%% file: thesis.tex
\documentclass[11pt,a4paper,twoside]{mythesis}

\usepackage{epsfig}
\usepackage{cite}
\usepackage{rotfloat}
\usepackage{amsmath}
\usepackage{theorem}
\usepackage{amssymb}
\usepackage{latexsym}
\usepackage{epic}
\usepackage{graphics}
\usepackage{rotating}
\usepackage{placeins}
\usepackage{hyperref}
\usepackage{lscape}

\newcommand{\be}{\begin{equation}}
\newcommand{\ee}{\end{equation}}

\newcommand{\Dslash}{\mbox{$\not \!\! D$}}
\newcommand{\delslash}{ \mkern-6mu \not \mkern-3mu \partial }
\newcommand{\del}{\partial}

\newcommand{\beq}{\begin{equation}}
\newcommand{\eeq}{\end{equation}}
\newcommand{\beqa}{\begin{eqnarray*}}
\newcommand{\eeqa}{\end{eqnarray*}}
\newcommand{\bea}{\begin{eqnarray}}
\newcommand{\eea}{\end{eqnarray}}
\newcommand{\bra}{\langle}
\newcommand{\ket}{\rangle}

\newcommand{\s}{\sigma}
\newcommand{\w}{\omega}
\newcommand{\fh}{c_1(u\bar{u}+d\bar{d})}
\newcommand{\sh}{c_2(s\bar{s})}
\newcommand{\half}{\frac{1}{2}}
\newcommand{\reci}[1]{\frac{1}{#1}}
\newcommand{\dbar}{d\mkern-6mu\mathchar'26}

\newcommand{\volint}[1]{\int \frac{d^4{#1}}{(2\pi)^4} \;}

\newcommand{\kslash}{\mbox{$\not \! k$}}
\newcommand{\pslash}{\mbox{$\not \! p$}}

\newcommand{\emdash}{\hspace{1pt}---\hspace{1pt}}

\newfont{\tapsmall}{tap scaled 1000}
\newfont{\tap}{tap scaled 1200}
\newfont{\tapsub}{tap scaled 1400}
\newfont{\tapmed}{tap scaled 2200}
\newfont{\tapbig}{tap scaled 3000}

\hypersetup{
    colorlinks=true,       
    linkcolor=red,          
    citecolor=green,        
    filecolor=magenta,      
    urlcolor=blue           
}

\begin{document}
\input{frontpage}

\pagenumbering{arabic}
\setcounter{page}{1}
\pagestyle{mystyle}
\include{Chapter1_intro}
\FloatBarrier
\include{Chapter2_particlephysics}

\FloatBarrier
\include{Chapter3_models}
\FloatBarrier
\include{Chapter4_methods}
\FloatBarrier
\include{Chapter5_results}
\FloatBarrier
\include{Chapter6_conclusions}
\FloatBarrier

\cleardoublepage
\addcontentsline{toc}{chapter}{\quad \ Bibliography}
\bibliographystyle{JHEP-2}
\bibliography{refs}
\cleardoublepage

\appendix
\FloatBarrier
\include{AppendixA_derivations}
\FloatBarrier
\include{AppendixB_particles}
\FloatBarrier
\include{AppendixD_papers}


\end{document}

%% file: frontpage.tex
\pagenumbering{roman}
 
\setcounter{page}{1}
\pagestyle{plain}

\newpage

\thispagestyle{empty}
\begin{center}
  \vspace*{0.5cm}
  {\Huge 
Applications of the Octet Baryon \\[1.5mm] 
Quark-Meson Coupling Model \\[4mm]
  to Hybrid Stars}

  \vspace*{1.5cm}
  {\huge
  Jonathan David Carroll } \\
  \vspace*{0.9cm}
  {
  \Large
  ( Supervisors:~~Prof.~~D.~B.~Leinweber,\
  Prof.~~A.~G.~Williams )
  } \\
  \vspace*{1.3cm}
  {\Large 
  A Thesis presented for the degree of\\[2mm]
  Doctor of Philosophy}
  \vspace*{1.0cm}
  \vfill

   \includegraphics[height=5.0cm]{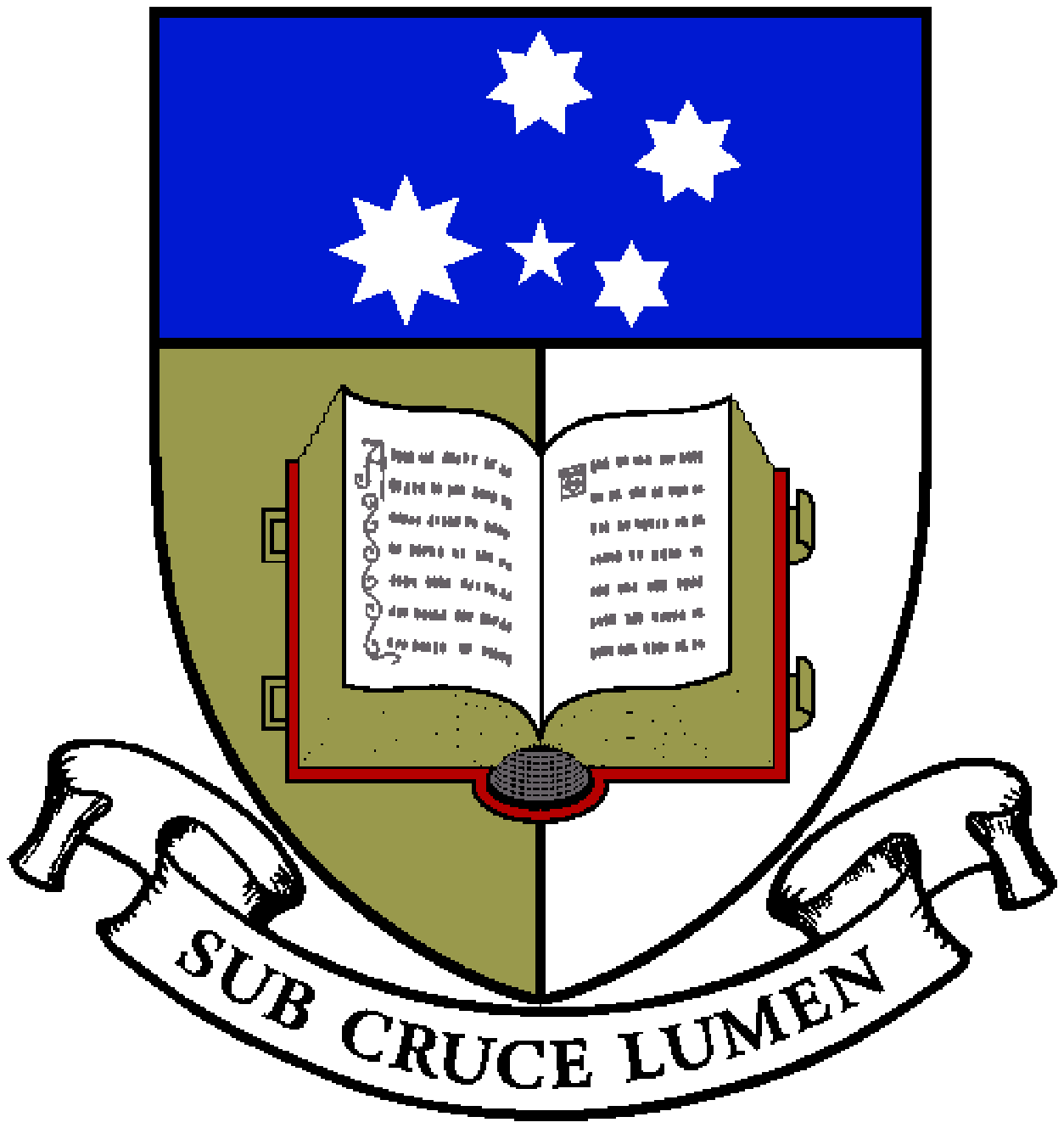}

  \vspace{0.5cm}

  {\large 
           Special Research Centre\\
	   for the Subatomic Structure of Matter\\
           School of Chemistry $\&$ Physics\\
           University of Adelaide\\
           South Australia\\[4mm]
           November 2009}

   \vspace{1cm}
   \phantom{endofpage}

\end{center}



\begin{mydedication}
Dedicated to my wife, Susanne;\\
for her love, sacrifice, and support of \\
a student for so many years.
\end{mydedication}
\cleardoublepage 

\originality{Abstract}
\begin{center}
        {\sf\Large Abstract}\\
\end{center}

The study of matter at extreme densities has been a major focus in
theoretical physics in the last half-century. The wide spectrum of
information that the field produces provides an invaluable
contribution to our knowledge of the world in which we live. Most
fascinatingly, the insight into the world around us is provided from
knowledge of the intangible, at both the smallest and largest scales
in existence.\par
Through the study of nuclear physics we are able to investigate the
fundamental construction of individual particles forming nuclei, and
with further physics we can extrapolate to neutron stars. The models
and concepts put forward by the study of nuclear matter help to solve
the mystery of the most powerful interaction in the universe; the
strong force.\par
In this study we have investigated a particular state-of-the-art model
which is currently used to refine our knowledge of the workings of the
strong interaction and the way that it is manifested in both neutron
stars and heavy nuclei, although we have placed emphasis on the former
for reasons of personal interest. The main body of this work has
surrounded an effective field theory known as Quantum Hadrodynamics
(QHD) and its variations, as well as an extension to this known as the
Quark-Meson Coupling (QMC) model, and variations thereof. We further
extend these frameworks to include the possibility of a phase
transition from hadronic matter to deconfined quark matter to produce
hybrid stars, using various models.\par
We have investigated these pre-existing models to deeply understand
how they are justified, and given this information, we have expanded
them to incorporate a modern understanding of how the strong
interaction is manifest.

\cleardoublepage 

\originality{Statement of Originality}

\vspace*{5cm}
\begin{center}
        {\sf\Large Statement of Originality}\\
\end{center}

This work contains no material which has been accepted for the award
of any other degree or diploma in any university or other tertiary
institution to \mbox{Jonathan David Carroll} and, to the best of my
knowledge and belief, contains no material previously published or
written by another person, except where due reference has been made in
the text.

I give consent to this copy of my thesis when deposited in the
University Library, being made available for loan and photocopying,
subject to the provisions of the Copyright Act 1968. The author
acknowledges that copyright of published works contained within this
thesis (as listed below$^*$) resides with the copyright holder(s) of
those works.

I also give permission for the digital version of my thesis to be made
available on the web, via the University's digital research
repository, the Library catalogue, the Australasian Digital Theses
Program (ADTP) and also through web search engines, unless permission
has been granted by the University to restrict access for a period of
time.

\vspace{4mm}

${}^*$Published article: Carroll {\it et al.} Physical Review C 79, 045810

\vspace{2cm}
\vspace{1mm}
\hfill Jonathan David Carroll
\cleardoublepage 

\originality{Acknowledgements}
\begin{center}
        {\sf\Large Acknowledgements}\\
\end{center}
%
The work contained herein would not have been possible without the
guidance, support, and co-operation of my
supervisors: \hbox{Prof.~Anthony~G.~Williams}
and \hbox{Prof.~Derek~B.~Leinweber}, nor that
of \hbox{Prof.~Anthony~W.~Thomas} at Jefferson Lab, Virginia,
USA. Their time and knowledge has been volunteered to seed my
learning, and I hope that the pages that follow are a suitable
representation of what has sprouted forth.\par
I would also like to acknowledge the help I have received from the
staff of\emdash and access I have been given to the facilities
at\emdash {\tapsmall eResearchSA} (formerly SAPAC). Many of the
calculations contained within this thesis have required computing
power beyond that of standard desktop machines, and I am grateful for
the use of supercomputer facilities, along with general technical
support.\par
In addition, I would also like to thank everyone else who has provided
me with countless hours of conversation both on and off topic which
have lead me to where I am
now: \hbox{Mr.~Mike~D'Antuoni}, \hbox{Dr.~Padric~McGee}, 
\hbox{Dr.~Sarah~Lawley}, \hbox{Dr.~Ping~Wang},
\hbox{Dr.~Sam~Drake}, \hbox{Dr.~Alexander~Kallionatis}, 
\hbox{Dr.~Sundance Bilson-Thompson}, and all of my teachers, lecturers and
fellow students who have helped me along the way.\par
I owe a lot of thanks to the CSSM office staff, who have played the
roles of `mums at work' for so many of
us: \hbox{Sara~Boffa}, \hbox{Sharon~Johnson}, \hbox{Silvana~Santucci},
\hbox{Rachel~Boutros} and \hbox{Bronwyn~Gibson}. Also, a special thanks to both
\hbox{Ramona~Adorjan} and \hbox{Grant~Ward}, for providing their inimitable support
with computers, and for their friendship; it is rare to find people
who can give both. \par
My sincere thanks go to \hbox{Benjamin~Menadue} for his thorough
proof-reading of this thesis, saving me from making several erroneous
and poorly worded statements.\par
Of course, I would like to thank my family (and family-in-law) for all
of their support and patience over the years, and for feigning
interest while I described something that I found particularly
remarkable.\par
Lastly, I would like to thank the caf\'es who have supported my
`passionate interest in coffee' over the past few years; I am now
known by name at several of these, and their abilities to make a great
latt\'e have kept me going through many a tired day (and night).

\cleardoublepage

\tableofcontents
\listoffigures
\listoftables
\cleardoublepage



%% file: Chapter1_intro.tex
\chapter{Introduction} \label{sec:introduction}

\font\dropcapfnt=cmr12 scaled \magstep5

\def\dropcap#1{\smallbreak \setbox0=\hbox{\dropcapfnt#1}
 \dimen0=\ht0 \setbox1=\hbox{t} \advance\dimen0 by-\ht1
 \dimen1=\wd0 \advance\dimen1 by 0.3em
 \hangindent\dimen1 \hangafter-2
 \noindent\smash{\llap{\lower\dimen0\box0\hskip0.3em}}
  \kern-0.0em }

\dropcap The world we live in is a strange and wonderful place. The vast
majority of our interactions with objects around us can be described
and predicted by relatively simple equations and relations, which have
been fully understood for centuries now. The concept of this tactile
world is however, a result of averaging a vast ensemble of smaller
effects on an unimaginably small scale, all conspiring together to
produce what we observe at the macroscopic level. In order to
investigate this realm in which particles can fluctuate in and out of
existence in the blink of an eye, we must turn to more sophisticated
descriptions of the players on this stage.\par
This research has the goal of furthering our understanding of the
interior of what are loosely called `neutron stars', though as shall
be shown herein, the contents are not necessarily only neutrons. With
this in mind, further definitions include `{\it hyperon stars}', `{\it
  quark stars}', and `{\it hybrid stars}', to describe compact stellar
objects containing hyperons\footnote{Baryons for which one or more of
  the three valence quarks is a strange quark. For example, the
  $\Sigma^+$ hyperon contains two up quarks and a strange quark. For a
  table of particle properties, including quark content, refer to
  Table~\ref{tab:particlesummary}}, quarks, and a mixture of each,
  respectively. In order to do this, we require physics beyond that
  which describes the interactions of our daily lives; we require
  physics that describes the individual interactions between
  particles, and physics that describes the interactions between
  enormous quantities of particles.\par
Only by uniting the physics describing the realm of the large and that
of the small can one contemplate so many orders of magnitude in scale;
from individual particles with a diameter of less than $10^{-22}~{\rm
  m}$, up to neutron stars with a diameter of tens of kilometers. Yet
the physics at each end of this massive scale are unified in this
field of nuclear matter in which interactions of the smallest
theorized entities conspire in such a way that densities equivalent to
the mass of humanity compressed to the size of a mere sugar cube
become commonplace, energetically favourable, and stable.\par
The sophistication of the physics used to describe the world of the
tiny and that of the enormous has seen much development over time. The
current knowledge of particles has reached a point where we are able
to make incredibly precise predictions about the properties of single
particles and have them confirmed with equally astonishing accuracy
from experiments. The physics describing neutron stars has progressed
from relatively simple (yet sufficiently consistent with experiment)
descriptions of neutron (and nucleon) matter to many more
sophisticated descriptions involving various species of baryons,
mesons, leptons and even quarks.\par
The outcome of work such as this is hopefully a better understanding
of matter at both the microscopic and macroscopic scales, as well as
the theory and formalism that unites these two extremes. The primary
methods which we have used to construct models in this thesis are
Quantum Hadrodynamics (QHD)\emdash which shall be described in
Sec.~\ref{sec:qhd}\emdash and the Quark-Meson Coupling
(QMC) model, which shall be described in Sec.~\ref{sec:qmc}.\par
In this thesis, we will outline the research undertaken in which we
produce a model for neutron star structure which complies with current
theories for dense matter at and above nuclear density and is
consistent with current data for both finite nuclei and observed
neutron stars. Although only experimental evidence can successfully
validate any theory, we hope to convey a framework and model that
possesses a minimum content of inconsistencies or unjustified
assumptions such that any predictions that are later shown to be
fallacious can only be attributed to incorrect initial conditions.\par
As a final defense of any inconsistencies that may arise between this
research and experiment, we refer the reader to one of the author's
favourite quotes:\par
\vspace{0.2cm}
\begin{quotation}
{\tap \raisebox{-2mm}{\Huge ``}\ 
There is a theory which states that if ever
anyone discovers exactly what the Universe is for and
why it is here, it will instantly disappear and be replaced by
something even more bizarre and inexplicable.\par \phantom{abc}}
\vspace{0.4mm}
{\tap There is another theory which states that this 
has already happened.\ \raisebox{-2mm}{\Huge ''}}
\begin{flushright}
-- Douglas Adams, The Hitchhiker's Guide To The Galaxy.
\end{flushright}
\end{quotation}
\par
Our calculations will begin at the particle interaction level, which
will be described in more detail in Section~\ref{sec:PPQFT}, from
which we are able to reproduce the bulk properties of matter at high
densities, and which we shall discuss in
Sections~\ref{sec:qhd}--\ref{sec:fockterms}. The methods for producing
our simulations of the interactions and bulk properties will be
detailed in Section~\ref{sec:methods}, along with a discussion on how
this is applied to the study of compact stellar objects. The results
of the simulations and calculations will be discussed in detail in
Section~\ref{sec:results}, followed by discussions on the
interpretation of these results in Section~\ref{sec:conclusions}. For
the convenience of the reader, and for the sake of completeness,
derivations for the majority of the equations used herein are provided
in Appendix~\ref{sec:derivations}, and useful information regarding
particles is provided in Appendix~\ref{sec:particleprops}. For now
however, we will provide a brief introduction to this field of
study.\par

\section{The Four Forces}\label{sec:fourforces}
Theoretical particle physics has seen much success and found many
useful applications; from calculating the individual properties of
particles to precisions that rival even the best experimental setups,
to determining the properties of ensembles of particles of greater and
greater scale, and eventually to the properties of macroscopic objects
as described by their constituents.\par
In order to do this, we need to understand each of the four
fundamental forces in Nature. The weakest of these forces\emdash
gravity\emdash attracts any two masses, and will become most important
in the following section. Slightly stronger is electromagnetism; the
force responsible for electric charge and magnetism. This force
provides an attraction between opposite electric charges (and of
course, repulsion between like charges), and thus helps to bind
electrons to nuclei. The mathematical description of this effect is
Quantum Electrodynamics (QED).\par
The `weak nuclear force'\emdash often abbreviated to `the weak
force'\emdash is responsible for the decay of particles and thus
radioactivity in general. At high enough energies, this force is
unified with electromagnetism into the `electro-weak force'. The
strongest of all the forces is the `strong nuclear force', abbreviated
to `the strong force'. This force is responsible for attraction
between certain individual particles over a very short scale, and is
responsible for the binding of protons within nuclei which would
otherwise be thrown apart by the repulsive electromagnetic force
between the positively charged protons. Each of these forces plays a
part in the work contained herein, but the focus of our study will be
the strong force.\par
A remarkably successful description of the strongest force at the
microscopic level is `Quantum Chromodynamics' (QCD) which is widely
believed to be the true description of strong interactions, relying on
quark and gluon degrees of freedom. The major challenge of this theory
is that at low energies it is non-perturbative\footnote{At high
  energies, QCD becomes asymptotically free~\cite{Gross:1973ju} and
  can be treated perturbatively. The physics of our world however is
  largely concerned with low energies.}, in that the coupling
constant\emdash which one would normally perform a series-expansion in
powers of\emdash is large, and thus is not suitable for such an
expansion. Regularisation techniques have been produced to create
perturbative descriptions of QCD, but perturbative techniques fail to
describe both dynamical chiral symmetry breaking and confinement; two
properties observed in Nature.\par
Rather than working directly with the quarks and gluons of QCD,
another option is to construct a model which reproduces the effects of
QCD using an effective field theory. This is a popular method within
the field of nuclear physics, and the route that has be taken for this
work. More precisely, we utilise a balance between attractive and
repulsive meson fields to reproduce the binding between fermions that
the strong interaction is responsible for.
\section{Neutron Stars}\label{sec:neutronstarhistory}
Although gravity may be the weakest of the four fundamental forces
over comparable distance scales, it is the most prevalent over
(extremely) large distances. It is this force that must be overcome
for a star to remain stable against collapse. Although a description
of this force that unifies it with the other three forces has not been
(satisfactorily) found, General Relativity has proved its worth for
making predictions that involve large masses.\par
At the time when neutron stars were first proposed by Baade and
Zwicky~\cite{Baade:1934}, neutrons had only been very recently proven
to exist by Chadwick~\cite{Chadwick:1932ma}. Nonetheless, ever
increasingly more sophisticated and applicable theories have
continually been produced to model the interactions that may lead to
these incredible structures; likely the most dense configuration
of particles that can withstand collapse.\par
The current lack of experimental data for neutron stars permits a wide
variety of
models~\cite{Lattimer:2000nx,Heiselberg:1999mq,Weber:2004kj,SchaffnerBielich:2004ch,Weber:1989hr,Chin:1974sa},
each of which is able to successfully reproduce the observed
properties of neutron stars, and most of which are able to reproduce
current theoretical and experimental data for finite nuclei and
heavy-ion collisions~\cite{Danielewicz:2002pu,Worley:2008cb}. The
limits placed on models from neutron star
observations~\cite{Podsiadlowski:2005ig,Grigorian:2006pu,Klahn:2006iw}
do not sufficiently constrain the models, so we have the opportunity
to enhance the models based on more sophisticated physics, while still
retaining the constraints above.\par
The story of the creation of a neutron star begins with a reasonably
massive star, with a mass greater than eight solar masses
($M>8~M_\odot$). After millions to billions of years or so (depending
on the exact properties of the star), this star will have depleted its
fuel by fusion of hydrogen into
${}^3$He, ${}^4$He, and larger elements up to iron (the most stable
element since has the highest binding energy per nucleon).\par
At this point, the core of the star will consist of solid iron, as the
heaviest elements are gravitationally attracted to the core of the
star, with successively lighter elements layered on top in accordance
with the traditional onion analogy. The core is unable to become any
more stable via fusion reactions and is only held up against
gravitational collapse by the degeneracy pressure of the
electrons\footnote{In accordance with the Pauli Exclusion Principle,
  no two fermions can share the same quantum state. This limits how
  close two fermions\emdash in this case, electrons\emdash can be
  squeezed, leading to the degeneracy pressure.}. The contents of the
upper layers however continue to undergo fusion to heavier elements
which also sink towards the core, adding to the mass of the lower
layers and thus increasing the gravitational pressure below.\par
This causes the temperature and pressure of the star to increase,
which encourages further reactions in the upper levels. Iron continues
to pile on top of the core until it reaches the Chandrasekhar limit of
$M = 1.4~M_\odot$, at which point the electron degeneracy pressure is
overcome. The next step is not fully understood\footnote{At present,
  models of supernova production have been unable to completely
  predict observations.}, but the result is a Type II supernova.\par
At the temperatures and pressures involved here, it is energetically
favourable for the neutrons to undergo $\beta$-decay into
protons, electrons (or muons), and antineutrinos according to
\be \label{eq:inversebeta}
n \to p^+ + e^- + \bar{\nu}_{e^-}.
\ee
These antineutrinos have a mean-free path of roughly
10~cm~\cite{Lattimer:2004} at these energies, and are therefore
trapped inside the star, causing a neutrino pressure bubble with
kinetic energy of order $10^{51}~{\rm erg} = 6.2\times 10^{56}~{\rm
  MeV}$~\cite{Lattimer:2004}. With the core collapsing (and producing
even more antineutrinos) even the rising pressure of the bubble cannot
support the mass of the material above and the upper layers begin
falling towards the core.\par
The sudden collapse causes a shock-wave which is believed to
`\emph{bounce}' at the core and expel the outer layers of the star in
a mere fraction of a second, resulting in what we know as a supernova,
and leaving behind the expelled material which, when excited by
radiation from another star, can be visible from across the galaxy as
a supernova remnant (SNR).\par
At the very centre of the SNR, the remaining core of the star
(na\"ively a sphere of neutrons, with some fraction of protons,
neutrons and electrons) retains the angular momentum of the original
star, now with a radius on the order of 10~km rather than $10^9$~km
and thus neutron stars are thought to spin very fast, with rotational
frequencies of up to 0.716~MHz~\cite{Hessels:2006ze}. Via a mechanism
involving the magnetic field of the star, these spinning neutron stars
may produce a beam of radiation along their magnetic axis, and if that
beam happens to point towards Earth to the extent that we can detect
it, we call the star a pulsar. For the purposes of this research, we
shall assume the simple case that the objects we are investigating are
static and non-rotating. Further calculations can be used to
extrapolate the results to rotating solutions, but we shall not focus
on this aspect here.\par
A further option exists; if the pressure and temperature (hence
energy) of the system become great enough, other particles can be
formed via weak reactions; for example, hyperons. The methods employed
in this thesis have the goal of constructing models of matter at
super-nuclear densities, and from these, models of neutron stars. The
outcome of these calculations is a set of parameters which describe a
neutron star (or an ensemble of them). Of these, the mass of a neutron
star is an observable quantity. Other parameters, such as radius,
energy, composition and so forth are unknown, and only detectable via
higher-order (or proxy) observations.\par
The ultimate goal would be finding a physically realistic model based
on the interactions of particles, such that we are able to deduce the
structure and global properties of a neutron star based only on an
observed mass. This however\emdash as we shall endeavor to show\emdash
is easier said than done.\par
%


%% file: Chapter2_particlephysics.tex
\chapter{Particle Physics $\&$ Quantum Field Theory}\label{sec:PPQFT}
In our considerations of the models that follow we wish to explore
ensembles of particles and their interactions. In order to describe
these particles we rely on Quantum Field Theory (QFT), which
mathematically describes the `rules' these particles obey. The
particular set of rules that are believed to describe particles
obeying the strong force at a fundamental level is Quantum
Chromodynamics (QCD), but as mentioned in the introduction, this
construction is analytically insolvable, so we rely on a model which
simulates the properties that QCD predicts.\par
In the following sections, we will outline the methods of calculating
the properties of matter from a field theoretic perspective.
\section{Lagrangian Density}\label{sec:lagrangiandensity}
The first step to calculating any quantity in a Quantum Field Theory
is to construct a Lagrangian density, which summarizes the dynamics of
the system, and from which the equations of motion can be
calculated. In order to do this, we must define precisely what it is
that we wish to calculate the properties of.\par
The classification schemes of particle physics provide several
definitions into which particles are identified, however each of these
provides an additional piece of information about those particles. We
wish to describe nucleons $N$ (consisting of protons $p$, and neutrons
$n$) which are hadrons\footnote{Bound states of quarks. In particular,
  bound states of three `valence' quarks plus any number of
  quark-antiquark pairs (the `sea' quarks, which are the result of
  particle anti-particle production via gluons) are called baryons.},
and are also fermions\footnote{Particles which obey Fermi--Dirac
  statistics, in which the particle wavefunction is anti-symmetric
  under exchange of particles; the property which leads to the Pauli
  Exclusion Principle.}.\par
We will extend our description to include the hyperons $Y$ (baryons
with one or more valence strange quarks) consisting of $\Lambda$,
$\Sigma^-$, $\Sigma^0$, $\Sigma^+$, $\Xi^-$, and $\Xi^0$ baryons. The
hyperons, together with the nucleons, form the octet of baryons (see
Fig.~\ref{fig:BaryonOctet}).\par
We can describe fermions as four-component spinors $\psi$ of
plane-wave solutions to the Dirac Equation (see later), such that
\be
\psi = u(\vec{p}\, ) e^{-ip_\mu x^\mu},
\ee
where $u(\vec{p}\, )$ are four-component Dirac spinors related to
plane-waves with wave-vector $\vec{p}$ that carry the spin information
for a particle, and which shall be discussed further in
Appendix~\ref{sec:qhdderiv}. For convenience, we can group the baryon
spinors by isospin group, since this is a degree of freedom that will
become important. For example, we can collectively describe nucleons
as a (bi-)spinor containing protons and neutrons, as
\be \psi_N = \begin{pmatrix} \psi_p(s) \\ \psi_n(s')
\end{pmatrix}.
\ee
Here we have used the labels for protons and neutrons rather than
explicitly using a label for isospin. We will further simplify this by
dropping the label for spin, and it can be assumed that this label is
implied. We will also require the Dirac Adjoint to describe the
antibaryons, and this is written as
\be \bar{\psi} = \psi^\dagger \gamma^0.  \ee
\par
Similarly, we can construct spinors for all the baryons. With these
spinors we can construct a Lagrangian density to describe the
dynamics of these particles. Since we are describing spin-$\half$
particles we expect the spinors to be solutions of the Dirac equation
which in natural units (for which $\hbar = c = 1$) is written as
\be
\label{eq:diraceq}
\left(i\gamma^\mu\del_\mu - M\right)\psi =
\left(i\delslash-M\right)\psi = 0, \ee
and similarly for the antiparticle $\bar{\psi}$. Feynman slash
notation is often used to contract and simplify expressions, and is
simply defined as $\not \!\!\! A = \gamma_\mu A^\mu$. Here, $\del_\mu$
is the four-derivative, $M$ is the mass of the particle, and
$\gamma^\mu$ are the (contravariant) Dirac Matrices, which due to the
anti-commutation relation of
\be \label{eq:cliff} \left\{\gamma^\alpha,\gamma^\beta\right\} =
\gamma^\alpha\gamma^\beta - \gamma^\beta\gamma^\alpha =
2\eta^{\alpha\beta}\mathbb{I}, \ee
(where $\eta = {\rm diag}(+1,-1,-1,-1)$ is the Minkoswki metric)
generate a matrix representation of the Clifford Algebra ${\it
  Cl}(1,3)$. They can be represented in terms of the $2\times 2$
identity matrix ${\mathbb I}$, and the Pauli Matrices $\vec{\sigma}$,
as
\be \gamma^0 =
\left(\begin{matrix}\mathbb{I}&0\\0&-\mathbb{I}\end{matrix}\right),
\qquad \gamma^i = \left(\begin{matrix}0&\sigma^i\;
  \\-\sigma^i&0\; \end{matrix}\right).  \ee
\par
Eq.~(\ref{eq:diraceq}) describes free baryons, so we can use this as
the starting point for our Lagrangian density, and thus if we include
each of the isospin groups, we have
\be {\cal L} = \sum_{k} \bar{\psi}_k\left(i\delslash-M_k\right)\psi_k
\ ; \ k \in\{N,\Lambda,\Sigma,\Xi\}, \ee
where the baryon spinors are separated into isospin groups, as
\be \label{eq:isospingroups} \psi_N = \begin{pmatrix}\psi_p\\\psi_n\end{pmatrix}, \quad
  \psi_\Lambda = \begin{pmatrix}\psi_\Lambda\end{pmatrix}, \quad
    \psi_\Sigma
    = \begin{pmatrix}\psi_{\Sigma^+}\\\psi_{\Sigma^0}\\\psi_{\Sigma^-}\end{pmatrix},
    \quad \psi_\Xi
    = \begin{pmatrix}\psi_{\Xi^0}\\\psi_{\Xi^-}\end{pmatrix}. \quad
    \ee
This implies that the mass term is also a diagonal matrix. In many
texts this term is simply a scalar mass term multiplied by a suitable
identity matrix, but that would imply the existence of a charge
symmetry; that the mass of the proton and of the neutron were
degenerate, and exchange of charges would have no effect on the
Lagrangian density. We shall not make this assumption, and will rather
work with the physical masses as found in Ref.~\cite{Amsler:2008zzb},
so $M_k$ will contain distinct values along the diagonal.\par
To this point, we have constructed a Lagrangian density for the
dynamics of free baryons. In order to simulate QCD, we require
interactions between baryons and mesons to produce the correct
phenomenology.  Historically, the scalar-isoscalar
meson\footnote{Despite it's dubious status as a distinct particle
  state, rather than a resonance of $\pi\pi$.} $\s$ and
vector-isoscalar meson $\w$ have been used to this end. Additionally,
the vector-isovector $\rho$ meson has been included (for asymmetric
matter) to provide a coupling to the isospin
channel~\cite{Serot:1984ey}.\par
\begin{figure}[!b]
\centering \includegraphics[width=0.65\textwidth]{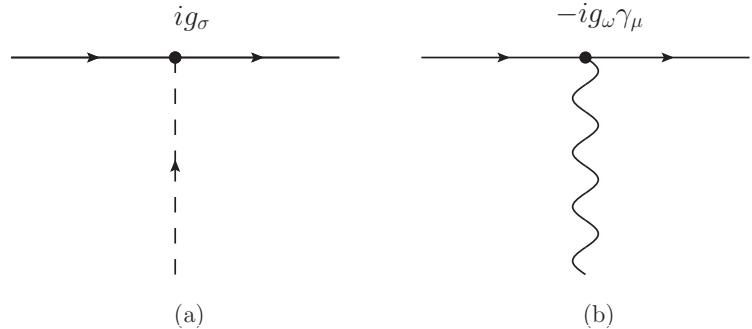}
\caption[Baryon-meson vertices]{Interaction vertex for the (a) scalar
  and (b) vector mesons, where the solid lines represent baryons
  $\psi$, the dashed line represents a scalar meson (e.g. $\s$), and
  the wavy line represents a vector meson (e.g. $\omega_\mu$).
  \protect\label{fig:vectorvertex}}
\end{figure}

In order to describe interactions of the baryons with mesons, we can
include terms in the Lagrangian density for various classes of mesons
by considering the appropriate bilinears that each meson couples
to. For example, if we wish to include the $\w$ meson, we first
observe that as a vector meson it will couple to a vector bilinear (to
preserve Lorentz invariance) as
\be -ig_{\w}\bar{\psi}\gamma_\mu\w^\mu\psi \ee
with coupling strength $g_\w$, which as we shall see, may be dependent
on the baryon that the meson is coupled to. The particular
coefficients arise from the Feynman rules for meson-baryon vertices
(refer to Appendix~\ref{sec:diagrams}). This particular vertex is
written in Feynman diagram notation as shown in
Fig.~\ref{fig:vectorvertex}(b).\par
This is not the only way we can couple a meson to a baryon. We should
also consider the Yukawa couplings of mesons to baryons with all
possible Lorentz characteristics; for example, the $\omega$ meson can
couple to a baryon $\psi$, with several different vertices:
\be \bar{\psi}\gamma_\mu\omega^\mu\psi,\ \ \bar{\psi}\s_{\mu\nu}q^\nu \!
\omega^\mu\psi,\ {\rm and}\ \ \bar{\psi}q_\mu\omega^\mu\psi, \ee
where $q_\mu$ represents the baryon four-momentum transfer $(q_f -
q_i)_\mu$. The latter two of these provides a vanishing contribution
when considering the mean-field approximation (which shall be defined
in Section~\ref{sec:mfa}), since $\s_{00}= 0$ and $q_\mu = 0$, as the
system is on average, static.\par
If we include the appropriate scalar and vector terms\emdash including
an isospin-coupling of the $\rho$ meson\emdash in our basic Lagrangian
density we have
\be \label{eq:interL} {\cal L} = \sum_{k}
\bar{\psi}_k\left(\gamma_\mu\left[i\del^\mu-g_{k \w}\w^\mu
  -g_{\rho}(\vec{\tau}_{(k)}\cdot\vec{\rho}^{\, \mu})\; \right]
-M_k+g_{k\s}\s\right)\psi_k \ ; \ k \in\{N,\Lambda,\Sigma,\Xi\}.  \ee
\par
The isospin matrices $\vec{\tau}_{(k)}$ are scaled Pauli matrices of
appropriate order for each of the isospin groups, the third components
of which are given explicitly here as
\be \label{eq:taus} \tau_{(N)3} = \tau_{(\Xi)3} = \reci{2}
\left[ \begin{matrix}\ 1\ & 0\ \\\ 0\ & -1\ \end{matrix} \right],
\quad
\tau_{(\Lambda)3} = 0, \quad \tau_{(\Sigma)3} =
\left[ \begin{matrix}\ 1\ &\ 0\ &\ 0\ \\\ 0\ &\ 0\ &\ 0\ \\\ 0\ &\ 0\ &
    -1\ \end{matrix} \right], \ee
for which the diagonal elements of $\tau_{(k)3}$ are the isospin
projections of the corresponding baryons within an isospin group
defined by Eq.~(\ref{eq:isospingroups}), i.e. $\tau_{(p)3} = I_{3p} =
+\half$.\par
At this point it is important that we remind the reader that the
conventions in this field do not distinguish the use of explicit
Einstein summation, and that within a single equation, indices may
represent summation over several different spaces. To make this
clearer, we will show an example of a term where all the indices are
made explicit; the interaction term for the $\rho$ meson in
Eq.~(\ref{eq:interL}) for which we explicitly state all of the indices
\bea \nonumber {\cal L}_{\rho} &=& \sum_{k}\, g_{\rho}\, \bar{\psi}_k
\gamma^\mu \vec{\tau}_{(k)}\cdot\vec{\rho\, }_\mu \psi_k\\ &=&
\sum_k\, \sum_{i,j=1}^{f_k}\, \sum_{\alpha,\beta=1}^{4}\,
\sum_{\mu=0}^{3}\, \sum_{a=1}^{3}\, g_{\rho}\,
\left(\bar{\psi}^i_k\right)_\alpha\; (\gamma^\mu)_{\alpha \beta}\;
\left(\tau_{(k)}^a\right)^{ij}\; \rho^a_\mu\;
\left(\psi_k^{j}\right)_\beta,\eea
where here $k$ is summed over isospin groups $N$, $\Lambda$, $\Sigma$,
and $\Xi$; $i$ and $j$ are summed over flavor space (within an isospin
group of size $f_k$, e.g. $f_N = 2$, $f_\Sigma = 3$); $\alpha$ and
$\beta$ are summed over Dirac space; $\mu$ is summed over Lorentz
space; and $a$ is summed over iso-vector space. The Pauli matrices, of
which $\left(\tau_k^a\right)^{ij}$ are the elements, are defined in
Eq.~(\ref{eq:taus}). This level of disambiguity is overwhelmingly
cluttering, so we shall return to the conventions of this field and
leave the indices as implicit.\par
In addition to the interaction terms, we must also include the free
terms and field tensors for each of the mesons, which are chosen with
the intent that applying the Euler--Lagrange equations to these terms
will produce the correct phenomenology, leading to
\bea \nonumber &\mathcal{L} = & \sum_k
\bar{\psi}_k\left(\gamma_\mu\left[i\del^\mu-g_{k\w}\w^\mu
  -g_{\rho}(\vec{\tau}_{(k)}\cdot\vec{\rho}^{\, \mu})\; \right]
-M_k+g_{k\s}\s\right)\psi_k \\ \nonumber &&
+\ \frac{1}{2}(\partial_{\mu}\sigma\partial^{\mu}\sigma-m^{2}_{\sigma}\sigma^{2})
+\frac{1}{2}m^{2}_{\omega}\omega_{\mu}\omega^{\mu}+
\frac{1}{2}m^{2}_{\rho}\rho_{\mu}\rho^{\mu}
-\frac{1}{4}\Omega_{\mu\nu}\Omega^{\mu\nu} -
\frac{1}{4}R^a_{\mu\nu}R_a^{\mu\nu}, \\ && \label{eq:fulllag} \eea
where the field tensors for the $\w$ and $\rho$ mesons are,
respectively,
\be \Omega_{\mu\nu} = \del_\mu\w_\nu - \del_\nu\w_\mu, \quad
R^a_{\mu\nu} = \del_\mu\rho^a_\nu - \del_\nu\rho^a_\mu - g_\rho
\epsilon^{abc}\rho^b_\mu\rho^c_\nu.  \ee
This is the Lagrangian density that we will begin with for the models
we shall explore herein. Many texts (for example,
Refs.~\cite{Alaverdyan:2009kv,Greco:2000dt,Uechi:2006pz,Menezes:2002cq})
include higher-order terms (${\cal O}(\s^3)$,\ ${\cal
  O}(\s^4)$,\ $\ldots$) and have shown that these do indeed have an
effect on the state variables, but in the context of this work, we
shall continue to work at this order for simplicity. It should be
noted that the higher order terms for the scalar meson can be included
in such a way as to trivially reproduce a framework consistent with
the Quark-Meson Coupling model that shall be described later, and thus
we are not entirely excluding this contribution.\par

\section{Mean-Field Approximation}\label{sec:mfa}
To calculate properties of matter, we will use an approximation to
simplify the quantities we need to evaluate. This approximation, known
as a Mean-Field Approximation (MFA) is made on the basis that we can
separate the expression for a meson field $\alpha$ into two parts: a
constant classical component, and a component due to quantum
fluctuations;
\be \alpha = \alpha_{\rm classical} + \alpha_{\rm quantum}.  \ee
If we then take the vacuum expectation value (the average value in the
vacuum) of these components, the quantum fluctuation term vanishes,
and we are left with the classical component
\be \bra\alpha\ket \equiv \bra\alpha_{\rm classical}\ket.  \ee
This component is what we shall use as the meson contribution, and we
will assume that this contribution (at any given density) is
constant. This can be thought of as a background `field' on top of
which we place the baryon components. For this reason, we consider the
case of \emph{infinite matter}, in which there are no boundaries to
the system. The core of lead nuclei (composed of over 200 nucleons)
can be thought of in this fashion, since the effects of the outermost
nucleons are minimal compared to the short-range strong nuclear
force.\par
Furthermore, given that the ground-state of matter will contain some
proportion of proton and neutron densities, any flavor-changing meson
interactions will provide no contribution in the MFA, since the
overlap operator between the ground-state $|\Psi\ket$ and any other
state $|\xi\ket$ is orthogonal, and thus
\be
\label{eq:overlap}
\bra\; \Psi\; |\; \xi\; \ket = \delta_{\Psi\xi}.
\ee
For this reason, any meson interactions which, say, interact with a
proton to form a neutron will produce a state which is not the ground
state, and thus provides no contribution to the MFA. We will show in
the next section that this is consistent with maintaining isospin
symmetry.

\section{Symmetries}\label{sec:symmetries}
In the calculations that will follow, there are several terms that we
will exclude from our considerations {\it ab initio} (including for
example, some that appear in Eq.~(\ref{eq:fulllag})) because they
merely provide a vanishing contribution, such as the quantum
fluctuations mentioned above. These quantities shall be noted here,
along with a brief argument supporting their absence in further
calculations.\par
\subsection{Rotational Symmetry and Isospin}\label{subsec:rotational}
The first example is simple enough; we assume rotational invariance of
the fields to conserve Lorentz invariance. In order to maintain
rotational invariance in all frames, we require that the spatial
components of vector quantities vanish, leaving only temporal
components. For example, in the MFA the vector-isoscalar meson
four-vector $\w_\mu$ can be reduced to the temporal component $\w_0$,
and for notational simplicity, we will often drop the subscript and
use $\bra\alpha\ket$ for the $\alpha$ meson mean-field
contribution.\par
A corollary of the MFA is that the field tensor for the rho meson
vanishes;
\be R^a_{\mu\nu} = \del_\mu\rho^a_\nu - \del_\nu\rho^a_\mu - g_\rho
\epsilon^{abc}\rho^b_\mu\rho^c_\nu \underset{\rm MFA}\longrightarrow
R^a_{00} = 0, \ee
since the derivatives of the constant terms vanish and
$(\vec{\rho}_0\times\vec{\rho}_0) = 0$. The same occurs for the omega
meson field tensor
\be \Omega_{\mu\nu} = \del_\mu\omega_\nu - \del_\nu\omega_\mu
\underset{\rm MFA}\longrightarrow \Omega_{00} = 0.  \ee
We also require rotational invariance in isospin space along a
quantization direction of $\hat{z}=\hat{3}$ (isospin invariance) as
this is a symmetry of the strong interaction, thus only the neutral
components of an isovector have a non-zero contribution. This can be
seen if we examine the general $2\times2$ unitary isospin
transformation, and the Taylor expansion of this term
\be \psi(x) \to \psi^\prime(x) = e^{i\vec{\tau}\cdot \vec{\theta}/2}
\psi(x)
\ \mathop{\longrightarrow}_{|\theta| \ll 1} \ \left(1 +
i\vec{\tau}\cdot\vec{\theta}/2 \right)\psi(x), \ee
where $\vec{\theta}=(\theta_1,\theta_2,\theta_3)$ is a triplet of real
constants representing the (small) angles to be rotated through, and
$\vec{\tau}$ are the usual Pauli matrices as defined in
Eq.~(\ref{eq:taus}). As for the $\rho$ mesons, we can express the
triplet as linear combinations of the charged states, as
\be \vec{\rho} = (\rho_1,\rho_2,\rho_3) = \left(
\reci{\sqrt{2}}(\rho_+ + \rho_-),\ \frac{i}{\sqrt{2}}(\rho_- -
\rho_+),\ \rho_0 \right).  \ee
The transformation of this triplet is then
\be \vec{\rho}\; (x) \to \vec{\rho\; }^\prime(x) =
e^{i\vec{T}\cdot\vec{\theta} } \vec{\rho}\; (x).  \ee
where $(T^i)_{jk} = -i\epsilon_{ijk}$ is the adjoint representation of
the SU(2) generators; the spin-1 Pauli matrices in the isospin basis,
a.k.a. SO(3). We can perform a Taylor expansion about
$\vec{\theta}=\vec{0}$, and we obtain
\be \rho_j(x) 
\ \mathop{\longrightarrow}_{|\theta| \ll 1} \ 
\left[\delta_{jk} + i
  (T^i)_{jk}\theta_i\right]\rho_k(x). \ee
We can therefore write the transformation as
\be \vec{\rho}\; (x) 
\ \mathop{\longrightarrow}_{|\theta| \ll 1} \ \vec{\rho}\; (x) -
\vec{\theta} \times \vec{\rho}\; (x).  \ee
Writing this out explicitly for the three isospin states, we
obtain the individual transformation relations
\bea \nonumber
&&\vec{\rho}\; (x) \to \vec{\rho}\; (x) -
  \vec{\theta} \times \vec{\rho}\; (x) =
  ( \rho_1 - \theta_2\rho_3 + \theta_3\rho_2,\ \rho_2 - \theta_1\rho_3
  + \theta_3\rho_1,\ \rho_3 - \theta_1\rho_2 + \theta_2\rho_1 ).\\[2mm]
&&
\eea
If we now consider the rotation in only the $\hat{z}=\hat{3}$
direction, we see that the only invariant component is $\rho_3$
\be \vec{\rho}\; (x) 
\xrightarrow[{}_{\stackrel{\theta1 = 0}{\theta_2 = 0}}]{} 
\left(\rho_1 + \theta_3\rho_2,\ \rho_2 +
\theta_3\rho_1,\ \rho_3\right).  \ee
If we performed this rotation along another direction\emdash i.e. $\hat{1}$,
 $\hat{2}$, or a linear combination of directions\emdash we would find
that the invariant component is still a linear combination of charged
states. By enforcing isospin invariance, we can see that the only
surviving $\rho$ meson state will be the charge-neutral state $\rho_3
\equiv \rho_0$.\par
\subsection{Parity Symmetry}\label{subsec:parity}
We can further exclude entire isospin classes of mesons from
contributing since the ground-state of nuclear matter (containing
equal numbers of up and down spins) is a parity eigenstate, and thus
the parity operator ${\cal P}$ acting on the ground-state produces
\be {\cal P}|{\cal O}\ket = \pm |{\cal O}\ket.  \ee
Noting that the parity operator is idempotent (${\cal P}^2 = {\mathbb
  I}$), inserting the unity operator into the ground-state overlap
should produce no effect;
\be \bra{\cal O}|{\cal O}\ket = \bra{\cal O}|{\mathbb I}|{\cal O}\ket
= \bra{\cal O}|{\cal P}{\cal P}|{\cal O}\ket = \bra{\cal
  O}|(\pm)^2|{\cal O}\ket = \bra{\cal O}|{\cal
  O}\ket.  \ee
We now turn our attention to the parity transformations for various
bilinear combinations that will accompany meson interactions. For
Dirac spinors $\psi(x)$ and $\bar{\psi}(x)$ the parity transformation
produces
\be \begin{array}{cc} \mathcal{P}\psi(t,\vec{x}\, )\mathcal{P} =
  \gamma^0\psi(t,-\vec{x}\, ), \\[2mm]
  \mathcal{P}\bar{\psi}(t,\vec{x}\, )\mathcal{P} =
  \bar{\psi}(t,-\vec{x}\, )\gamma^0, \end{array} \ee
where we have removed the overall phase factor $\exp(i\phi)$ since
this is unobservable and can be set to unity without loss of
generalisation. We can also observe the effect of the parity
transformation on the various Dirac field bilinears that may appear
in the Lagrangian density. The five possible Dirac bilinears are:
\be \bar{\psi}\psi,\quad \bar{\psi}\gamma^\mu\psi,\quad
i\bar{\psi}[\gamma^\mu,\gamma^\nu]\psi,\quad
\bar{\psi}\gamma^\mu\gamma^5\psi,\quad i\bar{\psi}\gamma^5\psi, \ee
for scalar, vector, tensor, pseudo-vector and pseudo-scalar meson
interactions respectively, where $\gamma_5$ is defined as
\be \label{eq:gam5} \gamma_5 = i\gamma_0\gamma_1\gamma_2\gamma_3 =
\left(\begin{array}{cc}0&{\mathbb I}\\{\mathbb I}&0\end{array}\right),
  \ee
in the commonly used Dirac basis. By acting the above transformation
on these bilinears we obtain a result proportional to the spatially
reversed wavefunction $\psi(t,-\vec{x}\, )$,
\bea \mathcal{P}\bar{\psi}\psi \mathcal{P} &=&
+\bar{\psi}\psi(t,-\vec{x}\, ),\\[3mm]
\mathcal{P}\bar{\psi}\gamma^\mu\psi \mathcal{P} &=& \left\{
   \begin{array}{ll}
     +\bar{\psi}\gamma^\mu\psi(t,-\vec{x}\, ) &\quad \mbox{for\ }
     \mu=0,\\ -\bar{\psi}\gamma^\mu\psi(t,-\vec{x}\, ) &\quad \mbox{for\ }
     \mu=1,2,3,
   \end{array}
\right.\\[3mm] \mathcal{P}\bar{\psi}\gamma^\mu\gamma^5\psi
\mathcal{P} &=& \left\{
   \begin{array}{ll}
     -\bar{\psi}\gamma^\mu\gamma^5\psi(t,-\vec{x}\, ) &\quad \mbox{for\ }
     \mu=0,\\ +\bar{\psi}\gamma^\mu\gamma^5\psi(t,-\vec{x}\, ) &\quad
     \mbox{for\ } \mu=1,2,3,
   \end{array}
\right.\\[3mm] \mathcal{P}i\bar{\psi}\gamma^5\psi \mathcal{P} &=&
-i\bar{\psi}\gamma^5\psi(t,-\vec{x}\, ).
\eea
By inserting the above pseudo-scalar and pseudo-vector bilinears into
the ground-state overlap as above, and performing the parity
operation, we obtain a result equal to its negative, and so the
overall expression \emph{must} vanish. For example
\be \langle \mathcal{O} | i\bar{\psi}\gamma^5\psi | \mathcal{O}
\rangle = \langle \mathcal{O} | \mathcal{P} i\bar{\psi}\gamma^5\psi
\mathcal{P} | \mathcal{O} \rangle = \langle \mathcal{O} |
-i\bar{\psi}\gamma^5\psi | \mathcal{O} \rangle 
= 0. \ee
Thus all pseudo-scalar and pseudo-vector meson contributions\emdash
such as those corresponding to $\pi$ and $K$\emdash provide no
contribution to the ground-state in the lowest order. We will show
later in Chapter~\ref{sec:fockterms} that mesons can provide higher
order contributions, and the pseudo-scalar $\pi$ mesons are able to
provide a non-zero contribution via Fock terms, though we will not
calculate these contributions here.\par

\section{Fermi Momentum}\label{sec:kf}
Since we are dealing with fermions that obey the Pauli Exclusion
Principle\footnote{That no two fermions can share a single quantum
  state.}, and thus Fermi--Dirac statistics\footnote{The statistics of
  indistinguishable particles with half-integer spin. Refer to
  Appendix~\ref{sec:chempotderivn}.}, there will be restrictions on the
quantum numbers that these fermions may possess. When considering
large numbers of a single type of fermion, they will each require a
unique three-dimensional momentum $\vec{k}$ since no two fermions may
share the same quantum numbers.\par
For an ensemble of fermions we produce a `Fermi sea' of particles; a
tower of momentum states from zero up to some value `at the top of the
Fermi sea'. This value\emdash the Fermi momentum\emdash will be of
considerable use to us, thus it is denoted $k_F$.\par
Although the total baryon density is a useful control parameter, many of
the parameters of the models we wish to calculate are dependent on the
density via $k_F$. The relation between the Fermi momentum and the
total density is found by counting the number of momentum states in a
spherical volume up to momentum $k_F$ (here, this counting is performed
in momentum space). The total baryon density\emdash a number density
in units of baryons/fm$^{3}$, usually denoted as just ${\rm
  fm}^{-3}$\emdash is simply the sum of contributions from individual
baryons, as
\be \label{eq:rho} \rho_{\rm total} = \sum_i \rho_i = \sum_i
\frac{(2J_i +1)}{(2\pi)^3}\int \theta(k_{F_i}-|\vec{k}|) \, d^3k =
\sum_i \frac{k_{F_i}^3}{3\pi^2}, \ee
where here, $i$ is the set of baryons in the model, $J_i$ is the spin
of baryon $i$ (where for the leptons and the octet of baryons, $J_i =
\half$), and $\theta$ is the Heaviside step function defined as
\be \label{eq:Heaviside} \theta(x) =
\left\{ \begin{array}{ll}1,\ \ {\rm if} & x > 0\\0,\ \ {\rm if} & x <
  0\end{array} \right. , \ee
which restricts the counting of momentum states to those between $0$
and $k_F$.\par
We define the species fraction for a baryon $B$, lepton $\ell$, or
quark $q$ as the density fraction of that particle, denoted by $Y_i$,
such that
\be 
\label{eq:Y}
Y_i = \frac{\rho_i}{\rho_{\rm total}}\ ;\quad i\in \{
B,\ell,q\} \, . \ee
Using this quantity we can investigate the relative proportions of
particles at a given total density.

\section{Chemical Potential}\label{sec:qftchempot}
In order to make use of statistical mechanics we must define the some
important quantities. One of these will be the chemical potential
$\mu$, also known as the Fermi energy $\epsilon_{F}$; the energy of a
particle at the top of the Fermi sea, as described in
Appendix~\ref{sec:chempotderivn}. This energy is the relativistic
energy of such a particle, and is the energy associated with a Dirac
equation for that particle. For the simple case of a non-interacting
particle, this is
\be \label{eq:muepsf} \mu_B = \epsilon_{F_B} = \sqrt{k_{F_B}^2 +
  M_B^2}. \ee
In the case that the baryons are involved in interactions with mesons,
we need to introduce scalar and temporal self-energy terms, which (for
example) for Hartree-level QHD using a mean-field approximation are
given by
\be \label{eq:selfenergy}
\Sigma^s_B = - g_{B\s} \bra\s\ket,\quad \Sigma^0_B = g_{B\w}\bra\w\ket
+ g_{\rho}I_{3B}\bra\rho\ket,
\ee
where $I_{3B}$ is the isospin projection of baryon $B$, defined by the
diagonal elements of Eq.~(\ref{eq:taus}), and where the scalar
self-energy is used to define the baryon effective mass as
\be \label{eq:effM}
M_B^* = M_B + \Sigma^s_B = M_B - g_{B\s} \bra\s\ket,
\ee
These self-energy terms affect the energy of a Dirac equation, and
thus alter the chemical potential, according to
\be
\label{eq:mu_sw}
\mu_B = \sqrt{k_{F_B}^2 + (M_B + \Sigma^s_B)^2}+\Sigma_B^0\ .  \ee
Eq~(\ref{eq:effM}) and Eq.~(\ref{eq:mu_sw}) define the important
in-medium quantities, and the definition of each will become
dependent on which model we are using.\par
For a relativistic system such as that which will consider here, each
conserved quantity is associated with a chemical potential, and we can
use the combination of these associated chemical potentials to obtain
relations between chemical potentials for individual species. In our
case, we will consider two conserved quantities: total baryon number
and total charge, and so we have a chemical potential related to each
of these. We can construct the chemical potential for each particle
species by multiplying each conserved charge by its associated
chemical potential to obtain a general relation. Thus
\be \label{eq:chempotrel} \mu_i = B_i \mu_n - Q_i \mu_e, \ee
where; $i$ is the particle species (which can be any of the baryons)
for which we are constructing the chemical potential; $B_i$ and $Q_i$
are the baryon number (`baryon charge', which is unitless) and
electric charge (normalized to the proton charge) respectively; and
$\mu_n$ and $\mu_e$ are the chemical potentials of neutrons and
electrons, respectively. Leptons have $B_\ell=0$, and all baryons have
$B_B=+1$. The relations between the chemical potentials for the octet
of baryons are therefore derived to be
\be \label{eq:allmus}
\begin{array}{rcrcrcl}
\mu_\Lambda &=& \mu_{\Sigma^0} &=& \mu_{\Xi^0} &=& \mu_n, \\
&&\mu_{\Sigma^-} &=& \mu_{\Xi^-} &=& \mu_n + \mu_e, \\
&&\mu_p &=& \mu_{\Sigma^+} &=& \mu_n - \mu_e, \\
&&&& \mu_\mu &=& \mu_e.
\end{array}
\ee
A simple example of this is to construct the chemical potential for
the proton (for which the associated charges are $B_p = +1$ and $ Q_p
= +1$);
\be \label{eq:betaeq} \mu_p = \mu_n - \mu_e.  \ee
This can be rearranged to a form that resembles neutron $\beta$-decay
\be
\label{eq:n0beta}
\mu_n = \mu_p + \mu_e.  \ee
\par
If we were to consider further conserved charges, such as lepton
number for example, we would require a further associated chemical
potential. In that example, the additional chemical potential would be
for (anti)neutrinos $\mu_{\bar{\nu}}$. The antineutrino would be
required to preserve the lepton number on both sides of the equation;
the goal of such an addition. Since we shall consider that neutrinos
are able to leave the system considered, we can ignore this
contribution {\it ab inito}. The removal of this assumption would
alter Eq.~(\ref{eq:n0beta}) to include the antineutrino, as would
normally be expected in $\beta$-decay equations
\be \mu_n = \mu_p + \mu_e + \mu_{\bar{\nu}}.  \ee

\section{Explicit Chiral Symmetry (Breaking)}\label{sec:chiral}
One of the most interesting symmetries of QCD is chiral symmetry. If
we consider the QCD Lagrangian density to be the sum of quark and
gluon contributions, then in the massless quark limit ($m_q = 0$);
%
%
%
\bea
\nonumber
{\cal L}_{\rm QCD} &=& {\cal L}_g + {\cal L}_q \\
\nonumber
&=& -\reci{4}G_{\mu\nu}^aG_{a}^{\mu\nu}
+ \bar{\psi}_i i \gamma^\mu (D_\mu)_{ij} \psi_j
\\
\label{eq:QCDm0lag}
&=& -\reci{4}G_{\mu\nu}^aG_{a}^{\mu\nu} +
\bar{\psi}_i i\gamma^\mu\del_\mu \psi_i -
gA^a_\mu\bar{\psi}_i\gamma^\mu T_{ij}^a\psi_j\ , \eea 
where here, $\psi_i(x)$ is a quark field of color $i \in \{r,g,b\}$,
$A_\mu^a(x)$ is a gluon field with color index $a \in \{1,\ldots,8\}$,
$T_{ij}^a$ is a generator\footnote{For example, $T^a = \lambda^a/2$
  using the Hermitian Gell-Mann matrices $\lambda_a$.} for SU(3), $g$
is the QCD coupling constant, and $G_{\mu\nu}^a$ represents the
gauge-invariant gluonic field strength tensor, given by
\be
G_{\mu\nu}^a = \left[\del_\mu,A_\nu^a\right] - gf^{abc}A_\mu^b
A_\nu^c\ ,
\ee
written with the structure constants $f^{abc}$.
%
Left- and right-handed components of Dirac fields can be
separated using the projection operators
\be \psi_{L\atop R} = \frac{1\mp \gamma_5}{2}\psi, \ee
using the definition of $\gamma_5$ of Eq.~(\ref{eq:gam5}), and so the
quark terms in the QCD Lagrangian density (the gluon terms are not
projected) can be written in terms of these components as
\be {\cal L}_{q}^{(f)} = i\bar{\psi}^{(f)}_L D_{\mu}\gamma^{\mu}
\psi^{(f)}_L + i\bar{\psi}^{(f)}_R D_{\mu}\gamma^{\mu} \psi^{(f)}_R.
\ee
This Lagrangian density is invariant under rotations in U(1) of the left- and
right-handed fields
\bea &{\rm U}(1)_L:& \psi_L \to e^{i\alpha_L}\psi_L, \quad
\psi_R \to \psi_R, \\ &{\rm U}(1)_R:& \psi_R \to e^{i\alpha_R}\psi_R,
\quad \psi_L \to \psi_L, \eea
where $\alpha_L$ and $\alpha_R$ are arbitrary phases. This invariance
is the chiral ${\rm U}(1)_L\otimes {\rm U}(1)_R$ symmetry. The Noether
currents associated with this invariance are then
\be J_L^\mu = \bar{\psi_L}\gamma^\mu \psi_L, \qquad J_R^\mu =
\bar{\psi_R}\gamma^\mu \psi_R, \ee
and as expected, these currents are conserved, such that $\del_\mu
J_L^\mu = \del_\mu J_R^\mu = 0$ according to the Dirac Equation. These
conserved currents can be alternatively written in terms of conserved
vector and axial-vector currents, as
\be J^\mu_{L} = \frac{V^\mu - A^\mu}{2}, \qquad J^\mu_{R} =
\frac{V^\mu + A^\mu}{2}, \ee
where here, $V^\mu$ and $A^\mu$ denote the vector and axial-vector
currents respectively\emdash the distinction of $A^\mu$ here from the
gluon fields in Eq.~(\ref{eq:QCDm0lag}) is neccesary\emdash and these
are defined by
\be V^\mu = \bar{\psi}\gamma^\mu\psi,\qquad A^\mu =
\bar{\psi}\gamma^\mu\gamma_5\psi, \ee 
and which are also conserved, thus $\del_\mu V^\mu = \del_\mu A^\mu =
0$. The chiral symmetry of ${\rm U}(1)_L\otimes {\rm U}(1)_R$ is
therefore equivalent to invariance under transformations under ${\rm
  U}(1)_V \otimes {\rm U}(1)_A$, where we use the transformations
\bea
\label{eq:U1V}
&{\rm U}(1)_V:& \psi \to e^{i\alpha_V}\psi,\qquad \bar{\psi} \to \psi^\dagger
e^{-i\alpha_V} \gamma_0,\\
\label{eq:U1A}
&{\rm U}(1)_A:& \psi \to e^{i\alpha_A\gamma_5}\psi,\quad \ \bar{\psi} \to
\psi^\dagger e^{-i\alpha_A\gamma_5} \gamma_0.
\eea
\par
Using the anticommutation relation
\be \label{eq:anticommutation} \left\{\gamma_5,\gamma_\mu\right\} =
\gamma_5\gamma_\mu + \gamma_\mu\gamma_5 = 0 \ee
we can evaluate the effect that the vector and axial-vector
transformations have on the QCD Lagrangian density, and we find that both
transformations are conserved.
If we now consider a quark mass term ${\cal L}_m$ in the QCD
Lagrangian density, the fermionic part becomes
\be {\cal L}^\psi_{\rm QCD} = {\cal L}_q + {\cal L}_m = 
\bar{\psi}_i\left(i\gamma^\mu
(D_\mu)_{ij}-m\delta_{ij}\right)\psi_j. \ee
%
For the purposes of these discussions, we can set the masses of the
quarks to be equal without loss of generality.
Although the massless Lagrangian density possesses both of the above
symmetries, the axial vector symmetry\emdash and hence chiral
symmetry\emdash is explicitly broken by this quark mass term;
\be
\label{eq:mbreaking}
{\cal L}_{m} = -\bar{\psi}m \psi \stackrel{{\rm U}(1)_A}{\longrightarrow}
-\bar{\psi}m e^{2i\alpha_A}\psi \neq -\bar{\psi}m \psi. 
\ee
The vector symmetry is nonetheless preserved when including this
term.\par

\section{Dynamical Chiral Symmetry (Breaking)}\label{sec:dynamicchiral}
Even with a massless Lagrangian density, it is possible that chiral
symmetry becomes dynamically broken, and we refer to this as
Dynamically Broken Chiral Symmetry, or DCSB.\par
\begin{figure}[!h]
\centering
\includegraphics[width=0.65\textwidth]{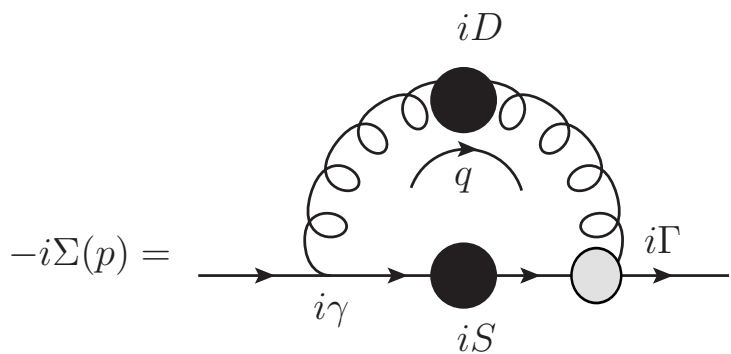}
\caption[Quark self-energy (DSE) in QCD]{Feynman diagram for the QCD
  self-energy for a quark, as given by the Dyson--Schwinger Equation
  (DSE). The full expression for this is given in
  Eq.~(\ref{eq:quarkDSE}). \protect\label{fig:quarkDSE}}
\end{figure}

Following the description of Ref.~\cite{Roberts:1994dr}, if we
consider the basic Lagrangian density of QCD to be
\be {\cal L}_{\rm QCD} = \bar{\psi}_i\left(i\gamma^\mu (D_\mu)_{ij} -
m \delta_{ij}\right) \psi_j - \reci{4}G_{\mu\nu}^aG^{\mu\nu}_a, \ee
with the definitions as in the previous section, of
\be
G^a_{\mu\nu} = \del_\mu A^a_\nu - \del_\nu A^a_\mu - g
f^{abc}A^b_\mu A^c_\nu, \quad 
D_\mu = \del_\nu + i g A^a_\mu T^a,
\ee
with standard definitions of other terms, then we can write the sum of
all QCD One-Particle Irreducible (1-PI) diagrams\footnote{Diagrams
  that cannot be made into two separate disconnected diagrams by
  cutting an internal line are called One-Particle Irreducible, or
  1-PI.} with two external legs as shown in Fig.~\ref{fig:quarkDSE};
illustrating the quark self-energy. The expression for the
renormalized quark self-energy in $d$ dimensions is
\be \label{eq:quarkDSE} -i\Sigma(p) = \frac{4}{3}Z_r\, g^2 \! \int
\frac{d^d q}{(2\pi)^d}
(i\gamma_\mu)(iS(q))(iD^{\mu\nu}(p-q))(i\Gamma_\nu(q,p)), \ee
where $Z_r$ is a renormalization constant, $g$ is the QCD
coupling, and $q$ is the loop momentum.\par
In the absence of matter fields or background fields (the
Lorentz-covariant case), we can write this self-energy as a sum of
Dirac-vector and Dirac-scalar components, as
\be \label{eq:LcovSE} \Sigma(p) = \pslash\; \Sigma_{\rm DV}(p^2) +
\Sigma_{\rm DS}(p^2).  \ee
where $\Sigma_{\rm DV}(p^2)$ is the Dirac-vector component, and
$\Sigma_{\rm DS}(p^2)$ is the Dirac-scalar component. These must both
be functions of $p^2$, since there are no other Dirac-fields to
contract with, and $\Sigma(p)$ is a Lorentz invariant quantity in this
case.\par
For the purposes of our discussion in this section, we will
approximate the Dirac-vector component of the self-energy to be
$\Sigma_{\rm DV} \sim 1$, in which case the self-energy is dependent
only on the Dirac-scalar component.\par
Even with a massless theory ($m = 0$) it is possible that the
renormalized self-energy develops a non-zero Dirac-scalar component,
thus $\Sigma_{\rm DS}(p^2) \neq 0$. This leads to a non-zero value for
the quark condensate $\bra\bar{\psi}_q\psi_q\ket$, and in the limit of
exact chiral symmetry, leads to the pion becoming a massless Goldstone
boson. Thus chiral symmetry can be dynamically broken. With the
addition of a Dirac-scalar component of the self-energy, the
Lagrangian density becomes
\be {\cal L}_{QCD} = \bar{\psi}_i\left(i\gamma^\mu (D_\mu)_{ij} - (m +
\Sigma_{\rm DS})\delta_{ij} \right) \psi_j -
\reci{4}G_{\mu\nu}^aG^{\mu\nu}_a, \ee
\vfill
\clearpage
%
and we can define a dynamic quark mass via the gap equation;
\be m^* = m + \Sigma_{\rm DS}. \ee
We will continue this discussion in Section~\ref{sec:njl}, in which we
will describe a particular model for $\Sigma_{\rm DS}$ in
order to describe DCSB.\par
\section{Equation of State}\label{sec:EoS}
In order to investigate models of dense matter, we need to construct
an Equation of State (EOS), which is simply a relation between two or
more state variables\emdash those which thermodynamically describe the
current state of the system, such as temperature, pressure, volume, or
internal energy\emdash under a given set of physical conditions. With
this, we will be able to investigate various aspects of a model and
compare differences between models in a consistent fashion.\par
For our purposes, we use the total baryon density $\rho_{\rm total}$
as the control parameter of this system, and so we need to obtain the
connection between, say, the energy density ${\cal E}$, the pressure
$P$, and this total baryon density, i.e.
\be {\cal E} = {\cal E}(\rho_{\rm total}), \qquad P = P(\rho_{\rm
  total}).  \ee
\par
State variables are important quantities to consider. Within any
transition between states the total change in any state variable will
remain constant regardless of the path taken, since the change is an
exact differential, by definition. For the hadronic models described
herein, the EOS are exact, in that they have an analytic form;
\be \label{eq:Panalytic} P(\rho_{\rm total}) = \rho_{\rm
  total}^2\frac{\del}{\del\rho_{\rm total}}\left(\frac{{\cal
    E}(\rho_{\rm total})}{\rho_{\rm total}}\right).  \ee
As simple as this exact form may seem, the derivative complicates
things, and we will find it easier to calculate the pressure
independently. Nonetheless, this expression will hold true. More
interestingly, this expression is equivalent to the first law of
thermodynamics in the absence of heat transfer; i.e.
\be PdV = - dE, \ee
(the proof of which can be found in Appendix~\ref{sec:firstlaw}) which
assures us that the theory is thermodynamically consistent.\par
A notable feature of each symmetric matter EOS we calculate is the
effect of saturation; whereby the energy per baryon for the system
possesses a global minimum at a particular value of the Fermi
momentum. This can be considered as a binding energy of the system. In
symmetric matter (in which the densities of protons and neutrons are
equal), the nucleon Fermi momenta are related via \mbox{$k_F = k_{F_n}
  = k_{F_p}$}, and the energy per baryon (binding energy) $E$ is
determined via
\be \label{eq:EperA} E = \left[ \reci{\rho_{\rm total}} \left( {\cal
    E} -\sum_B \rho_BM_B\right) \right].  \ee
In order to reproduce (a chosen set of) experimental results, this
value should be an extremum of the curve with a value of \mbox{$E_0 =
  -15.86~{\rm MeV}$} at a density of \mbox{$\rho_0 = 0.16~{\rm
    fm}^{-3}$} (or the corresponding Fermi momentum $k_{F_0}$).\par
The nucleon symmetry energy $a_{\rm sym}$ is approximately a measure
of the energy difference between the energy per baryon (binding
energy) of a neutron-only model and a symmetric nuclear model
(essentially a measure of the breaking of isospin symmetry). A more
formal expression (without assuming degeneracy between nucleon
masses, as derived in Appendix~\ref{sec:symenergy}) is
\be \label{eq:a4}
a_{\rm sym} = \frac{g_{\rho}^2}{3 \pi^2 m_\rho^2} k_{F}^3 
+ \reci{12} \frac{k_{F}^2}{\sqrt{k_{F}^2 + (M_p^*)^2}} 
+ \reci{12} \frac{k_{F}^2}{\sqrt{k_{F}^2 + (M_n^*)^2}} \, . 
\ee
At saturation, this should take the value of $(a_{\rm sym})_0 =
32.5~{\rm MeV}$ (for an analysis of values, see
Ref.~\cite{Tsang:2008fd}).\par
Another important aspect of an EOS is the compression modulus $K$
which represents the \emph{stiffness} of the EOS; the ability to
withstand compression. This ability is intimately linked to the Pauli
Exclusion Principle in that all other things being equal, a system
with more available states (say, distinguishable momentum states) will
have a softer EOS, and thus a smaller compression modulus. The
compression modulus itself is defined as the curvature of the binding
energy at saturation, the expression for which is
\be \label{eq:Kmod} K = \left[ k_F^2
  \frac{d^2}{dk_F^2}\left(\frac{\cal E}{\rho_{\rm
      total}}\right)\right]_{k_{F_{\rm sat}}} = 9 \left[ \rho_{\rm
    total}^2 \frac{d^2}{d\rho_{\rm total}^2}\left(\frac{\cal
    E}{\rho_{\rm total}}\right)\right]_{\rho=\rho_0}. \ee
The motivation for this is that by compressing the system, the energy
per baryon will rise. The curvature at saturation determines how fast
that rise will occur, and thus how resistant to compression the system
is. Experimentally, this is linked to the properties of finite nuclei,
particularly those with a large number of nucleons, and the binding of
these within a nucleus.\par
According to Ref.~\cite{Serot:1984ey} this should have a value in the
range $200$--$300~{\rm MeV}$, and we will calculate the value of
$K$ for each of the models to follow for comparison.\par

\section{Phase Transitions}\label{sec:phasetransitions}
In order to consider transitions between different phases of matter we
must use statistical mechanics. The simplest method of constructing a
phase transition\emdash known as a `Maxwell transition'\emdash is an
isobaric (constant pressure) transition constructed over a finite
density range. A transition of this form remains useful in
understanding the liquid--gas style phase transition that occurs
within QHD, which is a first-order transition (similar to that of ice
melting in a fluid) with the phases being separated by a non-physical
negative-pressure region. The inclusion of a Maxwell transition to
this simple model for QHD removes this unphysical region and replaces
it with a constant pressure phase.\par
The method for constructing a Maxwell transition will not be covered
here, though in-depth details can be found in
Ref.~\cite{Muller:1995ji}. We can however extract the transition
densities from Ref.~\cite{Serot:1984ey} to reproduce the results,
which are shown later in Fig.~\ref{fig:EOS_plusmaxwell} for the
various varieties of QHD. The more sophisticated method of
constructing a phase transition\emdash the `Gibbs
transition'~\cite{Glendenning:2001pe} that we have used for the
results produced herein\emdash relies on a little more statistical
mechanics. A comparison between the Maxwell and Gibbs methods for
models similar to those used in this work can be found in
Ref.~\cite{Bhattacharyya:2009fg}. For a full in-depth discussion of
this topic, see Ref.~\cite{Reif}.\par
If we consider a homogeneous (suitable for these mean-field
calculations) system with energy $E$, volume $V$, $N_m$ particles of
type $m$, and entropy $S$ which depends on these parameters such that
\be S=S(E,V,N_1,\ldots,N_m), \ee
then we can consider the variation of the entropy in the system as a
function of these parameters, resulting in
\be \label{eq:dSfullform} dS = \left(\frac{\del S}{\del
  E}\right)_{V,N_1,\ldots,N_m} \ dE + \left(\frac{\del S}{\del
  V}\right)_{E,N_1,\ldots,N_m} \ dV + \sum_{i=1}^m \ \left(\frac{\del
  S}{\del N_i}\right)_{V,N_{j\neq i}} \ dN_i, \ee
using standard statistical mechanics notation whereby a subscript $X$
on a partial derivative $\displaystyle{\left(\del A / \del
  B\right)_{X}}$ denotes that $X$ is explicitly held constant.
Eq.~(\ref{eq:dSfullform}) should be equal to the fundamental
thermodynamic relation when the number of particles is fixed, namely
\be \label{eq:fundtherm} dS = \frac{\dbar Q}{T} = \frac{dE + PdV}{T}.
\ee
Here, the symbol $\dbar$ denotes the inexact differential, since the
heat $Q$ is not a state function\emdash does not have initial and
final values\emdash and thus the integral of this expression is only
true for infinitesimal values, and not for finite values.
Continuing to keep the number of each type of particle $N_i$ constant,
a comparison of coefficients between
Eqs.~(\ref{eq:dSfullform})~and~(\ref{eq:fundtherm}) results in the
following relations:
\be \label{eq:invTdefs} \left(\frac{\del S}{\del
  E}\right)_{V,N_i,\ldots,N_m} = \reci{T}, \quad \left(\frac{\del
  S}{\del V}\right)_{E,N_i,\ldots,N_m} = \frac{P}{T}.  \ee
To provide a relation similar to Eq.~(\ref{eq:invTdefs}) for the case
where $dN_i \neq 0$, one defines $\mu_j$\emdash the chemical potential
per molecule\emdash as
\be \mu_i = -T\left(\frac{\del S}{\del N_i}\right)_{E,V,N_{j\neq i}}.
\ee
We can now re-write Eq.~(\ref{eq:dSfullform}) with the definitions in
Eq.~(\ref{eq:invTdefs}) for the case where the particle number can
change, as
\be 
\label{eq:dS}
dS = \reci{T}dE + \frac{P}{T}dV - \sum_{i=1}^m \frac{\mu_i}{T}dN_i,
\ee
which can be equivalently written in the form of the fundamental
thermodynamic relation for non-constant particle number,
\be dE = TdS - PdV + \sum_{i=1}^m \mu_i dN_i.  \ee
If we now consider a system $X$ of two phases $A$ and $B$, then we can
construct relations between their parameters by considering the
following relations:
\bea
\nonumber
E_X &=& E_A + E_B, \\
V_X &=& V_A + V_B, \\
\nonumber
N_X &=& N_A + N_B.
\eea
If we consider that these quantities are conserved between phases, we
find the following conservation conditions
\be
\label{eq:changesAB}
\begin{array}{rcrcr}
dE_X = 0 &\Rightarrow &dE_A + dE_B = 0 &\Rightarrow &dE_A = - dE_B, \\[2mm]
dV_X = 0 &\Rightarrow &dV_A + dV_B = 0 &\Rightarrow &dV_A = - dV_B,
\\[2mm]
%
dN_X = 0 &\Rightarrow &dN_A + dN_B = 0 &\Rightarrow &dN_A = - dN_B .
\end{array}
\ee
\par
The condition for phase equilibrium for the most probable situation is
that the entropy must be a maximum for
$S=S_X(E_X,V_X,N_X)=S(E_A,V_A,N_A;E_B,V_B,N_B)$, which leads to
\be dS_X = dS_A + dS_B = 0.  \ee
Thus, inserting Eq.~(\ref{eq:dS}) we find
\be dS = \left(\reci{T_A} dE_A + \frac{P_A}{T_A} dV_A -
\frac{\mu_A}{T_A} dN_A \right) + \left(\reci{T_B} dE_B +
\frac{P_B}{T_B} dV_B - \frac{\mu_B}{T_B} dN_B \right).  \ee
If we now apply the result of Eq.~(\ref{eq:changesAB}) we can simplify
this relation to
\be \label{eq:maxS} dS = 0 = \left(\reci{T_A} - \reci{T_B}\right) dE_A
+ \left(\frac{P_A}{T_A} - \frac{P_B}{T_B}\right) dV_A -\left(
\frac{\mu_A}{T_A} - \frac{\mu_B}{T_B}\right) dN_A \ee
and thus for arbitrary variations of $E_A$, $V_A$ and $N_A$, each
bracketed term must vanish separately, so that
\be \label{eq:eachvanishes} \reci{T_A} = \reci{T_B}, \quad
\frac{P_A}{T_A} = \frac{P_B}{T_B}, \quad \frac{\mu_A}{T_A} =
\frac{\mu_B}{T_B}.  \ee
Eq.~(\ref{eq:eachvanishes}) implies that at the phase transition the
system will be isentropic ($dS=0$), isothermal ($dT=0$), isobaric
($dP=0$), and isochemical ($d\mu=0$), where the terms
$S$,~$T$,~$V$,~and~$\mu$ now refer to the mean values rather than for
individual particles.\par
We only require two systems at any one time when considering a mixture
of phases, for example, a neutron (`neutron phase') can transition to
a proton and an electron (`proton and electron phase') provided that
the condition $\mu_n = \mu_p + \mu_e$ is met.\par
For a phase transition between hadronic- and quark-matter phases then,
the conditions for stability are therefore that chemical, thermal, and
mechanical equilibrium between the hadronic $H$, and quark $Q$ phases
is achieved, and thus that the independent quantities in each phase
are separately equal. Thus the two independent chemical potentials (as
described in Sec.~\ref{sec:qftchempot}) $\mu_n$~and~$\mu_e$ are each
separately equal to their counterparts in the other phase,
i.e. $\left[(\mu_n)_H=(\mu_n)_Q\right]$, and
$\left[(\mu_e)_H=(\mu_e)_Q\right]$ for chemical equilibrium;
$\left[T_H=T_Q\right]$ for thermal equilibrium; and $\left[P_H =
  P_Q\right]$ for mechanical equilibrium.\par
\begin{figure}[!t]
\centering
\includegraphics[width=0.9\textwidth]{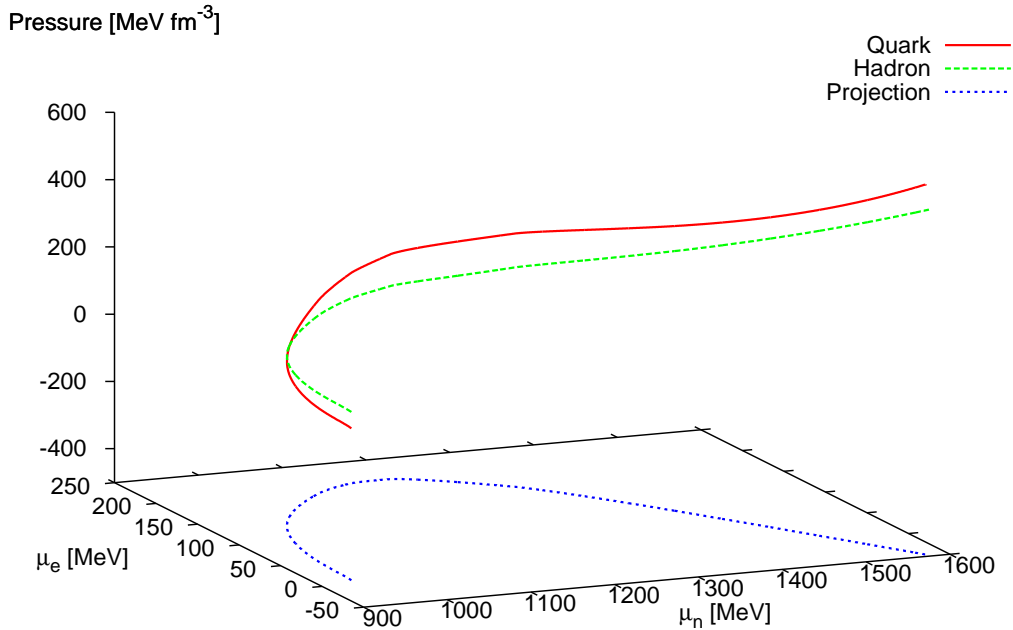}
\caption[Illustrative locus of phase transition variables]{(Color
  Online) Illustrative locus of values for the independent chemical
  potentials $\mu_e$ and $\mu_n$, as well as the pressure $P$ for
  phases of hadronic matter and deconfined quark matter. Note that
  pressure in each phase increases with density. and that a projection
  onto the $\mu_n\mu_e$ plane is a single line, as ensured by the
  chemical equilibrium condition. \protect\label{fig:3d}}
\end{figure}

An illustrative example of these relations is shown in
Fig.~\ref{fig:3d} in which the values of the independent chemical
potentials $\mu_n$ and $\mu_e$, as well as the pressure $P$ for a
hadronic phase and a quark phase are plotted for increasing values of
total density $\rho_{\rm total}$. In this case, the quark matter data
is calculated based on the hadronic matter data, using the chemical
potentials in the hadronic phase as inputs for the quark phase
calculations, and as such the chemical potentials are\emdash by
construction\emdash equal between the phases. As this is an
illustrative example of the relations between the phases, no
constraints have been imposed to reproduce a phase transition yet.\par
In this figure, the low-density points correspond to small values of
$\mu_n$, and we see that for densities lower than some phase
transition density $\rho_{\rm total} < \rho_{\rm PT}$ the hadronic
pressure is greater than the quark pressure and thus the hadronic
phase is dominant. At the transition the pressures are equal, and thus
both phases can be present in a mixed phase, and beyond the transition
the quark pressure is greater than the hadronic pressure indicating
that the quark phase becomes dominant.\par
Note that for all values of the total density, the chemical potentials
in each phase are equal, as shown by the projection onto the
$\mu_n$$\mu_e$ plane.\par
In our calculations, we will only
investigate these two phases independently up to the phase transition,
at which point we will consider a mixed phase, as shall be described
in the next section.\par
We consider both phases to be cold on the nuclear scale, and assume
$T=0$ so the temperatures are also equal, again by construction. We
must therefore find the point\emdash if it exists\emdash at which, for
a given pair of independent chemical potentials, the pressures in both
the hadronic phase and the quark phase are equal.\par
To find the partial pressure of any baryon, quark, or lepton species
$i$ we use
\be \label{eq:pressures} P_i = \frac{\left(2J_B + 1\right) {\cal
    N}_c}{3(2\pi)^3}\int \frac{\vec{k}^2\; \theta(k_{F_i} -
  |\vec{k}|)}{\sqrt{\vec{k}^2+(M_i^*)^2}}\; d^3k, \ee
where the number of colors is ${\cal N}_c = 3$ for quarks, ${\cal N}_c
= 1$ for baryons and leptons, and where $\theta$ is the Heaviside step
function defined in Eq.~(\ref{eq:Heaviside}). To find the total
pressure in each phase, we sum the pressures contributions in that
phase. The pressure in the hadronic phase $H$ is given by
%
%
%
\be \label{eq:Hpressure}
P_H = \sum_j P_j  + \sum_\ell P_\ell + \sum_m P_m,
\ee
in which $j$ represents the baryons, $\ell$ represents the leptons,
$m$ represents the mesons appearing in the particular model being
considered, if they appear, and the pressure in the quark phase $Q$ is
given by
%
\be \label{eq:Qpressure}
P_Q = \sum_q P_q + \sum_\ell P_\ell - B,
\ee
where $q$ represents the quarks, and $B$ denotes the bag energy
density, which we shall discuss further in Section~\ref{sec:MITbag}.\par
%
%
%
%
%
In order to determine the EOS beyond the point at which the pressures
are equal, we need to consider the properties of a mixed phase.\par

\section{Mixed Phase} \label{sec:MixedPhase}

Once we have defined the requirements for a phase transition between
two phases, we must consider the possibility of a mixed phase (MP)
containing proportions of the two phases. This adds a further degree
of sophistication to a model; we can not only find equations of state
for hadronic matter and quark matter and simply stitch them together,
but we can also allow the transition between these to occur
gradually.\par
To calculate the mixed phase EOS, we calculate the hadronic EOS with
control parameter $\rho_{\rm total}$, and use the independent chemical
potentials $\mu_n$ and $\mu_e$ as inputs to determine the quark matter
EOS, since we can determine all other Fermi momenta given these two
quantities. We increase $\rho_{\rm total}$ until we find a
density\emdash if it exists\emdash at which the pressure in the quark
phase is equal the pressure in the hadronic phase (if such a density
cannot be found, then the transition is not possible for the given
models).\par
Assuming that such a transition is possible, once we have the density
and pressure at which the phase transition occurs, we change the
control parameter to the quark fraction $\chi$ (which is an order
parameter parameterizing the transition to the quark matter phase)
which determines the proportions of hadronic matter and quark
matter. If we consider the mixed phase to be composed of some fraction
of hadronic matter and some fraction of quark matter, then the mixed
phase of matter will have the following properties: the total density
will be
\be 
\label{eq:mp_rho}
\rho_{\rm MP} = (1-\chi)\; \rho_{\rm HP} + \chi\; \rho_{\rm QP},
\ee
where $\rho_{\rm HP}$ and $\rho_{\rm QP}$ are the densities in the
hadronic and quark phases, respectively. A factor of three in the
equivalent baryon density in the quark phase,
\be 
\label{eq:equivrho}
\rho_{\rm QP} = \reci{3}\sum_q \rho_q = (\rho_u + \rho_d + \rho_s)/3,
\ee
arises because of the restriction that a baryon contains three
quarks.\par
According to the condition of mechanical equilibrium detailed earlier,
the pressure in the mixed phase will be
\be
\label{eq:mp_P}
P_{\rm MP} = P_{\rm HP} = P_{\rm QP}.
\ee
\par
We can step through values $0\! <\! \chi\! <\! 1$ and determine the
properties of the mixed phase, keeping the mechanical stability
conditions as they were above. In the mixed phase we need to alter our
definition of charge neutrality; while previously we have used the
condition that two phases were independently charge-neutral, such as
$n\to p^+ + e^-$, it now becomes possible that one phase is (locally)
charged, while the other phase carries the opposite charge, making the
system globally charge-neutral. This is achieved by enforcing
\be \label{eq:mp_charge}
0 = (1-\chi)\; \rho^c_{\rm HP} + \chi\; \rho^c_{\rm QP} + 
\rho^c_{\ell} \, ,
\ee
where this time we are considering charge-densities, which are simply
the sum of densities multiplying their respective charges
\be \rho^c_i = \sum_j Q_j \rho_j \ ; \quad i \in \{{\rm HP},{\rm
  QP},\ell\}, \ee
where $j$ are the all individual particles modelled within the
grouping $i$. For example, the quark charge-density in a non-interacting
quark phase is given by
\be \label{eq:qp_charge} \rho^c_{\rm QP} = \sum_q Q_q \rho_q =
\frac{2}{3}\rho_u - \reci{3}\rho_d - \reci{3}\rho_s.  \ee
\par
We continue to calculate the properties of the mixed phase for
increasing values of $\chi$ until we reach $\chi = 1$, at which point
the mixed phase is now entirely charge-neutral quark matter. This
corresponds to the density at which the mixed phase ends, and a pure
quark phase begins. We can therefore continue to calculate the EOS for
pure charge-neutral quark matter, once again using $\rho_{\rm total}$
as the control parameter, but now where the total density is the
equivalent density as defined in Eq.~(\ref{eq:equivrho}).\par

\section{Stellar Matter} \label{sec:stellarmatter}
The equations of state described above are derived for homogeneous
infinite matter. If we wish to apply this to a finite system we must
investigate the manner in which large ensembles of particles are held
together. The focus of this work is `neutron stars', and we must find
a way to utilise our knowledge of infinite matter to provide insight
to macroscopic objects. For this reason, we turn to the theory of
large masses; General Relativity.\par
The Tolman--Oppenheimer--Volkoff (TOV)
equation~\cite{Oppenheimer:1939ne} describes the conditions of
stability against gravitational collapse for an EOS, i.e. in which the
pressure gradient is sufficient to prevent gravitational collapse of
the matter. The equations therefore relate the change in pressure with
radius to various state variables from the EOS. To preserve
continuity, the equations are solved under the condition that the
pressure at the surface of the star \emph{must} be zero.\par
The TOV equation is given by
\be
\label{eq:TOV:full}
\ \ \frac{dP}{dr}=-\frac{ G \left( P / c^2 +\mathcal{E} \right)
  \left(M(r)+4 r^3 \pi P / c^2 \right)} {r(r-2 G M(r) / c^2)}, \ee
or, in Planck units\footnote{In which certain fundamental physical
  constants are normalized to unity, viz $\hbar = c = G = 1$.}
\be
\label{eq:TOV:nat}
\qquad \frac{dP}{dr}=-\frac{ \left( P+\mathcal{E} \right) \left(M(r)
  +4 \pi r^{3} P\right)}{r(r -2M(r))}, \ee
where the mass within a radius $R$ is given by integrating the energy
density, as
\be
\label{eq:TOV:massdef}
\qquad M(R) = \int_0^R 4 \pi r^{2} {\cal E}(r) \; dr.
\ee
For a full derivation of these equations, refer to
Appendix~\ref{sec:tovderiv}.\par
Supplied with these equations and a derived EOS, we can calculate
values for the total mass and total radius of a star\footnote{Stellar
  objects in these calculations are assumed to be static, spherically
  symmetric, and non-rotating, as per the derivation of this
  equation. For studies of the effect of rapid rotation in General
  Relativity see Refs.~\cite{Lattimer:2004nj,Owen:2005fn}.} for a given
central density. We refer to these values as the `stellar
solutions.'\par
This is particularly interesting, since the mass of a neutron star is
observable (either via observing a pair of objects rotating about a
barycenter\footnote{A common centre of mass for the system, the point
  about which both objects will orbit, which is the balance point of
  the gravitational force. In this case, the mass measurements are
  simplified.}, or some other indirect/proxy measurement), yet the
radius is not directly observable, as stars are sufficiently distant
that they all appear as `point-sources'. With these calculations, we
produce a relationship between two quantities: the stellar mass and
the stellar radius, of which only the mass is currently observable,
and even this is not always so. This provides useful data for further
theoretical work requiring both quantities, as well as an opportunity
to place theoretical bounds on future experimental observations.\par
In addition to this data, since we are able to solve our equations for
the radial distance from the centre of the star, we can provide data
that current experiments can not; we can investigate the interior of a
neutron star, by calculating the proportions of various particles at
successive values of internal radius and/or density. This allows us to
construct a cross-section of a neutron star, investigate the possible
contents, and examine the effects that various changes to the models
have on both the internal and external properties.\par
%

\section{SU(6) Spin-Flavor Baryon-Meson Couplings}\label{sec:su6deriv}

We have noted earlier that the coupling of baryons to mesons is
dependent on isospin group. The physics leading to this result is
highly non-trivial, but is often neglected in the literature. We will
therefore outline the process involved in determining the relations
between baryon-meson couplings.\par
In order to determine the normalized relations between the point
vertex couplings of various mesons to the full baryon octet $g_{Bm}$
it is common to express the octet as a $3\times 3$ matrix in flavor
space as
\be B = \left( \
\begin{matrix}
\frac{\Sigma^0}{\sqrt{2}}+\frac{\Lambda}{\sqrt{6}} & \Sigma^+ & p \\
\Sigma^- & -\frac{\Sigma^0}{\sqrt{2}}+\frac{\Lambda}{\sqrt{6}} & n \\
-\Xi^- & \Xi^0 & -\frac{2 \Lambda}{\sqrt{6}}
\end{matrix} \ \right).
\ee
This has been constructed as an array where rows and columns are
distinguished by rotations in flavor space, which can be seen if we
observe the quark content of these baryons;
\be B = \left( \
\begin{matrix}
uds & \mathbf{d \to u} & uus & \mathbf{s \to d} & uud \\ \mathbf{u \to
  d} & & & & \\ dds & & dus & & ddu \\ \mathbf{d \to s} & & & & \\ ssd
& & uss & & sud
\end{matrix} \ \right).
\ee
The vector meson octet ($J^P=1^-$) can be written in a similar fashion as
\be \label{eq:Pvec}
P^\textrm{vec}_\textrm{oct} = \left( \
\begin{matrix}
\frac{\rho^0}{\sqrt{2}}+\frac{\omega_8}{\sqrt{6}} & \rho^+ & K^{*+} \\
\rho^- & -\frac{\rho^0}{\sqrt{2}}+\frac{\omega_8}{\sqrt{6}} & K^{*0} \\
K^{*-} & \overline{K^{*0}} & -2\left(\frac{\omega_8}{\sqrt{6}} \right)
\end{matrix} \ \right),
\ee
which, along with the singlet state, $P_{\rm sing}^{\rm vec} =
\reci{\sqrt{3}}{\rm diag}(\omega_0,\omega_0,\omega_0)$ defines the
vector meson nonet
\be \label{eq:octplussing} P^{\rm vec} = P^{\rm vec}_{\rm oct} +
P^{\rm vec}_{\rm sing}.  \ee
Furthermore, the scalar meson octet ($J^P=0^+$) can be written as
\be \label{eq:Psca}
P^\textrm{sca}_\textrm{oct} = \left( \
\begin{matrix}
\frac{a_0^0}{\sqrt{2}}+\frac{\s_8}{\sqrt{6}} & a_0^+ & \kappa^+ \\
a_0^- & -\frac{a_0^0}{\sqrt{2}}+\frac{\s_8}{\sqrt{6}} & \kappa^0 \\
\kappa^- & \overline{\kappa^0} & -2\frac{\s_8}{\sqrt{6}}
\end{matrix} \ \right),
\ee
and along with singlet state $P^{\rm sca}_{\rm sing} =
\reci{\sqrt{3}}{\rm diag}(\s_0,\s_0,\s_0)$, these define the scalar
meson nonet. The meson octet matrices are constructed in a similar
fashion to the baryon octet matrix;
\be
P_\textrm{oct} = \left( \
\begin{matrix}
u\bar{u} & \mathbf{\bar{u} \to \bar{d}} & u\bar{d} & \mathbf{\bar{d}
  \to \bar{s}} & u\bar{s} \\ \mathbf{u \to d} & & & & \\ d\bar{u} & &
d\bar{d} & & d\bar{s} \\ \mathbf{d \to s} & & & & \\ s\bar{u} & &
s\bar{d} & & s\bar{s}
\end{matrix} \ \right).
\ee
The singlet and octet representations of both $\omega$ and $\sigma$
($\w_0, \w_8, \s_0, \s_8$, appearing in
Eqs.~(\ref{eq:Pvec})--(\ref{eq:Psca})) are not however the physical
particles which we wish to include in the model; these are linear
combinations of the physical particles. Due to explicit SU(3)
flavor-symmetry breaking ($m_s > m_u,\ m_d$), a mixture of the
unphysical $\omega_8$ and $\w_0$ states produces the physical $\omega$
and $\phi$ mesons, while a mixture of the unphysical $\sigma_8$ and
$\s_0$ states produces the physical $\sigma$ and $f_0$ mesons, the
properties of which we list in Appendix~\ref{sec:particleprops}.\par
The octet and singlet states are represented by linear combinations of
quark-antiquark pairs. The state vectors for these are
\be
\label{eq:quarklincombs}
|\omega_8\ket = |\sigma_8\ket = \reci{\sqrt{6}} \left( |\bar{u}u\ket +
|\bar{d}d\ket -2 |\bar{s}s\ket \right),\quad |\omega_0\ket =
|\sigma_0\ket = \reci{\sqrt{3}} \left( |\bar{u}u\ket + |\bar{d}d\ket
|+ |\bar{s}s\ket \right), \ee
where the normalizations arise by ensuring that 
\be
\bra\xi|\xi\ket = 1;\quad \xi \in \{\omega_8,\omega_0,\sigma_8,\sigma_0\}.
\ee
Since the quark contents of the physical states are predominantly 
\be \omega = \sigma = \reci{\sqrt{2}}\ (u\bar{u}+d\bar{d}),\quad {\rm
  and} \quad \phi = f_0 = -s\bar{s}, \ee
we can replace the octet and singlet combinations with the physical
states via the replacements of
\bea
&\omega_8 = \reci{\sqrt{3}}\ \omega +
\frac{2}{\sqrt{6}}\ \phi,\quad 
&\omega_0 = \sqrt{\frac{2}{3}}\ \omega
-\reci{\sqrt{3}}\ \phi,\\
&\sigma_8 = \reci{\sqrt{3}}\ \sigma +
\frac{2}{\sqrt{6}}\ f_0,\quad 
&\sigma_0 = \sqrt{\frac{2}{3}}\ \sigma
-\reci{\sqrt{3}}\ f_0. 
\eea
Using these definitions, we can express the octet and singlet states 
in terms of the physical states
\be P^\textrm{vec}_\textrm{oct} = \left( \
\begin{matrix}
\frac{\rho^0}{\sqrt{2}}+\frac{\omega}{\sqrt{18}}+\frac{\phi}{\sqrt{9}}
& \rho^+ & K^{*+} \\ \rho^- &
-\frac{\rho^0}{\sqrt{2}}+\frac{\omega}{\sqrt{18}}+\frac{\phi}{\sqrt{9}}
& K^{*0} \\ K^{*-} & \overline{K^{*0}} &
-2\left(\frac{\omega}{\sqrt{18}}+\frac{\phi}{\sqrt{9}}\right)
\end{matrix} \ \right),
\ee
\be P^\textrm{sca}_\textrm{oct} = \left( \
\begin{matrix}
\frac{a_0^0}{\sqrt{2}}+\frac{\sigma}{\sqrt{18}}+\frac{f_0}{\sqrt{9}}
& a_0^+ & \kappa^{+} \\ a_0^- &
-\frac{a_0^0}{\sqrt{2}}+\frac{\sigma}{\sqrt{18}}+\frac{f_0}{\sqrt{9}}
& \kappa^{0} \\ \kappa^{-} & \overline{\kappa^{0}} &
-2\left(\frac{\sigma}{\sqrt{18}}+\frac{f_0}{\sqrt{9}}\right)
\end{matrix} \ \right).
\ee
Each of the above mesons can interact with a pair of baryons in three
possible SU(3) invariant ways, which we shall identify as $F$-style
(anti-symmetric), $D$-style (symmetric) and $S$-style
(singlet). The singlet mesons $\omega_0$ and $\sigma_0$ are associated
with an $S$-style coupling, while the octet particles are associated
with $F$- and $D$-style couplings.\par
To determine these $F$-, $D$-, and $S$-style couplings for the vector
and scalar mesons we need to calculate the SU(3) invariant Lagrangian
density coefficients symbolically for each isospin group, with each
SU(3) invariant combination\footnote{Following the notation
  of Ref.~\cite{Rijken:1998yy}.} given by:
\bea \nonumber \left[ \bar{B}BP \right]_F &= &{\rm Tr}(\bar{B}PB)-{\rm
  Tr}(\bar{B}BP) \\ \nonumber &= &{\rm Tr}(\bar{B}P_{\rm oct}B)-{\rm
  Tr}(\bar{B}BP_{\rm oct}), \\[2mm] \left[ \bar{B}BP \right]_D &=
&{\rm Tr}(\bar{B}PB)+{\rm Tr}(\bar{B}BP)-\frac{2}{3}{\rm
  Tr}(\bar{B}BP){\rm Tr}(P) \\ \nonumber &= &{\rm Tr}(\bar{B}P_{\rm
  oct}B)+{\rm Tr}(\bar{B}BP_{\rm oct}), \\[2mm] \nonumber \left[
  \bar{B}BP \right]_S &= &{\rm Tr}(\bar{B}B){\rm Tr}(P) \\ \nonumber
&= &{\rm Tr}(\bar{B}B){\rm Tr}(P_{\rm sing}).\label{eq:SU3invcombs}
\eea
where we have expanded the meson matrices $P$ according to
Eq.~(\ref{eq:octplussing}), and we note that the octet matrices
$\bar{B}$, and $B$ are traceless.\par
Together, these terms can be combined to form an interaction
Lagrangian density for all possible SU(3) invariant interactions
involving each meson isospin group, with octet and singlet coupling
coefficients $F$, $D$, and $S$ (each defined separately for each
isospin group) as
\be
\label{eq:SU3intL}
{\cal L}^{\rm int} = -\sqrt{2}\left\{ F\left[ \bar{B}BP\right]_F +D
\left[ \bar{B}BP\right]_D \right\} - S\reci{\sqrt{3}} \left[
  \bar{B}BP\right]_S, \ee
where the remaining numerical factors are introduced for
convenience.\par
If we evaluate this Lagrangian density by matrix multiplication of
$B$, the octet and singlet matrices of $P_{\rm vec}$, and $\bar{B} =
B^{\dagger}\gamma^0$ in the combinations stated in
Eq.~(\ref{eq:SU3invcombs}), we can extract the coefficients of each
baryon-meson vertex in terms of $F$, $D$ and $S$ factors. These are
summarized in Table~\ref{tab:FDScouplings} for vertices involving the
physical vector mesons $\w$ and $\rho^0$ and $\phi$, for pairs of like
baryons\footnote{As discussed in Section~\ref{sec:mfa}, any
  flavor-changing meson-baryon interactions would produce a null
  overlap of ground-state operators, and as such we only focus on the
  like-baryon interactions of the form $g_{B\alpha}\bar{\psi}_B \alpha
  \psi_{B^\prime} \delta_{BB^\prime}$ for a meson $\alpha$ in this
  discussion.}.
The summary for the scalar mesons is the same under replacements of
$\omega \to \sigma$, $\vec{\rho} \to \vec{a}_0$, and $\phi \to
f_0$.\par
\vfill
\begin{table}[!h]
\caption[$F$-, $D$-, and $S$-style couplings for
  $\bar{B}BP$]{\protect\label{tab:FDScouplings}$F$-, $D$-, and
  $S$-style couplings of like baryon-baryon pairs to vector mesons
  used in these models, according to vertices of type $B + P \to
  \bar{B}$. The summary for the scalar mesons is the same under the
  replacements of $\w \to \s$, $\vec{\rho} \to \vec{a}_0$, and
  $\phi\to f_0$.}
\begin{center}
\hrule
\vspace{1mm}
\hrule
\vspace{2mm}
\hspace{0.2cm}
\centering
\begin{tabular}{cc}
\begin{minipage}[c]{0.5\textwidth}
$\begin{array}{lrl}
\overline{\Sigma^-}\Sigma^-\rho^0 \ & \propto & \ 2 F \\[2mm]
\overline{\Sigma^-}\Sigma^-\omega \ & \propto & \ \frac{1}{9} \left(-6 D-\sqrt{6}S\right)           \\[2mm]
\overline{\Sigma^-}\Sigma^-\phi   \ & \propto & \ \frac{1}{9} \left(-6 \sqrt{2}D+\sqrt{3} S\right)  \\[2mm]
\overline{\Sigma^0}\Sigma^0\omega \ & \propto & \ \frac{1}{9} \left(-6 D-\sqrt{6}S\right)           \\[2mm]
\overline{\Sigma^0}\Sigma^0\phi   \ & \propto & \ \frac{1}{9} \left(-6 \sqrt{2}D+\sqrt{3} S\right)  \\[2mm]
\overline{\Sigma^+}\Sigma^+\rho^0 \ & \propto & \ -2 F \\[2mm]
\overline{\Sigma^+}\Sigma^+\omega \ & \propto & \ \frac{1}{9} \left(-6 D-\sqrt{6}S\right)           \\[2mm]
\overline{\Sigma^+}\Sigma^+\phi   \ & \propto & \ \frac{1}{9} \left(-6 \sqrt{2}D+\sqrt{3} S\right)  \\[2mm]
 \end{array}$
 \end{minipage}
 & 
 \begin{minipage}[c]{0.5\textwidth}
$\begin{array}{lrl}
\overline{\Lambda }\Lambda\omega \ & \propto & \ \frac{1}{9} \left(6 D-\sqrt{6} S\right)           \\[2mm]
\overline{\Lambda }\Lambda \phi  \ & \propto & \ \frac{1}{9} \left(6 \sqrt{2} D+\sqrt{3}S\right)     \\[2mm]
\overline{p}p\rho^0              \ & \propto & \ -D-F  \\[2mm]
\overline{p}p\omega              \ & \propto & \ \frac{1}{9} \left(3 D-9 F-\sqrt{6} S\right)  \\[2mm]
\overline{p}p\phi                \ & \propto & \ \frac{1}{9} \left(3 \sqrt{2} D-9 \sqrt{2} F+\sqrt{3}S\right)  \\[2mm]
\overline{n}n\rho^0              \ & \propto & \ D+F  \\[2mm]
\overline{n}n\omega              \ & \propto & \ \frac{1}{9} \left(3 D-9 F-\sqrt{6} S\right)  \\[2mm]
\overline{n}n\phi                \ & \propto & \ \frac{1}{9} \left(3 \sqrt{2} D-9 \sqrt{2} F+\sqrt{3}S\right)  \\[2mm]
 \end{array}$
 \end{minipage}
\end{tabular} 
 \begin{minipage}[c]{0.5\textwidth}
$\begin{array}{lrl}
& & \\
\overline{\Xi^-}\Xi^-\rho^0      \ & \propto & \ -D+F  \\[2mm]
\overline{\Xi^-}\Xi^-\omega      \ & \propto & \ \frac{1}{9} \left(3 D+9 F-\sqrt{6}S\right) \\[2mm]
\overline{\Xi^-}\Xi^-\phi        \ & \propto & \ \frac{1}{9} \left(3 \sqrt{2} D+9 \sqrt{2}F+\sqrt{3} S\right)  \\[2mm]
\overline{\Xi^0}\Xi^0\rho^0      \ & \propto & \ D-F  \\[2mm]
\overline{\Xi^0}\Xi^0\omega      \ & \propto & \ \frac{1}{9} \left(3 D+9 F-\sqrt{6}S\right)  \\[2mm]
\overline{\Xi^0}\Xi^0\phi        \ & \propto & \ \frac{1}{9} \left(3 \sqrt{2} D+9 \sqrt{2}F+\sqrt{3} S\right)  \\[2mm]
 \end{array}$
\end{minipage}
\vspace{1mm}
\hrule
\vspace{1mm}
\hrule
\end{center}
\end{table}
The physical $\phi$ and $f_0$ states are purely strange quark
components. These do not couple to nucleons significantly (since
nucleons contain only up and down valence quarks) the only way to
produce these mesons is via gluons. Thus we set the (normalized or
not) couplings of these mesons to zero; $g_{B\phi}=g_{Bf_0}=0$. We are
then left with the physical $\sigma$ and $\omega$ mesons as the
effective meson degrees of freedom.\par
If we denote the \emph{total} (but not normalized) coupling (now
including all prefactors of Eq.~(\ref{eq:SU3intL})) of a (like) baryon
pair $\bar{B}B$ to a meson $\alpha$ by $f_{B\alpha}$, and we calculate
the SU(3) invariant combinations for the singlet $\w_0$ and mixed
state $\w_8$, we can use the relation between these normalizations
from Eq.~(\ref{eq:quarklincombs}) to relate the $S$-style couplings to
the remaining couplings via
\be -\frac{S}{3} = f_{N\w_0} = \sqrt{2}f_{N\w_8} =
\frac{\sqrt{2}}{\sqrt{3}}(D-3F), \ee
and so we can reduce the relation of the couplings to
\be
\label{eq:strangecoupling}
S = \sqrt{6}(3F-D),
\ee
and we can therefore find the couplings of the singlet $\w_0$ meson in
terms of just $F$ and $D$ factors. After removing the strange quark
components and substituting the result of
Eq.~(\ref{eq:strangecoupling}) we obtain a summary of couplings as
shown in Table~\ref{tab:FDcouplings}.\par
\begin{table}[!b]
\caption[$F$- and $D$-style couplings for
  $\bar{B}BP$]{\protect\label{tab:FDcouplings}Couplings of like
  baryon-baryon pairs to vector mesons used in these models, according
  to vertices of type $B + P \to \bar{B}$ using the relation of
  Eq.~(\ref{eq:strangecoupling})}
\begin{center}
\hrule
\vspace{1mm}
\hrule
\vspace{2mm}
\hspace{0.2cm}
\centering
\begin{tabular}{ccc}
\begin{minipage}[c]{0.3\textwidth}
$\begin{array}{lrr}
\overline{\Sigma^-}\Sigma^-\rho^0  \  & \propto &    2F                    \\[2mm]
\overline{\Sigma^-}\Sigma^-\omega  \  & \propto &    -\frac{2D}{3}         \\[2mm]
\overline{\Sigma^0}\Sigma^0\omega  \  & \propto &    -\frac{2D}{3}         \\[2mm]
\overline{\Sigma^+}\Sigma^+\rho^0  \  & \propto &    -2   F                \\[2mm]
\overline{\Sigma^+}\Sigma^+\omega  \  & \propto &    -\frac{2D}{3}         \\[2mm]
 \end{array}$
 \end{minipage}
 & 
 \begin{minipage}[c]{0.3\textwidth}
 $  \begin{array}{lrr}
\overline{p}p\rho^0              \  & \propto &   -D-F                   \\[2mm]
\overline{p}p\omega              \  & \propto &   \frac{D}{3}-F          \\[2mm]
\overline{n}n\rho^0              \  & \propto &   D+F                    \\[2mm]
\overline{n}n\omega              \  & \propto &   \frac{D}{3}-F          \\[2mm]
\overline{\Lambda}\Lambda\omega  \  & \propto &   \frac{2D}{3}           \\[2mm]
 \end{array}$
 \end{minipage}
 &
\begin{minipage}[c]{0.3\textwidth}
  $\begin{array}{lrr}
\overline{{{\Xi }}^-}{{\Xi}^-}{{\rho }^0}   \   & \propto &    -D+F          \\[2mm]
\overline{{{\Xi }}^-}{{\Xi}^-}\omega        \   & \propto &    \frac{D}{3}+F \\[2mm]
\overline{{{\Xi }}^0}{{\Xi}^0}{{\rho }^0}   \   & \propto &    D-F           \\[2mm]
\overline{{{\Xi }}^0}{{\Xi}^0}\omega        \   & \propto &    \frac{D}{3}+F \\[2mm]
  \end{array}$
\end{minipage}
\end{tabular}
\vspace{1mm}
\hrule
\vspace{1mm}
\hrule
\vspace{10mm}
\end{center}
\end{table}
We note however that the couplings in Tables~\ref{tab:FDScouplings}
and \ref{tab:FDcouplings} do not display isospin symmetry manifestly,
though our original Lagrangian density (refer to
Section~\ref{sec:lagrangiandensity}) was constructed in terms of
isospin groups only with common coefficients. This can be remedied by
considering a \emph{general} Lagrangian density constructed from
isospin groups, which we shall restrict to terms involving like
baryons and the mesons we are interested in, to give
\bea
\nonumber \mathcal{L}_{\rm int}^{\rm oct} &=&
-f_{N\rho}(\overline{N}\vec{\tau}^{\rm T}N)\cdot\vec{\rho}
+if_{\Sigma\rho}(\vec{\overline{\Sigma}}\times\vec{\Sigma})\cdot\vec{\rho}
-f_{\Xi\rho}(\overline{\Xi}\vec{\tau}^{\rm
  T}\Xi)\cdot\vec{\rho} \\
\label{eq:octintL}
&& - f_{N\omega}(\overline{N}N)\omega -
f_{\Lambda\omega}(\overline{\Lambda}\Lambda)\omega
-f_{\Sigma\omega}(\vec{\overline{\Sigma}}\cdot\vec{\Sigma})\omega -
f_{\Xi\omega}(\overline{\Xi}\Xi)\omega,
\eea
where the $N$, $\Lambda$, and $\Xi$ isospin groups are defined as
before as
\be N = \begin{pmatrix}p\\n\end{pmatrix}, \quad \Lambda
  = \begin{pmatrix}\Lambda\end{pmatrix},
 \quad \Xi = \begin{pmatrix}\Xi^0\\ \Xi^-\end{pmatrix}. \ee
%
%
The $\rho$ mesons terms are defined in isospin space as linear
combinations of the physical charged states (as we did in
Section~\ref{subsec:rotational}) as
\be \nonumber \rho^- = \reci{\sqrt{2}}(\rho_1 - i\rho_2), \quad \rho^+
= \reci{\sqrt{2}}(\rho_1 + i\rho_2), \quad \rho^0 = \rho_3, \ee
or, equivalently as
\be \rho_1 = \reci{\sqrt{2}}(\rho^+ + \rho^-), \quad \rho_2 =
\frac{i}{\sqrt{2}}(\rho^- - \rho^+), \quad \rho_3 = \rho^0, \ee
with the same convention for the replacement of $\vec{\rho} \to
\vec{\Sigma}$. This gives the expansion
\be \vec{\Sigma}\cdot\vec{\rho} = \Sigma^+\rho^- + \Sigma^0\rho^0 +
\Sigma^-\rho^+ .  \ee
We can expand the Lagrangian density term by term to find the
individual interactions
\bea \nonumber (\overline{N}\vec{\tau}^{\rm T}N)\cdot\vec{\rho} &=&
(\overline{p} \ \ \overline{n}) \ \tau^{\rm T}_i
\rho^i \begin{pmatrix}p\\n\end{pmatrix} \\ \nonumber &=&
  (\overline{p}n+\overline{n}p)\rho_1 +
  i(\overline{p}n-\overline{n}p)\rho_2 + (\overline{p}p -
  \overline{n}n)\rho_3 \\ \nonumber &=&\reci{\sqrt{2}}
  (\overline{p}n+\overline{n}p)(\rho_+ + \rho_-)
  -\frac{1}{\sqrt{2}}(\overline{p}n-\overline{n}p)(\rho_- - \rho_+) +
  (\overline{p}p - \overline{n}n)\rho_0 \\ &=& \overline{p}p\rho_0 -
  \overline{n}n\rho_0 +\sqrt{2}\overline{p}n\rho_+ +
  \sqrt{2}\overline{n}p\rho_-\ , \eea
where we note that a term $\bar{B}BP$ indicates the annihilation of a
baryon $B$ with a meson $P$, and the creation of a baryon $\bar{B}$
according to the reaction $B+P\to\bar{B}$. Continuing to expand terms,
for the $\Sigma$ baryons we have
\bea \nonumber
(\vec{\overline{\Sigma}}\times\vec{\Sigma})\cdot\vec{\rho} &=& -i
\rho^+ \left(\Sigma^-\overline{\Sigma^0}-\Sigma^0\overline{\Sigma^+}
\right) -i \rho^- \left(
\Sigma^0\overline{\Sigma^-}-\Sigma^+\overline{\Sigma^0} \right) 
-i \rho^0 \left(
\Sigma^+\overline{\Sigma^+}-\Sigma^-\overline{\Sigma^-} \right)\ ,\\[1mm]
&& \eea
and for the $\Xi$ baryons,
%
%
\bea \nonumber (\overline{\Xi}\vec{\tau}^{\rm T}\Xi)\cdot\vec{\rho}
&=& (\overline{\Xi^0} \ \ \overline{\Xi^-}) \ \tau^{\rm T}_i
\rho^i \begin{pmatrix}\Xi^0\\ \Xi^-\end{pmatrix}\\ \nonumber &=&
  (\overline{\Xi^0}\Xi^-+\overline{\Xi^-}p)\rho_1 +
  i(\overline{\Xi^0}\Xi^--\overline{\Xi^-}\Xi^0)\rho_2 +
  (\overline{\Xi^0}\Xi^0 - \overline{\Xi^-}\Xi^-)\rho_3 \\[2mm] &=&
  \overline{\Xi^0}\Xi^0\rho_0 - \overline{\Xi^-}\Xi^-\rho_0
  +\sqrt{2}\overline{\Xi^0}\Xi^-\rho_+ +
  \sqrt{2}\overline{\Xi^-}\Xi^0\rho_-. \\ \nonumber \eea
The iso-scalar terms are more straightforward;
\bea (\overline{N}N)\omega &=& \overline{p}p\omega +
\overline{n}n\omega, \\[1mm]
\overline{\Lambda}\Lambda\omega && (\textrm{requires no expansion}), \\[1mm]
( \vec{\overline{\Sigma}}\cdot\vec{\Sigma})\omega &=&
\overline{\Sigma^+}\Sigma^+\omega + \overline{\Sigma^0}\Sigma^0\omega
+ \overline{\Sigma^-}\Sigma^-\omega, \\[1mm]
(\overline{\Xi}\Xi)\omega &=& \overline{\Xi^0}\Xi^0\omega +
\overline{\Xi^-}\Xi^-\omega.  \eea
\par
Once we have calculated the full interaction Lagrangian density, and
the $F$ and $D$ coefficients of each interaction, we have factors of
the following form:
\be {\cal L}_{\rm int} = \sum_B \sum_m A_{Bm} f_{Bm} X_{Bm}; \quad
X_{Bm} = C_{Bm} \bar{B}Bm, \ee
where $A_{\Sigma \rho} = i$, and for all other interactions $A_{Bm}=
-1$. The term $X_{Bm}$ is the expanded interaction term (after
expanding cross products, etc.) arising from the \emph{general}
Lagrangian density, Eq.~(\ref{eq:octintL}), and contains factors of
$C_{Bm} = \pm 1,\pm \sqrt{2}$. We also require a term calculated from
the SU(3) invariant combinations $M_{Bm}$; the coefficient of the
interaction $\bar{B}Bm$ in terms of $F$ and $D$ factors as found in
Table~\ref{tab:FDcouplings}.\par
To calculate the values of $f_{Bm}$ we apply the following formula:
\be f_{Bm} = \frac{C_{Bm}}{A_{Bm}} M_{Bm}. \ee
For example, consider the interaction vertex $\omega + \Sigma^0 \to
\overline{\Sigma^0}$:
\be A_{\Sigma \w} = -1, \quad C_{\Sigma \w} = +1, \quad M_{\Sigma \w}
= -\frac{2D}{3}, \quad \Rightarrow \quad f_{\Sigma\omega} =
\frac{+1}{-1}(-\frac{2D}{3}) = \frac{2D}{3}.  \ee
Performing these calculations for every possible interactions provides
(consistently) the following couplings of the octet of baryons to the
octet of mesons:\par
\be 
\begin{array}{c}
f_{N\rho} = D+F,\quad f_{\Lambda\rho} = 0,\quad f_{\Sigma\rho} =
2F, \quad f_{\Xi\rho} = F-D, 
\\[2mm]
f_{N\w} = 3F-D, \quad f_{\Lambda\w} = -\frac{4}{3}D+2F, \quad
f_{\Sigma\w} = 2F, \quad f_{\Xi\w} = F-D. 
\end{array}
\ee
\par
We can further simplify our calculations by examining all the
different currents that one can form using a baryon, an antibaryon and
a meson. As discussed in Ref.~\cite{Sakita:1965qt}, all possible
couplings of baryons to vector mesons (denoted by $\bar{B}BV$) should
be considered when writing out the most general Lagrangian density. By
calculating the currents (prior to making any approximations or
assumptions that appear in earlier sections here) we are able to find
the $F$-, $D$-, and $S$-style couplings of the form $\bar{B}BX$ where
$X$ is a meson with either scalar (S), vector (V), tensor (T),
axial-vector (A) or pseudo-scalar (P) spin form.\par
Under an expanded SU(6) spin-flavor symmetry, the currents are shown
in Table~\ref{tab:currents}, where the various vector couplings are of
the forms
\be V_1 = \bar{\psi}\gamma_\mu\psi,\ V_2 =
\bar{\psi}\sigma_{\mu\nu}q^\nu\psi,\ V_3 = \bar{\psi}q_\mu\psi, \ee
and we use the convenience definitions of
\be \sigma_{\mu\nu}=\frac{i}{2}\left[\gamma_\mu,\gamma_\nu\right], \quad
H=\frac{4M^2+q^2}{2M^2}.  \ee
\par
\begin{table}[!t]
\centering
\caption[$F$-, $D$- and $S$-style baryon currents]{\protect $F$-, $D$-
  and $S$-style Baryon currents for all types of meson vertices of the
  form $\bar{B}BX$ where $X$ is a meson with either scalar (S), vector
  (V), tensor (T), axial(pseudo-) vector (A) or pseudo-scalar (P) spin
  form. Adapted from Ref.~\cite{Sakita:1965qt}.\label{tab:currents}}
\vspace{3mm}
\begin{tabular}{lccc}
\hline 
\hline 
&&&\\[-4mm]
   &     $F$    &    $D$    &    $S$ \\
&&&\\[-4mm]
\hline
&&&\\[-2mm]
S  & $\reci{3}H\bar{\psi}\psi$ & 0 & $\reci{3}H\bar{\psi}\psi$ \\
&&&\\[-2mm]
V${}_1$ & $\reci{3}\left(H-\reci{6}\frac{q^2}{M^2}\right)\bar{\psi}\gamma_\mu\psi$ &
 $\frac{q^2}{6M^2}\bar{\psi}\gamma_\mu\psi$ &
 $\reci{3}\left(H-\reci{3}\frac{q^2}{M^2}\right)\bar{\psi}\gamma_\mu\psi$ \\
&&&\\[-2mm]
V${}_2$ & $-\reci{9M}i\bar{\psi}\sigma_{\mu\nu}\psi$ &  
$ + \reci{3M}i\bar{\psi}\sigma_{\mu\nu}\psi$ &
$ - \frac{2}{9M}i\bar{\psi}\sigma_{\mu\nu}\psi$ \\
&&&\\[-2mm]
V${}_3$ &  0  &  0  &  0  \\
&&&\\[-2mm]
A & $ \frac{2}{9}H\bar{\psi}\gamma_5\gamma_\mu\psi $ &
    $ \reci{3}H\bar{\psi}\gamma_5\gamma_\mu\psi $    & 
    $ \reci{9}H\bar{\psi}\gamma_5\gamma_\mu\psi $ \\
&&&\\[-2mm]
P & $ \frac{2}{9}H\bar{\psi}\gamma_5\psi $ &
    $ \reci{3}H\bar{\psi}\gamma_5\psi $    & 
    $ \reci{9}H\bar{\psi}\gamma_5\psi $ \\
&&&\\[-2mm]
\hline
\hline
\end{tabular}
\end{table}
If we now consider this as a low energy effective field theory, we can
consider the case of $q^2=0$. We can also enforce rotational symmetry
due to lack of a preferred frame (or direction) and thus remove the
spatial components of both the mesons and the momenta, so that
$V^\mu=(V^0,\vec{0})$ and $q^\mu=(q^0,\vec{0})$, as per
Section~\ref{subsec:rotational}. Along with $\sigma_{00}=0$, all terms
proportional to $q^2$ vanish, and $H=2$. Using these assumptions, the
currents are reduced to those found in
Table~\ref{tab:currentassumptions}.\par
We can now observe the relations between the $F$- and $D$-style
couplings (with the $S$-style coupling now contributing to $F$ and
$D$); First, as a check, we observe that the ratio $D/F$ for the
pseudo-scalars (and the axial-vectors for that matter) is indeed
$\frac{3}{2}$ as commonly noted in the
literature~\cite{Gursey:1964,Aliev:2001} under SU(6)
symmetry~\cite{Ishida:1968}. Less commonly found in the literature is
that the $\gamma_\mu$-type vector coupling is purely $F$-style, thus
the vector analogy of the above relation is $D/F=0$, implying
$D=0$.\par
Using the couplings of Table~\ref{tab:FDcouplings}, we can evaluate
the couplings of the vector mesons to the entire baryon octet. This
provides us with a unified description of the couplings in terms of an
arbitrary parameter $F$. These couplings are thus
\be
\begin{array}{c}
f_{N\rho} = F,\quad f_{\Lambda\rho} = 0,\quad f_{\Sigma\rho} =
2F,\quad f_{\Xi\rho} = F, \\[2mm] f_{N\w} = 3F,\quad f_{\Lambda\w} =
2F,\quad f_{\Sigma\w} = 2F,\quad f_{\Xi\w} = F.
\end{array}
\ee
We can normalize these results to the nucleon-$\w$ coupling, since we
will fit this parameter to saturation properties (refer to
Section~\ref{sec:EoS}). Thus the normalized couplings are
\be g_{Bm} = g_{N\w}\frac{f_{Bm}}{f_{N\w}}. \ee
We can then separate the meson couplings, since the the normalization
above results in the following relations, using isospin $I_B$, and
strangeness $S_B$ of baryon $B$;
\be g_{B\w} =\ \frac{(3-S_B)}{3}\ g_{N\w}, \quad g_{B\rho}
=\ \frac{2I_B}{3}\ g_{N\w}.  \ee
These results are consistent with a commonly used na\"ive assumption
that the $\w$ meson couples to the number of light quarks, and that
the $\rho$ meson couples to isospin. To emphasize the isospin symmetry
in our models, we will include the isospin as a factor in our
Lagrangian densities in the form of the $\vec{\tau}$ matrices. In
doing so, rather than having an independent coupling for each isospin
group, we will have a global coupling for the $\rho$ meson,
$g_{\rho}$. \par
\begin{table}[!t]
\centering
\caption[$F$-, $D$- and $S$-style baryon currents with mean-field
  assumptions]{\protect $F$-, $D$- and $S$-style Baryon currents with
  mean-field assumptions $V^\mu=(V^0,\vec{0})$, $q^\mu=(q^0,\vec{0})$,
  and $q^2=0$. \label{tab:currentassumptions}}
\vspace{3mm}
\begin{tabular}{lccc}
\hline 
\hline 
&&&\\[-4mm]
   &     $F$    &    $D$    &    $S$ \\
&&&\\[-4mm]
\hline
&&&\\[-2mm]
S  & $\frac{2}{3}\bar{\psi}\psi$ & 0 & $\frac{2}{3}\bar{\psi}\psi$ \\
&&&\\[-2mm]
V${}_1$ & $\frac{2}{3}\bar{\psi}\gamma_\mu\psi$ & 0 & $\frac{2}{3}\bar{\psi}\gamma_\mu\psi$ \\
&&&\\[-2mm]
V${}_2$ & 0 & 0 & 0 \\
&&&\\[-2mm]
V${}_3$ &  0  &  0  &  0  \\
&&&\\[-2mm]
A & $ \frac{4}{9}\bar{\psi}\gamma_5\gamma_\mu\psi $ &
    $ \frac{2}{3}\bar{\psi}\gamma_5\gamma_\mu\psi $    & 
    $ \frac{2}{9}\bar{\psi}\gamma_5\gamma_\mu\psi $ \\
&&&\\[-2mm]
P & $ \frac{4}{9}\bar{\psi}\gamma_5\psi $ &
    $ \frac{2}{3}\bar{\psi}\gamma_5\psi $    & 
    $ \frac{2}{9}\bar{\psi}\gamma_5\psi $ \\
&&&\\[-2mm]
\hline
\hline
\end{tabular}
\end{table}
Similarly to the above relations for the vector mesons, we have the
same relation for the scalar mesons; that the coupling is purely
$F$-style ($D=0$). Therefore the couplings for the scalar mesons are
the same as for the vector mesons, under the replacements $\w \to \s$,
$\vec{\rho} \to \vec{a}_0$. In the calculations that follow, we shall
further neglect the contributions from the scalar iso-vector
$\vec{a}_0$ due to their relatively large mass (refer to
Table~\ref{tab:particlesummary}).\par
%
As an alternative to the SU(6) relations for the $\rho$ meson coupling
$g_\rho$, we can use an experimental constraint. As we have shown
above, the $\rho$ meson couples to isospin, and as we will show in
Section~\ref{sec:qhd} the isospin density is proportional to the
asymmetry between members of an isospin group; for example the
asymmetry between protons and neutrons. This asymmetry is measured by
the symmetry energy $a_4 \equiv a_{\rm sym}$ (derived in
Appendix~\ref{sec:symenergy}) which appears in the semi-empirical mass
formula (the connection is derived in Appendix~\ref{sec:SEMF}) which
in the absence of charge symmetry is defined by Eq.~(\ref{eq:a4}).
%
%
The coupling of $\rho$ to the nucleons is found such that the
experimental value of the asymmetry energy of $a_{\rm sym} = 32.5~{\rm
  MeV}$ is reproduced at saturation. The coupling of $\rho$ to the
remaining baryons follows the relations above.\par
\cleardoublepage

%% file: Chapter3_models.tex
\chapter{Models Considered}\label{sec:models}
As outlined in the introduction, QCD is widely believed to be an
accurate description of strong-interaction particle physics. As a
non-perturbative theory, it cannot be solved analytically. In order to
make any predictions for this theory we must simulate the physics of
QCD, and we choose to do so using a model.\par
As with most fields of research, there are several choices for models
to investigate. The validity of these models must always be
challenged, and there is always a tendency to have a preference for a
particular model. Our goal is to work with a model which does not
introduce any physics that is not manifest in Nature, and does not
make any assumptions that cannot be verified. To this end, we begin
with a model called Quantum Hadrodynamics (QHD) which\emdash although
it is not a quark-level model\emdash simulates the fundamental
interactions of QCD with effective meson interactions.
\section{Quantum Hadrodynamics Model (QHD)}\label{sec:qhd}
The origins of the mean-field approximation and QHD reach back to the
non-relativistic work of Johnson and Teller~\cite{Johnson:1955zz},
which was reformulated by Duerr in a relativistic
model~\cite{Duerr:1956zz}. Once Chin and
Walecka~\cite{Walecka:1974,Chin:1974sa} successfully reproduced
saturation properties (refer to section~\ref{sec:EoS}), QHD as it is
known today was born. The formalism for QHD used for this work is
expertly detailed by Serot and Walecka~\cite{Serot:1984ey} and by
Furnstahl and Serot~\cite{Furnstahl:2000in}. QHD was the first great
step towards a particle-physics understanding of nuclear matter,
particularly in the form of neutron stars.\par
QHD is an effective\footnote{Not involving fundamental particles as
  the degree of freedom, but rather composite particles which provide
  a useful approximation to the physics.}, fully
relativistic\footnote{Making use of and obeying relativity, both
  special and general.} field theory which makes use of a mean-field
approximation (MFA, refer to Section~\ref{sec:mfa}) to describe Dirac
nucleons interacting at the quantum level. The gluons of QCD are
simulated by a delicate balance between attractive interactions of
scalar mesons, and repulsive interactions of vector mesons which, when
added together, produce an effect which approximates the strong
nuclear interaction; the interaction responsible for holding protons
and neutrons together inside a nucleus where the the Coulomb
interaction would otherwise cause the protons to repel and prevent any
nuclei from existing.\par
In the original formulation of QHD (later dubbed QHD-I), interactions
between the degenerate\footnote{In which the particles share a common
  mass, and thus satisfy charge symmetry.} iso-doublet of nucleons
(protons $p$, and neutrons $n$) involved the scalar-isoscalar $\s$ and
vector-isoscalar $\w$ mesons in the zero temperature limit. This was
soon expanded to QHD-II by inclusion of the uncharged vector-isovector
$\rho_0$ meson. We extend this to include the full baryon octet by
including the hyperons;
$\Lambda$,$~\Sigma^+$,$~\Sigma^0$,$~\Sigma^-$,$~\Xi^0$,
and$~\Xi^-$. For historical purposes, and to emphasise the effects of
each of these advances, when we present results we shall do so for the
most sophisticated version of a particular model, but also for
conditions corresponding to these variations for comparison.\par
We further extend this to use the SU(6) spin-flavor symmetry to relate
the couplings of all the baryons to all the mesons using the $F$-style
couplings as detailed in Section~\ref{sec:su6deriv}.\par
We make our model more physically realistic by using the physical
masses for the baryons, since charge symmetry is violated in
Nature. This provides a mass difference between the protons and
neutrons, and the model becomes sufficiently precise that we may
include leptons, which have a considerably smaller mass than the
baryons. We may then consider the effects of $\beta$-equilibrium
between species; by including leptons\footnote{We can safely neglect
  the contribution of $\tau$ leptons\emdash which have a mass in
  excess of 1776~MeV~\cite{Amsler:2008zzb} which makes them more
  massive than the $\Xi$ hyperons\emdash since their chemical
  potential would be equal to that of the electrons, and thus one
  would require an extraordinarily large electron contribution in
  order to provide a non-negligible $\tau$ Fermi momentum.} ($\ell \in
\{e^-,\mu^-\}$) we allow the possibility of considering a balance
between various charged species, although we are not explicitly
modelling charge-conserving interactions, as this would require the
inclusion of photon terms in the Lagrangian density, photons being the
mediators of the electromagnetic force. If we did not include the
leptons, we would not be able to consider globally charge neutral
nucleonic matter, as no negative charges would be available to balance
the positive charge of the protons.\par
We will use the term `configuration of a model' to indicate
differences (such as types of particles included or neglected) within
a particular model. The following discussion will focus on the most
sophisticated configuration\emdash the octet of baryons in
$\beta$-equilibrium with leptons\emdash and less sophisticated
configurations can be obtained by restricting this description. The
Lagrangian density that describes such a configuration of QHD is
\bea \nonumber {\cal L} &=& \sum_k \bar{\psi}_k
\left[\gamma_{\mu}(i\del^{\mu}-g_{k\w}\omega^{\mu}-
  g_{\rho}\vec{\tau}_{(k)}\cdot\vec{\rho}^{\, \mu})-(M_k-{g_{k\s}}
  \s)\right]\psi_k \\[1mm]
\nonumber && +
\frac{1}{2}(\del_{\mu}\sigma\del^{\mu}\sigma-m^{2}_{\s}\sigma^{2}) -
\frac{1}{4}F_{\mu\nu}F^{\mu\nu} - \frac{1}{4}R^a_{\mu\nu}R_a^{\mu\nu}
\\[2mm] \label{eq:QHDlag} && + \frac{1}{2}m^{2}_{\omega}\omega_{\mu}\omega^{\mu}
+\frac{1}{2}m^{2}_{\rho}\rho^a_{\mu}\rho_a^{\mu} +
\sum_\ell \bar{\psi}_\ell\left[i \gamma_\mu \del^\mu - m_\ell \right] \psi_\ell
+\delta\mathcal{L}, \eea
where the indices $k\in\{N,\Lambda,\Sigma,\Xi\}$ and
$\ell\in\{e^-,\mu^-\}$ represent the isospin group of the baryon
states and the lepton states, respectively, $\vec{\tau}_{(k)}$ are the
isospin matrices for each isospin group (refer to
Eq.~(\ref{eq:taus})), and $\psi_k$ corresponds to the Dirac spinors
for these isospin groups, i.e.
\be \psi_N = \begin{pmatrix}\psi_p\\\psi_n\end{pmatrix}, \quad
  \psi_\Lambda = \begin{pmatrix}\psi_\Lambda\end{pmatrix}, \quad
    \psi_\Sigma
    = \begin{pmatrix}\psi_{\Sigma^+}\\\psi_{\Sigma^0}\\\psi_{\Sigma^-}\end{pmatrix},
    \quad \psi_\Xi
    = \begin{pmatrix}\psi_{\Xi^0}\\\psi_{\Xi^-}\end{pmatrix}. \quad
    \ee
The vector field strength tensors are
\be \label{eq:vecfieldtensors} F^{\mu\nu} = \del^\mu\w^\nu -
\del^\nu\w^\mu, \quad R_a^{\mu\nu} = \del^\mu\rho_a^\nu -
\del^\nu\rho_a^\mu - g_{\rho}\epsilon_{abc} \rho_b^\mu\rho_c^\nu, \ee
$\psi_\ell$ is a spinor for the leptons, and $\delta{\cal L}$ are
renormalization terms. The values of the baryon and meson masses (in
vacuum, as used in the calculations herein) are summarized later in
Table~\ref{tab:masses}. We have neglected nonlinear meson terms in
this description for comparison purposes, though it has been shown
that the inclusion of non-linear scalar meson terms produces a
framework consistent with the QMC model without the added hyperfine
interaction~\cite{Muller:1997re} (see Sec.~\ref{sec:qmc}).\par
Assuming that the baryon density is sufficiently large, we use a
Mean-Field Approximation (MFA, as described in Section~\ref{sec:mfa})
with physical parameters (breaking charge symmetry) in which the meson
fields are replaced by their classical vacuum expectation values,
\mbox{$\alpha \to \bra\alpha\ket_{\rm classical}$}. With this
condition, the renormalization terms $\delta {\cal L}$ can be
neglected.\par
By enforcing rotational symmetry (refer to
Section~\ref{subsec:rotational}) and working in the frame where the
matter as a whole is at rest, we set all of the three-vector
components of the vector meson fields to zero, leaving only the
temporal components. Furthermore, we remove all charged meson states
as per the discussion in
Section~\ref{subsec:rotational}. Consequently, because the mean-fields
are constant, all meson derivative terms vanish, and thus so do the
vector field tensors. The only non-zero components of the vector meson
mean fields are then the temporal components, $\bra\w^\mu\ket =
\bra\w\ket\delta^{\mu 0}$ and $\bra\vec{\rho\, }^\mu\ket =
\bra\vec{\rho\, }\ket\delta^{\mu 0}$. Similarly, only the third
isospin component of the $\rho$ meson mean-field is non-zero,
corresponding to the uncharged $\rho_0$ meson.\par
The couplings of the mesons to the baryons are found via SU(6)
spin-flavor symmetry~\cite{Rijken:1998yy}. This produces the following
relations for the $\s$ and $\w$ couplings to each isospin group (and
hence each baryon $B$ in that isospin group) as per
Section~\ref{sec:su6deriv}:
\be
\reci{3}\; g_{N\s} = \reci{2}\; g_{\Lambda\s} = \reci{2}\; g_{\Sigma\s} = g_{\Xi\s},
\quad
\reci{3}\; g_{N\w} = \reci{2}\; g_{\Lambda\w} = \reci{2}\; g_{\Sigma\w} = g_{\Xi\w}.
\ee
Using the formalism of Eq.~(\ref{eq:QHDlag}) with isospin expressed
explicitly in the Lagrangian density, the couplings of the $\rho$
meson to the octet baryons are unified, and thus we can calculate the
coupling of any baryon to either the $\s$, $\w$ or $\rho$ meson.\par
By evaluating the equations of motion from the Euler--Lagrange
equations,
\be \label{eq:EL}
\frac{\del\mathcal{L}}{\del\phi_i} 
- \del_{\mu}\frac{\del\mathcal{L}}{\del(\del_{\mu}\phi_i)} = 0,
\ee
we find the mean-field equations for each of the mesons, as well as
the baryons. Prior to applying the MFA for the mesons, the $\s$
equation of motion produces a Klein--Gordon equation, while the $\w$
and $\rho$ equations of motion produce Maxwell equations. This is by
construction, and these terms can be found in
Appendix~\ref{sec:qhdderiv}. Returning to the use of the MFA, the
equations of motion for the meson fields are
\bea \label{eq:MFsigma} \bra\s\ket &=& \sum_B
\frac{g_{B\s}}{m^{2}_{\s}}\bra\bar{\psi}_B\psi_B\ket, \\
\label{eq:MFomega}
\bra\w\ket &=& \sum_B \frac{g_{B\w}}{m^{2}_\w}\bra\bar{\psi}_B\gamma^{0}\psi_B\ket = \sum_B
\frac{g_{B\w}}{m^{2}_{\w}}\bra\psi_B^\dag \psi_B\ket, \\
\label{eq:MFrho}
\bra\rho\ket &=& \sum_k
\frac{g_{\rho}}{m^{2}_{\rho}}\bra\bar{\psi}_k\gamma^{0} \tau_{(k)3}
\psi_k\ket = \sum_k \frac{g_{\rho}}{m^{2}_{\rho}}\bra\psi_k^\dag
\tau_{(k)3} \psi_k\ket = \sum_B
\frac{g_{\rho}}{m^{2}_{\rho}}\bra\psi_B^\dag I_{3B} \psi_B\ket , \eea
where the sum over $B$ corresponds to the sum over the octet of
baryons, and the sum over $k$ corresponds to the sum over isospin
groups. $I_{3B}$ is the third component of the isospin of baryon $B$,
as found in the diagonal elements of $\tau_{(k)3}$ in
Eq.~(\ref{eq:taus}). $\bra\omega\ket$, $\bra\rho\ket$, and
$\bra\sigma\ket$ are proportional to the conserved baryon density,
isospin density and scalar density respectively, where the scalar
density is calculated self-consistently.\par
\begin{figure}[!b]
\centering
\includegraphics[angle=90,width=0.9\textwidth]{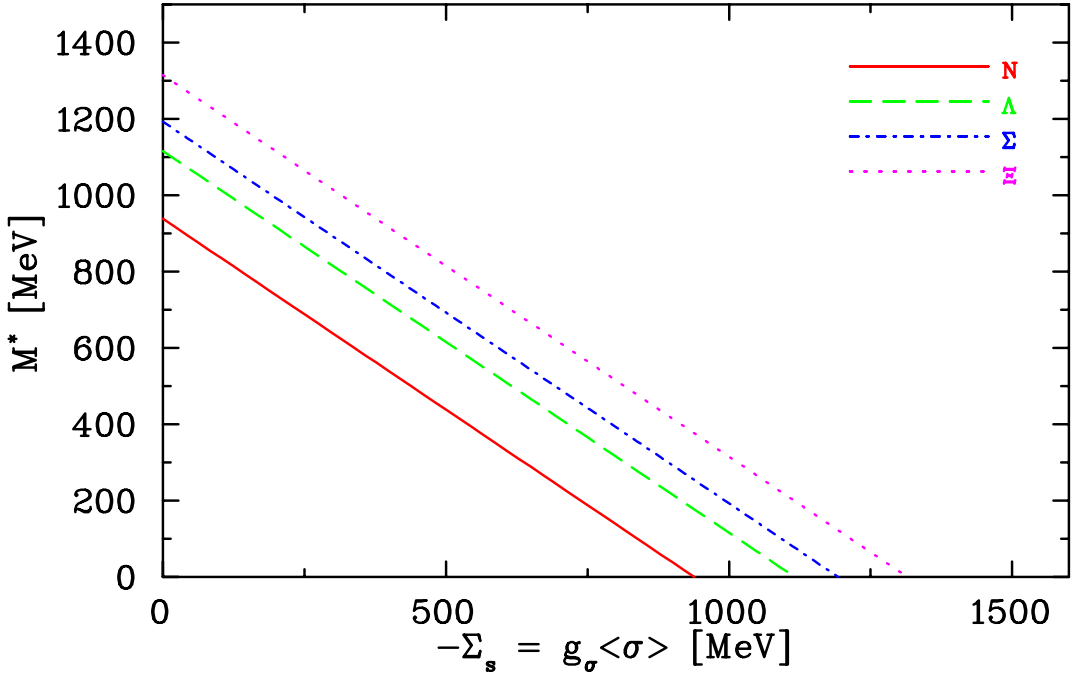}
\caption[Octet baryon effective masses in QHD]{(Color Online) Baryon
  effective masses $M^*$ as defined by the scalar self-energy
  $\Sigma^s$ for QHD. Note that it becomes possible that at some
  densities these effective masses may become negative. Negative
  effective masses in the Dirac equation imply the presence of
  antibaryons, which this model has neglected, so at this point the
  model becomes unreliable.  \protect\label{fig:Mstar_QHD}}
\end{figure}

The Euler--Lagrange equations also provide a Dirac equation for the
baryons
\be \label{eq:dirac} \sum_B
\left[i\!\!\not\!\partial-g_{B\w}\gamma^0\bra\w\ket-
  g_{\rho}\gamma^0I_{3B}\bra\rho\ket - M_B + g_{B\s} \bra\s\ket
  \right]\psi_B = 0,  \ee
in which we have inserted the expressions for the self-energies in
QHD. The effective mass (as defined by the scalar self-energy inserted
above, as per Eq.~(\ref{eq:effM})) is given by
\be \label{eq:QHDeffM} M_B^* = M_B + \Sigma^s_B = M_B - g_{B\s}
\bra\s\ket, \ee
and the evaluation of this is shown in Fig.~\ref{fig:Mstar_QHD} for
the octet of baryons. We note here that this definition of the
effective mass\emdash being linear in the scalar field\emdash has the
possibility of being negative for particular values of
$\bra\s\ket$. We will discuss this issue (and a remedy to it) in
more detail later, but for now we shall acknowledge that this
definition of the effective mass includes only the first term of many
as a linear approximation.\par
The baryon chemical potential (Fermi energy) is defined here as the
energy associated with the Dirac equation Eq.~(\ref{eq:dirac}), which
involves the above self-energies as
\be \label{eq:mu} \mu_B = \epsilon_{F_B} = \sqrt{k_{F_B}^2 +
  (M_B^*)^2}+g_{B\w}\bra\w\ket + g_{\rho}I_{3B}\bra\rho\ket.  \ee
The chemical potentials for the leptons are simply
\be \label{eq:ellmu} \mu_\ell = \sqrt{k_{F_\ell}^2 + m_\ell^2}, \ee
since the leptons do not interact with the mesons.\par

The energy density ${\cal E}$ and pressure $P$ for the EOS can be
obtained using the relations for the energy-momentum tensor for a
perfect fluid at rest, where $u^\mu$ is the four-velocity
\be \label{eq:SET} \bra T^{\mu \nu}\ket = \left({\cal
  E}+P\right)u^{\mu}u^{\nu} + P {\rm g}^{\mu\nu},\quad \Rightarrow
\quad P = \reci{3} \bra T^{ii} \ket,\quad {\cal E} = \bra T^{00} \ket,
\ee
since $u^i=0$ and $u_0 u^0=-1$, where ${\rm g}^{\mu\nu}$ here is the
inverse metric tensor having a negative temporal component, ${\rm g} =
{\rm diag}(-1,+1,+1,+1)$, in contrast to the Minkowski metric of
Eq.~(\ref{eq:cliff}). In accordance with Noether's Theorem, the
relation between the energy-momentum tensor and the Lagrangian density
is
\be \label{eq:EML} T^{\mu \nu} = -{\rm g}^{\mu
  \nu}\mathcal{L}+\del^{\mu}\psi \frac{\del \mathcal{L}}{\del
  (\del_{\nu}\psi)}.  \ee
We find the energy density and pressure for QHD to be the following
sum of contributions from baryons $B$, leptons $\ell$, and mesons $m$
to be
\bea \label{eq:E_H} \nonumber {\cal E} &=& \sum_{j=B,\ell,m} {\cal
  E}_j \\
&=& \sum_{i=B,\ell} \frac{\left(2J_i+1\right)}{(2 \pi)^3} \int \theta
(k_{F_i} - |\vec{k}|) \sqrt{\vec{k}^2+(M_i^*)^2} \; d^3k +
\sum_{\alpha = \s,\w,\rho} \reci{2} m_\alpha^2\bra\alpha\ket^2,
\\
\nonumber
P &=& \sum_{j=B,\ell,m} P_j \\
&=& \sum_{i=B,\ell} \frac{\left(2J_i+1\right)}{3(2 \pi)^3} \int
\frac{\vec{k}^2 \; \theta (k_{F_i} -
  |\vec{k}|)}{\sqrt{\vec{k}^2+(M_i^*)^2}} \; d^3k + \sum_{\alpha =
  \w,\rho} \reci{2} m_\alpha^2\bra\alpha\ket^2 - \reci{2}
m_\s^2\bra\s\ket^2,
\label{eq:P_H}
\eea
where $J_i$ is the spin of particle $i$ ($J_i=\frac{1}{2}\ \forall\ i
\in \{B,\ell\}$) which in this case accounts for the availability of
both up and down spin-states, and $\theta$ is the Heaviside Step
Function (see Eq.~(\ref{eq:Heaviside})). Note that the pressure
arising from the vector mesons is positive, while it is negative for
the scalar meson. For a full derivation of these terms, refer to
Appendix~\ref{sec:qhdderiv}.\par
The expression for the self-consistent scalar field $\bra\s\ket$ is
determined by the derivative of the energy density with respect to the
effective mass. In the case of QHD this expression is given by
\be \label{eq:QHDscalarfield} \bra\s\ket = \sum_B
\frac{g_{B\s}}{m_\s^2} \frac{\left(2J_B+1\right)}{(2\pi)^3} \int
\frac{M_B^*\; \theta(k_{F_B} - |\vec{k}|)}{\sqrt{\vec{k}^2 +
    (M_B^*)^2}} \; d^3k , \ee
which is solved self-consistently since the effective mass (appearing in the
integral) is defined in terms of this quantity, as per
Eq.~(\ref{eq:QHDeffM}).\par
The couplings $g_{N\s}$ and $g_{N\w}$ are determined such that
symmetric nuclear matter (in which $\rho_p = \rho_n = 0.5\rho_{\rm
  total}$) saturates with the appropriate minimum in the (binding)
energy per baryon, as per Section~\ref{sec:EoS}. The couplings for QHD
which provide a fit to saturated nuclear matter are shown in
Table~\ref{tab:couplings}.\par
The EOS for QHD can be obtained by finding solutions to
Eqs.~(\ref{eq:MFsigma})--(\ref{eq:MFrho}) subject to
charge-neutrality, conservation of a chosen total baryon number, and
equivalence of chemical potentials. These conditions can be summarized
as
\be 
\label{eq:equilconds}
\left.
\begin{array}{rcl}
0 &=& \sum_i Q_i \rho_i\\[2mm]
\rho &=& \sum_i B_i \rho_i\\[2mm] 
\mu_i &=& B_i \mu_n - Q_i \mu_e
\end{array}
\quad \right\} \quad i \in \{
p,n,\Lambda,\Sigma^+,\Sigma^0,\Sigma^-,\Xi^0,\Xi^-,e^-,\mu^- \}.  \ee
Once these equations are solved, the energy density and pressure can
be calculated.\par
It should be noted that, as with many relativistic models for baryonic
matter, once we include more than one species of baryon this model
eventually predicts the production of baryons with negative effective
masses at sufficiently high densities ($\rho > 1~{\rm fm}^{-3}$). This
is a direct result of the linear nature of the effective mass as shown
in Eq.~(\ref{eq:QHDeffM}). As the Fermi energy (see Eq.~(\ref{eq:mu}))
approaches zero, the cost associated with producing baryon-antibaryon
pairs is reduced and at this point the model breaks down. From a more
physical point of view, as the density rises one would expect that the
internal structure of the baryons should play a role in the
dynamics. Indeed, within the QMC model, the response of the internal
structure of the baryons to the applied mean-scalar-field ensures that
no baryon mass ever becomes negative.\par

\section{Quark-Meson Coupling Model (QMC)}\label{sec:qmc}
Up to this point we have only considered QHD as an effective field
theory; we have considered baryons as the fundamental degrees of
freedom for this scale. It may however be the case that further
internal degrees of freedom are more significant at high-densities. We
therefore wish to extend our model to include the effect of baryon
structure in the form of quarks\footnote{Some portions of this section
  are adapted from Carroll {\it et. al.}~\cite{Carroll:2008sv}.}.\par
Deep inelastic scattering experiments have shown that nucleons do
indeed have internal structure~\cite{Bloom:1969kc}, and few would
discount quarks as the fundamental particles involved. To include this
degree of freedom, one needs to solve the boundary condition equations
for the Dirac particles and include the energy contribution from a
`{\em bag}' similar to the MIT bag model parameter, as will be
shown.\par
%
%
%
%
The work here will focus on the latest development of the QMC
model~\cite{RikovskaStone:2006ta} which includes a quadratic term with
numerical factor $d$ in the effective mass which accounts for the
scalar polarizability. Like QHD, QMC is a relativistic quantum field
theory formulated in terms of the exchange of scalar and vector
mesons. However, in contrast with QHD these mesons couple not to
structureless baryons but to clusters of confined quarks. As the
density of the medium grows and the mean-scalar and mean-vector fields
grow, the structure of the clusters adjusts self-consistently in
response to the mean-field coupling.\par
While such a model would be extremely complicated to solve in general,
it has been shown by Guichon {\it et al.}~\cite{Guichon:1995ue} that
in finite nuclei one should expect the Born--Oppenheimer approximation
(in which we are able to isolate and distinguish the effects of the
quarks, in the same way that one is able to separate the net effects
of electrons in an atomic calculation) to be good at the 3\% level. Of
course, in nuclear matter it is exact at mean-field level as a result
of the constant meson fields.\par
Within the Born--Oppenheimer approximation, the major effect of
including the structure of the baryon is that the internal quark wave
functions respond in a way that opposes the applied scalar field. To a
very good approximation this physics is described through the `scalar
polarizability' $d$, which in analogy with the electric
polarizability\footnote{In QED, the expression for the energy shift
  due to the quadratic Stark effect is $\Delta E = -\half \alpha
  |E_z|^2$, in which $\alpha$ is the electric polarizability and $E_z$
  is the external electric field, taken to point along the
  $\hat{z}$-axis~\cite{Friedrich}.}, describes the term in the baryon
effective mass quadratic in the applied scalar
field~\cite{Guichon:1987jp,Thomas:2004iw,Ericson:2008tv,Massot:2008pf,Chanfray:2003rs}.
Recent explicit calculations of the equivalent energy functional for
the QMC model have demonstrated the very natural link between the
existence of the scalar polarizability and the many-body forces, or
equivalently the density dependence, associated with successful,
phenomenological forces of the Skyrme
type~\cite{Guichon:2004xg,Guichon:2006er}. In nuclear matter, the
scalar polarizability is the {\it only} effect of the internal
structure in the mean-field approximation. On the other hand, in
finite nuclei the variation of the vector field across the hadronic
volume also leads to a spin-orbit term in the nucleon
energy~\cite{Guichon:1995ue}.\par
Once one chooses a quark model for the baryons and specifies the quark
level meson couplings, there are no new parameters associated with
introducing any species of baryon into the nuclear matter. Given the
well known lack of experimental constraints on the forces between
nucleons and hyperons (let alone hyperons and hyperons) which will be
of great practical importance as the nuclear density rises above
(2--3)$\rho_0$, this is a particularly attractive feature of the QMC
approach and it is crucial for our current investigation. Indeed, we
point to the very exciting recent results~\cite{Guichon:2008zz} of the
QMC model\emdash modified to include the effect of the scalar field on
the hyperfine interaction\emdash which led to $\Lambda$ hypernuclei
being bound in quite good agreement with experiment and $\Sigma$
hypernuclei being unbound because of the modification of the hyperfine
interaction, thus yielding a very natural explanation of this observed
fact. We note the success that this description has generated for
finite nuclei as observed in Ref.~\cite{Guichon:2006er}.\par
While we will use the QMC model for our considerations of baryon
structure here, we note that there has been a parallel
development~\cite{Bentz:2001vc} based upon the covariant, chiral
symmetric NJL model~\cite{Nambu:1961tp}, with quark confinement
modelled using the proper time regularization proposed by the
T\"ubingen group~\cite{Hellstern:1997nv,Ebert:1996vx}. The latter
model has many advantages for the computation of the medium
modification of form factors and structure functions, with the results
for spin structure functions~\cite{Cloet:2005rt,Cloet:2006bq} offering
a unique opportunity to test the fundamental idea of the QMC model
experimentally. However, in both models it is the effect of quark
confinement that leads to a positive polarizability and a natural
saturation mechanism.\par
Comparisons between the QHD and QMC derivations {\it a posteriori}
reveal that although the underlying physics of QHD and QMC is rather
different, at the hadronic level the equations to be solved are very
similar. Full discussions and derivations can be found in
Refs.~\cite{Guichon:1987jp,Saito:1996yb,Saito:2005rv,Muller:1997re}. We
shall rather focus on the changes to QHD which are required to produce
the QMC model: \par
\vspace{3mm}
{\bf 1.} Because of the scalar polarizability of the hadrons,
which accounts for the self-consistent response of the internal quark
structure of the baryon to the applied scalar
field~\cite{Guichon:2006er}, the effective masses appearing in QMC are
non-linear in the mean-scalar field. We write them in the general form
\be \label{eq:effMQMC}
M_B^* = M_B - w_B^\s \; g_{N\s} \bra\s\ket 
+ \frac{d}{2} \tilde{w}_B^\s \; (g_{N\s}\bra\s\ket)^2
\, , 
\ee
where the weightings $w_B^\s,\ \tilde{w}_B^\s,$ and the scalar
polarizability of the nucleon $d$, must be calculated from the
underlying quark model. Note now that only the coupling to the
nucleons $g_{N\s}$, is required to determine all the effective
masses.\par
The most recent calculation of these effective masses, including the
in-medium dependence of the spin dependent hyperfine
interaction~\cite{Guichon:2008zz}, yields the explicit expressions:
\be \label{eq:MstarsinQMC1}
M_{N}(\bra\s\ket) = M_{N}-g_{N\s}\bra\s\ket
+\left[0.0022+0.1055R_{N}^{\rm free}-
0.0178\left(R_{N}^{\rm free}\right)^{2}\right]
\left(g_{N\s}\bra\s\ket\right)^{2},
\ee
\bea
M_{\Lambda}(\bra\s\ket) &=& M_{\Lambda}-\left[0.6672+0.0462R_{N}^{\rm free}-
0.0021\left(R_{N}^{\rm free}\right)^{2}\right]g_{N\s}\bra\s\ket
\nonumber \\ 
 &  & +\left[0.0016+0.0686R_{N}^{\rm free}-0.0084\left(R_{N}^{\rm free}\right)^{2}
\right]\left(g_{N\s}\bra\s\ket\right)^{2},
\eea
\bea
\nonumber
M_{\Sigma}(\bra\s\ket) &=& M_{\Sigma}-\left[0.6706-0.0638R_{N}^{\rm free}-
0.008\left(R_{N}^{\rm free}\right)^{2}\right]g_{N\s}\bra\s\ket
 \\
 &  & +\left[-0.0007+0.0786R_{N}^{\rm free}-0.0181\left(R_{N}^{\rm free}\right)^{2}
\right]\left(g_{N\s}\bra\s\ket\right)^{2},
\eea
\bea
\label{eq:MstarsinQMC4}
M_{\Xi}(\bra\s\ket) &= & M_{\Xi}-\left[0.3395+0.02822R_{N}^{\rm free}-
0.0128\left(R_{N}^{\rm free}\right)^{2}\right]g_{N\s}\bra\s\ket
\nonumber \\
 &  & +\left[-0.0014+0.0416R_{N}^{\rm free}-0.0061\left(R_{N}^{\rm free}\right)^{2}
\right]\left(g_{N\s}\bra\s\ket \right)^{2}\, . 
\eea
We take $R_N^{\rm free}=0.8~{\rm fm}$ as the preferred value of the
free nucleon radius, although in practice the numerical results depend
only very weakly on this parameter~\cite{Guichon:2006er}.\par
\begin{figure}[!b]
\centering
\includegraphics[angle=90,width=0.8\textwidth]{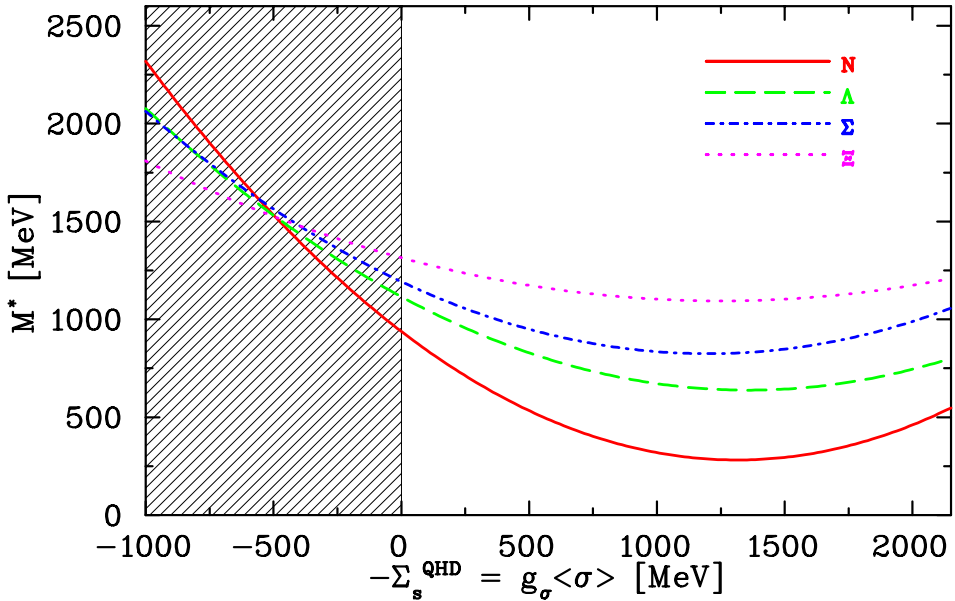}
\caption[Octet baryon effective masses in QMC]{(Color Online) Baryon
  effective masses for QMC as a function of the QHD (linear) scalar
  self-energies. The values at $\Sigma^{\rm QHD}_s=0$ are the vacuum
  masses as found in Table~\ref{tab:masses}. In order to emphasise the
  curve we have shown the effective masses beyond $-\Sigma^{\rm QHD}_s
  = 2000~{\rm MeV}$, though in practice we only require values up to
  $-\Sigma^{\rm QHD}_s = 800~{\rm MeV}$ which corresponds to densities
  of $\sim 2~{\rm fm}^{-3}$ (6--8 $\rho_0$), beyond which higher order
  terms not shown in Eq.~(\ref{eq:effMQMC}) become significant. The
  (shaded) unphysical region of $\Sigma^{\rm QHD}_s > 0$ contains a
  point for which all of the effective masses become unified; this is
  the point at which SU(3) symmetry is an accurate
  symmetry. \protect\label{fig:effMvsS}}
\end{figure}

Given the parameters in
Eqs.~(\ref{eq:MstarsinQMC1})--(\ref{eq:MstarsinQMC4}), all the
effective masses for the baryon octet are entirely determined. They
are plotted as functions of the Hartree scalar self-energy $\Sigma^s =
-g_{N\s}\bra\s\ket$ in Fig.~\ref{fig:effMvsS} and we see clearly that
they never become negative (note that the range of $\Sigma^s$ covered
here corresponds to densities well above (6--8)$\rho_0$ in QMC).\par
\vfill
%
%
\clearpage
{\bf 2.} The mean-scalar field $\bra\s\ket$ is derived
self-consistently by taking the derivative of the energy density with
respect to $\bra\s\ket$, thus the scalar field equation
\be \label{eq:scalarfield} \bra\s\ket = \sum_B \frac{g_{N\s}}{m_\s^2}
C(\bra\s\ket) \frac{\left(2J_B+1\right)}{(2\pi)^3} \int \frac{M_B^*\;
  \theta(k_{F_B} - |\vec{k}|)}{\sqrt{\vec{k}^2 + (M_B^*)^2}} \; d^3k ,
\ee
has an extra factor when compared to Eq.~(\ref{eq:QHDscalarfield}),
denoted by
\be \label{eq:Csigma} C(\bra\s\ket) = w_B^\s - \tilde{w}_B^\s d
g_{N\s}\bra\s\ket.  \ee
Note that the the scalar polarizability $d$ term in $C(\bra\s\ket)$
does not have the factor of $\reci{2}$ that is found
Eq.~(\ref{eq:effMQMC}), because of the differentiation of the squared
term.\par
Given this new term in the equation for the mean-scalar field, we can
see that this allows feedback of the scalar field which is modelling
the internal degrees of freedom of the baryons. This feedback
prevents certain values of $\bra\s\ket$ from being accessed.\par
\vspace{3mm}
{\bf 3.} The couplings to the nucleons are re-determined by the fit to
saturation properties (as per Section~\ref{sec:EoS}) with the new
effective masses for the proton and neutron. The couplings for QMC
which provide a fit to saturated nuclear matter are shown in
Table~\ref{tab:couplings}.\par
Given these changes alone, QHD is transformed into QMC. The
implications of these changes however will be profound.\par
%

\section{MIT Bag Model}\label{sec:MITbag}
We consider two models for a deconfined quark matter phase, both of
which model free quarks. The first model, the MIT bag
model~\cite{Chodos:1974je}, is commonly used to describe quark matter
(and, by extension, hadronic matter) because of its simplicity.\par
In this model we construct baryons as three quarks confined within a
`bag,' a region defined by the local energy density which is greater
than the energy density of the surrounding vacuum by a factor of
$B$\emdash which is commonly quoted to a power of one-quarter, thus
$B^{1/4} \sim 180~{\rm MeV}$\emdash and which is lower in pressure by
the same factor. The motivation for this is that the quarks reside in
a high-energy, low pressure region which confines them.\par
This is the observed property of confinement\emdash that no quarks
have ever been observed, either directly or indirectly, in
isolation\emdash that perturbative approaches to QCD are unable to
reproduce. This property is only of concern for hadronic matter
though; at high energies, QCD predicts that the quarks become
`asymptotically free', and at this point we can consider quark
matter.\par
The simplest Lagrangian density that can be constructed for such a
scenario is
\be {\cal L}_{\rm MIT} = \bar{\psi}_q \left( i\delslash - m_q
\right)\psi_q + B, \ee
where $\psi_q$ is a spinor for quarks with mass $m_q$ where we
typically use masses of $m_u = 3~{\rm MeV}$, $m_d = 7~{\rm MeV}$, and
$m_s = 95~{\rm MeV}$ for the up, down, and strange quarks respectively,
and $B$ is the aforementioned bag energy density. In a similar fashion
to the derivations for QHD and QMC, the energy density and pressure
can be calculated to be
\be {\cal E} = B + \sum_q \frac{(2J_q + 1){\cal N}_c}{(2\pi)^3}\int
\theta(k_{F_q} - |\vec{k}|)\sqrt{\vec{k}^2+(m_q)^2}\; d^3k\ , \ee
\be P = -B + \sum_q \frac{(2J_q + 1){\cal N}_c}{3(2\pi)^3}\int
\frac{\vec{k}^2 \; \theta(k_{F_q} -
  |\vec{k}|)}{\sqrt{\vec{k}^2+(m_q)^2}}\; d^3k\ .  \ee
\par
In order to utilise statistical mechanics and explore phase
transitions involving this scenario, we model point objects with fixed
masses which possess chemical potentials related to the independent
chemical potentials of Eq.~(\ref{eq:chempotrel}) via
\be \label{eq:qchempot}
\mu_u = \reci{3}\mu_n - \frac{2}{3}\mu_e,\qquad
\mu_d = \reci{3}\mu_n + \frac{1}{3}\mu_e,\qquad
\mu_s = \mu_d,
\ee
where quarks have a baryon charge of $\reci{3}$ since baryons contain
3 quarks. Because the current quarks do not interact with mesons in
this model, the quark chemical potential has the same form as the
lepton chemical potential (no meson terms) and thus
\be \label{eq:quarkmu}
\mu_q = \sqrt{k_{F_q}^2 + m_q^2}\ ; \quad q\in\{u,d,s\}.
\ee
The EOS can therefore be solved under the conditions of
Eq.~(\ref{eq:equilconds}). In our calculations of this model, we use
the current quark masses as found in Ref.~\cite{Amsler:2008zzb}, which
represent the physical, bare quark masses.\par

\section{Nambu--Jona-Lasinio Model (NJL)}\label{sec:njl}
As an alternative model for deconfined quark matter, we consider the
Nambu--Jona-Lasinio (NJL) model~\cite{Nambu:1961tp}, in which the
quarks have dynamically generated masses, ranging from constituent
quark masses at low densities to current quark masses at high
densities.\par
Dynamical breaking of the chiral symmetry produces Nambu--Goldstone
bosons\emdash the triplet of pions in the two-flavor case\emdash which
are massless while the symmetry is preserved, but which possess a
non-zero mass when the symmetry is broken. According to the
Gell-Mann--Oaks--Renner relation (for example, see
Ref.~\cite{Weinberg:1977hb}), in the leading order of the chiral
expansion,
\be m_\pi^2 = \reci{f_\pi^2}\left(m_u \bra\bar{u}u\ket + m_d
\bra\bar{d}d\ket\right), \ee
or more commonly reduced to the fact that the quark mass scales as the
square of the pion mass,
\be m_q\sim m_\pi^2. \ee
\par
At large densities, manifest chiral symmetry is expected to be
partially restored. If it was fully restored, the quarks would be
massless (in which case, $m_\pi = m_q = 0$). The symmetry is however
not exact, and the quarks retain a very small mass. This small mass is
the current quark mass as noted in Ref.~\cite{Amsler:2008zzb}, whereas
the dynamically generated quark masses under broken chiral symmetry
are the constituent masses that together na\"ively sum to the mass of
a baryon. By using the NJL model rather than a simple treatment for
the quark masses we endeavor to make our phase transition models more
realistic and more sophisticated by including more physics believed to
represent Nature. The NJL model is a simple construction that displays
the correct phenomenology, namely DCSB.\par
If we consider a particular choice for a massless Lagrangian
density\footnote{Following the considerations of Nambu and
  Jona-Lasinio~\cite{Nambu:1961tp}.} with strong coupling $G$, in a
Lorentz-covariant frame (with no matter or background fields) to be
\be \label{eq:NJLlag} {\cal L} = - \bar{\psi}i\delslash\psi +
G\left[(\bar{\psi}\psi)^2 - (\bar{\psi}\gamma_5\psi)^2\right] \ee
we can see that this is invariant under the vector and axial vector
symmetries of Eqs.~(\ref{eq:U1V})--(\ref{eq:U1A}). For example,
\bea \nonumber (\bar{\psi}\psi)^2 = \bar{\psi}\psi\bar{\psi}\psi
\stackrel{{\rm U}(1)_A}{\longrightarrow}
\bar{\psi}e^{2i\alpha_A\gamma_5}\psi\ \bar{\psi}e^{2i\alpha_A\gamma_5}\psi
&=& \bar{\psi}\psi e^{2i\alpha_A\gamma_5}
\ e^{-2i\alpha_A\gamma_5}\bar{\psi}\psi \\
&=& (\bar{\psi}\psi)^2, 
\eea
where we recall Eq.~(\ref{eq:anticommutation}), and that $\bar{\psi} =
\psi^\dagger \gamma_0$.\par
\begin{figure}[!h]
\centering
\includegraphics[width=0.4\textwidth]{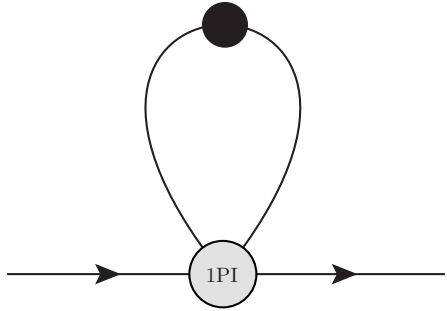}
\caption[NJL self-energy]{Feynman diagram for the NJL quark
  self-energy, where the four-fermion vertex is 1-PI, and the loop
  corresponds to the full quark
  propagator.\protect\label{fig:NJLfullselfenergy}}
\end{figure}

The self-energy of this particular model is shown in
Fig.~\ref{fig:NJLfullselfenergy} in which the four-fermion vertex is
the 1-PI vertex (refer to Section~\ref{sec:dynamicchiral}), and the
loop is the full quark propagator.\par
\begin{figure}[!t]
\centering
\includegraphics[width=0.4\textwidth]{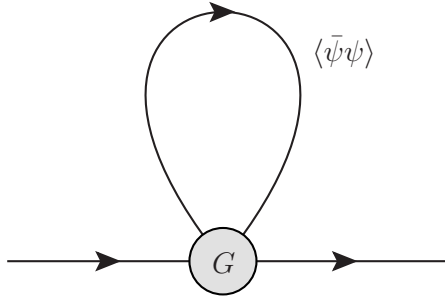}
\caption[NJL self-energy (approximated)]{Feynman diagram for the NJL quark
  self-energy, where the four-fermion vertex is the bare vertex with
  coupling $G$, and the loop corresponds to quark
  condensate.\protect\label{fig:NJLselfenergy}}
\end{figure}

If we then approximate the four-fermion vertex to be the bare vertex
with coupling $G$, and the loop to correspond to the bare propagator,
then the loop now corresponds to the quark condensate
$\bra\bar{\psi}\psi\ket$ and the self-energy is given by the Feynman
diagram as shown in Fig.~\ref{fig:NJLselfenergy}. The Lagrangian
density for NJL then becomes
\be {\cal L} = - \bar{\psi}\left(i\delslash +
2G\bra\bar{\psi}\psi\ket\right)\psi = - \bar{\psi}\left(i\delslash +
\Sigma_s\right)\psi.  \ee
where we can see from Fig.~\ref{fig:NJLselfenergy} that the
self-energy is a Dirac-scalar term. We can now identify the dynamical
quark mass as a scalar self-energy term in this Lagrangian
density.\par
If we include a further constant mass term $\bar{\psi}m_0\psi$ in
Eq.~(\ref{eq:NJLlag}) that explicitly breaks the chiral symmetry, we
can define the effective (dynamic) quark mass as
\be
\label{eq:dynamicalqmass}
m^*_q = m_0 + \Sigma_s = m_0 - 2G\bra\bar{\psi}\psi\ket.  \ee
This is called the `gap equation' in analogy to
superconductivity\footnote{The BCS theory of
  superconductivity~\cite{Bardeen:1957mv} is a primary motivation for
  this model, in which electrons in a metal may become a paired
  bosonic state\emdash a Cooper pair\emdash by possessing a lower
  energy than the Fermi energy. In that case, a temperature-dependent
  energy gap exists, and electron excitations must be of a minimum
  energy, as opposed to the continuous spectrum that the electrons
  would normally have.}.\par
\vfill
To connect with Section~\ref{sec:dynamicchiral}, we can understand
the NJL self-energy of Fig.~\ref{fig:NJLselfenergy} in analogy to that
of Fig.~\ref{fig:quarkDSE} in the (albeit, unphysical) case that we
shrink the gluon loop in Fig.~\ref{fig:quarkDSE} to a point, and in
which case the Lagrangian densities are related via the (arbitrary)
replacement of
\be
\label{eq:ctfl}
(\bar{\psi}\Gamma\psi)^2 \to
2\bar{\psi}\Gamma\psi\; \bra\bar{\psi}\Gamma\psi\ket, \ee
where $\Gamma$ is any gamma matrix appearing in the interaction
Lagrangian density. The term $\bra\bar{\psi}\Gamma\psi\ket$ denotes
the ground-state (vacuum) expectation value of $\bar{\psi}\Gamma\psi$,
and as noted in Section~\ref{subsec:parity}, the ground-state is a
parity eigenstate, thus the only term in Eq.~(\ref{eq:NJLlag}) that
has a non-zero vacuum expectation value is
$\bra\bar{\psi}\psi\ket$.\par
The expression for the Feynman diagram in Fig.~\ref{fig:NJLselfenergy}
is simply
\be \label{eq:LCSF} \Sigma(p) = 2G\bra\bar{\psi}\psi\ket = 2iG\, {\rm Tr}S_F(0) \ee
where $S_F(0) = S_F(x-y)\delta(x-y)$ is the Feynman propagator for a
quark (refer to Appendix~\ref{sec:propagators}) which starts and ends
at the same space-time point.\par
As with Section~\ref{sec:dynamicchiral}, we have described the
above model for a Lorentz-covariant frame, in which there are no matter
or background fields. In this case, the Feynman propagator in
Eq.~(\ref{eq:LCSF}) is the free-space propagator. We make a further
approximation by replacing this propagator with the in-medium
propagator, one which is in the presence of matter fields, and thus we
introduce a Fermi momentum. The expression for the quark condensate in
this case becomes
\be \label{eq:condensate} \bra\bar{\psi}_q\psi_q\ket = i {\rm Tr}
S_F(0) = -4\ \frac{{\cal N}_c}{(2\pi^3)} \int \frac{m_q^*\;
  \theta(k_{F}-|\vec{k}|)\; \theta(\Lambda -
  k_F)}{\sqrt{\vec{k}^2+(m_q^*)^2}}\; d^3k, \ee
where we have introduced a momentum cutoff of $k_F < \Lambda$ to
regularize this integral, at which point we expect to recover current
quark masses, and ${\cal N}_c$ is the number of color degrees of
freedom of quarks.
In order to calculate the effective quark mass at each density, we
must first find the coupling $G$ which yields the appropriate
constituent quark masses in free space ($k_F=0$). The coupling is
assumed to remain constant as the density rises. Given a (free-space)
constituent quark mass $m_q^{\rm free}$, we can solve
Eqs.~(\ref{eq:dynamicalqmass}) and (\ref{eq:condensate}) to find the
coupling
\be \label{eq:qcoupling} G = \frac{(m_q^{\rm free}-m_0)}{4\ }
\left[\frac{{\cal N}_c}{(2\pi)^3} \int \frac{m_q^{\rm free} \;
    \theta(|\vec{k}| - k_{F})\; \theta(\Lambda -
    k_F)}{\sqrt{\vec{k}^2+(m_q^{\rm free})^2}}\; d^3k \right]^{-1}
\Bigg|_{k_F = 0} \, .  \ee
We evaluate Eq.~(\ref{eq:qcoupling}) for ${\cal N}_c = 3$ to produce
constituent quark masses of \mbox{$m_{u,d}^{\rm free} = 350~{\rm
    MeV}$} using current quark masses of $m_0^{u,d} = 10~{\rm MeV}$
for the light quarks, and a constituent quark mass of \mbox{$m_s^{\rm
    free} = 450~{\rm MeV}$} using a current quark mass of $m_0^s =
160~{\rm MeV}$ for the strange quark (both with a momentum cutoff of
$\Lambda = 1~{\rm GeV}$) as per the phenomenology of this field. At
$k_F = 0$ we find the couplings to be
\be
\label{eq:NJLcouplings}
G_{u,d} = 0.148~{\rm fm}^2,\quad G_{s} = 0.105~{\rm fm}^2.
\ee
\par
\begin{figure}[!b]
\centering
\includegraphics[angle=90,width=0.8\textwidth]{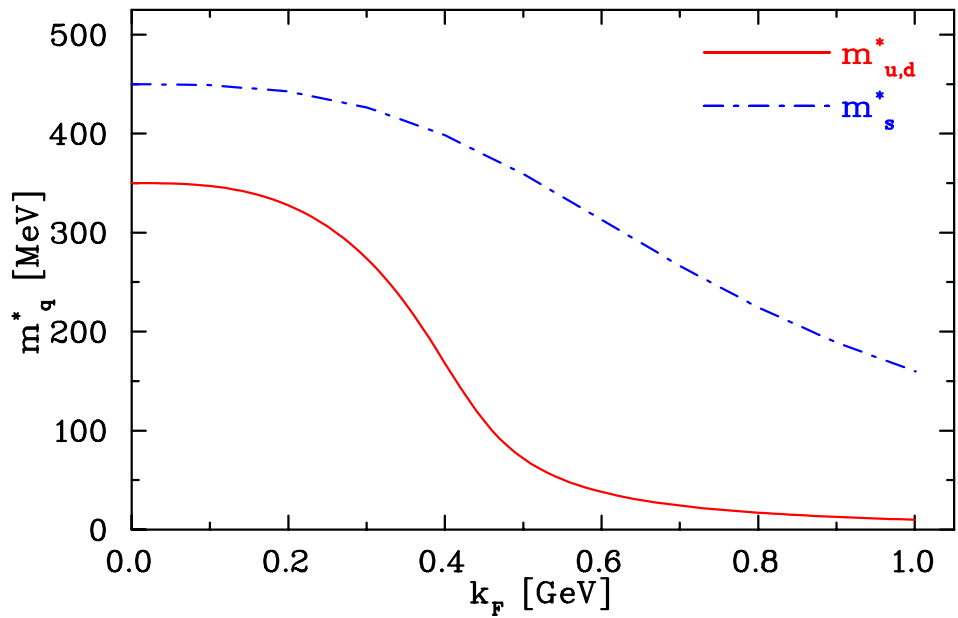}
\caption[Quark masses in NJL]{(Color Online) Dynamic quark masses in
  the NJL model. The mass at $k_F = 0$ is the constituent quark mass,
  and the mass at the cutoff of $k_F = \Lambda = 1~{\rm GeV}$ is the
  current quark mass. This model successfully reproduces the behavior
  found within the Schwinger--Dyson formalism for dynamical chiral
  symmetry breaking.  \protect\label{fig:mkf}}
\end{figure}

We can now use these parameters to evaluate the dynamic quark mass
$m_q^*$ for varying values of $k_F$ by solving
Eqs.~(\ref{eq:dynamicalqmass}) and (\ref{eq:condensate})
self-consistently. The resulting density dependence of $m_q^*$ is
illustrated in Fig.~\ref{fig:mkf} where we observe that the quark
masses\emdash particularly the light quark masses\emdash eventually
saturate and are somewhat constant above a certain Fermi momentum
(hence, density). For densities corresponding to $k_F > \Lambda$, the
dynamic quark mass is constant.\par
%
%
We can now construct the quark matter EOS in the same way as we did
for the MIT bag model, but with density-dependent masses rather than
fixed masses.\par
\clearpage

\section{Fock Terms}\label{sec:fockterms}
As an extension to the models described above, we now introduce terms
which are of a higher order\emdash the Fock terms\emdash leading us to a
Hartree--Fock description of matter. As we wish to work to the next
leading order, we require a perturbative description of the models in
which we can identify the next terms. To this end, we utilise Feynman
diagrams to illustrate which terms shall be calculated, and which shall
be neglected, since we can identify the order of a diagram by the
number of baryon-meson vertices, each of which contributes a factor of
the baryon-meson coupling.\par
The derivation of QHD as described in Appendix~\ref{sec:qhdderiv} does
not lend itself to such an expansion, since all the contributing terms
must appear in the Lagrangian density {\it ab initio}, but we can
re-derive QHD perturbatively which will allow the identification of
the next-order terms. The diagrammatic derivation for QHD at Hartree
level can be found in Appendix~\ref{sec:hartreeEderiv}.\par
\begin{figure}[!t]
\centering
\includegraphics[width=0.5\textwidth]{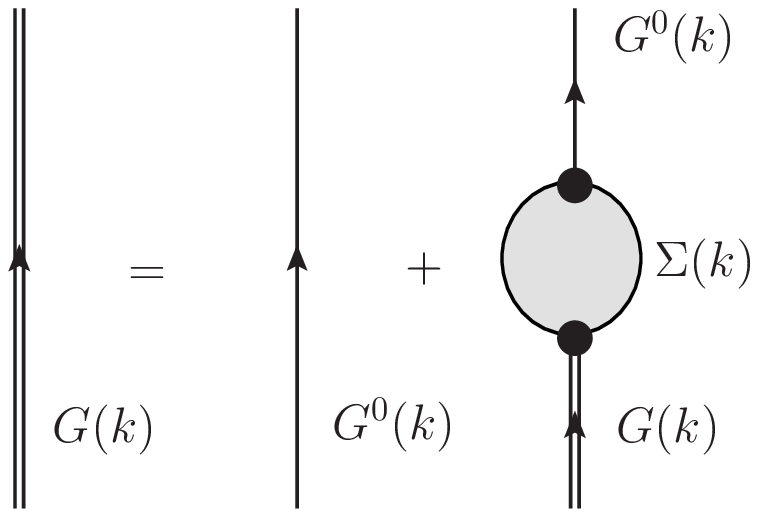}
\caption[Dyson's Equation: Feynman Diagram]{Feynman diagram for
  Dyson's Equation, Eq.~(\ref{eq:dysonseq}), providing a
  self-consistent description for the full baryon propagator (double
  line) involving the bare propagator (single line) and self-energy
  (filled blob).
  \protect\label{fig:DysonsEqFeynman}}
\end{figure}

The first Fock terms that we wish to calculate are the next-to-leading
order contributions to the baryon self-energies $\Sigma_B(k)$, which
as we shall see are momentum dependent. The full derivation of this
Fock contribution can be found in
Appendix~\ref{sec:selfenergyderiv}. In order to calculate the
Hartree--Fock EOS, we consider Dyson's Equation which
self-consistently relates the full momentum-dependent baryon Green's
function (propagator) $G(k)$ to the bare (vacuum) baryon propagator
$G^0(k)$ and the self-energy, as
\be
\label{eq:dysonseq}
G(k) = G^0(k) + G^0(k)\Sigma(k)G(k).
\ee
This can be represented with Feynman diagrams as shown in
Fig.~\ref{fig:DysonsEqFeynman}.\par
We can write the as-yet undefined self-energy as a sum of terms by
defining components of the self-energy proportional to the identity
matrix ${\mathbb I}$, or gamma matrices $\gamma^0$ and
$\vec{\gamma}$ for scalar ($\Sigma^s$), temporal
($\Sigma^0$), and vector ($\Sigma^i \equiv \Sigma^v$) self-energies,
such that the full self-energy (in the case where the following terms
are defined in-medium) is expanded as
\bea
\label{eq:STVcomponents}
\Sigma(k) &=& \Sigma^s(k) - \gamma_\mu \Sigma^{\mu}(k) \\ &=&
\Sigma^s(|\vec{k}|,k^0) - \gamma_0\Sigma^0(|\vec{k}|,k^0) +
\vec{\gamma}\cdot\vec{k}\; \Sigma^v(|\vec{k}|,k^0).  \eea
Note the differences between Eq.~(\ref{eq:STVcomponents}) and
Eq.~(\ref{eq:LcovSE}); the components of the self-energy in the
Lorentz-covariant case (free-space). In this in-medium case, we now
have dependence on $\vec{k}$, which will lead to a dependence on Fermi
momentum.\par
Dyson's Equation can be solved formally to give the baryon propagator
\be \left[G(k)\right]^{-1} = \gamma_\mu(k^\mu + \Sigma^\mu(k)) -
    [M+\Sigma^s(k)].  \ee
\par
For convenience, we define the following quantities;
\bea M^*(k) &=& M + \Sigma^s(k), \\[2mm] \label{eq:kstar} \vec{k}^*
&=& \vec{k}+\Sigma^v(k), \\[1mm] E^*(k) &=& \sqrt{(\vec{k}^*)^2 +
  (M^*)^2}, \\[2mm] k^{*\mu} &=& k^\mu + \Sigma^\mu(k) = [k^0 +
  \Sigma^0(k),\vec{k}^*]. \eea
\par
We can express the solution to Dyson's Equation as a sum of Dirac and
Fermi components, where the Dirac contribution is responsible for
Pauli blocking, and the Fermi contribution accounts for
antibaryons~\cite{Serot:1984ey}, which shall be neglected
here\footnote{In their absence we will use a momentum cutoff to
  regularize the integrals.}
\bea G(k) &=& G_F(k) + G_D(k); \\[2mm]
%
%
G_F(k) &=& [\gamma^\mu k_\mu^* + M^*(k)]\left(k^{*\mu}k_\mu^* -
(M^*(k))^2 +i\epsilon\right)^{-1}, \\ G_D(k) &=& [\gamma^\mu k_\mu^* +
  M^*(k)] \,
\frac{i\pi}{E^*(k)}\ \delta(k_0-E(k))\theta(k_F-|\vec{k}|), \eea
where the energy $E(k)$ is the self-consistent single-particle energy,
calculated `on-shell'\footnote{In which case $p_\mu p^\mu = M^2$, and
  thus $p^0 = E(p)$.}
\be \label{eq:singleparticle} E(k) = [E^*(k) - \Sigma^0(k)]_{k_0=
  E(k)}. \ee
\par
\begin{center}
\begin{figure}[!t]
\centering
\includegraphics[width=0.7\hsize]{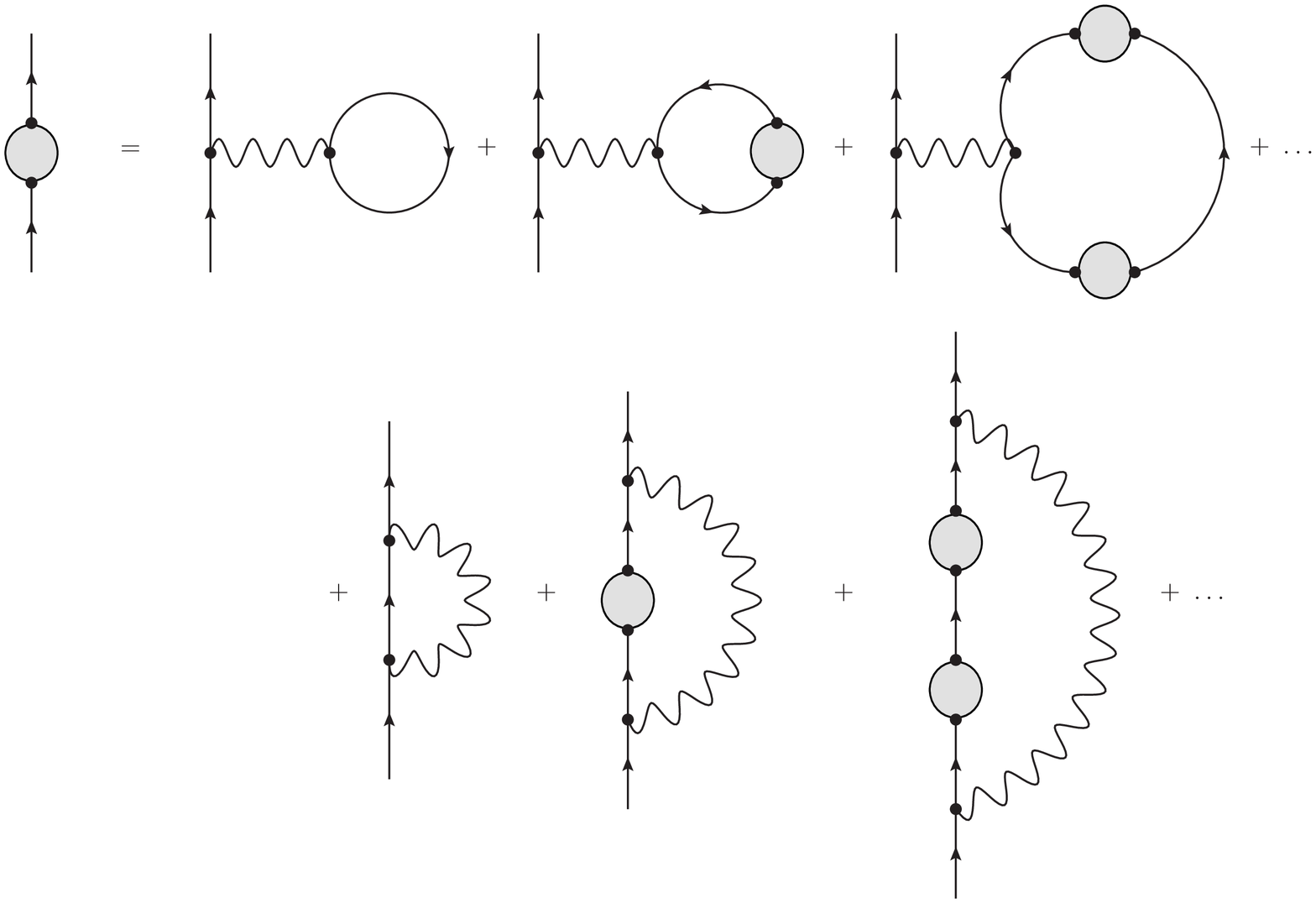}
\caption[Feynman diagrams for Hartree--Fock self-energy]{Summation of
  Feynman diagrams for all possible (vector) interaction terms
  contributing to the self-energy. The first row of diagrams are the
  tadpole terms, and the second row are the exchange terms, where in
  both cases, each term is derived by inserting the full baryon Green's
  function into the previous term, which self-consistently includes
  the self-energy. Similar diagrams exist for the scalar meson
  interactions.\protect\label{fig:body_alldiagrams}}
\end{figure}
\end{center}

In the Hartree case, the tadpole diagrams define the self-energy. If
we now include the exchange contributions, the diagrams for which (for
the vector meson terms) are shown in Fig.~\ref{fig:body_alldiagrams},
the expressions for the total (including all Lorentz forms) $\s$ and
$\w$ contributions to the self-energy become
\bea \Sigma(k) &=& \sum_B \Sigma_{B\s}(k) + \Sigma_{B\w}(k), \\[2mm]
\Sigma_{B\s}(k) &=& ig_{B\s} \volint{q}
\left[ e^{iq^0\eta} \sum_{B'} g_{B'\s} \frac{{\rm Tr}[G(q)]}{m_\s^2} +
  \frac{g_{B\s}\; G(q)}{(k-q)_\mu^2 - m_\s^2 + i\epsilon}\right], \\[2mm]
\Sigma_{B\w}(k) &=& ig_{B\w} \volint{q}
\left[e^{iq^0\eta} \sum_{B'} g_{B'\w}  \frac{{\rm Tr}[\gamma_\mu G(q)]}{m_\w^2} +
  \frac{g_{B\omega}\; \gamma_\mu {\rm g}^{\mu\nu} G(q)\gamma_\nu}{(k-q)_\lambda^2 -
    m_\w^2+i\epsilon} \right], \eea
where for each case in the sum, $G$ refers to the baryon propagator of
baryon $B$. Here ${\rm g}^{\mu\nu}$ is the Minkowski metric tensor
(a.k.a. $\eta^{\alpha\beta}$ in Eq.~(\ref{eq:cliff})).\par
For simplicity, we will neglect the contribution from the $\rho$ meson
in these Fock calculations. The self-energy can now be separated into
terms proportional to those in Eq.~(\ref{eq:STVcomponents}).
\par
Since we will be using a momentum cutoff to regularize the integrals,
we\footnote{Following the procedure of Ref.~\cite{Serot:1984ey}} drop
the antibaryon components of the baryon propagator $G_F(k)$, leaving
$G_D(k)$ as the full propagator. Some of the integrals can then be
performed to produce the mean-field results plus the Fock additions
\bea
\nonumber
\Sigma_B^s(k,E(k)) &=& g_{B\s} \sum_{B'} \frac{-(2J_{B'}+1)}{(2\pi)^3}
\frac{g_{B'\s}}{m_\s^2} \int_0^{k_{F_{B'}}} d^3q
\frac{M_{B'}^*(q)}{E_{B'}^*(q)} \\[2mm]
&& + \reci{4\pi^2k}\int_0^{k_{F_B}}q\; dq \frac{M_B^*(q)}{E_B^*(q)}
\left[\reci{4}g_{B\s}^2\Theta_s(k,q)-g_{B\w}^2\Theta_\w(k,q)\right], \eea
\bea \nonumber \Sigma_B^0(k,E(k)) &=& g_{B\w} \sum_{B'} \frac{-(2J_{B'}+1)}{(2\pi)^3}
\frac{g_{B'\w}}{m_\w^2} \int_0^{k_{F_{B'}}} d^3q \\[2mm] && -
\reci{4\pi^2k}\int_0^{k_{F_B}}q\; dq
\left[\reci{4}g_{B\s}^2\Theta_\s(k,q)+\reci{2}g_{B\w}^2\Theta_\w(k,q)\right],
\eea
\bea
\Sigma_B^v(k,E(k)) &=& - \reci{4\pi^2k^2}\int_0^{k_{F_B}}q\; dq \frac{q^*}{E_B^*(q)}
\left[\reci{2}g_{B\w}^2\Phi_\s(k,q)+g_{B\w}^2\Phi_\w(k,q)\right],
\eea
with the definitions (for convenience) of
\bea \Theta_i(k,q) &=& \ln \left| \frac{A_i(k,q) + 2kq}{A_i(k,q)-
  2kq}\right|, \\[2mm] \Phi_i(k,q) &=& \reci{4kq}
A_i(k,q)\Theta_i(k,q) - 1, \\[2mm] A_i(k,q) &=& \vec{k}^{\; 2} +
\vec{q}^{\; 2} + m_i^2 - [E(q)-E(k)]^2, \eea
for which $q = |\vec{q}\, |$, $k = |\vec{k}|$.\par
All of these self-energies are evaluated on-shell at the
self-consistent single-particle energies defined in
Eq.~(\ref{eq:singleparticle}). To further simplify our calculations,
we will use the approximation $\Sigma_B^v = 0$, since the
momentum-dependence of this term is a power weaker than $\Sigma^s_B$
and $\Sigma^0_B$. The only consequence of this is that in
Eq.~(\ref{eq:kstar}), $\vec{k}^* \to \vec{k}$.\par
These contributions to the self-energy will affect the effective
masses of the baryons in QHD as additional terms in
Eq.~(\ref{eq:QHDeffM}), and once we have made this change we can once
again calculate the energy density. In this perturbative description
of QHD, although the energy density is still derived using the
energy-momentum tensor, the expression for this now involves the meson
propagators. Using the Hartree meson propagators (since these are not
affected by the Fock addition to the self-energy) we can evaluate the
energy density, and we must once again re-fit the baryon-meson
couplings such that the saturation properties are reproduced. The
couplings which provide this are given in Table~\ref{tab:couplings}
for this case of Fock term additions to the self-energy.\par
\begin{figure}[!h]
\centering
\includegraphics[width=0.45\textwidth]{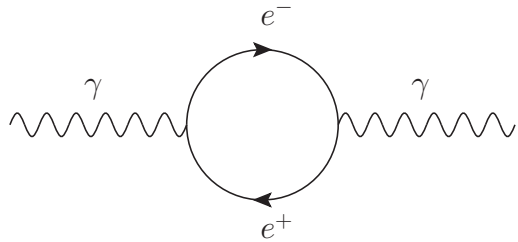}
\caption[QED vacuum polarization]{Feynman diagram for vacuum
  polarization in QED. Here the photon creates two fermions which are
  off-shell (virtual particles) which annihilate back to a photon. The
  charged pair acts as an electric dipole, and thus are able to
  partially screen an external electromagnetic field.
  \protect\label{fig:VacPol}}
\end{figure}

The second Fock terms we can calculate are additional terms in the
meson propagators\emdash the medium pol\-arisation\emdash which will
alter the definition of the energy density. The medium polarization is
named in analogy to the vacuum polarization, the simplest example of
which is in Quantum Electrodynamics (QED); in vacuum, a photon can
spontaneously create a virtual\footnote{These particles are
  `off-shell', meaning that $p_\mu p^\mu \neq m^2$.}
particle-antiparticle pair\emdash namely an electron and a
positron\emdash which annihilate back to a photon, as shown in
Fig.~\ref{fig:VacPol}. While they temporarily exist, the charged pair
acts as an electric dipole and can partially screen an external
electromagnetic field. Similarly, the medium polarization is a
contribution from the creation of a baryon loop which affects the
matter field.\par
\vfill
\begin{figure}[!b]
\centering
\includegraphics[width=\textwidth]{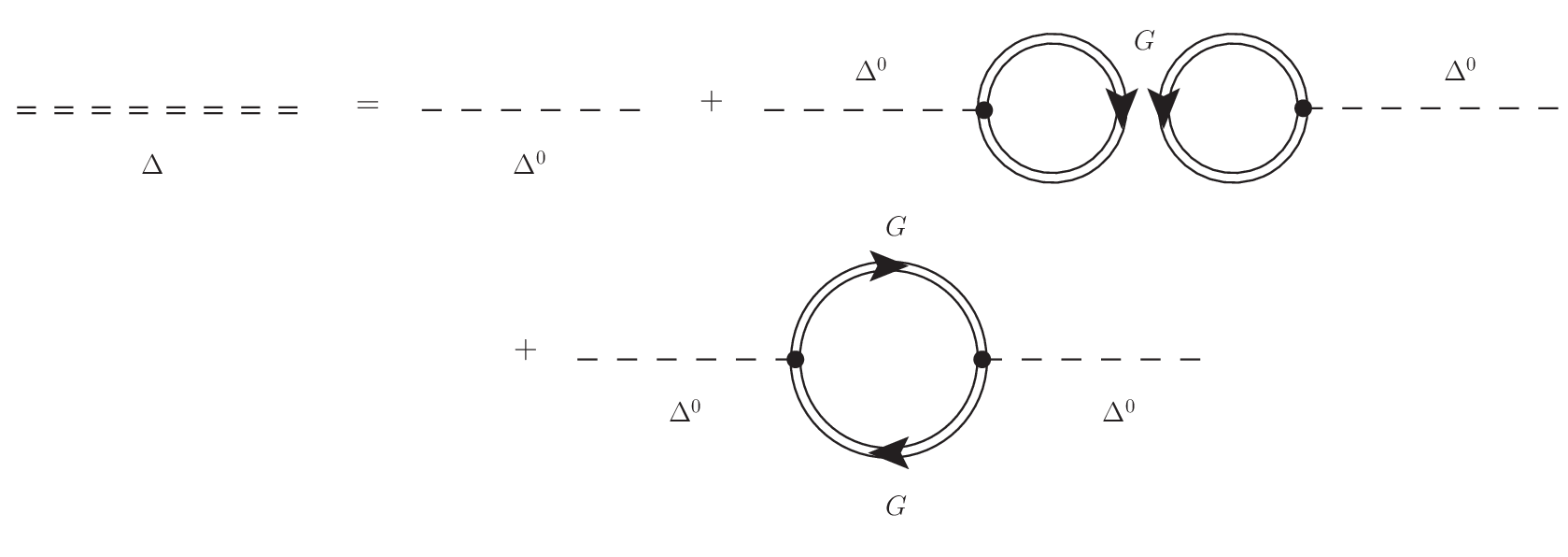}
\caption[3 Terms in the Meson Propagator]{The three leading terms
  contributing to the full meson propagator (double, dashed line) in
  this case of the scalar meson; the bare propagator $\Delta^0$
  (single, dashed line), the (squared) tadpole diagram, and the medium
  polarization diagram. Similar terms can be written for the vector
  meson also. \protect\label{fig:body_3PartMesonPropagator}}
\end{figure}

\clearpage
The three terms contributing to the scalar propagator are shown in
Fig.~\ref{fig:body_3PartMesonPropagator}. The first of these is the
bare meson propagator. The second is the Hartree contribution,
essentially the square of a tadpole term. The last of these is the
medium polarization, and the scalar and vector contributions to this
term are given by
\bea
\Delta'(k) &=& \Delta^0 \Pi_\s(k) \Delta^0(k) \\
\Pi_\s(k) &=& -i\sum_B g_{B\s}^2 \volint{q} {\rm Tr}\left[ G_D(q)G_D(k+q)\right] 
\eea
\bea D_{\mu\nu}'(k) &=& D_{\mu\lambda}^0 \Pi_\w^{\lambda\sigma}(k)
D_{\sigma\nu}^0(k) \\ \Pi_\w^{\lambda\sigma}(k) &=& -i\sum_B g_{B\w}^2
\volint{q} {\rm Tr}\left[ \gamma^\lambda G_D(k+q)\gamma^\sigma
  G_D(q)\right] \eea
where $\Pi_\s$ and $\Pi_\w^{\lambda\sigma}$ are the medium
polarizations.\par
%
%
By evaluating the temporal components of the energy-momentum tensor
$T^{00}$ with the Hartree--Fock propagator as shown in
Fig.~\ref{fig:body_3PartMesonPropagator} we find the full expression
for the Hartree--Fock energy density
\bea \nonumber {\cal E}_{\rm HF} &=& \sum_B \frac{(2J_B+1)}{(2\pi)^3}
\int_0^{k_F} d^3k \ E(k) + \frac{1}{2}m_\s^2 \bra\s\ket^2 -
\frac{1}{2}m_\w^2 \bra\w\ket^2 \\[2mm] \nonumber &&+ \ \reci{2}
\frac{(2J_B+1)}{(2\pi)^6} \int_0^{k_{F_B}} \int_0^{k_{F_B}}
\frac{d^3k \, d^3q}{E^*(k)E^*(q)} \\[2mm] \nonumber &&\times \ \left\{
g_{B\s}^2
D_\s^0(k-q)\left[\reci{2}-[E(k)-E(q)]^2D_\s^0(k-q)\right]\right.
\left[k^{*\mu}q_\mu^* + M^*(k)M^*(q)\right] \\[2mm] \nonumber
&&+\ 2g_{B\w}^2D_\w^0(k-q)\left.\left[\reci{2}-[E(k)-E(q)]^2D_\w^0(k-q)\right]
     [k^{*\mu}q_\mu^* - 2M^*(k)M^*(q)]\right\} 
     . \label{eq:HFENERGY} \\[2mm] \eea
The first line of terms in Eq.~(\ref{eq:HFENERGY}) are of the same
form as the Hartree contributions, but these terms are now defined
using the second-order self-energies. The double integral arises from
the medium polarization terms. Eq.~(\ref{eq:HFENERGY}) is fully derived
in Section~\ref{sec:selfenergyderiv}.\par
For consistency, once we include Fock terms to the energy density, we
must once again recalculate the couplings to fit saturation
properties. The values that best reproduce the data for this
Hartree--Fock description of QHD are summarized in
Table~\ref{tab:couplings}.\par
While we have presented the results for the $\s$ and $\w$ mesons, it
should be noted that other meson contributions are possible. We could
add the contribution from the $\rho$ meson with the appropriate
coupling factors, and furthermore since the Fock terms are of a higher
order than mean-field, we could now include the pion contribution, and
all pseudo-scalars not otherwise excluded. For now we shall continue
to neglect these terms for simplicity.\par
An alternative additional contribution to the energy density is the
second-order exchange contribution as per Ref.~\cite{Krein:1998vc}
which shall be neglected here for simplicity.
\cleardoublepage


%% file: Chapter4_methods.tex
\chapter{Methods of Calculation}\label{sec:methods}
As is often the case in theoretical physics, for many of the
quantities mentioned in the preceding chapters we lack all the
information required to evaluate them. Although we may know the vacuum
mass constants and couplings, there are several quantities which are
self-consistently defined\emdash where the quantity of interest
appears on both the left- and right-hand sides of an equation\emdash
for example the equation for the effective baryon mass in QHD,
\be \label{eq:egMstar} M_{B}^* = M_{B} - g_{B\s}\bra\s\ket = M_{B} -
g_{B\s} \sum_{B'}\frac{g_{B'\s}}{m_\s^2}\int
\frac{(2J_{B'}+1)}{(2\pi)^3} \frac{M^*_{B'}\,
  \theta(k_{F_{B'}}-|\vec{k\, }|)}{\sqrt{\vec{k}^2+M_{B'}^{*2}}}\, d^3
k\ , \ee
is defined self-consistently\footnote{In that the effective masses of
  \emph{all} the baryons are required in order to calculate the
  effective mass of a single baryon.}. Technically,
Eq.~(\ref{eq:egMstar}) describes $N_B$ coupled, self-consistent
equations, where $N_B$ is the number of baryons in the model
considered. Furthermore, although we use the total baryon density
$\rho_{\rm total}$ as the control parameter, for non-trivial cases the
integration limit of $k_{F_B}$ is reliant on knowing the density of
each species separately, which in turn can be reliant on $M_B^*$ via
the chemical potentials (refer to Eq.~(\ref{eq:mu_sw})). Equations of
this form cannot be solved analytically, so we turn our attention to
numerical calculations.\par
All of the calculations performed in this thesis have been computed in
Fortran~90 using methods inspired by `Numerical Recipes in
Fortran~90'~\cite{NumRec}. Rather than publish the particular code
used to perform the exact calculations of this work, in the following
sections we will briefly outline some of the challenges that have been
met in performing these calculations, and the major steps required to
reproduce the simulations.
\section{Newton's Method}\label{sec:newtonsmethod}
One of the most widely renowned methods for iteratively solving
non-linear equations is the Newton--Raphson method, also known simply
as `Newton's Method'. It has the additional benefit that it is
suitable for solving self-consistent equations, as above.\par
\begin{figure}[!t]
\centering
\includegraphics[width=0.6\textwidth]{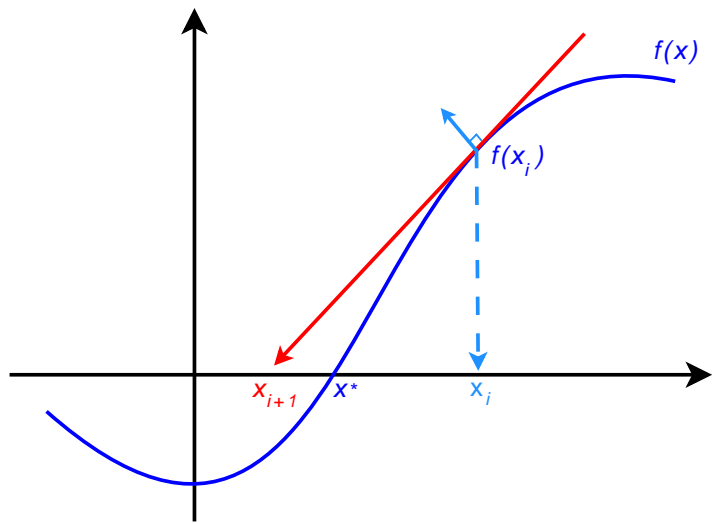}
\caption[Diagram: Newton's Method]{(Color Online) Diagram for Newton's
  method as described in the text. Here, the starting value $x_i$ lies
  within the radius of convergence which for this example is bounded
  somewhere between the two extrema. For example, a tangent at or
  beyond these points will have an $x$-intercept of $x_{i+1}$ that is
  further from $x^*$ than $x_i$ is.  \protect\label{fig:newtons}}
\end{figure}

Newton's method is elegant in its simplicity; In order to find the
value $x^*$ which is a solution of $f(x) = 0$ where $f$ is a general
function, we begin by considering the gradient $f'$ of a tangent to
$f$ at a point $x_i$ that is sufficiently close to $x^*$ (within the
radius of convergence). The radius of convergence is defined in this
case by the region in which the tangent to $f$ intersects the $x$-axis
at a point $x_{i+1}$ such that $|x^* - x_{i+1}| < |x^* - x_i|$, as
shown in Fig.~\ref{fig:newtons}.\par
The gradient of the tangent is given by
\be f'(x_i) = \frac{\Delta f(x)}{\Delta x} = \frac{f(x_i)}{x_i -
  x_{i+1}}, \ee
where we have used the fact that the tangent to $f$ intercepts the
$x$-axis at $x_{i+1}$, and thus $f(x_{i+1}) = 0$. Rearranging this
gives the iterative procedure for Newton's method;
\be
\label{eq:newtons}
x_{i+1} = x_i - \frac{f(x_i)}{f'(x_i)}.
\ee
If the function is convergent, for some value of $i$ it must be true
that $|x_{i+1} - x_i| < \epsilon$ for some tolerance $\epsilon$, at
which point the solution $x^*$ is considered to be found. To make use
of this method, we need to define our functions, and the equations
that we wish to find the solutions to.\par
For QHD, QMC, and in general, we use the total baryon density as the
control parameter, allowing us to calculate the properties of matter at
a specific density. At a given density, the total baryon density is
therefore conserved (constant), and equal to the sum of individual
baryon densities;
\be
\rho_{\rm total} = \sum_B\rho_B.
\ee
We also require that the total charge of the system, defined by
\be
Q_{\rm total} = \sum_BQ_B\rho_B
\ee
is constant, and vanishing. With two constraining equations we can use
Newton's method to find the values of (at most) two quantities
constrained by these equations. For the purpose of simplicity, we use
the proton and neutron Fermi momenta $k_{F_p}$ and $k_{F_n}$ since
these appear in the integration limits of many quantities.\par
In keeping with the discussion of chemical potentials in
Section~\ref{sec:qftchempot}, we could equivalently use the neutron
and electron chemical potentials; the choice is arbitrary, though
using only the nucleon Fermi momenta allows us to neglect conservation
of total charge, such as the case of nuclear matter where we only need
a single constraining equation\footnote{The nucleon Fermi momenta are
  constrained such that the densities are equal in that case.}, without
loss of generality.\par
Newton's method is a root-solving method, so our constraint equations
need to be written in a form of $\vec{f}(\vec{x\, }) = \vec{0}$. This is
achieved simply as
\bea
\label{eq:constraint1}
f_1(k_{F_n}) = \rho_{\rm total} - \sum_B\rho_B &=& 0, \\
\label{eq:constraint2}
f_2(k_{F_p}) = Q_{\rm total} - \sum_BQ_B\rho_B &=& 0.
\eea
Lastly, in order to solve the self-consistent equations for the baryon
effective masses in QHD and QMC, we require an additional quantity
that we can allow to vary. Considering QHD, since (at Hartree level at
least) the scalar field $\bra\s\ket$ is independent of baryon species
(it is a sum of contributions from \emph{all} baryons) we use this
quantity in a further constraint equation
\be \label{eq:constraint3} f_3(\bra\s\ket) = \bra\s\ket -
\sum_B\frac{g_{B\s}}{m_\s^2}\int \frac{(2J_B+1)}{(2\pi)^3}\frac{M^*_B\,
  \theta(k_{F_B}-|\vec{k}|)}{\sqrt{\vec{k}^2+M_B^{*2}}}\, d^3 k = \ 0.  \ee
Note that since 
$\rho_B = k_{F_B}^3 /3\pi^2$, each of the above constraint equations
contains $k_{F_p}$ and $k_{F_n}$, so these equations are not only
non-linear, but are all also highly correlated.\par
Although the above derivation of Newton's method is for a function of
a single variable, we can extend this concept to an arbitrary number
of $m$ functions in $n$ variables by replacing the derivative by a
Jacobian
\be \label{eq:jacobian} J=\begin{bmatrix} \dfrac{\partial f_1}{\partial x_1} & \cdots &
\dfrac{\partial f_1}{\partial x_n} \\ \vdots & \ddots & \vdots
\\ \dfrac{\partial f_m}{\partial x_1} & \cdots & \dfrac{\partial
  f_m}{\partial x_n} \end{bmatrix}.  \ee
\par
In order to include the hyperon contributions to these calculations,
we do not require any further constraint equations, since all of the
hyperon Fermi momenta can be derived from that of the nucleons, by
rearranging Eq.~(\ref{eq:allmus}). The constraint equations above,
Eqs.~(\ref{eq:constraint1})--(\ref{eq:constraint3}) do however change
as they now include terms for the hyperons in the sums.\par
We can use the same method to perform the quark matter calculations in
this thesis, with only slight changes to the constraint equations. The
conserved baryon density is now defined for quark matter as per
Eq.~(\ref{eq:equivrho}), and the total charge remains zero, though the
individual charges $Q_i$ now reflect the quark charges; $Q_u = +2/3$,
$Q_d = Q_s = -1/3$. For the calculations of the MIT Bag Model, only
these two constraint equations are required, since the quark masses
are constant. For the NJL Model calculations, the quark masses are
calculated self-consistently in the same manner as the baryon
effective masses above, so the equation for a quark condensate,
Eq.~(\ref{eq:condensate}) can take the place of the scalar field in
Eq.~(\ref{eq:constraint3}).\par
This method is used to produce all of the calculations for this work
at Hartree level, but it is not suitable for calculations involving
the Fock terms; the additional factors that are included at
Hartree--Fock level alter the self-consistent equations for the
effective masses (via the self-energy) such that the effective masses
become baryon- and momentum-dependent. Because of this, we can no
longer rely on the mean-scalar field to produce a constraint equation,
since we would require $2+n_k\times n_B$ equations ($n_k$ = number of
momentum samples, $n_B$ = number of baryons in the
calculation). Clearly this is not a feasible route to take, since for
a reasonable number of momentum samples, say $n_k = 100$ we would
require 202 equations to just calculate properties of nucleonic
matter.\par
The alternative is to find a new method of evaluating our
self-consistent equations. The method we have used is described next.

\section{Steffensen's Method}\label{sec:impnewtons}
Until now we have described solving the self-consistent equations
within Newton's method, but as stated above the Fock terms would
introduce too many new equations for this to be workable. To make
matters worse, the addition of the Fock terms makes the effective
masses momentum dependent, and so the equations for these become
integral equations
\bea \nonumber M^*_B(k) &=& M_B - g_{B\s} \sum_{B'}
\frac{g_{B\s}}{m_\s^2} \int_0^{q_{F_{B'}}}
\frac{(2J_{B'}+1)}{(2\pi)^3} \frac{M_{B'}^*(q)}{E_{B'}^*(q)}\, d^3q
\\[1mm] &&+ \reci{4\pi^2k}\int_0^{q_{F_B}}q\;
\frac{M_{B}^*(q)}{E_{B}^*(q)}
\left[\reci{4}g_{B\s}^2\Theta_\s(k,q)-g_{B\w}^2\Theta_\w(k,q)\right]
dq.\label{eq:HFmstar} \eea
where the terms in this equation are defined in
Section~\ref{sec:fockterms}.\par As a solution to this problem, we
find a new method of solving the self-consistency. Many improvements
to Newton's method are available, and we have chosen to investigate
one in particular; Steffensen's Method. The advantage that this method
has over Newton's method is that it doesn't require a calculation of
the derivative of the function $f$.\par
\begin{figure}[!b]
\centering
\includegraphics[width=0.6\textwidth]{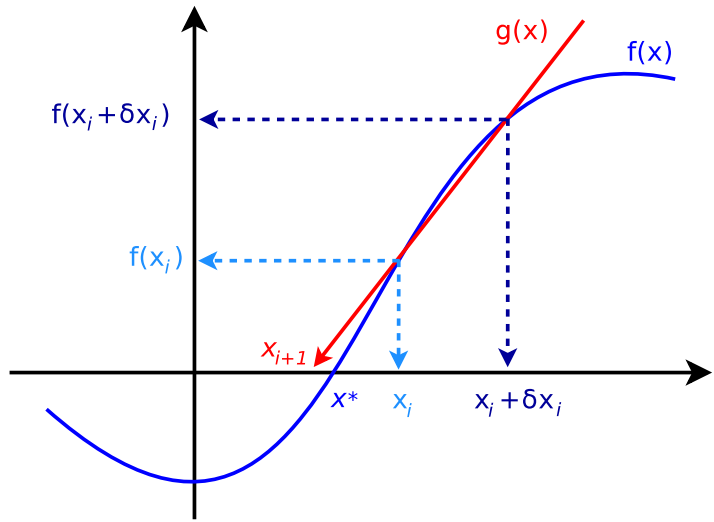}
\caption[Diagram: Steffensen's Method]{(Color Online) Diagram for
  Steffensen's method as described in the text. This method has the
  numerical advantage that it does not require the evaluation of a
  derivative, though the requirement of a `good' starting value is now
  somewhat more strict.
  \protect\label{fig:steffensens}}
\end{figure}

As an improvement over Newton's method, we replace the derivative (the
slope of the tangent to $f$) in the denominator of
Eq.~(\ref{eq:newtons}) with the slope $g(x)$ of a line joining the
point $(x_i,f(x_i))$ with an auxiliary point $(x_i+\delta
x_i,f(x_i+\delta x_i))$ as shown in Fig.~\ref{fig:steffensens}. The
choice of $\delta x_i$ is somewhat arbitrary, and the choice of
$\delta x_i = f(x_i)$ merely ensures that the step size is scaled
appropriately according to the distance from the solution, and in fact
$\delta x_i$ vanishes as $f(x_i)$ approaches 0.\par
The slope of this new line is then
\be
g(x) = \frac{f(x_i+f(x_i))-f(x_i)}{f(x_i)}.
\ee
If we insert this into Eq.~(\ref{eq:newtons}) in place of the
derivative we obtain the iterative procedure for Steffensen's Method,
where the $x$-intercept of $g(x)$ defines the next approximation to
$f(x^*) = 0$;
\be
\label{eq:steffensens}
x_{i+1} = x_i - \frac{f(x_i)}{g(x_i)} = x_i -
\frac{f(x_i)^2}{f(x_i+f(x_i))-f(x_i)}.  \ee
%
%
This method can be applied to solving integral equations such as
Eq.~(\ref{eq:HFmstar}).\par
An issue that the reader may notice is that the effective masses in
that equation have a momentum-dependence in the form of a $1/k$ term,
as well as $k$-dependence in the $\Theta(k,q)$ terms; this makes the
definition of $M^*(0)$ very critical, since the $k$-dependence of the
$\Theta$ term (as defined in Section~\ref{sec:fockterms}) produces
$\Theta_i(0,q) = \ln(1) = 0$, thus we have a $0/0$ issue. Since we
require the value of $M^*(0)$ in order to calculate the integrals
above, we must define this value very carefully. Since the function is
fairly smooth, we use a linear interpolation of the function evaluated
at the next two momentum points to define the limit of the function at
$k=0$.\par
This gives us all the information we need in order to solve the
self-consistent equations, but we can improve the method by
accelerating the convergence, as we shall show in the next section.\par
\subsection{Aitken's $\Delta^2$ Process}\label{sec:aitken}
We can further accelerate the rate of convergence for Steffensen's
Method by using Aitken's delta-squared process. If the function we are
attempting to find the roots of is $f(x)$ we can iterate Steffensen's
method to produce intermediate values $h_0$, $h_1$, and $h_2$ such
that
\bea \nonumber h_0 &=& x_i, \\[2mm] \nonumber h_1 &=&
\displaystyle{h_0 - \frac{f(h_0)}{g(h_0)}}, \\[1mm] h_2 &=&
\displaystyle{h_1 - \frac{f(h_1)}{g(h_1)}}.  \eea
%
With these definitions, Aitken's delta-squared process defines the
next approximation to the root of the function $f$ as
\be
x_{i+1} = h_0-\frac{(h_{2}h_0-h_{1})^2}{h_{2}-2h_{1}+h_0}.
\ee
If the process has failed to converge at this stage by satisfying the
condition $|x_{i+1} - x_{i}| < \epsilon$ then the entire procedure is
repeated for the next value of $x$. If convergence is achieved, it is
trivial to confirm that the convergence has produced the correct value
of $f(x_{i+1}) = 0$.\par
Once again, this description has involved only a function of a single
variable. The method can be modified to handle functions of several
variables, but we shall not discuss this modification here as we will
only be using this method for nucleon calculations. With this method
we are able to solve (at least) two self-consistent integral
equations, and we shall do so for the nucleon effective masses for a
suitable number of momentum points.\par
We nest the solving of the self-consistent effective masses within the
Newton's method in the previous section which still varies the proton
and neutron Fermi momenta to find the roots of the equations. The
functions $\vec{f}$ that we wish to find the roots of via Steffensen's
Method are simply rearrangements of Eq.~(\ref{eq:HFmstar}) for each of
the baryons such that $\vec{f}(M_p^*,M_n^*) = \vec{0}$. Once we have
solved these self-consistencies, we no longer require the third
constraint equation Eq.~(\ref{eq:constraint3}) in Newton's method.

\subsection{Example Integral Equations}\label{sec:exampleintegral}

As an example, we consider the fixed-point integral equation
\be \label{eq:inteq} u(x) = F(u(x)) = e^x + e^{-1} \int_0^1 u(t)\; dt, \ee
which has the solution
\be u^*(x) = e^x + 1. \ee
Being such a simple case, we can solve Eq.~(\ref{eq:inteq}) in several
ways. For the sake of comparison, we will compare a na\"ive method; in
which a previous function value is used as the next guess, the
iterative procedure for which is $u_{i+1}(x) = F(u_{i})$, where for
this example the minimized function is the difference between the
left- and right-hand sides of Eq.~(\ref{eq:inteq}), viz $f(x_i) =
u(x_i) - F(u(x_i))$; with the Aitken-improved Steffensen's method used
for our actual calculations.\par
\begin{figure}[!t]
\begin{minipage}[c]{0.85\textwidth}
\centering
\includegraphics[angle=90,width=0.8\textwidth]{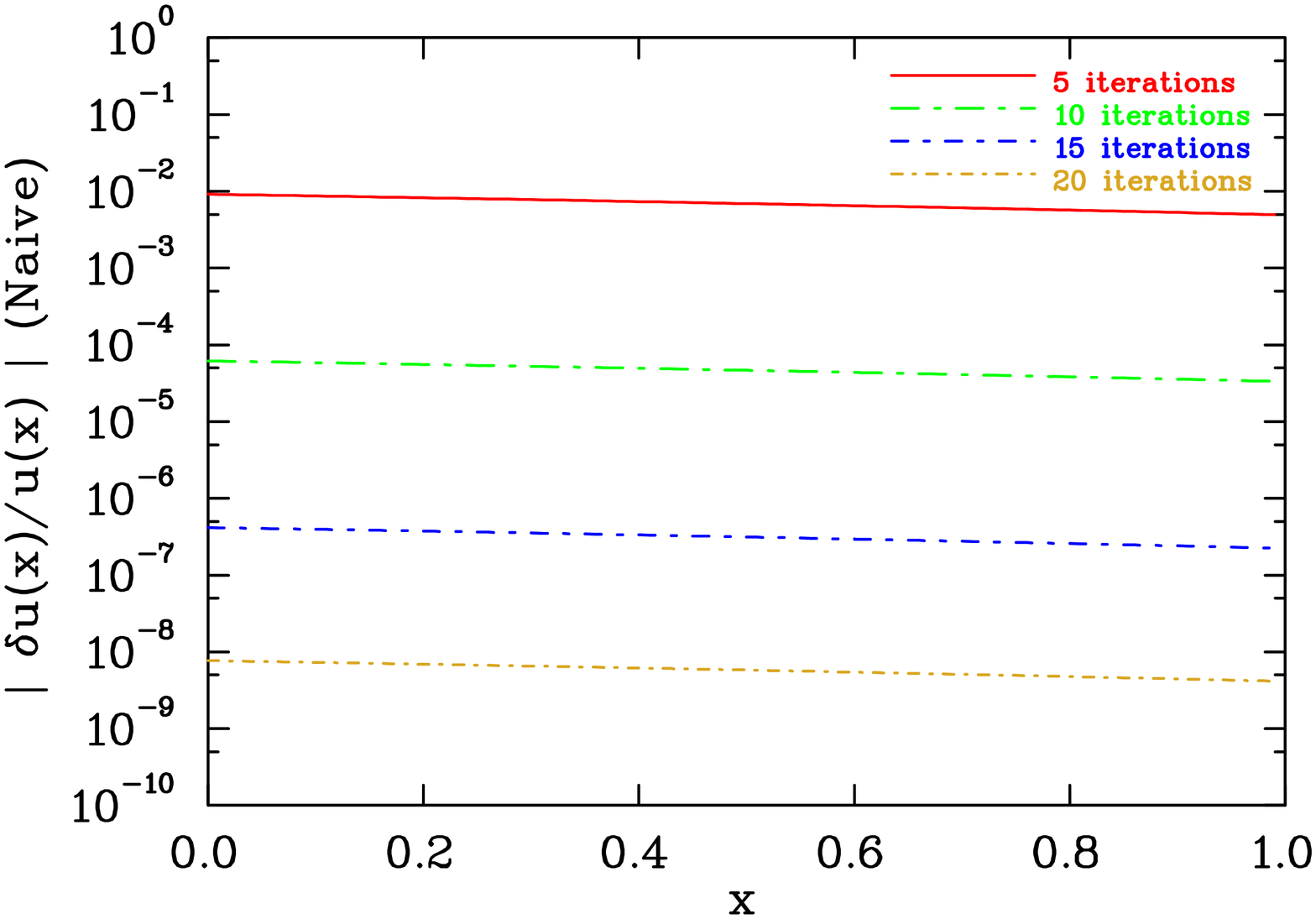}
\includegraphics[angle=90,width=0.8\textwidth]{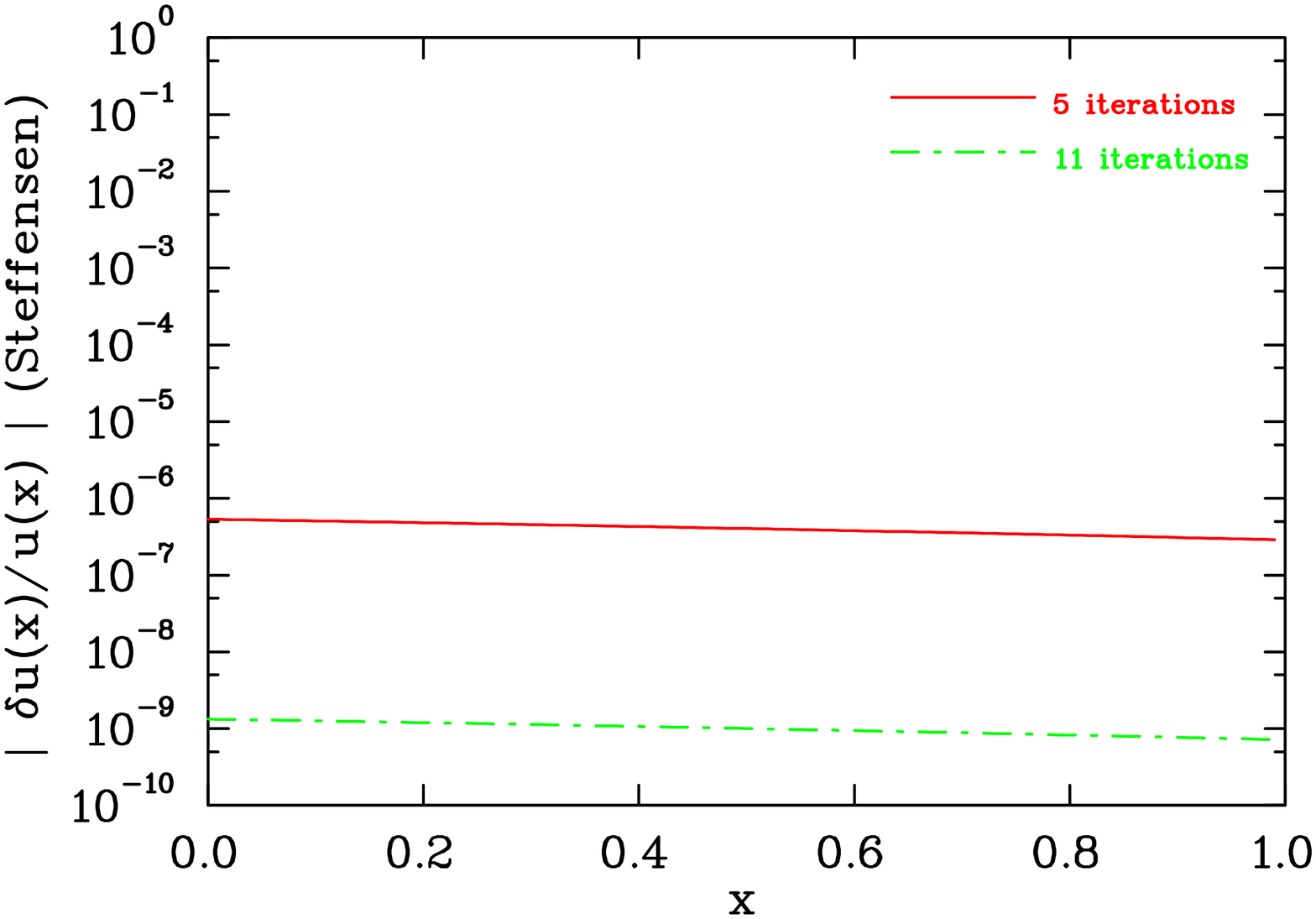}
\end{minipage}
\caption[Relative errors for convergence of iterative methods] {(Color
  Online) Errors for convergence of a na\"ive method (top) and
  Steffensen's method (bottom) relative to the exact solution
  $u^*(x)$. The na\"ive method converges in 20 iterations, while
  Steffensen's method converges in 11 iterations.}
\label{fig:converge}
\end{figure}

In both cases, we supply an identical initial guess of $u_0(x) = 0$,
sampled over 1000 $x$-values, and set a tolerance for convergence (via
the relative error) at each sampled value of $x_j$ (where
$j=1,\ldots,1000$) of
\be \epsilon_j = \frac{u_{i+1}(x_j)-u_i(x_j)}{u_{i+1}(x_j)} = 1\times
10^{-8}. \ee
In both cases the procedures converge to within the above tolerance,
and the actual errors relative to $u^*(x)$ are shown in
Fig.~\ref{fig:converge}. The na\"ive method converges with 20
iterations, while Steffensen's method converges with just 11, a vast
improvement.\par
The speed of convergence is one advantage of the Aitken-improved
Steffensen's method, but the more important advantage comes when we
try to solve a more complicated integral equation, such as
Eq.~(\ref{eq:HFmstar}) for a single baryon, which is a fixed-point
integral equation.\par
With all constants defined the integral equation reduces to $M^* =
f(M^*,k_F)$, where the value of $k_F$ is given a fixed value before
the integral equation is to be solved. For values of $k_F$ up to some
critical value of $(k_F)_{\rm critical}$, the na\"ive method
successfully solves the integral equation within a finite number of
iterations.\par
\begin{figure}[!t]
\centering
\includegraphics[width=0.6\textwidth,angle=90]{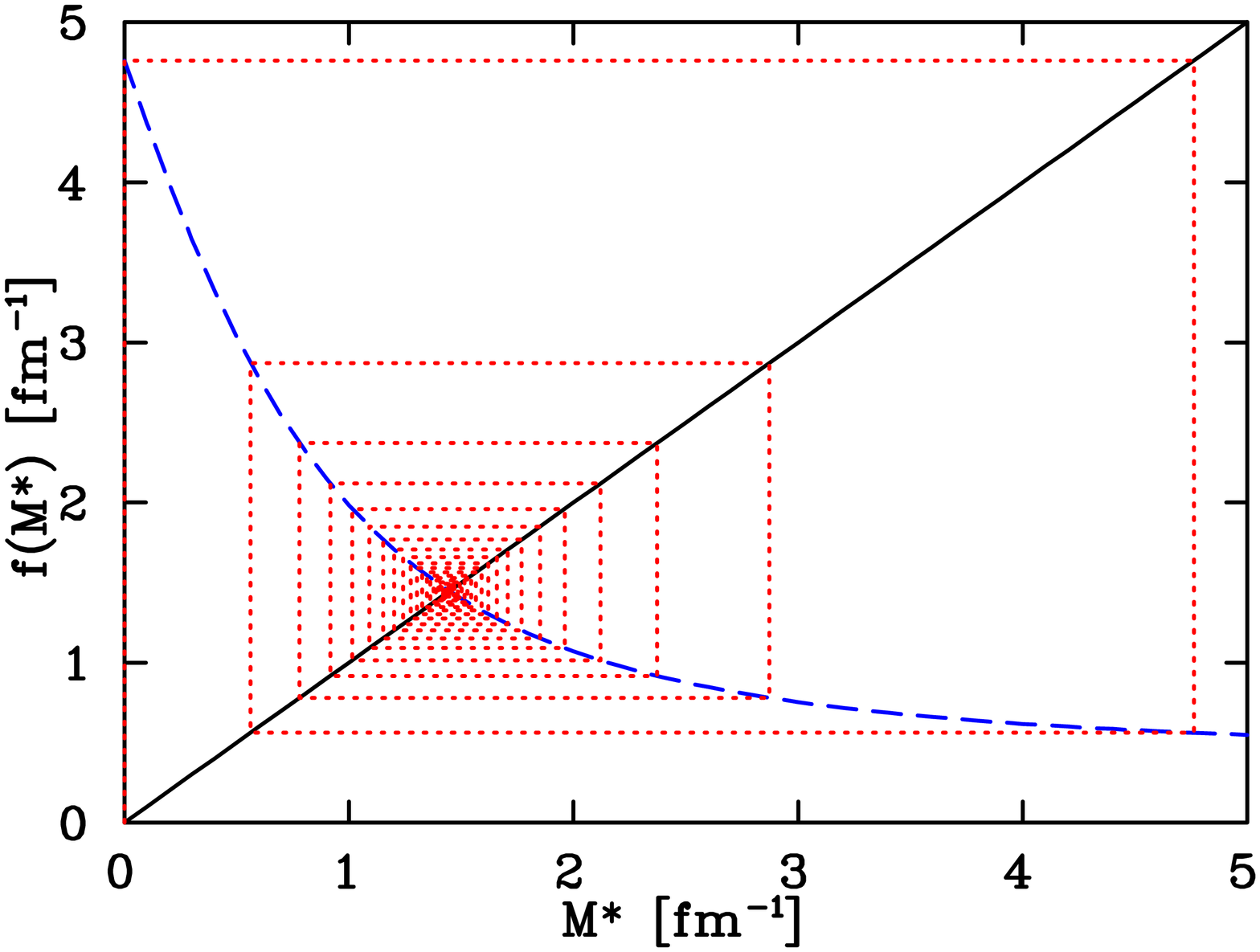}
\caption[Cobweb Diagram: Na\"ive method, $k_F < (k_F)_{\rm
    critical}$]{(Color Online) Cobweb diagram for a na\"ive method of
  solving Eq.~(\ref{eq:HFmstar}) for a single baryon, in which $k_F <
  (k_F)_{\rm critical}$. The short-dashed line traces the guesses and
  function evaluations from the initial guess of $M^*=0$ to the value
  that satisfies the integral
  equation. \protect\label{fig:cobweb_kfunder}}
\end{figure}

In order to investigate the convergence we use a cobweb diagram, as
shown in Fig.~\ref{fig:cobweb_kfunder} in which we `join-the-dots'
between successive guesses and function evaluations. For clarity, the
function $f(M^*,k_F)$ is shown, as is the value of $M^*$; the point at
which these lines intersect represents the solution to the integral
equation. In this way we can trace the convergence of the guesses from
an initial guess of $M^* = 0$ to the value that satisfies the integral
equation.\par
\begin{figure}[!b]
\begin{minipage}[c]{0.85\textwidth}
\centering
\includegraphics[angle=90,width=0.8\textwidth]{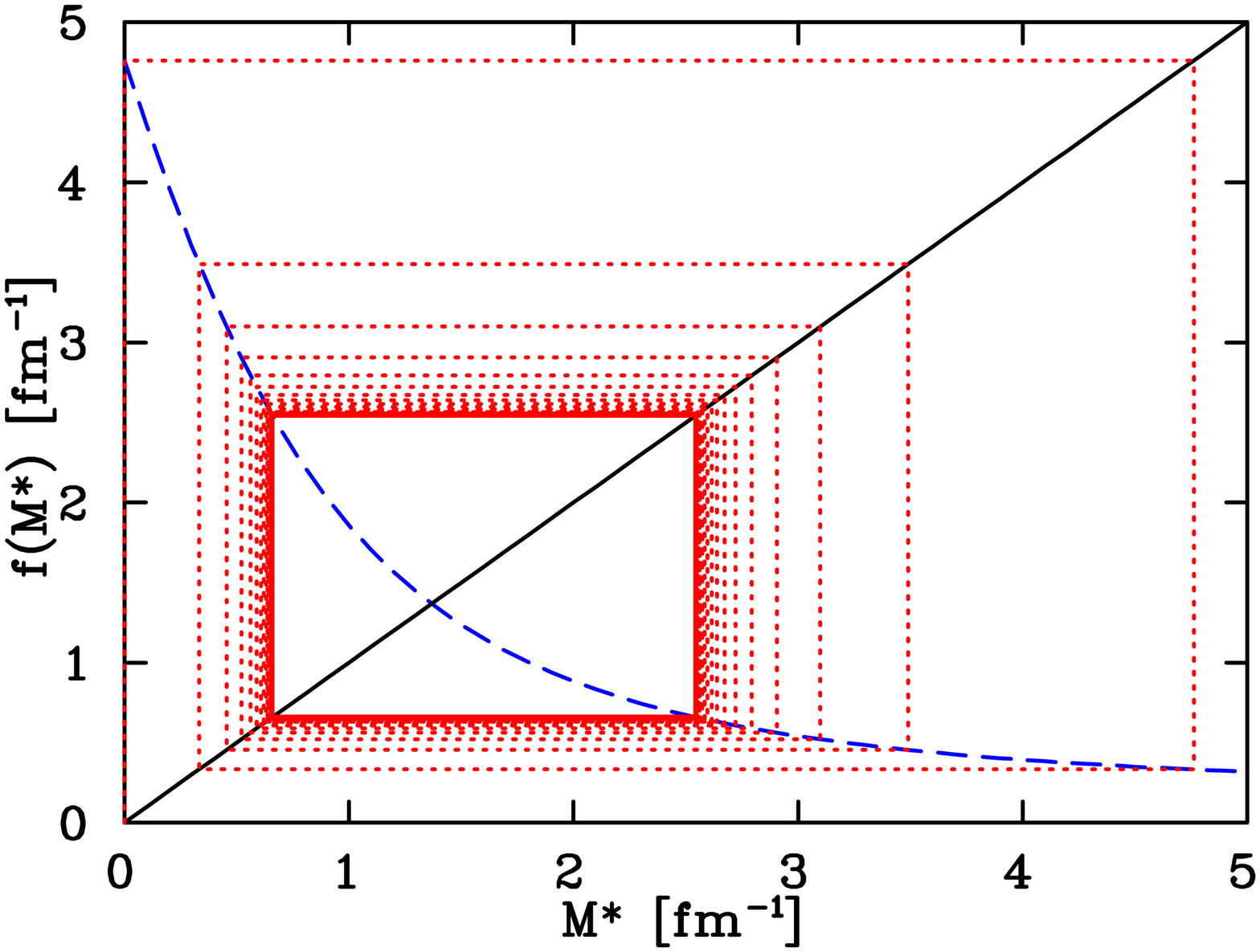}
\caption[Cobweb Diagram: Na\"ive method, $k_F > (k_F)_{\rm
    critical}$]{(Color Online) Cobweb diagram for a na\"ive method of
  solving Eq.~(\ref{eq:HFmstar}) for a single baryon, in which $k_F >
  (k_F)_{\rm critical}$. In this case, the procedure does not converge
  to the a single solution, but alternates between a pair of
  points. \protect\label{fig:cobweb_kfover}}
\vspace{3mm}
\includegraphics[angle=90,width=0.8\textwidth]{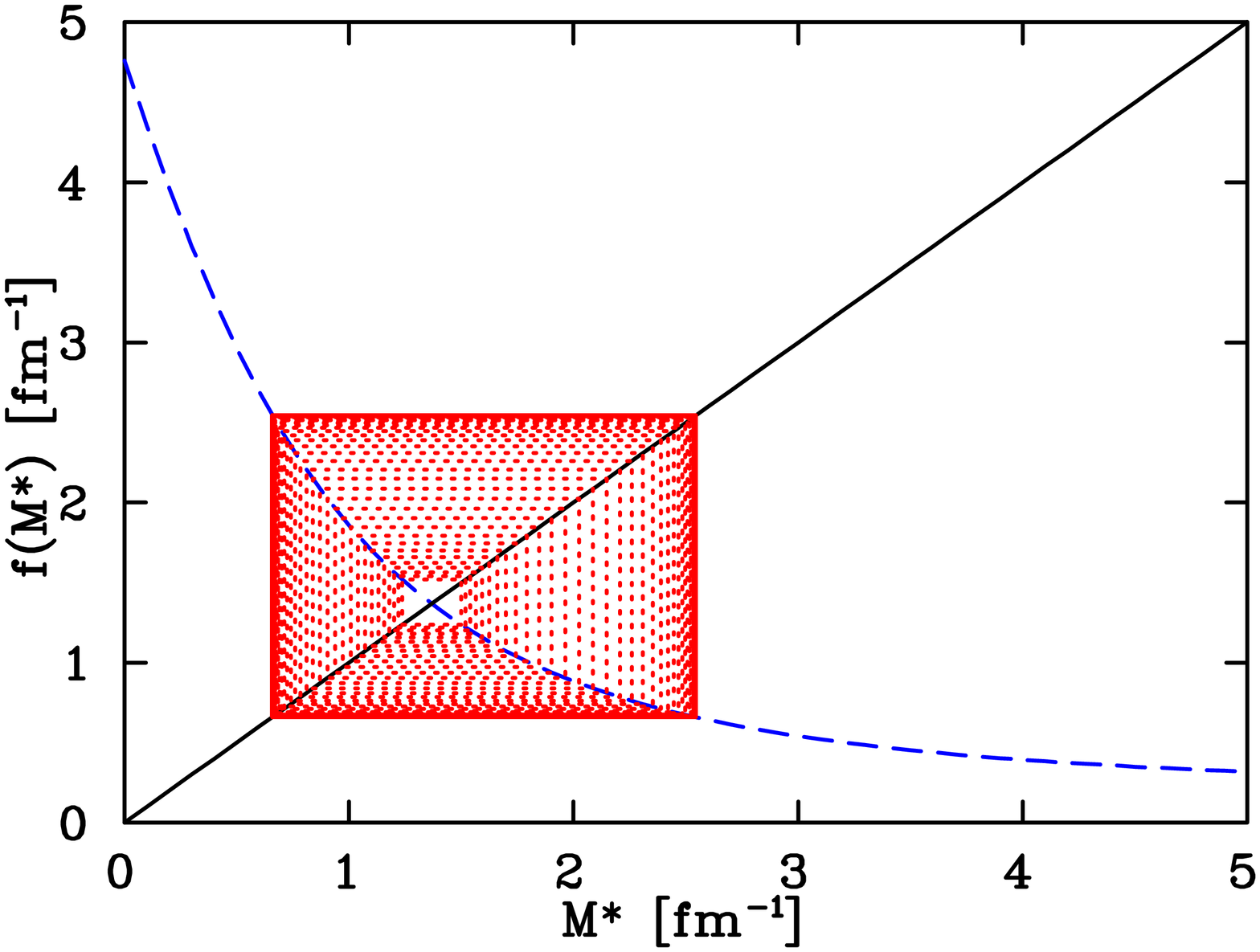}
\caption[Cobweb Diagram: Na\"ive method, $k_F > (k_F)_{\rm critical}$,
  $M^* \sim$ exact solution]{(Color Online) Cobweb diagram for a
  na\"ive method of solving Eq.~(\ref{eq:HFmstar}) for a single
  baryon, in which $k_F > (k_F)_{\rm critical}$ and the starting point
  is deliberately chosen to be near the exact solution. In this case,
  the procedure does not converge to the a single solution, but in
  fact diverges, then alternate between a pair of
  points. \protect\label{fig:cobweb_kfclose}}
\end{minipage}
\end{figure}

If however, we wish to solve the integral equation for a value of $k_F
> (k_F)_{\rm critical}$, the na\"ive method fails to converge, and
rather `flip-flops' between a pair of points. The cobweb diagram for
this case is shown in Fig.~\ref{fig:cobweb_kfover}, and we see that
the solution is never reached.\par
The significance of the value of $(k_F)_{\rm critical}$ is that the
slope of the function at the fixed-point becomes too negative; in
particular, $f^\prime(M^*,(k_F)_{\rm critical}) < -1$. Proving that
this will cause this method to fail to converge is as easy as creating
a cobweb diagram for any function with $f^\prime($fixed-point$) <
-1$. Even if we select a starting point particularly close to the
actual fixed-point solution and perform this iterative procedure, if
$k_F > (k_F)_{\rm critical}$ the iterations diverge to a pair of
points, as shown in Fig.~\ref{fig:cobweb_kfclose}. Although we have
only shown the case of the na\"ive substitution iterative procedure,
we note that Newton's method suffers from the same limitation.\par
\begin{figure}[!t]
\centering
\includegraphics[width=0.6\textwidth,angle=90]{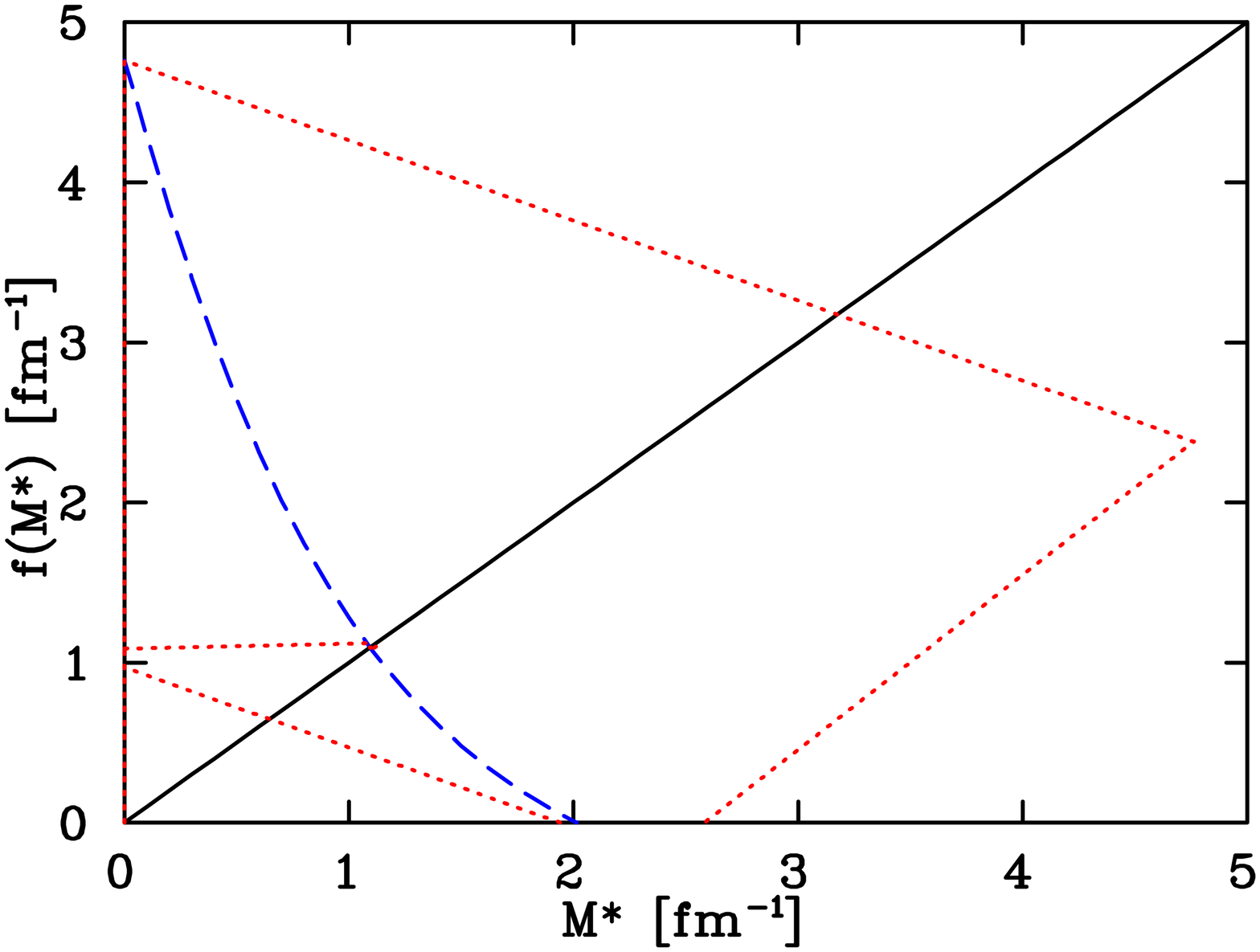}
\caption[Cobweb Diagram: Steffensen's method, $k_F \gg (k_F)_{\rm
    critical}$]{(Color Online) Cobweb diagram for a Aitken-improved
  Steffensen's method of solving Eq.~(\ref{eq:HFmstar}) for a single
  baryon, in which $k_F \gg (k_F)_{\rm critical}$, with initial guess
  $M^* = 0$. This shows that even for cases in which the na\"ive
  method (and Newton's method for that matter) would fail, this method
  is sufficient to solve the integral
  equation. \protect\label{fig:cobweb_stef}}
\end{figure}

If however, we use Aitken-improved Steffensen's method for solving the
same integral equation with the same initial value of $M^* = 0$, the
procedure converges to the correct result, irrespective of the choice
of $k_F$. Fig.~\ref{fig:cobweb_stef} shows an example of this
convergence for a value of $k_F \gg (k_F)_{\rm critical}$, and the
solution is not only found, but is done so in a small number of
iterations. We shall therefore rely on Aitken-improved Steffensen's
method to solve the integral equations in the work that follows.\par
While we have shown an example of a single integral equation, the
equations we will need to solve for our calculations will be a series
of coupled integral equations, $M^*_B =
f(M^*_p,M^*_n,k_{F_p},k_{F_n})$ and this complicates matters. To be
precise, we would require an equivalent of the Jacobian,
Eq.~(\ref{eq:jacobian}) but derived in the manner of Steffensen's
method. For our purposes, we are able to use the linear version to
solve two coupled integral equations; one for protons and one for
neutrons. This may be due to the similarities between the values of
$M^*$ for each of these. Further research will be undertaken to
investigate the roles that the octet baryons take in these
calculations.\par
\clearpage

\section{Infinite Matter}\label{sec:infmattermethod}
Utilising the above techniques we are able to compute simulations of
infinite matter. The entire procedure is coded in Fortran~90. The
method by which we do this is as follows;
\begin{enumerate}
\item Define all vacuum masses, coupling constants, and internal
  parameters.
\item Provide suitable initial guesses for variables $k_{F_p}$ and
  $k_{F_n}$, and if applicable, $\bra\s\ket$.
\item Select a value for the control parameter $\rho_{\rm total}$. We
  select a small non-zero value to begin with. When we are computing
  models involving the leptons, we also select the value of the total
  charge, which is always $Q_{\rm total} = 0$.
\item Call a subroutine {\tt SolveConstraints} which solves the
  constraint equations
  Eqs.~(\ref{eq:constraint1})--(\ref{eq:constraint2}) using Newton's
  Method. Eq.~(\ref{eq:constraint3}) can be excluded here even if not
  calculating Fock terms, but this self-consistency must then be
  solved independently. This subroutine takes initial guesses for
  $k_{F_p}$ and $k_{F_n}$ as well as as required tolerances as inputs,
  and produces the Fermi momentum values that satisfy the constraint
  equations above as outputs.
\item As part of the evaluation of
  Eqs.~(\ref{eq:constraint1})--(\ref{eq:constraint2}) within {\tt
    SolveConstraints}, the self-consistent effective masses must be
  calculated by calling another subroutine {\tt
    SolveEffectiveMasses}. This subroutine takes initial guesses for
  the effective masses and self-energies, as well as required
  tolerances as inputs, and produces the self-consistent effective
  masses that satisfy Eq.~(\ref{eq:constraint3}) as outputs.
\item With the Fermi momenta and effective masses determined, we
  finally have all the information required to evaluate quantities of
  interest, such as the energy density (e.g. Eq.~(\ref{eq:E_H}))
  and pressure (e.g. Eq.~(\ref{eq:P_H})) of the system. These
  values are evaluated and saved to a data-file.
\item Finally, the control parameter $\rho_{\rm total}$ is increased by
  some small step size $\delta \rho$ and the process repeated, where
  the initial guesses for the subroutines are now the final values
  from the previous iteration.
\end{enumerate}
\par
This of course assumes that the couplings constants are known
quantities. In each of the models we use we require that the
properties of saturated nuclear matter are reproduced, as per
Eq.~(\ref{eq:EperA}). In order to find the appropriate values of
$g_{N\s}$ and $g_{N\w}$ that reproduce the correct results we perform
the above method with initial guesses for these values, and calculate
the energy per baryon at saturation (as defined by the lower extrema
of the data) and the Fermi momentum at which this occurs. By
performing this several times for several choices of couplings, we can
refine our choices for initial guesses in order to better reproduce
the saturation properties. Of course, in practice we perform these
steps using a Fortran~90 subroutine which minimises a function of the
saturation property variables for values of the couplings, such that
the couplings found using this method produce saturation properties
for any EOS that match the constraints with great precision.\par
This process is suitable for solving all the models presented in this
thesis; the model-specific details do not require changes to the
method of solving the equations. This provides us with enough data to
investigate fully the properties of infinite matter. To investigate
compact stellar objects however, we require a further calculation.
\section{Runge--Kutta Integration}\label{sec:rungekutta}
In order to calculate the properties of compact stellar objects, we
need to solve the Tolman--Oppenheimer--Volkoff Equations, as described
in Section~\ref{sec:stellarmatter} and derived in
Appendix~\ref{sec:tovderiv}. These equations provide a connection
between the infinite matter equation of state and the properties of
macroscopic stellar objects which can support the enormous
gravitational pressures involved without collapsing.\par
The relation between the state variables of the EOS; the energy
density ${\cal E}$ and the pressure $P$; and the the mass-radius
relationship of a compact stellar object is determined by the pressure
gradient
%
%
\be \label{eq:RKOV} \frac{dP}{dr} = -\frac{G\left(P/c^2+{\cal
    E}\right)\left(M(r)+4 \pi r^{3}P/c^2\right)}{r(r-2G M(r)/c^2)}. \ee
In order to make use of this, we will need to integrate this equation
to find the pressure at some radius $R'$. We use fourth-order
Runge--Kutta integration, which provides a useful scheme for
integrating since it only requires a small number of function
evaluations.\par
Fourth-order Runge--Kutta integration is defined by the following
iterative procedure; If the differential equation $y' = G(x,y)$ is to
be integrated to give
\be y(b) = y_0 + \int_{0}^{b} G(x,y)dx\ee
with initial values and intervals of 
\be \label{eq:init} y_0 = y(0), \quad x_0 = 0, \quad x_i = i\, \delta
x, \ee
where $b = n\delta x$, then we can define the Runge--Kutta steps as
\bea
\nonumber
RK1_i &=& G\left(x_i,y_i\right), \\[2mm] \nonumber
RK2_i &=&  G\left(x_i + \frac{\delta x}{2},\ y_i+\frac{\delta x}{2}\,
RK1_i\right), 
\\[2mm] \nonumber
RK3_i &=& G\left(x_i + \frac{\delta x}{2},\ y_i + \frac{\delta x}{2}\,
RK2_i\right), \\[2mm]
RK4_i &=& G\left(x_i + \delta x, \ y_i+\delta x\, RK3_i\right),
\eea
where $\delta x$ is a small shift in $x$ that will determine the
resolution to which we wish to know $y(x)$. The values $RK1_i$ and
$RK4_i$ are the slopes of tangents to the function $y$ at the points
$x_i$ and $x_i + \delta x$ respectively, while $RK2_i$ and $RK3_i$ are
the slopes of $y$ at the midpoint $x_i + \delta x/2$, where the $y$
value of the former is defined by the Euler method\footnote{Euler's
  method provides a very simple approximation, utilising the first two
  terms of a Taylor expansion of $y$ to solve \mbox{$y'(t) =
    f(t,y(t))$}, thus \mbox{$y_{n+1} = y_n + h f(t_n,y_n)$}.} using
$RK1_i$, and the latter defined by the Euler method using $RK2_i$.\par
The next value of the function, located at $x_{i+1} = (i+1)\, \delta
x$, is then defined using a modification of Simpson's Rule, as
\be y_{i+1} = y_i + \frac{\delta x}{6}\, \left(RK1_i + 2\, RK2_i + 2\,
RK3_i + RK4_i\right).  \ee
By performing this procedure iteratively we can integrate a
differential equation until after $n$ steps we reach the upper
integration limit of $x_n = b$.\par
For our purposes, we select a central density of a stellar object
$\rho_{\rm central}$ as the control parameter. The EOS data calculated
for a particular model then provides the energy density and pressure
at the centre of the star.\par
Assuming spherical symmetry, we require that the pressure gradient
$dP/dr$ at the centre of the star ($r=0$) be zero, thus $y'(0) =
0$. By using Runge--Kutta integration, we can evaluate the pressure at
the boundary of a sphere of matter with a radius $r=R'$. By relating
this pressure to the EOS we can also find the energy density at this
boundary, ${\cal E}(R')$.\par
By performing this integration for larger and larger values of $r=R'$
(since Eq.~(\ref{eq:RKOV}) describes a negative pressure gradient,
this results in smaller and smaller values of pressure $P$) we
eventually find some value $r=R$ for which the integral of
Eq.~(\ref{eq:RKOV}) becomes negative, i.e. $y_{i+1} < 0$; thus at a
distance of approximately $r=R$ (to within the resolution of $\delta
x$) the pressure is zero, and we define this point as the edge of the
star, and hence the radius.\par
Since we have also collected data for ${\cal E}(r)$, we can calculate
the mass residing within some radius $R'$ as
\be M(R') = \int_0^{R'} 4 \pi r^{2} {\cal E}(r) \; dr, \ee
and thus the total mass of the star is defined by $M(R)$, where $R$ is
the radius of the star. With this information we can investigate the
mass-radius relation for a star with a given central
density. Repeating this calculation for several values of $\rho_{\rm
  central}$ provides a locus of values as shall be shown, for example,
in Section~\ref{sec:QHDSTARS}. We can also investigate other
properties of a star as a function of internal radius by calculating
them at various values of $x_i = r$ in Eq.~(\ref{eq:init}).
\section{Phase Transitions}\label{sec:phasetransmethod}
In order to combine the EOS for hadronic and quark matter (since we
will be using a Gibbs transition, refer to
Section~\ref{sec:phasetransitions}) we require a method of calculating
the phase transitions from hadronic matter to a mixed phase, and from
a mixed phase to quark matter. Following the requirements described in
Section~\ref{sec:MixedPhase} we calculate the EOS of hadronic matter
with control parameter $\rho_{\rm total}$, while at each density we
use $\mu_n$ and $\mu_e$ as inputs to the quark matter EOS (for a given
model of quark matter). At each density we can calculate the pressure
for the hadronic phase $P_H$ and the quark phase $P_Q$ and compare
these values. At the point (if it exists) where $P_H = P_Q$ we
consider the models to be in equilibrium.\par
We then change the control parameter in our code from $\rho_{\rm
  total}$ to $\chi$ (parameterizing the quark fraction in the mixed
phase, and acting as the order parameter between the phases) at which
point we have $\chi=0$, still purely hadronic matter. We can then
increase $\chi$ by $\delta\chi$ and calculate properties of the mixed
phase such as the total baryon density and energy density with
weightings of $\chi$ and $(1-\chi)$ for the quark and hadronic phases
respectively, with the condition that the chemical potentials are
still equal for each phase, and the pressures are equal for each
phase.\par
We continue to increase $\chi$ until we reach $\chi=1$ at which point
the EOS is described entirely by quark matter, and thus the mixed
phase is in equilibrium with a pure quark phase. From this point on,
we return to using $\rho_{\rm total}$ as the control parameter and
increasing this value by $\delta\rho$ up to some arbitrary density.
\cleardoublepage

%% file: Chapter5_results.tex
\chapter{Results}\label{sec:results}
Within this chapter we will once again use the term `configuration of
a model' to indicate differences (such as types of particles included
or neglected) within a particular model. Some of the most interesting
and important numerical results are shown, though results for all
calculations  for all configurations will not be shown, due to the
overwhelmingly large number of possibilities that exist. A summary
table is provided in Section~\ref{sec:summary} which contains many
numerical results of interest.\par
\section{QHD Equation of State}\label{sec:QHDEOS}
Though QHD has been studied extensively, and many excellent summaries
exist (e.g. Ref.~\cite{Serot:1984ey}), we will present the results of
our QHD calculations for the purpose of comparison in later sections,
and to verify that our results do indeed reproduce the established
results.\par
\subsection{QHD Infinite Matter}\label{sec:QHDINF}
\begin{figure}[!t]
\centering
\includegraphics[angle=90,width=0.9\textwidth]{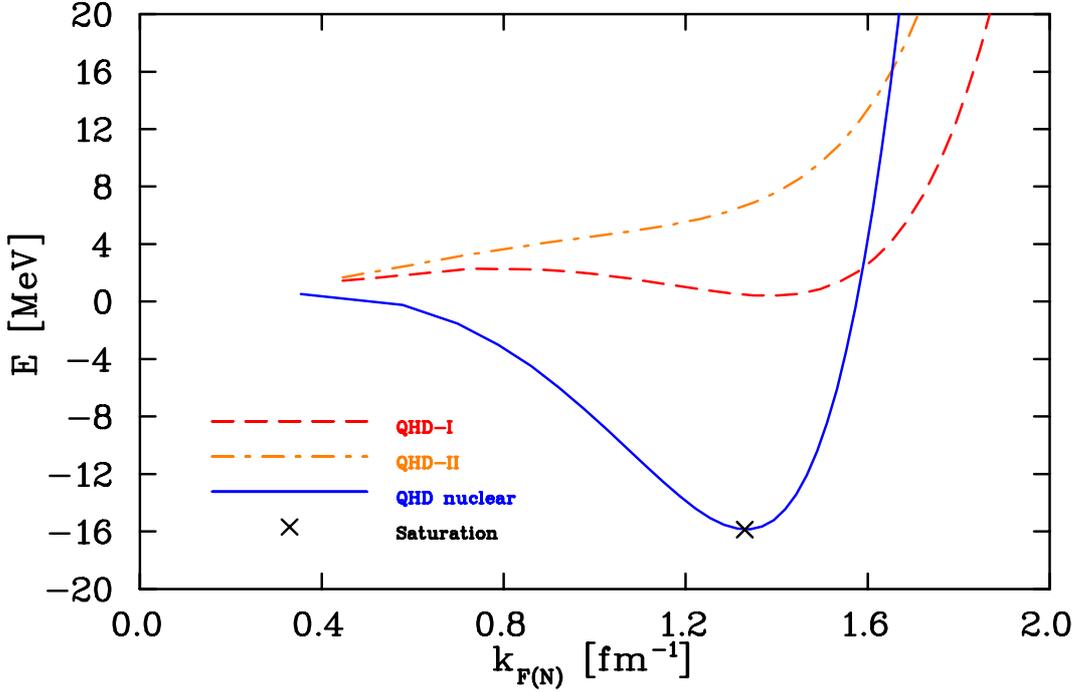}
\caption[Energy per baryon for nucleonic QHD]{(Color Online) Energy
  per baryon for nucleonic QHD-I, QHD-II, and nuclear QHD which
  demonstrates saturation at the correct value of $E$ as defined in
  Eq.~(\ref{eq:EperA}), and value of $k_F$ (here we use the neutron
  Fermi momentum, as it is common to all three configurations)
  corresponding to $\rho_0$ as per Eq.~(\ref{eq:rho}). The
  reproduction of this value occurs by construction via the use of
  appropriate couplings $g_{N\s}$ and $g_{N\w}$ as described in
  Section~\ref{sec:infmattermethod}. \protect\label{fig:EperA_QHD}}
\end{figure}

To obtain numerical results, we solve the meson field equations,
Eqs.~(\ref{eq:MFsigma})--(\ref{eq:MFrho}), with the conditions of
charge neutrality and fixed baryon density for various configurations
of models. For configurations involving leptons, we include the
condition of equivalence of chemical potentials given by
Eq.~(\ref{eq:equilconds}). The energy per baryon given by
Eq.~(\ref{eq:EperA}) for various configurations of nucleonic QHD (in
which QHD-I neglects contributions from $\rho$ mesons, QHD-II includes
these contributions, nuclear QHD contains equal proportions of protons
and neutrons, and in each case we do not model leptons) are shown in
Fig.~\ref{fig:EperA_QHD} and we see that the saturation of nuclear
matter occurs at the correct value of $k_F$ (here we use the neutron
Fermi momentum, as it is common to all three configurations; for the
relation between Fermi momentum and density, refer to
Section~\ref{sec:kf}) corresponding to $\rho_0$ as per
Eq.~(\ref{eq:rho}) by construction via the use of appropriate
couplings $g_{N\s}$ and $g_{N\w}$.\par
The compression modulus for symmetric nuclear matter, as defined by
the curvature at saturation, refer to Eq.~(\ref{eq:Kmod}) is found to
be $K = 525~{\rm MeV}$, which is in agreement with
Ref.~\cite{Serot:1984ey}, but as stated in that reference, not with
experiment. This point will be discussed further in
Sec.~\ref{sec:QMCINF}. We further note that the pressure of nuclear
matter calculated via Eq.~(\ref{eq:P_H}) is zero at the saturation
point, as predicted by Eq.~(\ref{eq:Panalytic}).\par
The case of `symmetric' nuclear matter\emdash in which we include
protons, but not $\beta$-equilibrium with leptons\emdash is of course
purely academic, since this is infinite matter, and we are therefore
considering an infinite charge. The saturation that occurs in this
case and the binding (a negative energy per baryon indicates a binding
energy) is considered to resemble heavy finite nuclei.\par
\begin{figure}[!t]
\centering
\includegraphics[angle=90,width=0.9\textwidth]{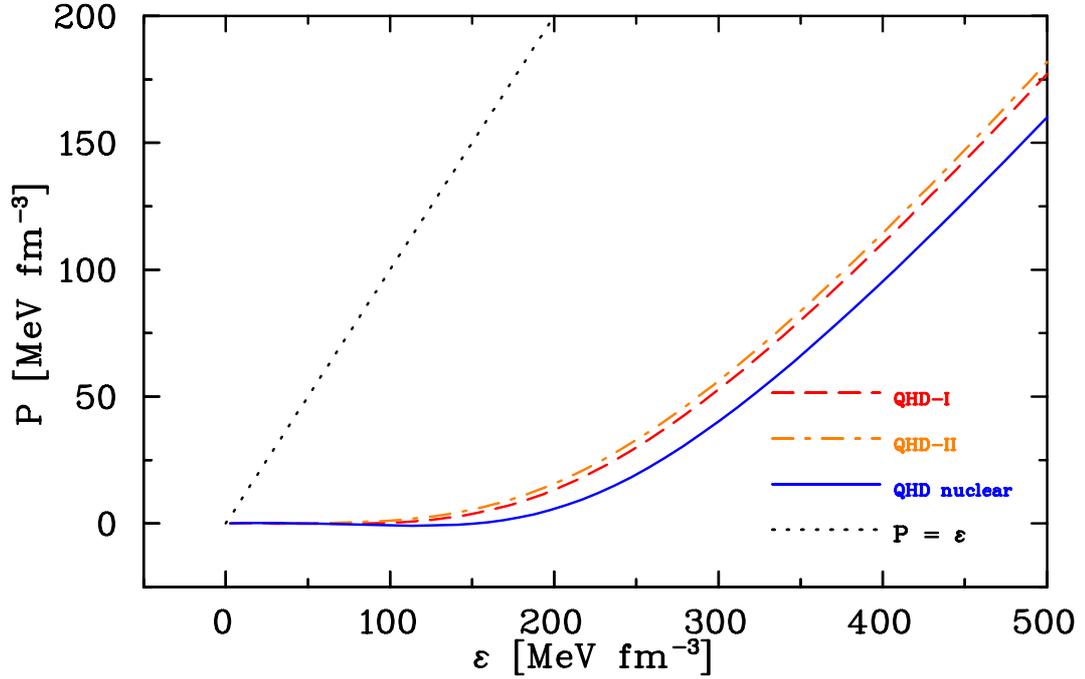}
\caption[Low density Equation of State]{(Color Online) Low density EOS
  for nucleonic QHD, as well as the causal limit of $P={\cal E}$ in
  which the speed of sound in-medium equals the speed of light. Here
  we see that the low density EOS are rather similar, independent of
  which configuration of QHD is considered. The common feature however
  is that the low density EOS are rather soft, in that it does not yet
  approach the limit of $P = {\cal E}$. The EOS for all densities for
  these configurations are shown in
  Fig.~\ref{fig:EOS_plusmaxwell}. \protect\label{fig:EOSpe}}
\end{figure}

The low density EOS for nucleonic QHD is shown in Fig.~\ref{fig:EOSpe}
in which we see that the low density EOS is rather soft (does not
approach the limit of $P = {\cal E}$; the stiffest possible EOS in
which the speed of sound equals that of light), and nearly
configuration independent in that QHD-I produces similar results to
that of nuclear QHD. At higher densities, the EOS for all
configurations approach the limit.\par
\begin{figure}[!t]
\centering
\includegraphics[angle=90,width=0.9\textwidth]{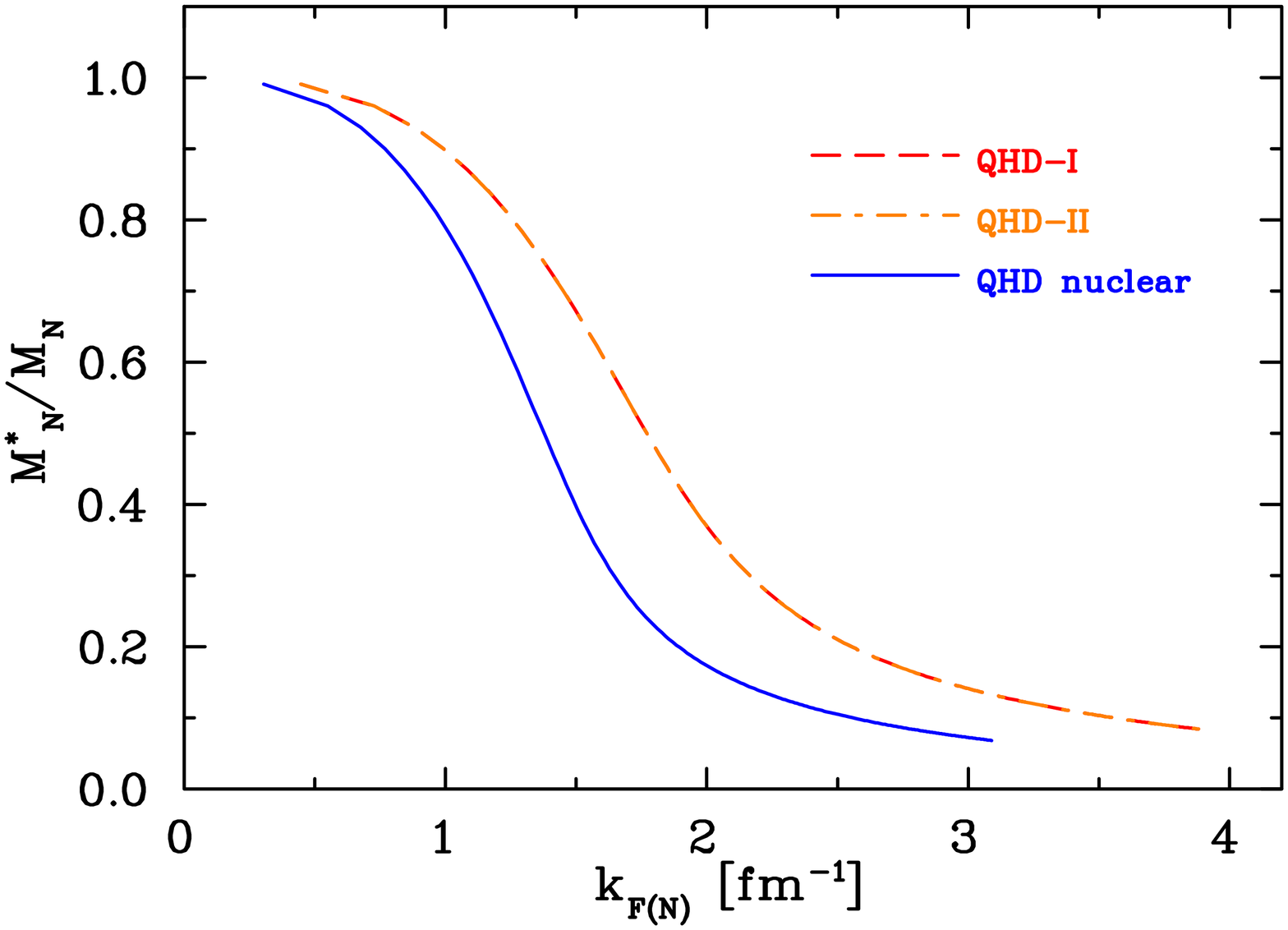}
\caption[Neutron effective masses in nucleonic QHD]{(Color Online)
  Neutron effective masses for the various configurations of nucleonic
  QHD. Note that the curves for neutron matter (QHD-I and QHD-II) are
  identical; the addition of the $\rho$ meson contribution (a vector
  meson) does not contribute to the effective mass, as this is a
  purely scalar effect. Although the form of the effective mass is
  linear in Fig.~\ref{fig:Mstar_QHD}, in that case the effective mass
  is plotted against the scalar self-energy, which is shown in
  Fig~\ref{fig:HartreeSelfEnergies}, and we note that this quantity is
  non-linear, which determines the shape of the curve
  here. \protect\label{fig:MstarQHDN}}
\end{figure}

\begin{figure}[!b]
\centering
\includegraphics[angle=90,width=0.9\textwidth]{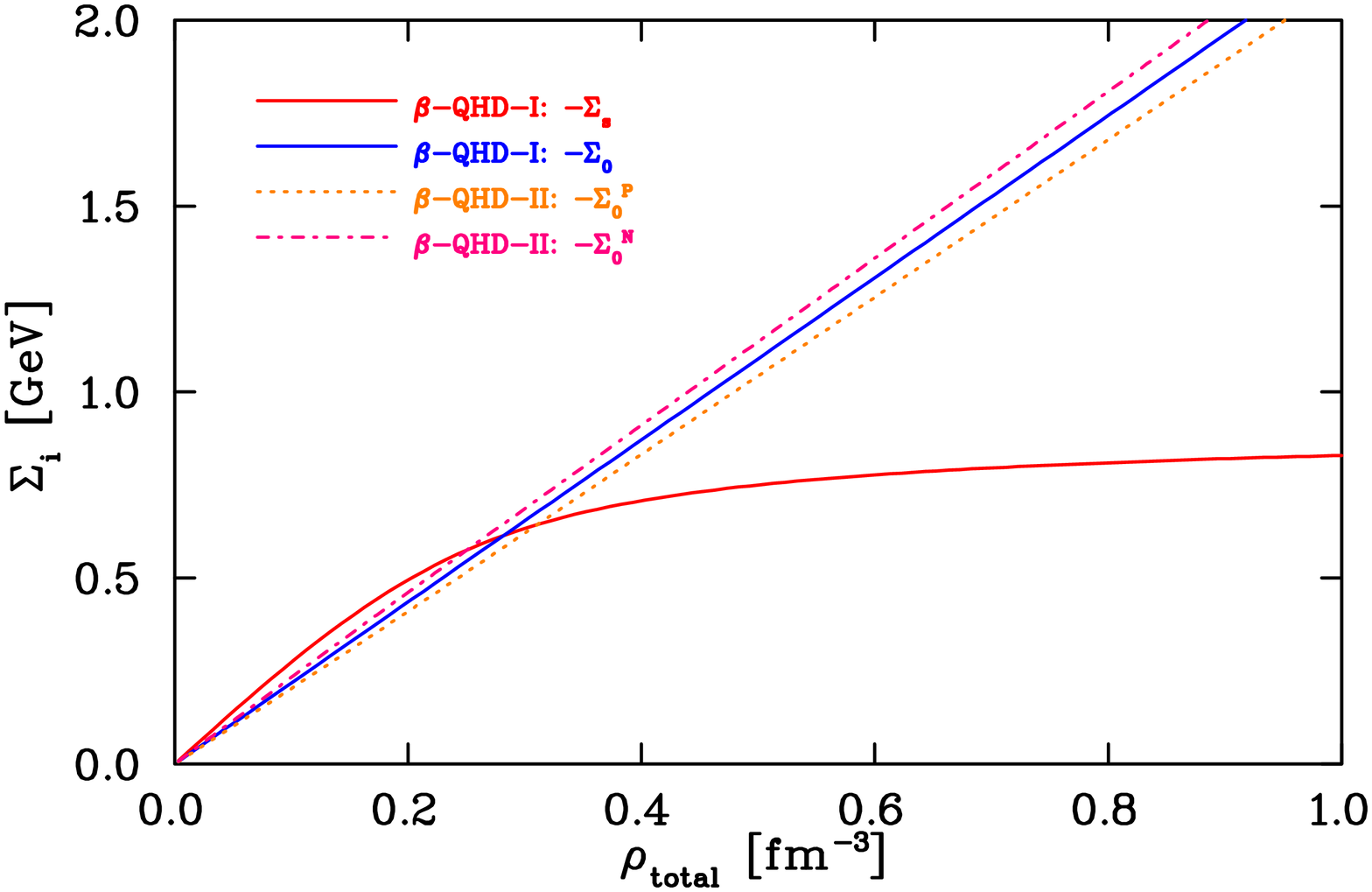}
\caption[$\beta$-equilibrium QHD-I and QHD-II self-energies ]{(Color
  Online) Self-energies for $\beta$-equilibrium QHD-I and QHD-II.
  Nuclear QHD has the same self-energies as QHD-I since the $\rho$
  meson does not contribute; it is proportional to the asymmetry
  between proton and neutron densities. Here the density is shown up
  to $1.0~{\rm fm}^{-3}$ after which the curves become linear. Note
  the significance of the shape of the $\Sigma_s$ curve to that of the
  effective mass as shown in Fig.~\ref{fig:MstarQHDN}. Also note that
  in Hartree-level QHD, $\Sigma_v = 0$ for all baryons.
  \protect\label{fig:HartreeSelfEnergies}}
\end{figure}

The effective masses for the various configurations of nucleonic QHD
are shown in Fig.~\ref{fig:MstarQHDN}, for which we note that the
neutron matter curves are identical, as the $\rho$ meson (a vector
meson) provides no contribution to the effective mass, which is a
purely scalar effect. Although the form of the effective mass is
linear in Fig.~\ref{fig:Mstar_QHD}, in that case the effective mass is
plotted against the scalar self-energy, which is shown in
Fig~\ref{fig:HartreeSelfEnergies}, and we note that this quantity is
non-linear, which determines the shape of the curve in
Fig.~\ref{fig:MstarQHDN}. For the case of nuclear QHD we find values
of the effective mass at saturation of the neutrons to be
$(M_n^*/M_n)_{\rm sat} = 0.56$, corresponding to an effective mass of
$M_n^* = 526~{\rm MeV}$.\par
We can investigate the balance between the various meson fields by
examining the self-energy contributions, which are defined explicitly
for QHD in Eq.~(\ref{eq:selfenergy}). These are shown in
Fig.~\ref{fig:HartreeSelfEnergies} for QHD-I and QHD-II both in
$\beta$-equilibrium (the self-energies for nuclear QHD are the same as
those for QHD-I, since the $\rho$ meson does not contribute to nuclear
QHD as its contribution is proportional to the asymmetry between
proton and neutron densities). From this figure it is clear that there
exists an important balance between the scalar and vector
interactions. Note the significance of the shape of the $\Sigma_s$
curve to that of the effective mass as shown in
Fig.~\ref{fig:MstarQHDN}. Also note that in Hartree-level QHD,
$\Sigma_v = 0$ for all baryons.\par
\begin{figure}[!t]
\centering
\includegraphics[angle=90,width=0.9\textwidth]{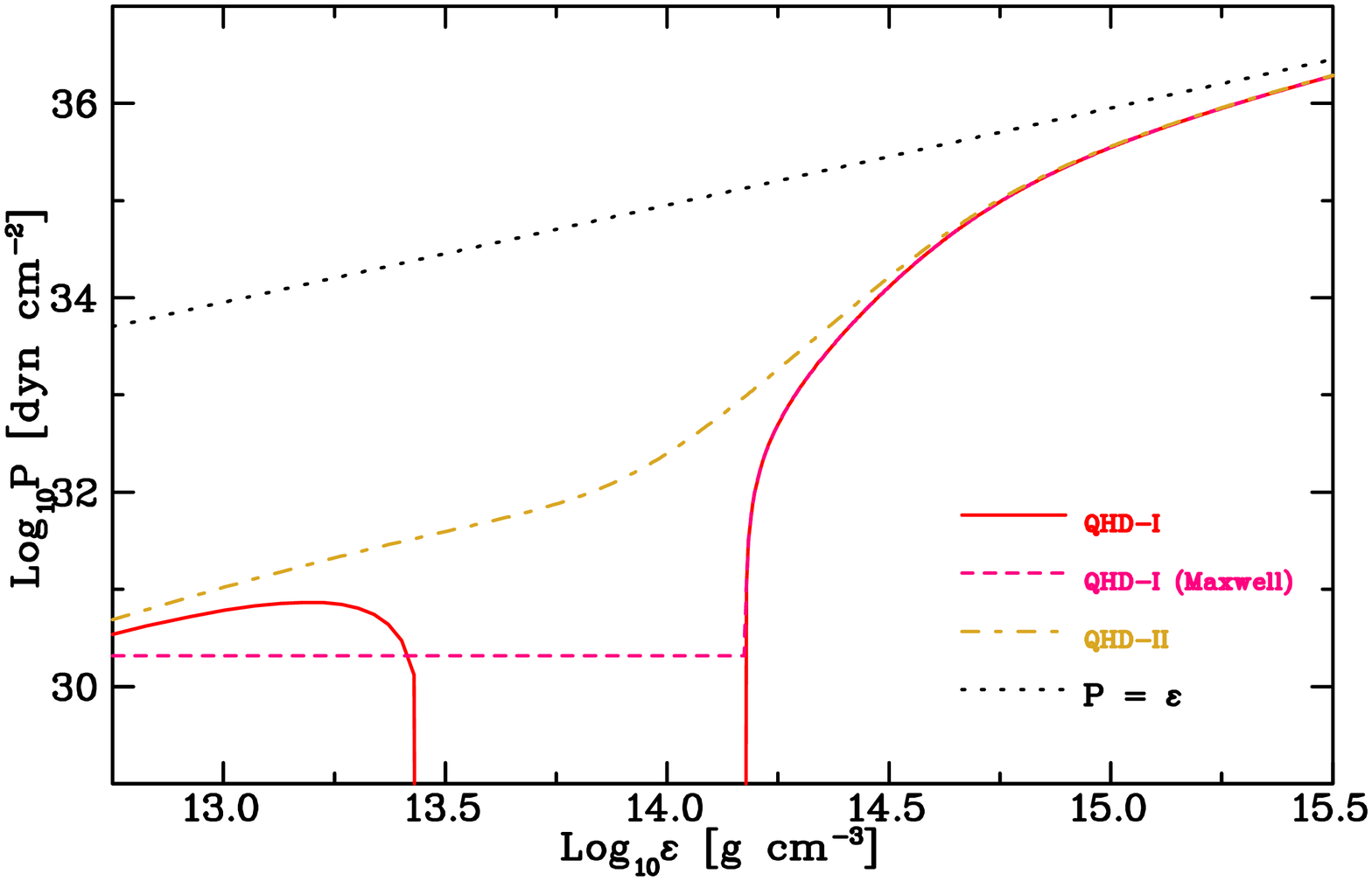}
\caption[Log-log EOS for nucleon QHD]{(Color Online) Log-log EOS for
  nucleonic QHD-I and QHD-II. QHD-I produces features similar to that
  of a van der Waals EOS, which can be interpreted as a liquid-gas
  phase transition. The addition of the $\rho$ meson contribution
  removes this transition. A Maxwell construction has been performed
  (using the transition pressure from Ref.~\cite{Serot:1984ey}) to
  produce a constant pressure transition between the phases. Note that
  at high densities (large values of ${\cal E}$) all the
  configurations approach the limit of $P={\cal
    E}$. \protect\label{fig:EOS_plusmaxwell}}
\end{figure}

A log-log graph of the EOS for nucleonic QHD in which we show a
Maxwell transition for the van der Waals style liquid-gas phase
transition in QHD-I is shown in Fig.~\ref{fig:EOS_plusmaxwell}. The
transition pressure is taken from Ref.~\cite{Serot:1984ey}. In this
form of a transition (in contrast to that of
Section~\ref{sec:phasetransitions}) the pressure is constant between
the phases (isobaric transition) and the density of the combined EOS
becomes disjoint. This example refers to a liquid-gas style phase
transition, where later we will discuss phase transitions between
phases with different degrees of freedom. The inclusion of the $\rho$
meson removes this transition.\par
\begin{figure}[!b]
\begin{tabular}[h!tb]{cc}
\begin{minipage}[c]{0.45\textwidth}
\centering
\includegraphics[angle=90,width=\textwidth]{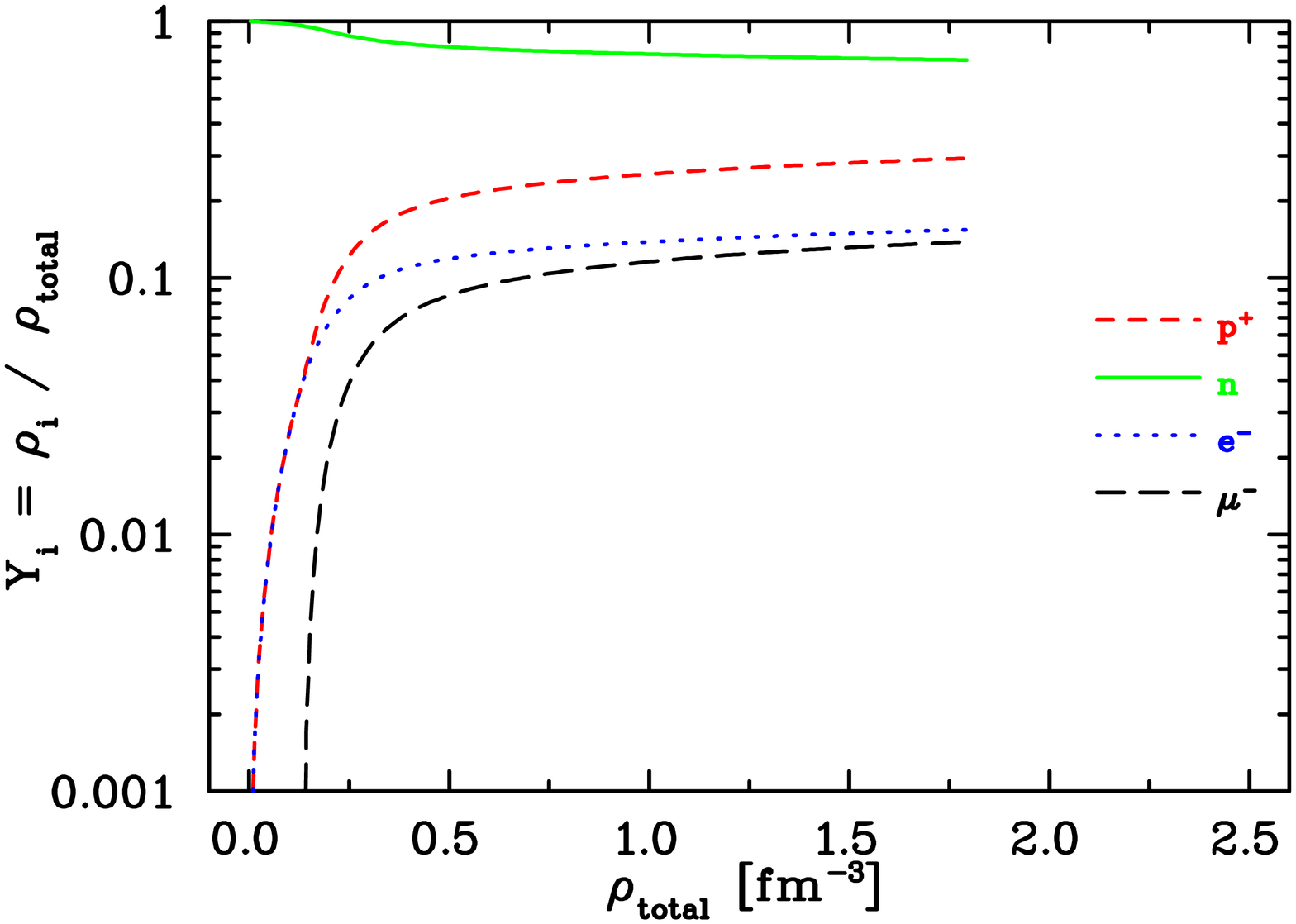}
\end{minipage}
&
\begin{minipage}[c]{0.45\textwidth}
\centering
\includegraphics[angle=90,width=\textwidth]{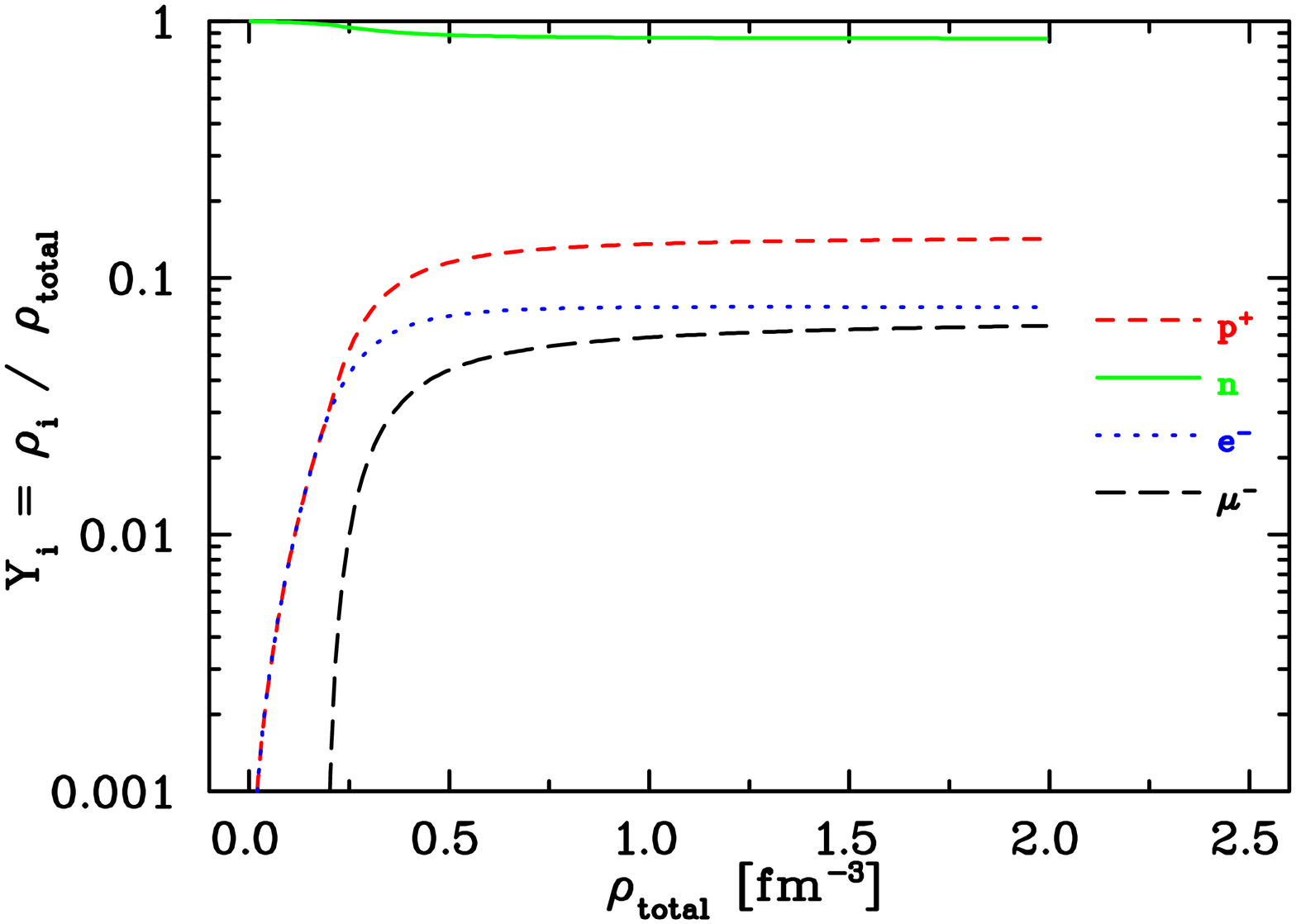}
\end{minipage}
\end{tabular}
\caption[Species fractions for nucleonic QHD in $\beta$-equilibrium
  with leptons] {(Color Online) Species fractions $Y_i$ for nucleonic
  QHD in $\beta$-equilibrium with leptons $\ell=e^-,\mu^-$, including
  (left) and lacking (right) a contribution from the $\rho$ meson. In
  both cases, at low densities the system is composed of nearly
  entirely neutron matter, but at higher densities the contribution
  from protons become non-negligible.}
\label{fig:CNem_rhonorho}
\end{figure}

For nuclear QHD, the species fractions (refer to Eq.~(\ref{eq:Y})) are
by definition equal for the protons and neutrons, i.e. $Y_p = Y_n =
0.5$. For QHD-I and QHD-II the neutrons provide the only baryonic
contribution. If we include protons and leptons ($\ell = e^-, \mu^-$)
we can investigate the effects of $\beta$-equilibrium matter with
global charge neutrality. The species fractions $Y_i$ are shown in
Fig.~\ref{fig:CNem_rhonorho} for a configuration with (and without)
contributions from $\rho$ mesons. In each case, at low densities the
system is composed of nearly entirely neutron matter, but we can see
that at higher densities the contributions from protons become
non-negligible.\par
We do not show any results for the addition of hyperons to the QHD
EOS. The QHD model as described in Section~\ref{sec:qhd} is derived
assuming a weak-field limit (in order to define the MFA; refer to
Section~\ref{sec:mfa}) in order to define the Dirac equation for the
baryons. In this limit, where the meson fields are assumed to be small
in magnitude (weak), we can ignore the effects of antibaryons, since
it is not possible in this limit for the creation of
particle--antiparticle pairs via vacuum fluctuations\footnote{This is
  a well known limitation of the Dirac equation, and has been
  discussed in detail, for example in the case of electron scattering
  from a potential barrier, resulting in the `Klein
  Paradox'~\cite{Klein}.}. The inclusion of hyperons causes the
magnitudes of the fields\emdash particular the scalar field, as can be
seen by examining the form of this in Eq.~(\ref{eq:QHDscalarfield}) in
which we sum over \emph{all} baryons\emdash to become large, and thus
the weak-field assumption becomes violated. We can observe that this
is indeed the case by examining the effective masses. We observe that
when the hyperons are included, the baryon effective masses for
several of the baryons become negative, indicating that the scalar
self-energy (defined by the mean-scalar field as per
Eq.~(\ref{eq:QHDeffM})) has become larger than the vacuum baryon
mass. We consider this to be a breakdown of the model, and thus we do
not perform calculations where this occurs. Calculations for QHD can
be performed prior to this breakdown, but we elect to not present any
of these results, as they represent a model which we consider to be
inaccurate. We shall therefore consider it not possible to include
hyperons into QHD in an interesting manner for our purposes. The issue
of negative effective masses is however remedied in QMC as will be
shown later.\par
\clearpage

\subsection{QHD Stars}\label{sec:QHDSTARS}
Having found the QHD EOS by evaluating the energy density,
Eq.~(\ref{eq:E_H}), and pressure, Eq.~(\ref{eq:P_H}), we can calculate
properties of stellar objects (which we shall refer to as 'stellar
solutions') based on this EOS, using the TOV equation (refer to
Section~\ref{sec:stellarmatter}). The radius of a star is defined as
the radius at which the pressure is zero and is calculated using a
fourth-order Runge--Kutta integration method (refer to
Section~\ref{sec:rungekutta}).\par
\begin{figure}[!h]
\centering
\includegraphics[angle=90,width=0.9\textwidth]{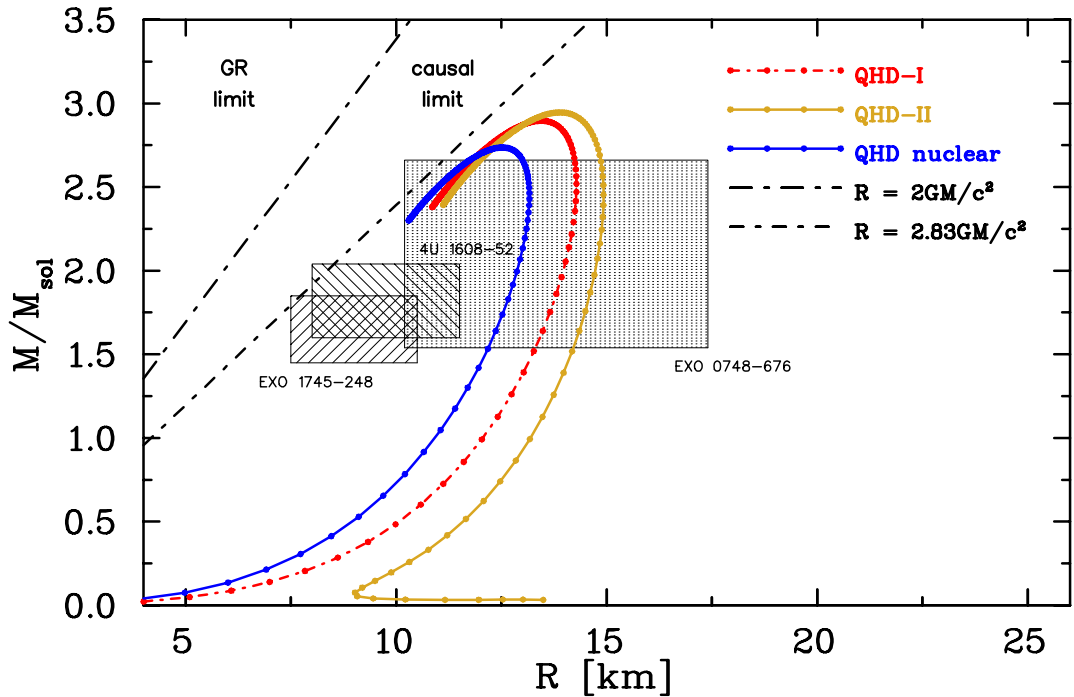}
\caption[Mass-radius relations for nucleonic QHD]{(Color Online)
  Mass-radius relation for various configurations of nucleonic
  QHD. Also shown are the theoretical limits due to General Relativity
  and causality (where $G$ is the gravitational constant, $c$ is the
  speed of light), as well as the $2\s$ error bounds from various
  experiments~\cite{Ozel:2006bv,Guver:2008gc,Ozel:2008kb} which
  illustrate how loosely the requirements of an EOS are currently
  defined. Note that the low-density shape of the curve changes with
  the addition of the $\rho$ meson in
  QHD-II.\protect\label{fig:MvsRQHDN}}
\end{figure}

The mass-radius relations for various configurations of nucleonic QHD
are shown in Fig.~\ref{fig:MvsRQHDN} along with the $2\s$
results\footnote{The shaded areas for \mbox{EXO 1745-248} and \mbox{4U
    1608-52} represent a conservative reproduction of the $2\s$ data
  from the relevant reference. The shaded area for \mbox{EXO 0748-676}
  represents the central data point plus error bars expanded to a
  rectangular area.} of various
experiments~\cite{Ozel:2006bv,Guver:2008gc,Ozel:2008kb}\emdash which
detail the only measurements to date of the radii of neutron stars
along with their masses, namely for the stars denoted by \mbox{EXO
  0748-676}, \mbox{4U 1608-52}, and \mbox{EXO 1745-248}\emdash for
comparison. The authors of Ref.~\cite{Ozel:2006bv} claim that their
findings (corresponding to \mbox{EXO 0748-676}) rule out soft
equations of state, though as noted in a response to that
paper~\cite{Alford:2006vz} and in this work, the $2\s$ data admits a
wide variety of EOS. The results of Ref.~\cite{Ozel:2006bv} should
likely be discounted, since developments in the field have shown
possible large errors with that experiment~\cite{Galloway:2007dn}.\par
\vfill
We note that the shape of the low-central-density relation changes
with the addition of the $\rho$ meson producing QHD-II, which should
be expected, given the differences between the QHD-I and QHD-II EOS as
shown in Fig.~\ref{fig:EOS_plusmaxwell}.\par
\begin{figure}[!b]
\centering
\includegraphics[angle=90,width=0.9\textwidth]{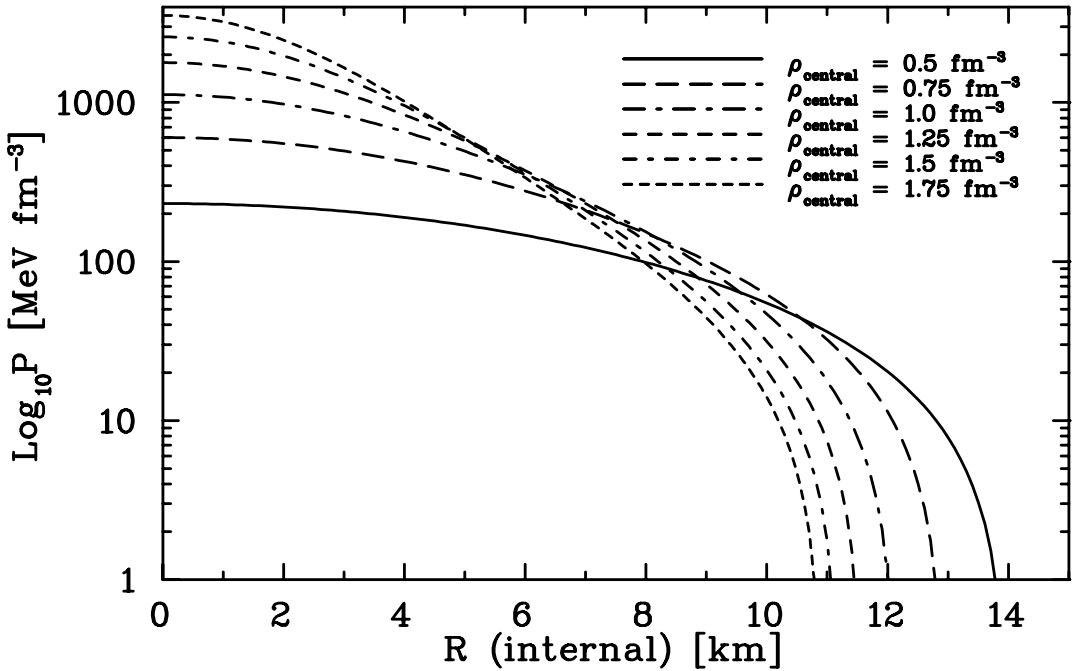}
\caption[Pressure vs Radius for QHD-I]{(Log) pressure vs. internal
  radius for stars with various central densities for QHD-I. Note that
  the stars with a larger central density have a larger central
  pressure, and a smaller total radius. The data for $\rho_{\rm
    central} = 0.5~{\rm fm}^{-3}$ corresponds to a star with $M =
  2.84~M_\odot$ and $R = 13.96~{\rm km}$, while the data for
  $\rho_{\rm central} = 1.75~{\rm fm}^{-3}$ corresponds to a star with
  $M = 2.39~M_\odot$ and $R = 10.9~{\rm km}$, both of which correspond
  to the high-central-density positive-gradient section of the
  relevant curve in Fig.~\ref{fig:MvsRQHDN}.
  \protect\label{fig:PvsRadQHDI}}
\end{figure}

We show (as an example, for QHD-I) the relation between pressure and
internal radius for a variety of stars with different central
densities in Fig.~\ref{fig:PvsRadQHDI}. We note the trend that a star
with a larger central density has a greater central pressure, and a
smaller total radius. Each of the curves in that figure correspond to
the high-central-density positive-gradient section of the relevant
curve in Fig.~\ref{fig:MvsRQHDN}.\par
\begin{figure}[!b]
\centering
\includegraphics[angle=90,width=0.9\textwidth]{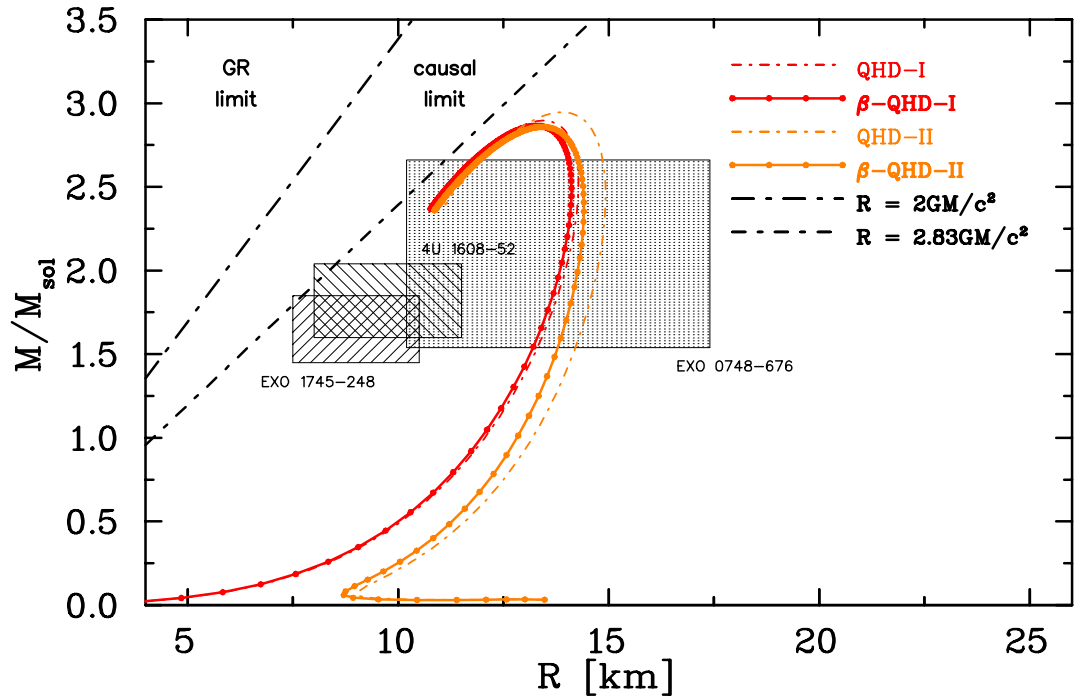}
\caption[Mass-radius relations for $\beta$-equilibrium nucleonic QHD
  matter]{(Color Online) Mass-radius relations for QHD-I and QHD-II in
  $\beta$-equilibrium, with the neutrons-only relations (as shown in
  Fig.~\ref{fig:MvsRQHDN}) for comparison. Note that the effect of
  including protons and leptons does not have such a pronounced effect
  on QHD-I, since in that case the protons do not provide such a large
  contribution, as can be seen in Fig.~\ref{fig:CNem_rhonorho}. The
  similarities between the stellar solutions for each configuration
  emphasises the difficulty of using the mass and radius alone to
  constrain a model.\protect\label{fig:TOV_leptons}}
\end{figure}

If we now include the leptons to our calculations, we can observe the
effect that these have on the mass-radius relations, as shown in
Fig.~\ref{fig:TOV_leptons} for $\beta$-equilibrium nucleonic
matter. While the QHD-I and $\beta$ QHD-I configurations may not be
dramatically different, the similarities of the stellar solutions for
these configurations demonstrates the difficulties that exist in
determining the content of stars from these two parameters alone.\par
\clearpage

\section{QMC Equation of State}\label{sec:QMCEOS}
By including the effect of the quark content of baryons we can
investigate the impact on various quantities, using the QMC model as
described in Sec.~\ref{sec:qmc}. We will see that this has profound
consequences for the EOS. Many of the results presented in this
section are published by the author~\cite{Carroll:2008sv} as a unique
investigation of the octet QMC model.\par
\subsection{QMC Infinite Matter}\label{sec:QMCINF}
\begin{figure}[!b]
\centering
\includegraphics[angle=90,width=0.9\textwidth]{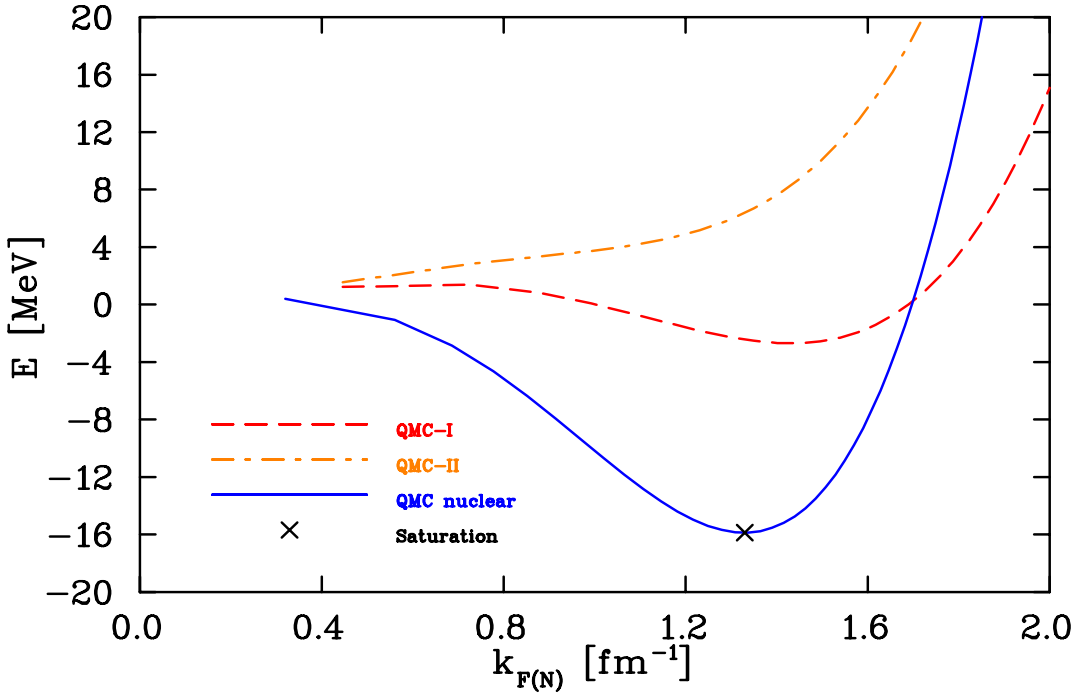}
\caption[Energy per baryon for nucleonic QMC]{(Color Online) Energy
  per baryon for nucleonic QMC-I, QMC-II, and nuclear QMC which
  demonstrates saturation at the correct value of $E$ as defined in
  Eq.~(\ref{eq:EperA}), and correct value of $k_F$. For comparison to
  Fig.~\ref{fig:EperA_QHD} we use the same notation to differentiate
  between the configurations. The reproduction of the saturation value
  occurs by construction via the use of appropriate couplings
  $g_{N\s}$ and $g_{N\w}$ which have been re-calculated for
  QMC. \protect\label{fig:EperA_QMC}}
\end{figure}

The energy per baryon curves for QMC are shown in
Fig.~\ref{fig:EperA_QMC} where, for the sake of comparison to
Fig.~\ref{fig:EperA_QHD} we have used the same notation for
distinguishing the configurations, thus QMC-I neglects the
contribution of the $\rho$ meson, QMC-II includes it, and QMC nuclear
restricts the species fractions of protons and neutrons to be
equal. We neglect the contributions of leptons for these simple
configurations. The couplings of the baryons to the mesons are found
such that once again the appropriate saturation properties are
reproduced.\par
\begin{figure}[!b]
\centering
\includegraphics[angle=90,width=0.9\textwidth]{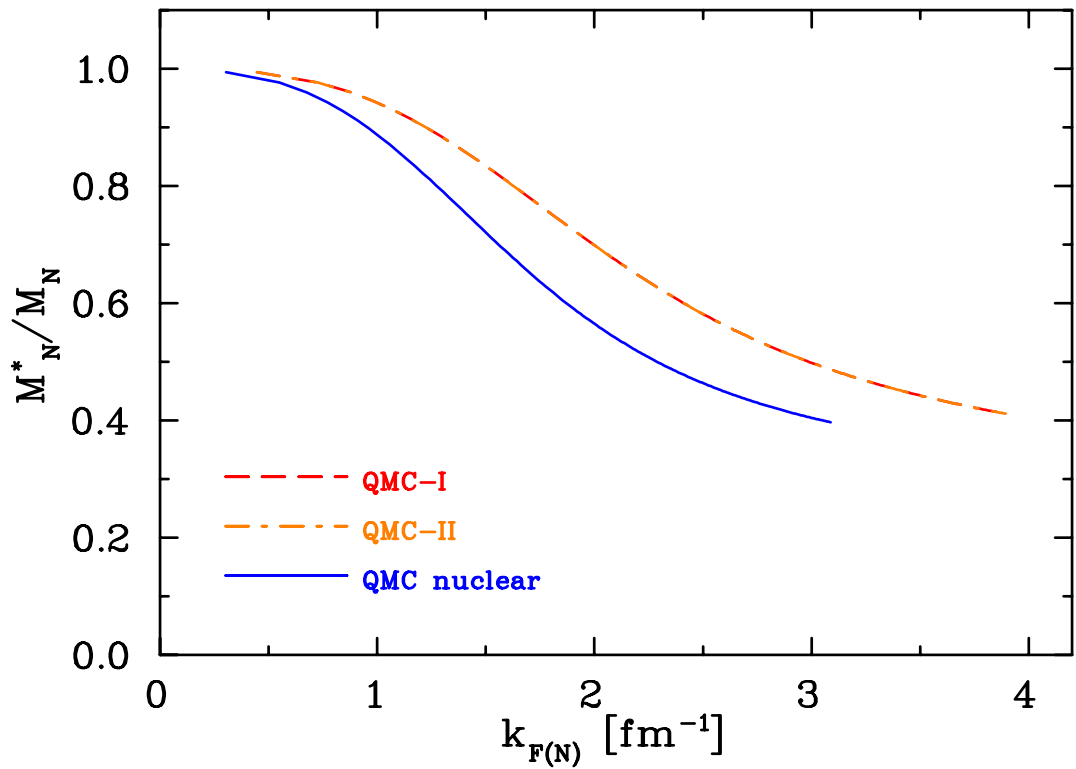}
\caption[Effective neutron masses in nucleonic QMC]{(Color Online)
  Effective neutron masses for the various configurations of nucleonic
  QMC in which the effective masses have a quadratic form. Note the
  subtle differences in shape between these curves and those of
  Fig.~\ref{fig:MstarQHDN}, particularly the value of the effective
  mass at saturation. \protect\label{fig:MstarQMCN}}
\end{figure}

An interesting feature to note for Fig.~\ref{fig:EperA_QMC} is that
the curvature for nuclear QMC at saturation is less than that of the
same curve in Fig.~\ref{fig:EperA_QHD} indicating that the compression
modulus for QMC takes a smaller value. This is of particular interest,
since the value of the compression modulus for QHD is known to be too
large~\cite{Serot:1984ey}, which is typical of models that neglect
quark level interactions. For QMC we find a significant improvement in
the compression modulus; $K = 280~{\rm MeV}$ which lies at the upper
end of the experimental range. The nucleon effective mass at
saturation for QMC is found to be $(M^*)_{\rm sat} = 735~{\rm MeV}$,
producing $(M^*/M)_{\rm sat} = 0.78$. The effective masses for the
various configurations of nucleonic QMC are shown in
Fig.~\ref{fig:MstarQMCN} and we see that these effective masses do
indeed remain positive, as opposed to the negative effective masses
we encountered in QHD.\par
\begin{figure}[!t]
\centering
\includegraphics[angle=90,width=0.9\textwidth]{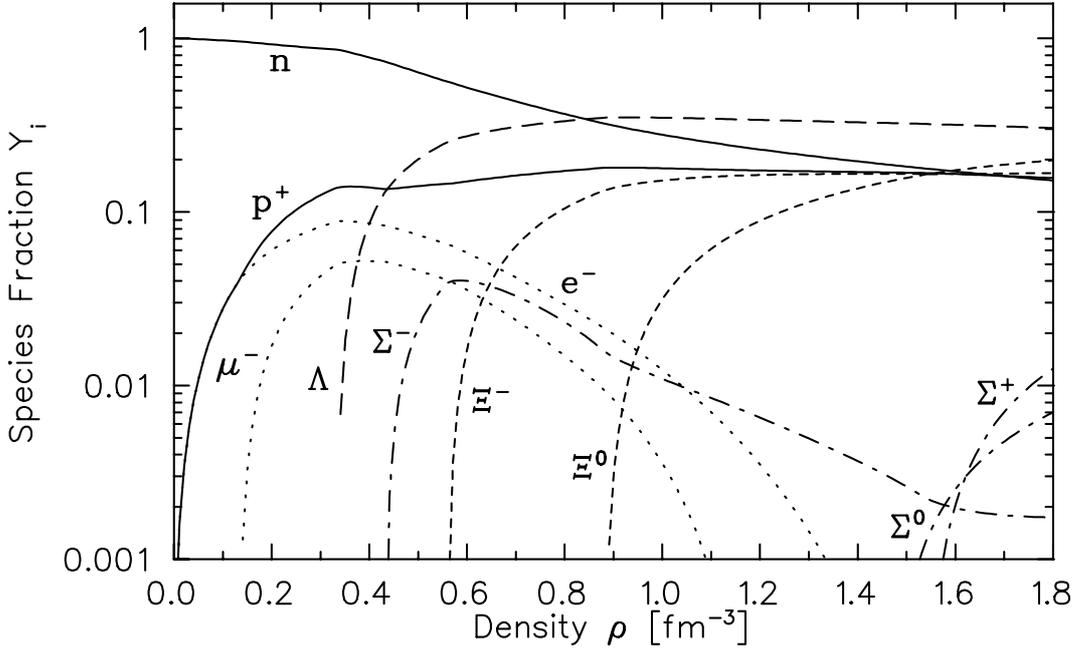}
\caption[Species fractions for octet QMC]{Species fractions $Y_i$ for
  octet QMC, in which the $\rho$ meson contribution is included. The
  density here is the total baryon density, as per
  Eq.~(\ref{eq:rho}). Note that in this case, all of the octet baryons
  contribute at some density, and that the species fractions of
  $\Sigma$ hyperons are suppressed\emdash in particular that of the
  $\Sigma^-$\emdash compared to the other baryons. When compared to
  the nucleon-only data (as in Fig.~\ref{fig:QMCCNem_rhonorho}) we
  observe that in this case the lepton densities do not plateau, but
  rather decrease once the hyperons appear, as explained in the
  text. Note that the $\Lambda$ hyperon is the first hyperon to
  appear, and that at high-densities the relative species fractions
  plateau for the nucleons, $\Lambda$, and $\Xi$ hyperons. The
  relative baryon proportions at high density are ordered by isospin,
  in that $\rho_\Lambda > \rho_N \sim \rho_\Xi > \rho_\Sigma$. The
  parameters used here are shown in Table~\ref{tab:couplings}.
  \protect\label{fig:specfrac_QMC}}
\end{figure}

The most interesting aspects of QMC are found when we include the
remainder of the octet of baryons\emdash the hyperons\emdash which was
not possible in QHD due to the Klein Paradox. As discussed in
Sec.~\ref{sec:qmc}, the baryon effective masses in QMC do not become
negative at any density, and thus we do not encounter the same issues
as we do when trying to calculate octet QHD. The species fractions for
octet QMC are shown in Fig.~\ref{fig:specfrac_QMC}, where we note that
the $\Lambda$ species fraction is significantly larger than that of
the $\Sigma$. The investigations by Rikovska--Stone {\it et
  al.}~\cite{RikovskaStone:2006ta} (in which alternative Fock terms
were introduced to the QMC model) lead us to expect that the $\Sigma$
would disappear entirely from the system if we were to include Fock
terms for the hyperons.\par
\begin{figure}[!b]
\begin{tabular}[h]{cc}
\begin{minipage}[c]{0.45\textwidth}
\centering
\includegraphics[angle=90,width=\textwidth]{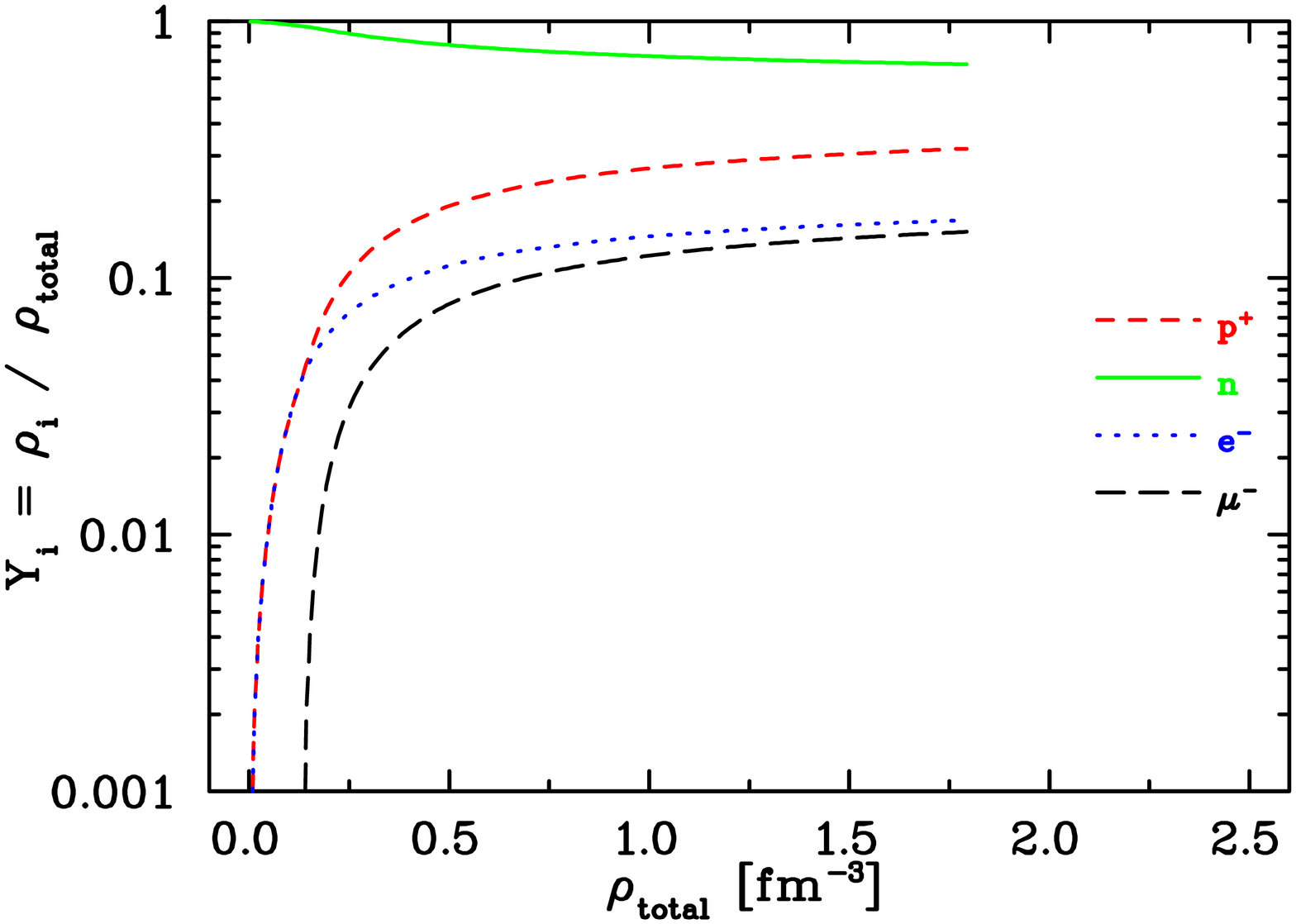}
\end{minipage}
&
\begin{minipage}[c]{0.45\textwidth}
\centering
\includegraphics[angle=90,width=\textwidth]{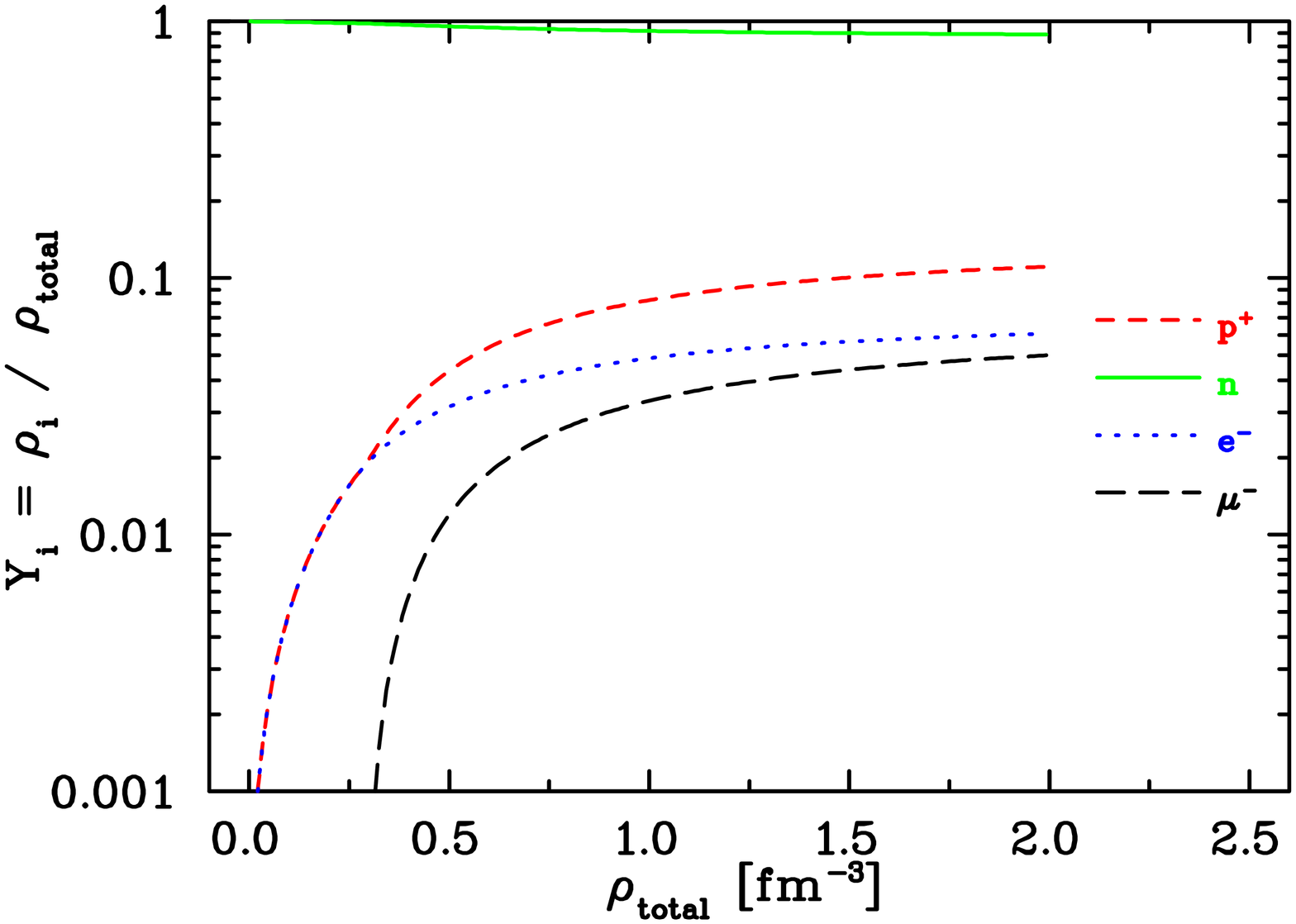}
\end{minipage}
\end{tabular}
\caption[Species fractions for nucleonic QMC in $\beta$-equilibrium
  with leptons] {(Color Online) Species fractions $Y_i$ for nucleonic
  QMC in $\beta$-equilibrium with leptons $\ell=e^-,\mu^-$, including
  (left) and lacking (right) a contribution from the $\rho$
  meson. These results are very similar to those of
  Fig.~\ref{fig:CNem_rhonorho}, though the differences are most
  pronounced when lacking the $\rho$ meson (right).}
\label{fig:QMCCNem_rhonorho}
\end{figure}
The species fractions in Fig.~\ref{fig:specfrac_QMC} are particularly
interesting when compared to the $\beta$-equilibrium nucleonic
configurations (Fig.~\ref{fig:QMCCNem_rhonorho}) as the leptonic
contributions \emph{decrease} at higher densities, in contrast to the
plateau observed in Fig.~\ref{fig:QMCCNem_rhonorho}. This is a result
of the chemical potential equilibria which provide preference to
hadronic charged states with large masses, rather than those with
smaller masses. In this case, the system is more stable with the
negative charge (to counter that of the protons) being provided by the
$\Xi^-$ hyperon rather than the leptons.\par
%
%
One may reasonably ask then why the $\Sigma^-$ meson does not play the
role of balancing positive charge, as it is the lightest baryon with a
negative electric charge. The relation between the chemical potentials
in Eq.~(\ref{eq:allmus}); $\mu_{\Sigma^-} = \mu_{\Xi^-}$ involves the
full interacting chemical potentials of Eq.~(\ref{eq:mu_sw}) which
includes scalar meson terms that affect the mass, and vector meson
terms that effect the energy. It is a balance of these effects which
dictates which particle will have the greater Fermi momentum (hence,
density) when considering Eq.~(\ref{eq:allmus}) above. For the case
shown in Fig.~\ref{fig:specfrac_QMC} the balance dictates that the
$\Xi$ hyperons will have a greater species fraction at the densities
shown.\par
\begin{figure}[!b]
\centering
\includegraphics[angle=90,width=0.85\textwidth]{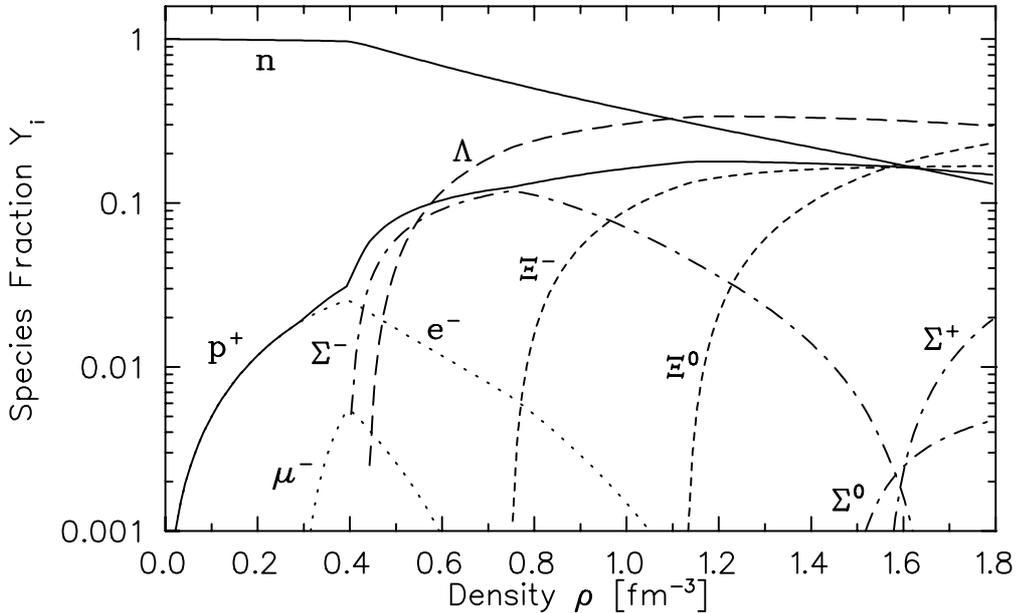}
\caption[Species fractions for octet QMC (neglecting $\rho$)]{Species
  fractions $Y_i$ for octet QMC, in which the $\rho$ meson
  contribution is neglected, but otherwise using the same parameters
  as used in Fig.~\ref{fig:specfrac_QMC}. We note that at
  high-densities, the baryons are no longer sorted by isospin as
  distinctly, and that the $\Sigma^-$ hyperon is now the first to
  appear. Furthermore, the hyperon threshold now occurs at a higher
  density than that of Fig.~\ref{fig:specfrac_QMC} where the $\Lambda$
  was the first to appear. In this case, the $\Sigma^-$ species
  fraction is not suppressed at high density, indicating that the
  $\rho$ meson plays a vital role in
  this.\protect\label{fig:specfrac_QMCnorho}}
\end{figure}

If however we exclude the contribution of the $\rho$ meson, we can
observe the effects that are attributed to this meson. The species
fractions for octet QMC in which we neglect the $\rho$ meson are shown
in Fig.~\ref{fig:specfrac_QMCnorho}, and we note that the distribution
is remarkably different. First, we note that the $\Sigma^-$ hyperon is
now the first to appear, in stark contrast to
Fig~\ref{fig:specfrac_QMC} in which the $\Sigma$ hyperons were largely
suppressed. The effect on the chemical potentials (in particular the
vector potential terms) due to neglecting the $\rho$ meson is such
that this is now possible. We also note that the high-density plateau
of Fig.~\ref{fig:specfrac_QMC} is less evident, and the baryons are
not sorted by isospin as distinctly. The lack of suppression for the
$\Sigma^-$ is made clear in this case, and indicates the strong link
between this meson and baryon, which is reasonable given that the
$\rho$ meson couples to isospin, and the $\Sigma$ baryons have the
largest magnitude isospin ($I_\Sigma = 1$).
%
%
\clearpage

\subsection{QMC Stars}\label{sec:QMCSTARS}
\begin{figure}[!b]
\centering
\includegraphics[angle=90,width=0.9\textwidth]{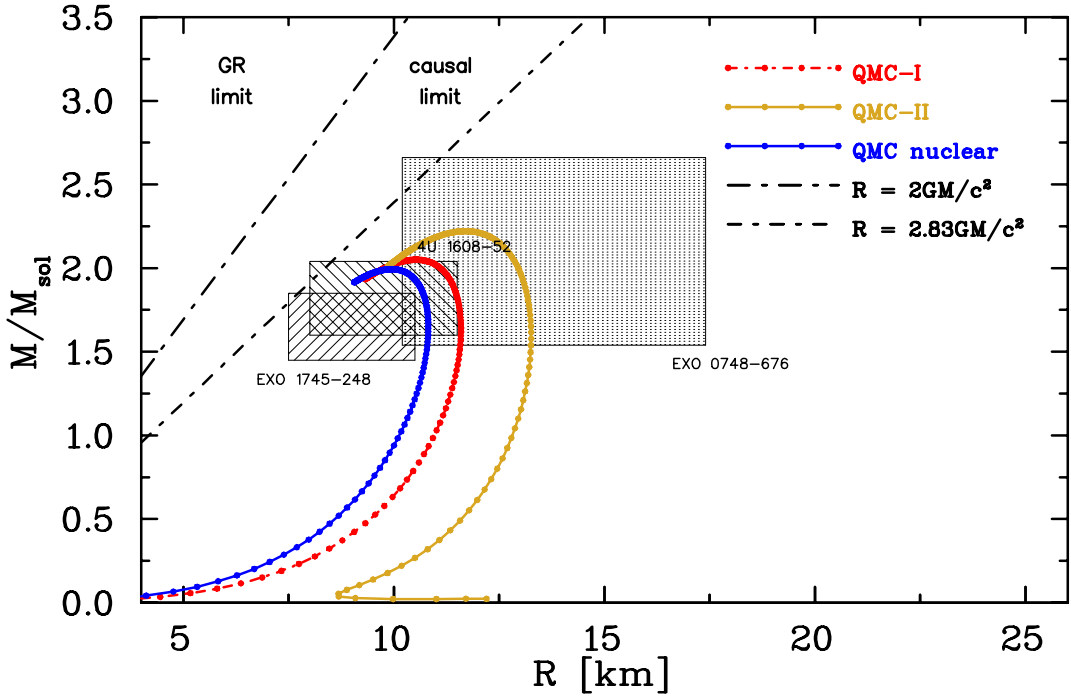}
\caption[Mass-radius relations for nucleonic QMC]{(Color Online)
  Mass-radius relations for various configurations of nucleonic QMC,
  with additional features the same as Fig.~\ref{fig:MvsRQHDN}. Note
  that the maximum masses of QMC-I and nuclear QMC stars are much
  lower than for QHD-I and nuclear QHD, indicating that the QMC EOS is
  much softer that that of QHD as evidenced by the smaller value of
  the compression modulus $K$ in QMC. \protect\label{fig:MvsRQMCN}}
\end{figure}

By solving the TOV equation, we can once again investigate `stellar
solutions'. The mass-radius relations for nucleonic QMC EOS are shown
in Fig.~\ref{fig:MvsRQMCN}. The maximum masses for the QMC-I and
nuclear QMC EOS are much smaller than their counterparts in QHD, since
the QMC EOS is much softer than that of QHD as evidenced by the
smaller value of the compression modulus $K$ in QMC. The softening is
due to the effective inclusion of the response to the scalar field via
the scalar polarizability, as discussed in
Section~\ref{sec:qmc}. Inclusion of the $\rho$ meson appears to undo
much of this softening. We do note however that the QMC-I curves still
lie within the $2\s$ bounds.\par
\begin{figure}[!t]
\centering
\includegraphics[angle=90,width=0.9\textwidth]{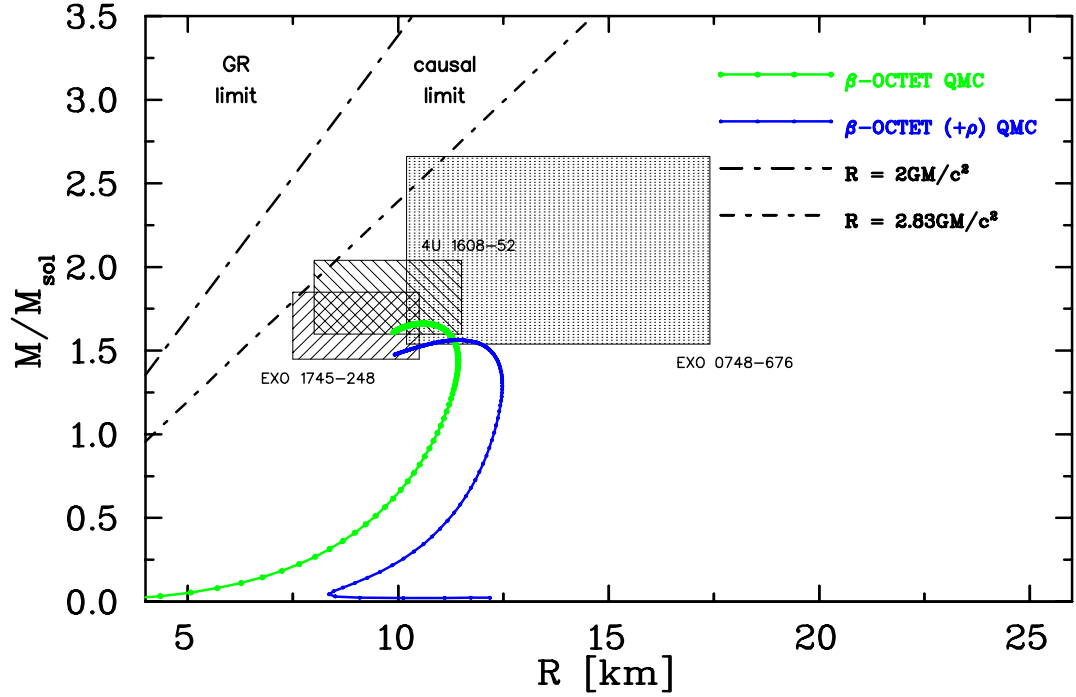}
\caption[Mass-radius relations for octet QMC]{(Color Online)
  Mass-radius relations for octet QMC-I (as per
  Fig.~\ref{fig:specfrac_QMCnorho}) and QMC-II (as per
  Fig.~\ref{fig:specfrac_QMC}), with additional features the same as
  Fig.~\ref{fig:MvsRQHDN}. Note that the maximum masses of octet QMC-I
  and octet QMC-II stars are lower than nucleonic QHD-I and nucleonic
  QHD-II stars, indicating that the octet QMC EOS is softer that that
  of the nucleonic QHD. \protect\label{fig:MvsRQMCY}}
\end{figure}

If we now introduce hyperons to the QMC EOS, as shown in
Fig.~\ref{fig:MvsRQMCY} we see a further softening of the EOS. In this
case, the addition of the $\rho$ meson does not stiffen the EOS back
to the level of the nucleonic EOS, since the $\rho$ contribution is
much smaller in the hyperonic case. As shown in
Fig.~\ref{fig:specfrac_QMC}, the asymmetry between the components of
the isodoublets and those of the isotriplet is small, and the $\rho$
contribution is proportional to this asymmetry (refer to
Eq.~(\ref{eq:MFrho})). This is an important distinction between the
nucleonic and hyperonic EOS.\par
While the mass-radius relations for these configurations may be
incapable of explaining the given observed data of~\cite{Ozel:2006bv},
it is clearly a great step forward in that we may now model the
effects of hyperons, which was not possible in QHD.\par
\vfill
\begin{figure}[!b]
\centering
\includegraphics[angle=90,width=0.9\textwidth]{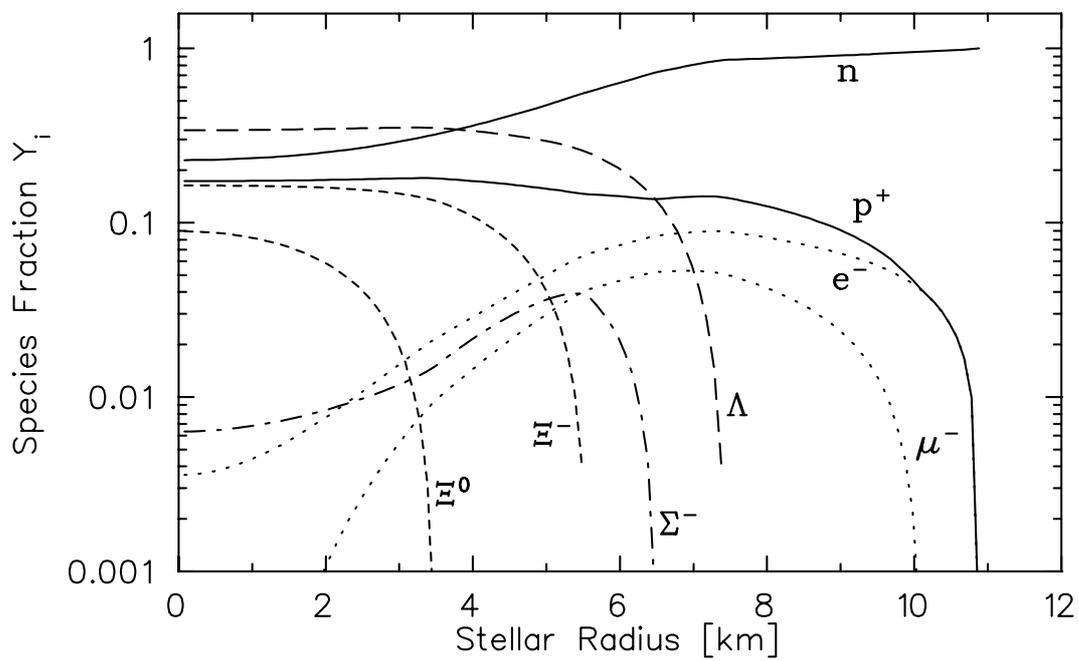}
\caption[Species fractions for octet QMC in a star]{Species fractions
  for octet QMC in $\beta$-equilibrium as a function of stellar radius
  for a stellar solution with a central density of $\rho_{\rm central}
  = 1.2~{\rm fm}^{-3}$. The parameters used here are the same as those
  used to produce Fig.~\ref{fig:specfrac_QMC}. We note that the
  outermost $3~{\rm km}$ of this star contains only nucleons and
  leptons, with hyperon effects occurring much deeper.
  \protect\label{fig:TOVRAD_noPT}}
\end{figure}
If we consider the case of a star with a central density of $\rho_{\rm
  central} = 1.2~{\rm fm}^{-3}$ (an arbitrary choice) as shown in
Fig.~\ref{fig:TOVRAD_noPT} for the case of hyperonic QMC, we observe
that the outermost $3~{\rm km}$ of this $11~{\rm km}$ star contains
only nucleonic matter (in $\beta$-equilibrium). The core of this star
contains roughly equal proportions of nucleons, $\Lambda$ and $\Xi$
hyperons, with a notable lack of $\Sigma$ hyperons. This is the
simplest case for which we produce a stellar object with hyperons, and
comprises our simplest method for including a strangeness degree of
freedom.\par
\clearpage 

\section{Hybrid Equation of State}\label{sec:HYBRIDEOS}
A focus of this work has been to calculate the properties of hybrid
stars; involving contributions from both baryons and deconfined quarks
in a statistical mechanics method. The process by which we obtain
these results has been described in
Section~\ref{sec:phasetransitions}. The results presented in this
section have been published by the author~\cite{Carroll:2008sv} as a
novel extension of the octet QMC model.\par

\subsection{Hybrid Infinite Matter}\label{sec:HYBRIDINF}
In Section~\ref{sec:MITbag} we discussed the MIT bag model for quark
matter, in which we model three deconfined free quarks in a Fermi gas
possessing constant masses consistent with current quark
phenomenology. It is this model that we shall refer to (unless stated
otherwise) for hybrid models involving quark matter. The hadronic
model we will use in this section is the octet QMC model, unless
otherwise specified.\par
\begin{figure}[!t]
\centering
\includegraphics[angle=90,width=0.9\textwidth]{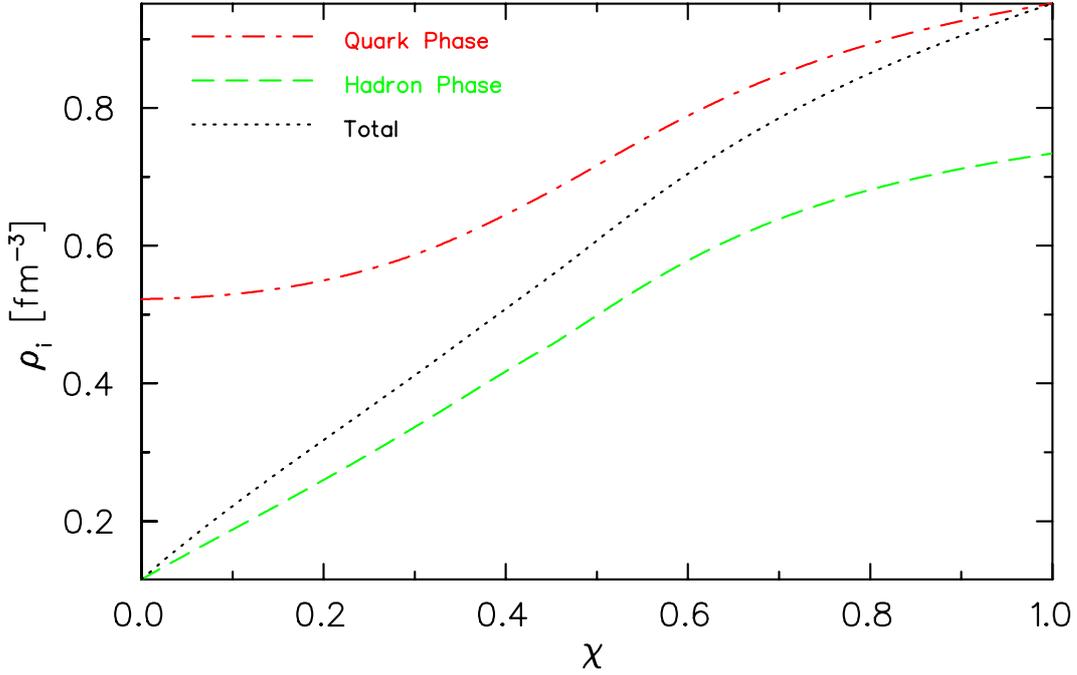}
\caption[Densities in a mixed phase]{(Color online) Densities in the
  mixed phase for octet QMC mixed with three-flavor quark matter
  modelled with the MIT bag model. Note that at all values of $\chi$
  (the mixing parameter according to Eq.~(\ref{eq:mp_rho}) which also
  defines the total density), the equivalent baryon density of quarks
  is greater than the hadronic baryon density, allowing the total
  baryon density to increase monotonically with increasing
  $\chi$. Clearly, at $\chi=0$ the density corresponds entirely to the
  hadronic phase, and at $\chi=1$ the density corresponds entirely to
  the quark phase. Between these points, the density is that of the
  mixed phase.  \protect\label{fig:densities}}
\end{figure}

The conditions for a Glendenning-style mixed phase as discussed in
Section~\ref{sec:MixedPhase} require that for a given pair of $\mu_n$
and $\mu_e$ (common to the hadronic and quark phases) at any value of
the mixing parameter $\chi$, the quark density is greater than the
hadronic density. This condition ensures that the total baryon density
increases monotonically within the range $\rho_{\rm QP} > \rho_{\rm
  MP} > \rho_{\rm HP}$, as can be seen in Eq.~(\ref{eq:mp_rho}). An
example of this is illustrated in Fig.~\ref{fig:densities} for a mixed
phase of octet QMC and three-flavor quark matter modelled with the MIT
bag model.\par
Recall that the method for evaluating the properties of a
mixed phase involves calculating the neutron and electron chemical
potentials of the hadronic phase and using these as inputs to the
quark phase calculations, providing the quark Fermi momenta via
Eq.~(\ref{eq:quarkmu}). In this case, with the dynamic quark masses of
NJL and no vector potentials (self-energies) due to non-interacting
quarks, the calculated quark Fermi momentum corresponds a quark
density that is lower than the hadronic density, and as a result there
are no configurations for a mixed phase in which the proportion of
quarks increases while at the same time the total baryon density
increases. It may be possible that with smaller constituent quark
masses at low density, the Fermi momenta would provide sufficiently
high quark densities, but we feel that it would be unphysical to use
smaller constituent quark masses. This result implies that\emdash
at least for the models we have investigated\emdash dynamical chiral
symmetry breaking (in the production of constituent quark masses at
low density) prevents a phase transition from a hadronic phase to a
mixed phase involving quarks.\par
\begin{figure}[!t]
\centering
\includegraphics[angle=90,width=0.9\textwidth]{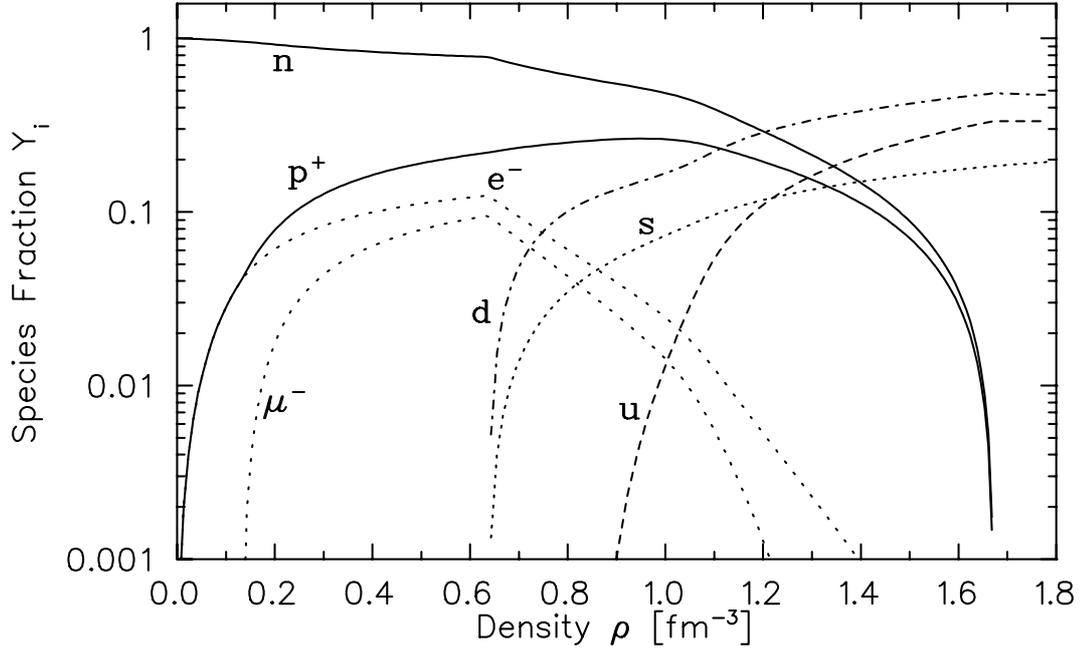}
\caption[Species fractions for hybrid nucleonic QMC matter]{Species
  fractions for a phase transition from nucleonic QMC matter to
  three-flavor quark matter modelled with NJL. Note that in this
  unphysical (we see no physical reason for which hyperons should be
  neglected) case, a phase transition is possible, and occurs at a
  value of $\rho = 0.64~{\rm fm}^{-3}$. In this case the $u$ quark
  first appears at a higher density than the $d$ or $s$ quarks due to
  its positive charge which can only be balanced by the other two
  quarks since the leptons provide a decreasing contribution as per
  Eq.~(\ref{eq:betaeq}).\protect\label{fig:QMCnuclear}}
\end{figure}

We do note, however, that if we restrict consideration to nucleons
only within the QMC model (with the same parameters\footnote{The
  couplings of mesons to baryons are defined by the energy per baryon,
  which saturates at a lower density than the hyperon threshold. We
  can therefore safely assume that the inclusion or removal of
  hyperons plays no part in defining the couplings of mesons to the
  nucleons.} as octet QMC), and represent quark matter with the NJL
model, we do in fact find a possible mixed phase. More surprisingly,
the density at which a phase transition to the mixed phase occurs for
this combination is significantly larger than the case where hyperons
are present (in which case we model a transition to MIT bag model
quark matter). An example of this is shown in
Fig.~\ref{fig:QMCnuclear}, the parameters for which can be found in
Table~\ref{tab:results}. This produces a mixed phase at about $4
\rho_0$ ($\rho = 0.64~{\rm fm}^{-3}$) and a pure quark matter phase
above about $10 \rho_0$ ($\rho = 1.67~{\rm fm}^{-3}$). In this case
the $u$ quark first appears at a higher density than the $d$ or $s$
quarks due to its positive charge which can only be balanced by the
other two quarks since the leptons provide a decreasing contribution
as per Eq.~(\ref{eq:betaeq}), since the proton and neutron densities
become more and more similar. Although this example does show a phase
transition, the omission of hyperons is certainly unrealistic. This
does however illustrate the importance and significance of including
hyperons, in that their inclusion alters the chemical potentials which
satisfy the equilibrium conditions in such a way that the mixed phase
is no longer produced.\par
\begin{figure}[!t]
\centering
\includegraphics[angle=90,width=0.9\textwidth]{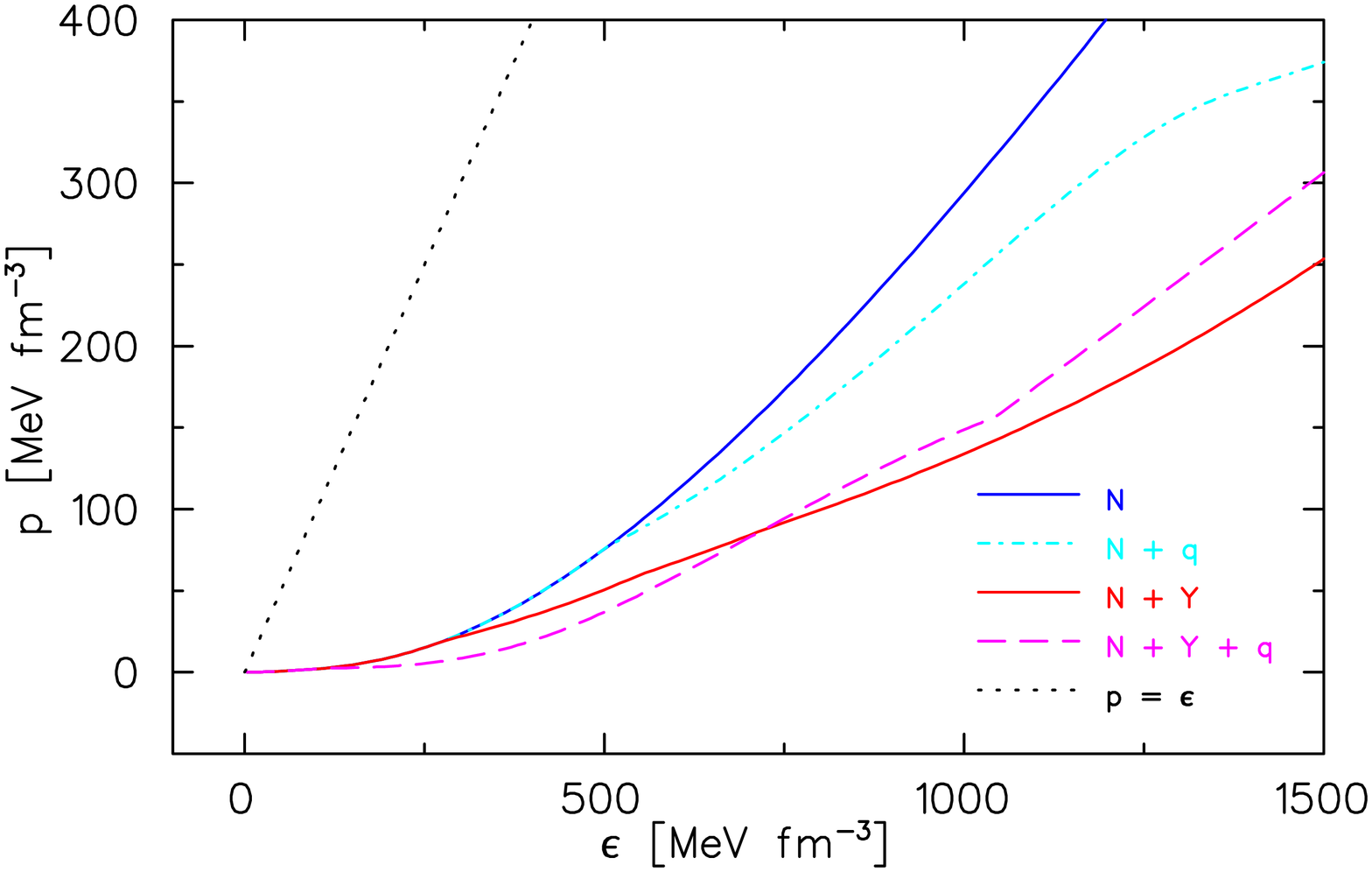}
\caption[EOS for various models]{(Color online) Equation of State for;
  nucleonic `N' matter modelled with QMC; nucleonic matter where a
  phase transition to NJL modelled quark matter is permitted; baryonic
  `N+Y' matter modelled with octet QMC; and baryonic matter where a
  phase transition to MIT bag modelled quark matter is permitted. The
  line $P = {\cal E}$ represents the causal limit, $v_{\rm sound} =
  c$. The bends in these curves indicate a change in the composition
  of the EOS, such as the creation of hyperons or a transition to a
  mixed or quark phase. Note that at very low energies (densities) the
  curves are identical, where only nucleonic matter in
  $\beta$-equilibrium is present. \protect\label{fig:EOSlots}}
\end{figure}

Using the MIT bag model, we can calculate the hybrid EOS, which is
shown in Fig.~\ref{fig:EOSlots}. We can see that at low densities (low
energies) all of the configurations of EOS are fairly soft. At the
lowest densities, each of the configurations of EOS are approximately
equal; at this point they all represent nucleons in
$\beta$-equilibrium. \par
\begin{figure}[!b]
\centering
\includegraphics[angle=90,width=0.9\textwidth]{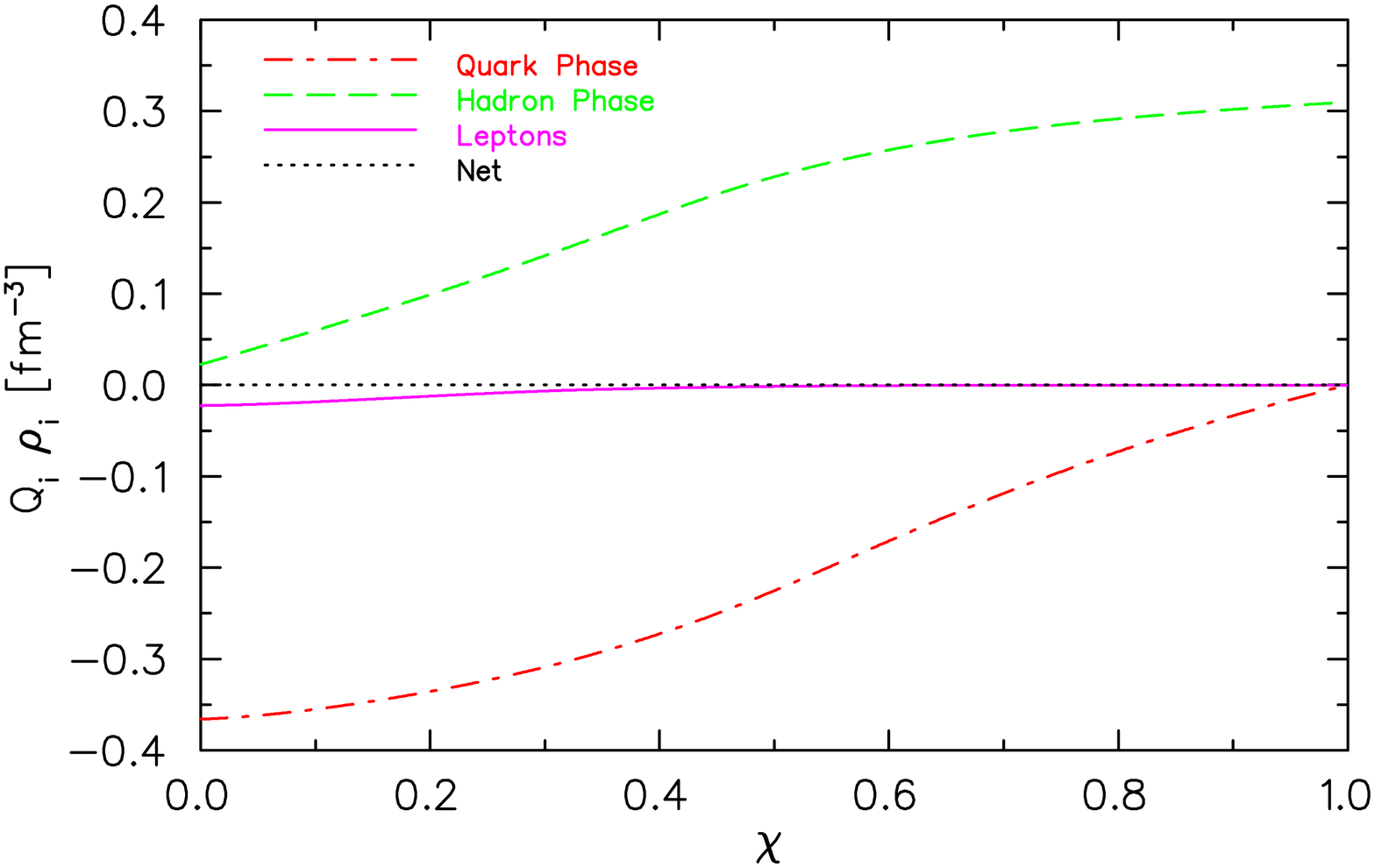}
\caption[Charge-densities in a mixed phase]{(Color online)
  Charge-densities (in units of the proton charge per ${\rm fm}^3$) in
  the mixed phase for a transition from octet QMC to three-flavor
  quark matter modelled with the MIT bag model. The net charge of the
  mixed phase is exactly zero, as per the constraining equation of
  Eq.~(\ref{eq:mp_charge}). Note that following the mixed phase, the
  quarks are able to satisfy charge neutrality with no leptons, in
  contrast to the nucleonic cases of $\beta$-equilibrium in which the
  lepton contribution remains stable. $\chi$ is the mixing parameter
  within the mixed phase according to Eq.~(\ref{eq:mp_rho}).
  \protect\label{fig:chargedensities}}
\end{figure}

For each of the configurations in which we find a phase transition
from baryonic matter to quark matter, the EOS consists of negatively
charged quark matter, positively charged hadronic matter, and a small
proportion of leptons, to produce globally charge-neutral matter. The
proportions of hadronic, leptonic and quark matter throughout the
mixed phase (for example, during a transition from octet QMC matter to
three-flavor quark matter modelled with the MIT bag model) are
displayed in Fig.~\ref{fig:chargedensities} in which we note that the
quarks are able to satisfy charge neutrality without lepton
contributions, in contrast to the cases of nucleonic
$\beta$-equilibrium in which the lepton contributions remain stable at
increasing densities. A summary of the results of interest is given in
Table~\ref{tab:results}.\par
Results of calculations for larger quark masses are not shown, as they
require a much lower bag energy density to satisfy the pressure
equilibrium conditions. For constituent quark masses, we find that no
phase transition is possible for any value of the bag energy density,
as the quark pressure does not rise sufficiently fast to overcome the
hadronic pressure. This is merely because the masses of the quarks do
not allow a sufficiently large Fermi momentum at a given chemical
potential, according to Eq.~(\ref{eq:quarkmu}).\par
\begin{figure}[!t]
\begin{tabular}[!tb]{c}
\begin{minipage}[c]{0.95\textwidth}
\centering
\includegraphics[angle=90,width=0.8\textwidth]{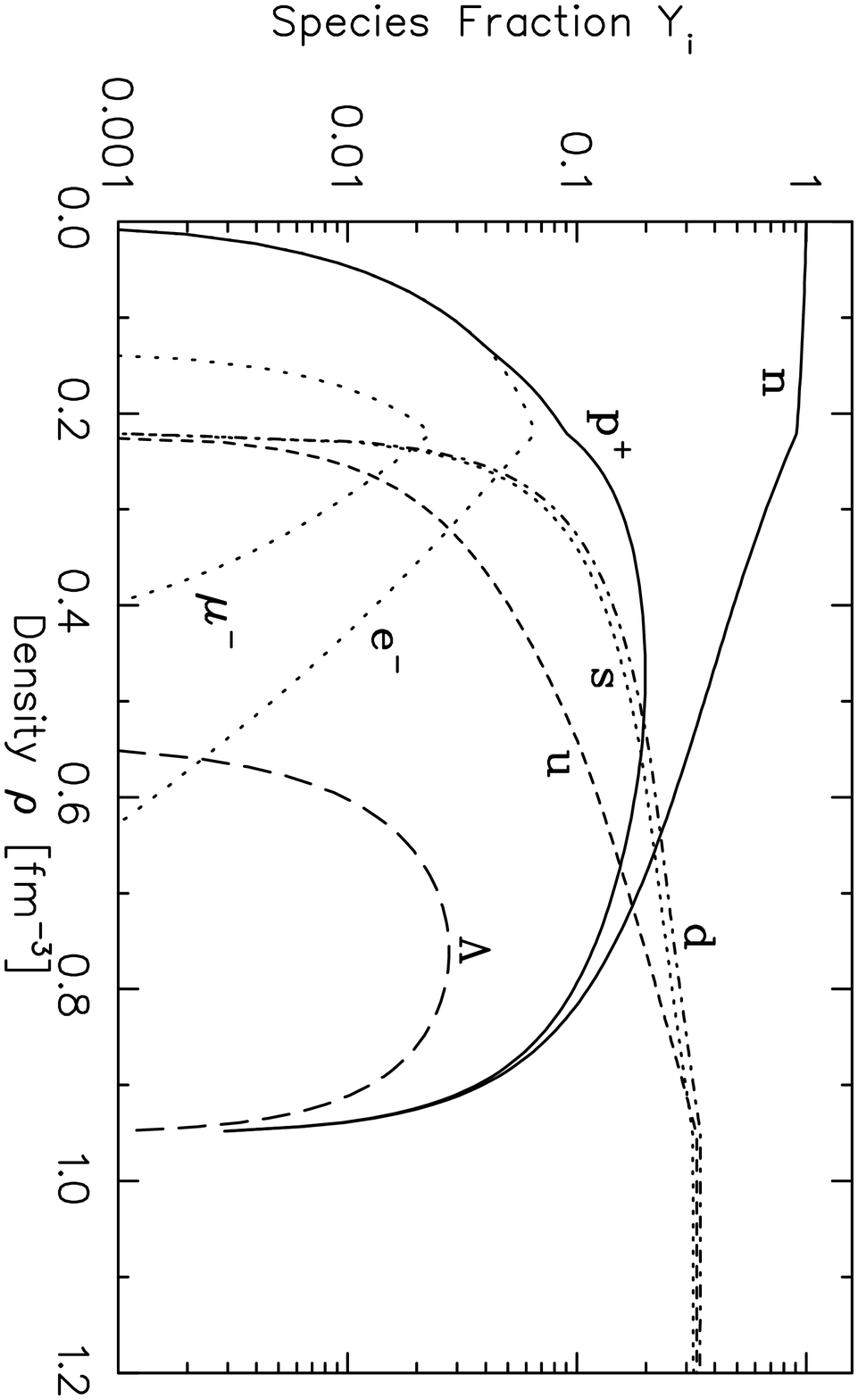}
\caption[Species fractions for octet QMC with phase transition to
  quark matter]{Species fractions $Y_i$ for octet QMC (the same as in
  Fig.~\ref{fig:specfrac_QMC}) but where now we allow the phase
  transition to a mixed phase involving quark matter modelled with the
  MIT bag model. Note that the $\Lambda$ hyperon is the only hyperon
  to appear in the mixed phase, and does so at a much higher density
  than the configuration where the transition to a mixed phase is
  forbidden. A similar transition from nucleonic QMC matter to
  three-flavor quark matter modelled with the MIT bag model (as shown
  in Fig.~\ref{fig:QMCnuclear}) produces results almost identical to
  these, except of course that in that case there is no contribution
  from the $\Lambda$ hyperon. We note the significant differences in
  threshold densities when comparing these
  results. \protect\label{fig:SpecFrac_PTQMC}}
\end{minipage}
\\[3mm]
\phantom{blah}\\[1mm]
\begin{minipage}[c]{0.95\textwidth}
\centering
\includegraphics[angle=90,width=0.8\textwidth]{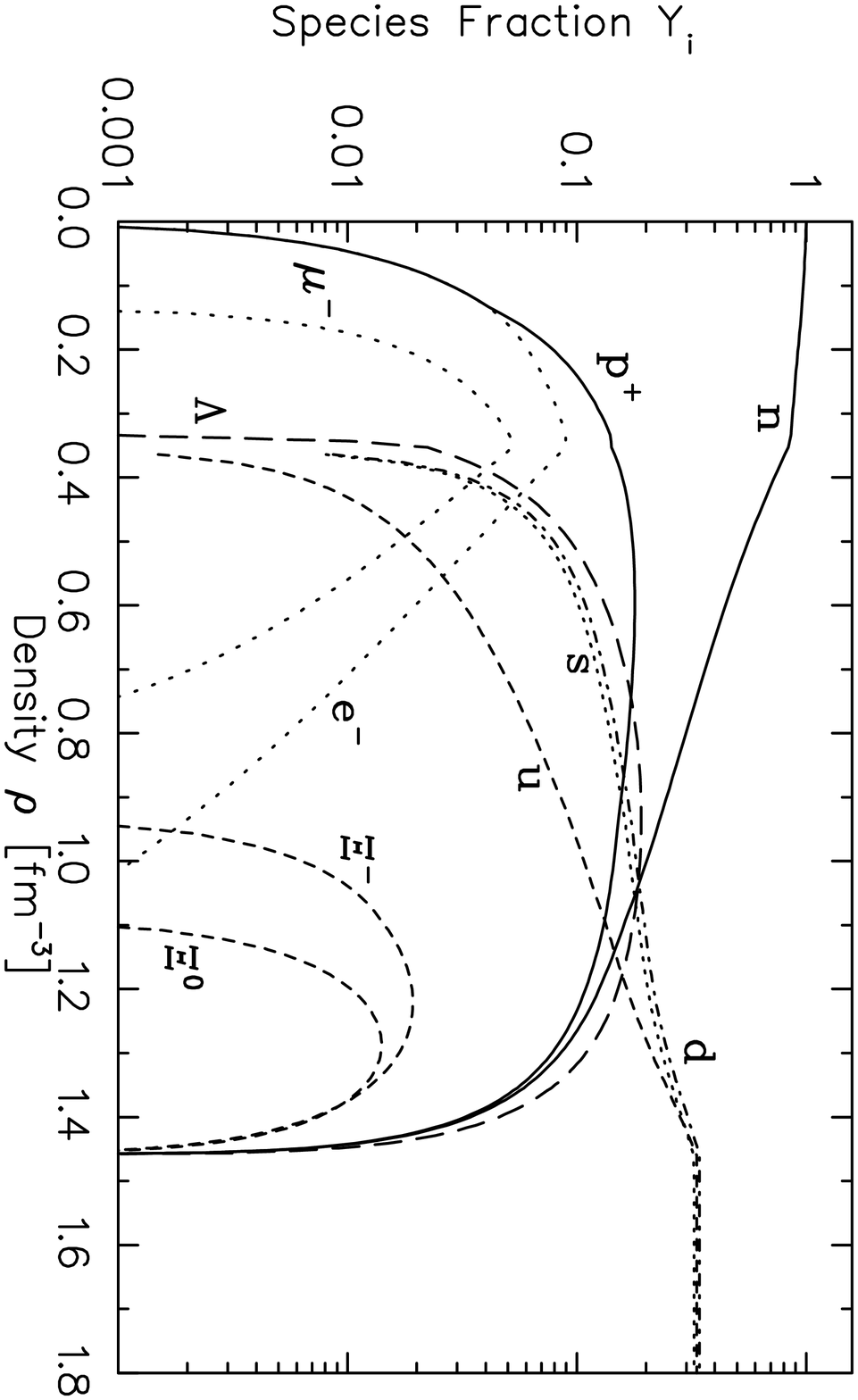}
\caption[Species fractions for octet QMC with $B^{1/4}=195~{\rm
    MeV}$]{Species fractions, $Y_i$, for octet QMC (the same as in
  Fig.~\ref{fig:SpecFrac_PTQMC} but now where the bag energy density
  has been increased to $B^{1/4}=195~{\rm MeV}$). Note that now the
  appearance of hyperons occurs at a smaller density than in the case
  of Fig.~\ref{fig:SpecFrac_PTQMC}, the transition to a mixed phase
  occurs at a slightly larger density, and that now $\Xi$ hyperons are
  present in the mixed phase.  \protect\label{fig:SpecFrac_PTQMC_195}}
\end{minipage}
\end{tabular}
\end{figure}

When we calculate the EOS including a mixed phase and subsequent pure
quark phase, we find that small changes in the parameters can
sometimes lead to very significant changes. In particular, the bag
energy density $B$, and the quark masses in the MIT bag model have the
ability to both move the phase transition points, and to vary the
constituents of the mixed phase. We have investigated the range of
parameters which yield a transition to a mixed phase and these are
summarized in Table~\ref{tab:results}. For illustrative purposes we
show an example of species fractions for a reasonable set of
parameters ($B^{1/4}=180~{\rm MeV}$ and $m_{\rm u,d,s} = 3,7,95~{\rm
  MeV}$) in Fig.~\ref{fig:SpecFrac_PTQMC}. Note that in this case the
$\Lambda$ hyperon enters the mixed phase briefly (and at a low species
fraction). Note that the transition density of $\rho_{\rm MP} \sim
0.22~{\rm fm}^{-3}$ produced by the combination of the octet QMC and
MIT bag models in that case seems unlikely to be physical as it
implies the presence of deconfined quarks at densities less than
$2\rho_0$, which would contradict the results of searches for such
entities. A similar transition from nucleonic QMC matter to
three-flavor quark matter modelled with the MIT bag model
(Fig.~\ref{fig:QMCnuclear}) produces results almost identical to those
of Fig.~\ref{fig:SpecFrac_PTQMC}, except of course that in that case
there is no contribution from the $\Lambda$ hyperon. We note the
significant differences in threshold densities between these two
cases.\par
%
%
With small changes to parameters, such as those used to produce
Fig.~\ref{fig:SpecFrac_PTQMC_195} in which the bag energy density is
given a slightly higher value from that used in
Fig.~\ref{fig:SpecFrac_PTQMC} ($B^{1/4}$ increased from $180~{\rm
  MeV}$ to $195~{\rm MeV}$, but the quark masses remain the same), it
becomes possible for the $\Xi$ hyperons to also enter the mixed phase,
albeit in that case with small species fractions, $Y_\Sigma, Y_\Xi
\leq 0.02$. This highlights the need for strict tolerances for `known'
values of model parameters.\par
%
%
%
%
\clearpage 

\subsection{Hybrid Stars}\label{sec:HYBRIDSTARS}
\begin{figure}[!b]
\begin{tabular}[!tb]{c}
\begin{minipage}[c]{0.95\textwidth}
\centering
\includegraphics[angle=90,width=0.9\textwidth]{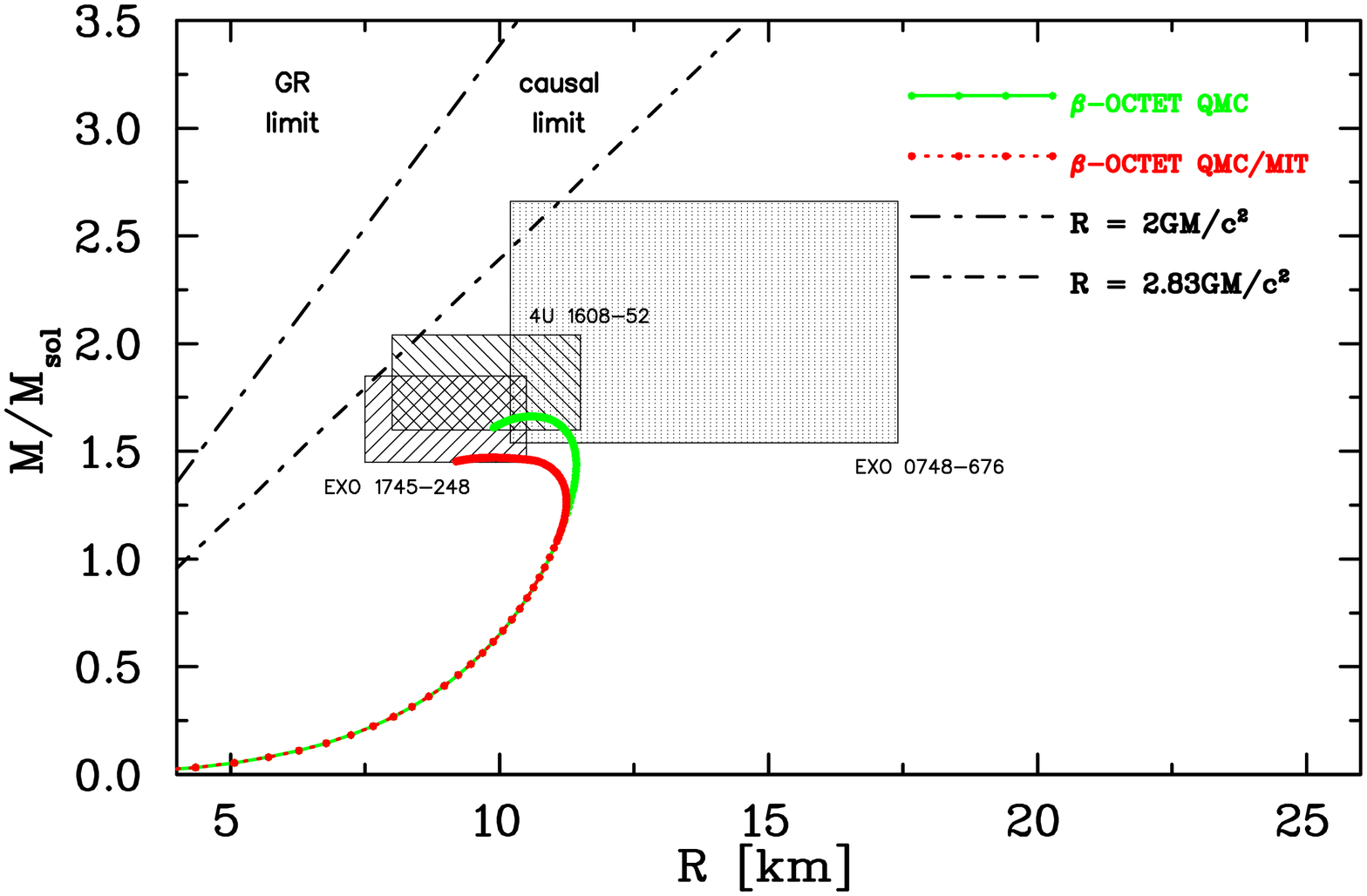}
\end{minipage}
\\[3mm]
\phantom{blah}\\[1mm]
\begin{minipage}[c]{0.95\textwidth}
\centering
\includegraphics[angle=90,width=0.9\textwidth]{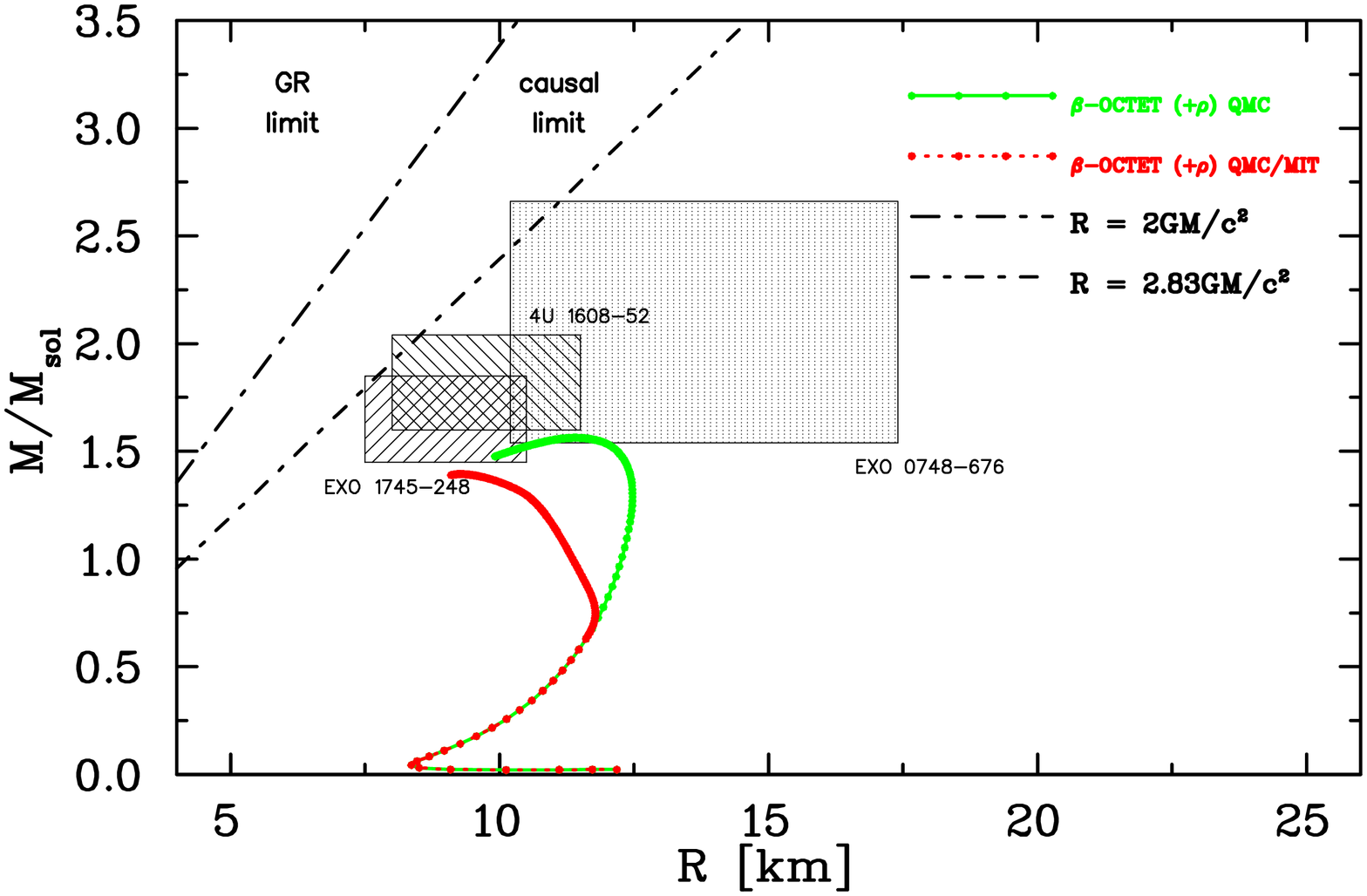}
\end{minipage}
\end{tabular}
\caption[Mass-radius relations for octet QMC baryonic and hybrid
  stars] {(Color Online) Mass-radius relations for octet QMC in
  $\beta$-equilibrium (neglecting the $\rho$ meson, upper; and
  including it, lower) for baryonic and hybrid stars. Also shown are
  the additional features described in Fig.~\ref{fig:MvsRQHDN}. Note
  that the stellar solutions for configurations involving a phase
  transition are identical to those where the phase transition is
  neglected, up to some value of the central density at which the
  solutions diverge. This point corresponds to the lowest density
  stars which contain quark matter. The stars containing quark matter
  generally have a lower mass than those which only contain baryonic
  matter due to a softening of the EOS.}
\label{fig:QMChybrid}
\end{figure}

The stellar solutions for octet QMC hybrid stars (in which a phase
transition to a mixed phase is allowed) are shown in
Fig.~\ref{fig:QMChybrid} along with the baryon-only stellar solutions
for comparison. We note that the solutions for configurations
involving a phase transition are identical to those where a transition
is neglected, up to some value of the central density at which point
the solutions diverge. This indicates the lowest mass stars that
contain quark matter.\par
We note that the stellar masses for configurations involving a phase
transition to quark matter are lower than those in which the
transition is neglected, for the same value of central density. This
is due to the softening of the EOS by the introduction of quarks (see
Section~\ref{sec:EoS}). Overall, the stellar masses for these
configurations are similar to observed neutron star masses, though
notably lower than the masses of the most massive observed neutron
stars. This could be attributed to an over-softening of the EOS.\par
\begin{figure}[!t]
\begin{tabular}[!tb]{c}
\begin{minipage}[c]{0.95\textwidth}
\centering
\includegraphics[angle=90,width=0.9\textwidth]{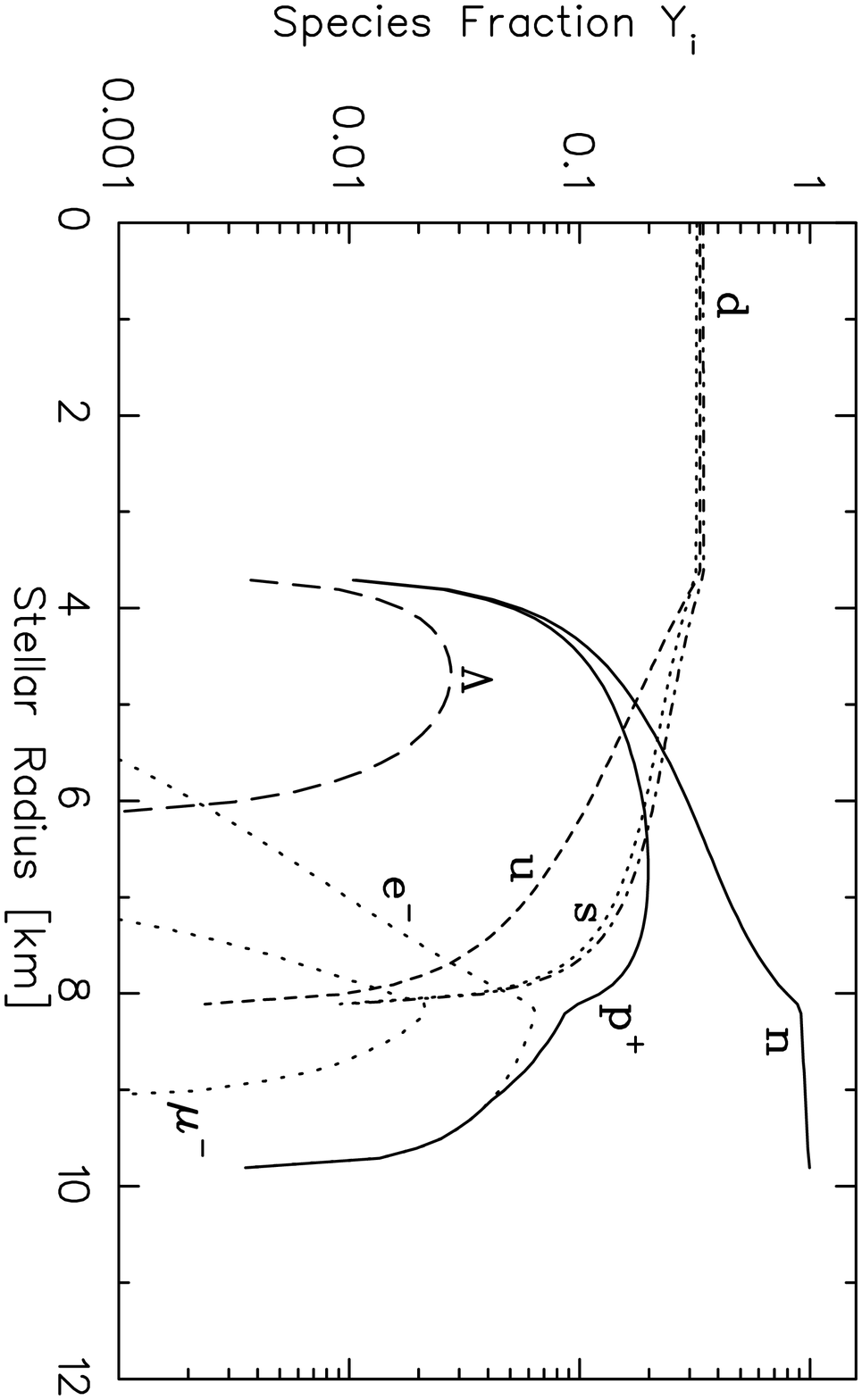}
\caption[Species fractions for hybrid octet QMC vs Stellar Radius]{Species
  fractions for octet QMC with a phase transition to three-flavor quark
  matter modelled with the MIT bag model, as a function of stellar
  radius for a stellar solution with a central density of $\rho_{\rm
    central} = 1.2~{\rm fm}^{-3}$. The parameters used here are the
  same as those used to produce Fig.~\ref{fig:SpecFrac_PTQMC}. Note
  that in this case one finds pure deconfined three-flavor quark matter at
  the core (all of some 3.5~km) of this star, and a small proportion
  of $\Lambda$ in the mixed phase.  \protect\label{fig:TOVRAD_PT}}
\end{minipage}
\\[3mm]
\phantom{blah}\\[1mm]
\begin{minipage}[c]{0.95\textwidth}
\centering
\includegraphics[angle=90,width=0.9\textwidth]{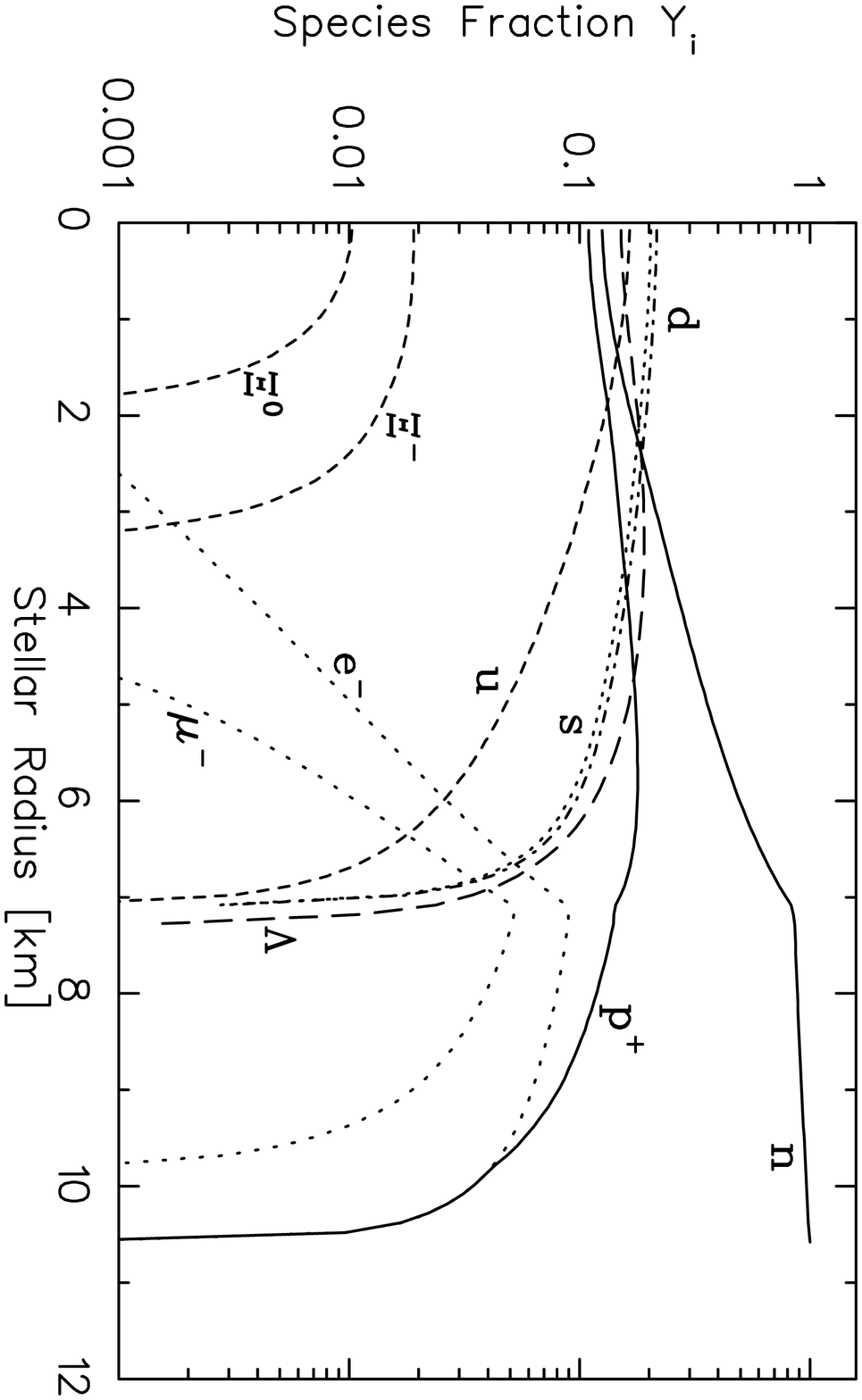}
\caption[Increasing the bag energy density to $B^{1/4} = 195~{\rm
    MeV}$]{Species fractions for the interior of a star with central
  density $\rho_{\rm central} = 1.2~{\rm fm}^{-3}$ where the bag
  energy density is given a slightly higher value from that used in
  Fig.~\ref{fig:TOVRAD_PT} (increased from $B^{1/4} = 180~{\rm MeV}$
  to $195~{\rm MeV}$), but the quark masses remain the same.
  \protect\label{fig:TOVRAD_PT195}}
\end{minipage}
\end{tabular}
\end{figure}

If we consider a star with a central density of $\rho_{\rm central} =
1.2~{\rm fm}^{-3}$ for octet QMC with a phase transition to quark
matter modelled with the MIT bag model (using the same parameters as
used in Fig.~\ref{fig:SpecFrac_PTQMC}) we observe the prediction of
$3.5~{\rm km}$ of quark core for a star with a radius of $10~{\rm
  km}$, with only a very small contribution from the hyperons, via
$\Lambda$, as shown in Fig.~\ref{fig:TOVRAD_PT}.\par
%
%
If we increase the value of the bag energy density from
$B^{1/4}=180~{\rm MeV}$ as used in Fig.~\ref{fig:TOVRAD_PT} to
$B^{1/4}=195~{\rm MeV}$ we find that the radius of a star with a
central density of $\rho_{\rm central} = 1.2~{\rm fm}^{-3}$ increases
by around $1~{\rm km}$, but the baryonic contribution (in the mixed
phase) is more prolific, as we now find contributions from $\Lambda$
and $\Xi$ hyperons in the core of the star, along with quark
contributions. The effect of this increase in the bag energy can be
seen by comparing Fig.~\ref{fig:TOVRAD_PT} with
Fig.~\ref{fig:TOVRAD_PT195}, where the bag energy densities are
$B^{1/4}=180~{\rm MeV}$ for the former, and $B^{1/4}=195~{\rm MeV}$
for the latter, and this is the only change that has been made.\par
%
%
\clearpage

\section{Hartree--Fock QHD Equation of State}\label{sec:HYBRIDFOCK}
As discussed in Section~\ref{sec:fockterms}, we can further extend our
models by including Fock terms. For comparison purposes, we show the
effects of including \emph{only} the Fock contribution to the
self-energy, as well as the effects of the full Hartree--Fock
calculations (including the medium polarization energy), as compared
to the Hartree calculations presented earlier.\par
The results presented in this section extend the calculations of
Ref.~\cite{Serot:1984ey} in which the properties of Hartree--Fock
nuclear matter were calculated \emph{without} $\beta$-equilibrium. To
the extent of the author's knowledge, the following calculations have
not been performed or published elsewhere, and as such are further
novel calculations.\par
For simplicity, we present the extension from Hartree to Hartree--Fock
for QHD, though similar calculations can be performed for extending
QMC similarly. We leave this as work for the future.\par

\subsection{Hartree--Fock QHD Infinite Matter}\label{sec:HYBRIDFOCKINF}
\begin{figure}[!b]
\centering
\includegraphics[angle=90,width=0.9\textwidth]{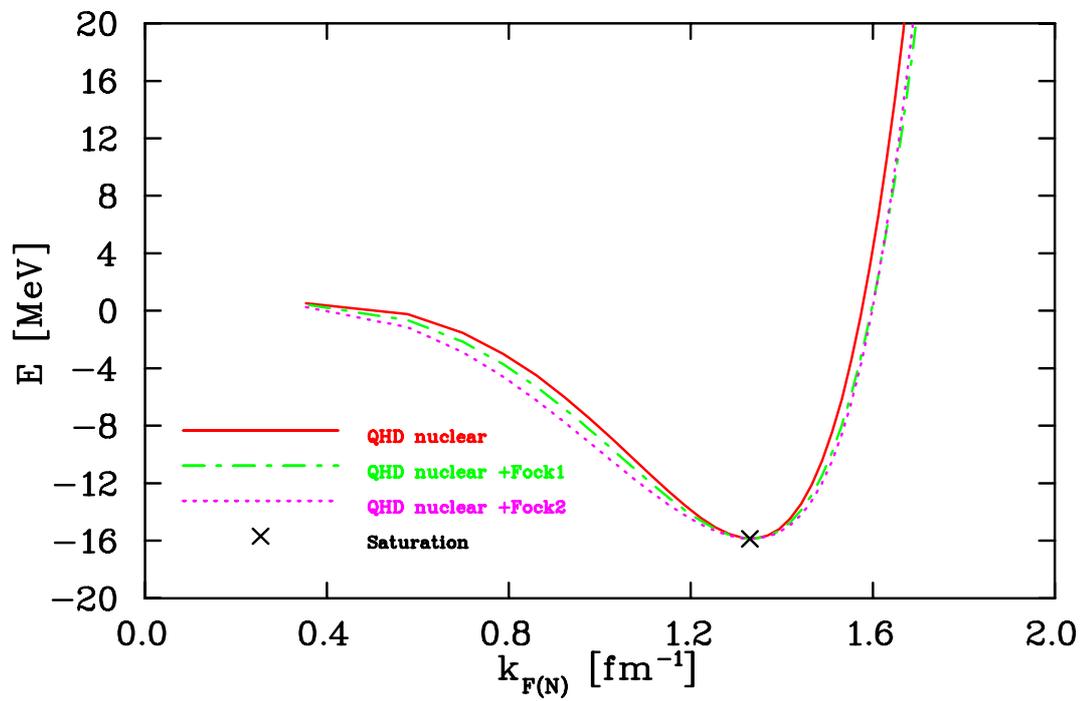}
\caption[EOS for nuclear QHD including Fock terms]{(Color Online) EOS
  for nuclear QHD matter calculated for three configurations; at
  Hartree level, including only Fock self-energy terms, and
  additionally including the medium polarization energy
  contribution. Couplings for each configuration are found such that
  saturation properties are reproduced. The values of these couplings
  can be found in Table~\ref{tab:couplings}.
  \protect\label{fig:QHDsymEOSFock}}
\end{figure}

In Fig.~\ref{fig:QHDsymEOSFock} we show the saturation curves for
nuclear QHD, in which we observe that the saturation properties for
each configuration are reproduced accurately, in order to calculate
the couplings to be used for each configuration.\par
We also note from that figure that the curvature for each
configuration is different at saturation, implying differences between
the compression modulii for each configuration, with Hartree QHD
having the largest value for $K$, and the full Hartree--Fock (Fock2)
having the smallest. For comparison, the values are shown in
Table~\ref{tab:couplings}. The softening of this EOS brings the
compression modulus for Hartree--Fock QHD closer to the experimental
range, though we note that this may further over-soften the EOS since
the maximum masses in Hartree QMC appear too small already.\par
\begin{figure}[!t]
\centering
\includegraphics[angle=90,width=0.9\textwidth]{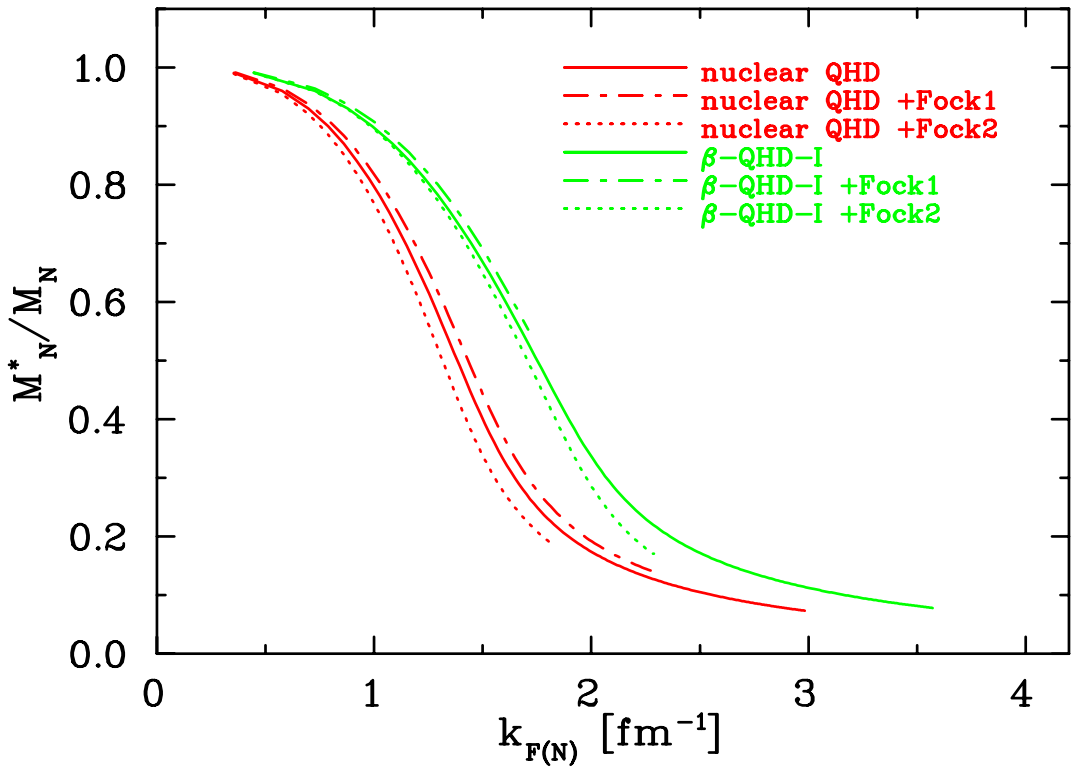}
\caption[Nucleon effective masses for QHD including self-energy Fock
  terms]{(Color Online) Nucleon effective masses for nuclear QHD and
  $\beta$-equilibrium nucleon QHD-I including Fock contributions to
  the self-energies (Fock1) and including this term as well as the
  medium polarization energy contribution (Fock2). The Hartree--Fock
  contributions are plotted for a limited range of Fermi momenta
  corresponding to densities up to $\sim 0.5~{\rm
    fm}^{-3}$. \protect\label{fig:QHDFockSE_mstar}}
\end{figure}

%
%
%
We can examine the effective masses for Hartree--Fock QHD for the case
of only including the self-energy Fock terms, and compare these with
those for Hartree QHD as shown in Fig.~\ref{fig:MstarQHDN}. The
comparison is shown in Fig.~\ref{fig:QHDFockSE_mstar} and we note that
the differences are subtle at this stage. The Hartree--Fock data
corresponds to two configurations; Fock1 denotes the inclusion of the
self-energy Fock contribution, while Fock2 denotes the additional
inclusion of the medium polarization energy Fock contribution. In each
case, the effective mass is plotted for a limited range of Fermi
momenta corresponding to densities up to $\sim 0.5~{\rm fm}^{-3}$.\par
\clearpage

\subsection{Hartree--Fock QHD Stars}\label{sec:HYBRIDFOCKSTARS}
\begin{figure}[!b]
\centering
\includegraphics[angle=90,width=0.9\textwidth]{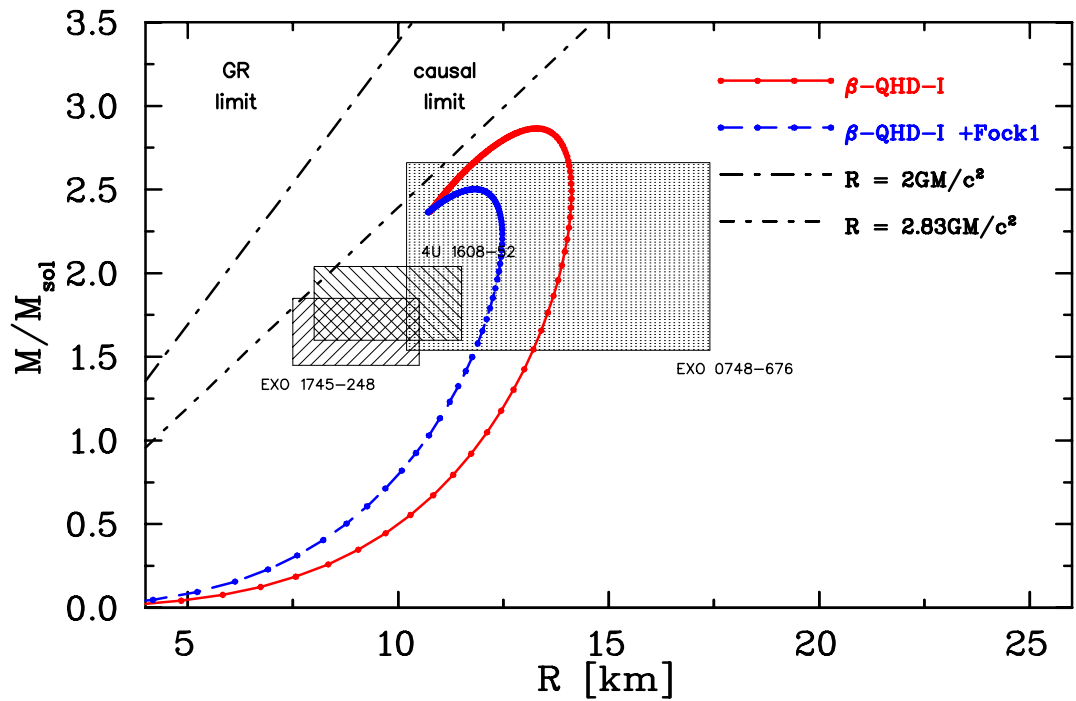}
\caption[Mass-radius relations for Hartree--Fock QHD-I]{(Color Online)
  Mass-radius relations for Hartree--Fock QHD-I, in which we have
  included the self-energy Fock terms. Also shown for comparison are
  the mass-radius relations of Hartree QHD-I as per
  Fig.~\ref{fig:MvsRQHDN}. We observe that the inclusion of this Fock
  term softens the EOS, resulting in a reduced maximum stellar
  mass. \protect\label{fig:QHDCNFockTOV}}
\end{figure}

To investigate the effects that the addition of the various Fock terms
have on stellar solutions for these EOS, we can solve the TOV
equations in each case. Fig.~\ref{fig:QHDCNFockTOV} shows the
mass-radius relations for nucleonic QHD-I with the additional
self-energy Fock terms (Fock1) included, in which the couplings used
provide a fit to the saturation properties of nuclear matter. For
comparison purposes, the same stellar solutions with Fock terms
neglected (as per Fig.~\ref{fig:MvsRQHDN}) are shown, and we note that
the maximum mass is lower when the Fock terms are included, indicating
a softening of the EOS.\par
Since the compression modulus of the full Hartree--Fock configuration
(Fock2) for QHD is smaller than than of the Fock1 configuration, we
expect that maximum mass calculated with the Fock2 configuration would
be lower than those in Fig.~\ref{fig:QHDCNFockTOV}.\par
\clearpage

\section{Summary Tables}\label{sec:summary}
The vacuum masses of the baryons, leptons, and mesons used in the
calculations for this work are shown in Table~\ref{tab:masses}, as per
Ref.~\cite{Amsler:2008zzb}.\par
The couplings used in each configuration of each model are found such
that the nuclear matter configuration reproduces saturation properties
(as per Section~\ref{sec:EoS}). The couplings used for all of the
individual configurations of QHD, QMC, and Hartree--Fock QHD are shown
in Table~\ref{tab:couplings}. In the case of Hartree--Fock QHD we show
the couplings that reproduce saturation properties for the case of
only including the self-energy Fock contribution (denoted
Hartree--Fock1), and for the case of including that contribution and
additionally the medium-polarization Fock contribution (denoted
Hartree--Fock2). The couplings are given to three decimal places to
allow comparison between investigations, though more significant
figures would be required to exactly reproduce these results.\par
Table~\ref{tab:results} contains comparison values for each of the
configurations of each model; The input parameters used to calculate
each configuration, such as species modelled, and\emdash where
applicable\emdash the quark masses and bag energy density; as well as
the calculated values for phase transition densities, compression
modulii, and effective mass at saturation.\par
%
%
\begin{table}[!h]
\centering
\caption[Baryon, lepton, and meson
  masses]{\protect\label{tab:masses}The vacuum (physical) baryon,
  lepton, and meson masses (in units of MeV) as used throughout this
  work~\cite{Amsler:2008zzb}.\vspace{2mm}}
\begin{tabular}{cccccccc}
\hline
\hline\\[-3mm]
$M_{p}$         & 
$M_n$           &
$M_\Lambda$     & 
$M_{\Sigma^-}$  & 
$M_{\Sigma^0}$  & 
$M_{\Sigma^+}$  & 
$M_{\Xi^-}$     &
$M_{\Xi^0}$     \\[1mm]
938.27          &
939.57          &
1115.68         &
1197.45         &
1192.64         &
1189.37         &
1321.31         &
1314.83         \\[1mm]
\hline \\[-3.5mm]
\multicolumn{3}{c}{$m_\s$}   &
\multicolumn{2}{c}{$m_\w$}   & 
\multicolumn{3}{c}{$m_\rho$} \\[1mm]
\multicolumn{3}{c}{$550.0$}  &
\multicolumn{2}{c}{$782.6$}  & 
\multicolumn{3}{c}{$775.8$}  \\[1mm]
\hline
 & 
\multicolumn{3}{c}{$m_{e^-}$} &
\multicolumn{3}{c}{$m_{\mu^-}$} 
 & \\[1mm]
 &
\multicolumn{3}{c}{$0.51$} &
\multicolumn{3}{c}{$105.66$} 
 & \\[1mm]
\hline
\hline
\end{tabular}
\end{table}
\begin{table}[!h]
\centering
\caption[Baryon-meson couplings and compression modulii]{Baryon-meson
  couplings for all configurations used in this work, found such that
  the saturation properties of nuclear matter (as described in
  Section~\ref{sec:EoS}) are reproduced. Also shown here are the
  resultant compression modulii for each configuration (we only need
  present the isospin symmetric value, as per the definition of $K$.
\protect\label{tab:couplings}\vspace{4mm}}
\begin{tabular}{c|ccccc}
\hline
\hline \\[-3.5mm]
Configuration & $g_{N\s}$ & $g_{N\w}$ & $g_{\rho}$ && $K~({\rm MeV})$ \\[1mm]
\hline\\[-3mm]
Hartree QHD & 10.644 & 13.179 & 6.976 && 525  \\[2mm]
Hartree QMC & 8.268 & 8.417 & 4.167 && 281  \\[2mm]
Hartree--Fock1 QHD & 10.001 & 11.819 & -- && 456  \\[1mm]
Hartree--Fock2 QHD & 11.289 & 14.170 & -- && 366  \\[1mm]
\hline
\hline
\end{tabular}
\end{table}
%
%
\clearpage
%
\begin{landscape}
\begin{table}[!h]
\centering
\caption[Summary of results]{\protect\label{tab:results}Table of
  species content ($N$ = nucleons, $Y$ = hyperons, $\ell$ = leptons,
  $q$ = quarks); inputs ($B^{1/4}$, $m_q$); and results for octet QMC
  and quark models presented in this thesis. $\rho_{\rm Y}$,
  $\rho_{\rm MP}$ and $\rho_{\rm QP}$ represent the density at which
  hyperons first appear ($\Lambda$ is the first hyperon to enter in
  these calculations in all but one configuration; octet QMC-I
  (Fig.~\ref{fig:specfrac_QMCnorho} where the $\rho$ meson is
  neglected, in which case the $\Sigma^-$ is the first hyperon to
  enter); the density at which the mixed phase begins; and the density
  at which the quark phase begins, respectively. Figures for selected
  parameter sets are referenced in the final column. Dynamic NJL quark
  masses are determined by
  Eqs.~(\ref{eq:condensate})--(\ref{eq:qcoupling}).
\vspace{10mm}}
\begin{tabular}{lcccccl}
\hline
\hline
&&&&&& \\[-2mm]
Particles: & 
$B^{1/4}~({\rm MeV})$ & 
$\{m_u,m_d,m_s\}~({\rm MeV})$ & 
$\rho_{\rm Y}~({\rm fm}^{-3})$ & 
$\rho_{\rm MP}~({\rm fm}^{-3})$ & 
$\rho_{\rm QP}~({\rm fm}^{-3})$ &Figure:\\
&&&&&& \\[-3mm]
\hline
&&&&&& \\[-2mm]
N, Y, $\ell$, \phantom{q,} ($\s$, $\w$, $\rho$) & --- & ---               & 0.33 & ---  & ---  & Fig.~\ref{fig:specfrac_QMC} \\
N, Y, $\ell$, \phantom{q,} ($\s$, $\w$\; $\phantom{\rho}$)         & --- & ---               & 0.39 & ---  & ---  & Fig.~\ref{fig:specfrac_QMCnorho} \\
N, Y, $\ell$, q, ($\s$, $\w$, $\rho$) & 180 & \{3, 7, 95\}      & 0.54 & 0.22 & 0.95 & Fig.~\ref{fig:SpecFrac_PTQMC} \\
N, Y, $\ell$, q, ($\s$, $\w$\; $\phantom{\rho}$)       & 180 & \{3, 7, 95\}      & 0.51 & 0.36 & 0.95 & --- \\
N, Y, $\ell$, q, ($\s$, $\w$, $\rho$) & 195 & \{3, 7, 95\}      & 0.33 & 0.35 & 1.46 & Fig.~\ref{fig:SpecFrac_PTQMC_195} \\
N, Y, $\ell$, q, ($\s$, $\w$, $\rho$) & 170 & \{30, 70, 150\}   & 0.55 & 0.20 & 0.87 & --- \\
N, Y, $\ell$, q, ($\s$, $\w$, $\rho$) & 175 & \{100, 100, 150\} & 0.41 & 0.28 & 1.41 & --- \\
N, \phantom{Y,} $\ell$,    q, ($\s$, $\w$, $\rho$) & 180 & Dynamic (NJL)     & ---  & 0.64 & 1.67 & Fig.~\ref{fig:QMCnuclear} \\
&&&&&& \\[-2mm]
\hline
\hline
\end{tabular}
\end{table}
\end{landscape}

%% file: Chapter6_conclusions.tex
\chapter{Conclusions}\label{sec:conclusions}
We have derived and performed calculations for various Quantum
Hadrodynamics (QHD) equation of state (EOS) configurations
(preliminarily at Hartree level) as a foundation for further work,
both reproducing the results of a well-known reference, viz
Ref.~\cite{Serot:1984ey} and extending the model to include
$\beta$-equilibrium between baryons and leptons. In this model the
baryons are treated as effective degrees of freedom and interact with
a mean-field of mesons. We have shown that this model is not capable
of modelling hyperons due to a breakdown in the assumptions\emdash in
particular that the meson potentials are small, an assumption made in
order to require the Dirac equation to be defined for this
model\emdash and thus we have not pursued this model any further,
apart from calculating the stellar solutions; the mass-radius
relations.\par
We have investigated the latest (to date) manifestation of the
quark-meson coupling (QMC) model to produce an EOS for nucleonic
matter as well as investigations of the inclusion of hyperons to this
model, in $\beta$-equilibrium with leptons. In this model we include
the self-consistent response of the internal quark degrees of freedom
to the applied scalar field, allowing the hyperons to be modelled
without violating the above assumptions of the Dirac equation that
prevent us from calculating hyperon effects in QHD, as evidenced by
the fact that the effective baryon masses in QMC\emdash which now
include a quadratic scalar-field term\emdash remain positive at all
densities. We do note however that other models for hadronic matter
exist in which hyperon degerees of freedom become accessible, and in
which case a transition from hadronic matter to quark matter becomes
possible, e.g. Ref~\cite{Burgio:2002sn}. We have calculated the EOS
for various configurations of the models described in this thesis with
various values for variable quantities, and calculated constituents of
infinite matter for several of these configurations, providing
information on the density fractions of various particles in
equilibrium. We have investigated the effects that changing various
parameters of this model have on these density fractions of particles,
as well as the effects of restricting the types of particles modelled
with QMC. We note that qualitatively, the relative proportions of the
hyperons are in good agreement with other treatments (for example,
Ref.~\cite{Schaffner:1996}) and that we observe expected
phenomenology, in particular the suppression of $\Sigma$ hyperons
consistent with hypernuclei studies as noted in
Ref.~\cite{RikovskaStone:2006ta}. We have calculated and investigated
the stellar solutions for this EOS and compared the mass-radius
relations to the current state-of-the-art experimental
observations. We have noted the softening of the EOS and the
corresponding decrease in the maximum stellar mass, which we attribute
to the increase in the number of degrees of freedom (both due to the
hyperons, and more fundamentally, to the inclusion of baryon
structure) over which the Fermi momenta may be shared.\par
By performing the above calculations for both QHD and QMC, we have
investigated the effect of introducing the quadratic term in the QMC
effective masses, as compared to the linear form of the effective mass
in QHD. The effect is highly non-trivial, as can be inferred by the
change to the couplings that reproduce saturation properties for
nucleonic matter alone. More importantly, and as noted above, the
inclusion of strangeness degrees of freedom via the hyperons is a
profound difference between these models.\par
We have produced several EOS that simulate a phase transition from
octet QMC modelled hadronic matter, via a continuous Glendenning style
mixed phase to a pure, deconfined quark matter phase via a mixed phase
containing some fractions of hadronic and quark matter. We believe
that this should correspond to a reasonable description of the
relevant degrees of freedom in each density region. We have performed
investigations using the MIT bag model as a preliminary model for
deconfined quark matter, as well as investigations of the
Nambu--Jona-Lasinio (NJL) model for quark matter. The NJL model is
more sophisticated than the MIT bag model, in that it incorporates a
mechanism for simulating dynamical chiral symmetry breaking (DCSB),
while the MIT bag model assumes constant current quark masses. We have
shown that in the NJL scenario of DCSB the relevant chemical potential
relations prevent a transition from hadronic matter modelled with
either QHD or QMC to quark matter modelled with NJL in which the total
baryon density increases with increasing quark content. Further
investigations need to be performed in order to make any final
conclusions from this, but in any case this is a particularly
interesting result. Both cases of phase transitions from octet QMC
matter to quark matter are novel calculations not otherwise published
by any other researcher to the best of the author's knowledge.\par
We have thus shown that the models considered here reveal some
important clues to the possible nature of dense nuclear matter. It
appears that if dynamical chiral symmetry does indeed result in
typical constituent quark masses in low density quark matter, then a
phase transition from hadronic matter to quark matter is
unlikely. This result invites further investigation, particularly in
the quark matter phase with the goal of describing deconfined quarks
in a manner fundamentally consistent with QCD, but also in the
hadronic phase where\emdash as we have shown\emdash too-wide a range
of model parameters are consistent with experimental
observations. Once again, a future goal should involve a description
of hadronic matter fundamentally consistent with QCD.\par
As a further extension to QHD, we have derived and calculated
properties of Hartree--Fock QHD infinite matter and stellar
solutions. We have investigated the effects of including only the
corrections to the baryon self-energies as a method of increasing the
sophistication of the model, as well as including also the corrections
due to the medium polarization of baryons. Further work would include
extensions of QMC to this sophistication, as well as investigations of
the effects of the $\rho$ and $\pi$ mesons in Hartree--Fock QHD and
QMC. The calculations performed for Hartree--Fock QHD in
$\beta$-equilibrium are to the extent of the author's knowledge novel
calculations not otherwise published, and will be published by the
author in an upcoming article.\par
Of particular interest, the methods used here to solve the non-linear
self-consistent integral equations required in Hartree--Fock QHD (viz,
Aitken-improved Steffensen's Method) appear to be of a much greater
sophistication than previous efforts, in particular that the equations
are solved in full without introducing approximations such as
linearization, as was the case in Ref.~\cite{Horowitz:1983}. This is
of academic interest in that we have produced a suitable, reliable
method for performing these calculation in full, particularly given
that\emdash as we have shown\emdash a simpler treatment (Newton's
Method) fails for this system of equations.\par
The multiple-phase EOS demonstrate the complexity and intricacy of the
models, as well as the dependence on small changes in parameters. The
mass-radius relations predicted by the multiple-phase EOS calculations
provide overlap with the current experimentally acceptable range,
though the range of masses predicted is not yet able to reproduce all
currently observed stellar masses. The range of predicted stellar
radii are consistent with the current observational data. This is a
non-trivial result, especially when one considers the scale over which
we are investigating; the range of predicted radii for the most
sophisticated models in this work (for example, that used to produce
Fig.~\ref{fig:QHDCNFockTOV}) provides no predictions above twice the
observed range of radii of the stars in question, yet the radius of an
average main sequence star, such as our sun for example, is eight
orders of magnitude larger. While this is some concession,
unfortunately it is one shared by most radius predictions of stellar
matter, most models for which produce stellar masses and radii that
are at least order-of-magnitude consistent with observation.\par
The couplings of the baryons to mesons used throughout this work (as
summarized in Table~\ref{tab:couplings}) have been found such that the
EOS in which they are used reproduces the properties of saturated
nuclear matter with the appropriate value of the energy per baryon,
and occurring at the appropriate value of saturation density. These
couplings are dependent on the exact specifications of the model in
question, and as such are difficult to compare between treatments. In
the defining study of QHD~\cite{Serot:1984ey} the authors utilise
charge symmetry for the nucleons, and thus the nucleons have a
degenerate mass. This is not the case in our work, and the distinction
is an important one, as neglecting charge symmetry allows us to
appropriately model leptons as well as baryons. We remind the reader
that these couplings are only applicable to reproducing the values of
saturation properties quoted in this text, and that the use of
different values will result in different coupling constants. The
saturation properties used in the above reference are slightly
different to those used in this work, but nonetheless, our values for
the couplings of the nucleons to the mesons are suitably consistent
with that reference. \par
According to the SU(6) quark model described in
Sec.~\ref{sec:su6deriv}, the coupling of the nucleons to the $\rho$
meson should relate to the coupling of the nucleons to the $\w$ meson
in a ratio of $g_{\rho}:g_{N\w} = 1:3$. As stated in that section
though, rather than using the SU(6) relations for the $\rho$ meson we
determine the $\rho$ coupling to the nucleons via the symmetry energy
$a_4$ as per Eq.~(\ref{eq:a4}). For both QHD and QMC, the $\rho$
coupling $g_{\rho}$ (when determined such that the particular value of
$a_4 = 32.5~{\rm MeV}$ is reproduced) is closer to $g_{N\w}/2$. This
is intimately linked to the particular value of $a_4$ that is to be
reproduced, as well as the values of saturation properties used to
determine $g_{N\w}$. As can be seen by comparing the values in
Table~\ref{tab:couplings}, the model in question has considerable
impact on the couplings that reproduce the desired properties, the
cause of which is that the couplings appear in many times in many
equations in highly non-linear fashions. A small change to the value
of a baryon-meson coupling alters the meson potential, which in turn
alters the effective masses and chemical potentials, which in turn
alter the energy per baryon, and thus the saturation properties that
constrain the couplings. Likewise, a change to the functional form of
the effective mass (such as the difference between QHD and QMC)
introduces similar alterations, and thus model-dependent couplings are
essential.\par
\begin{figure}[!b]
\centering
\includegraphics[width=0.7\textwidth]{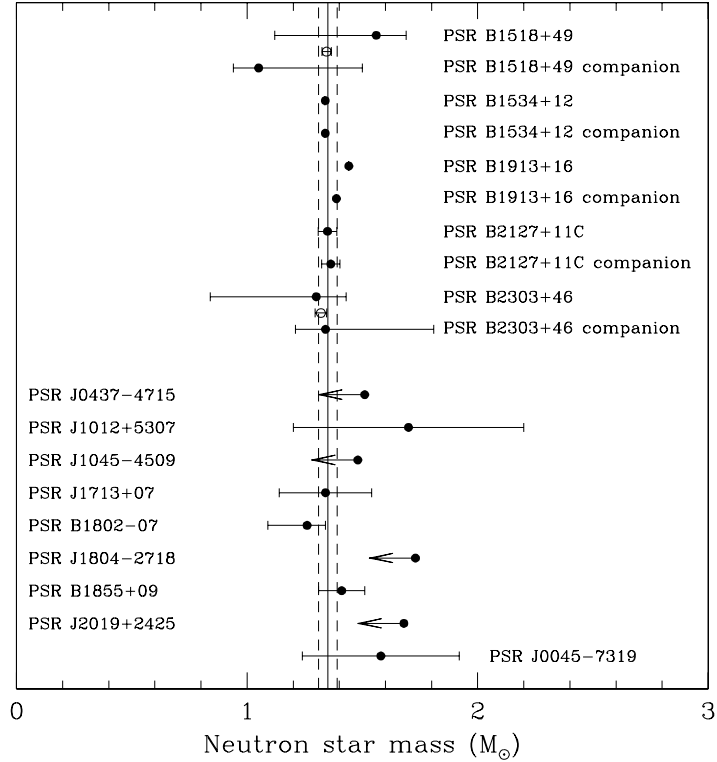}
\caption[Observed pulsar masses]{Observed pulsar masses from
  Ref.~\cite{Thorsett:1998uc} in which we note that the largest
  observed mass is $2.2~M_\odot$. Many results of our calculations are
  consistent with much of this data, though we caution a direct
  comparison between the models used in this work and observed compact
  stellar objects, for reasons discussed in the
  text. \protect\label{fig:observedmasses}}
\end{figure}

A comparison between the compression modulii of various models and
configurations has been made, and the range of values compared to the
literature; viz the experimental range noted in
Ref.~\cite{Serot:1984ey} of $K=200$--$300~{\rm MeV}$. While Hartree
QHD appears to produce a value corresponding to an EOS which is too
stiff ($K$ too large), QMC produces one at the upper bound of this
range. This is somewhat in contradiction with experiment due to the
concept that stiffness of an EOS is related to the maximum mass of a
compact object modelled with that EOS; although the QHD stellar
solutions include compact objects with masses in excess of much of the
observed data
of~\cite{Ozel:2006bv,Guver:2008gc,Ozel:2008kb}\footnote{As examples
  only; none of the models here are able to reproduce the masses of
  the largest observed neutron stars/pulsars, though we do not
  necessarily conclude that this invalidates any of our calculations},
the QMC EOS\emdash while adequately reproducing the value for
compression modulus\emdash results in stellar solutions with a maximum
mass that are only just consistent with the range of observed
data. The hybrid star stellar solutions (while these configurations do
not alter the value of compression modulus) provide even smaller
maximum masses, and one might conclude from this that the hybrid QMC
EOS is `too soft'. Herein lies our contradiction; how can any EOS
simultaneously reproduce the data for compression modulus and yet
produce stellar solutions with masses approaching $2.2~M_\odot$, the
masses of the largest observed pulsars? As an example of the
observational data for pulsars we show the range from
Ref.~\cite{Thorsett:1998uc} of observed pulsar masses in
Fig.~\ref{fig:observedmasses}. We note that in this case the largest
observed mass is roughly $2.2~M_\odot$, while many of these
observations are consistent with our results. We do however reiterate
that the predictions of our calculations correspond to static,
spherically symmetric (non-rotating) compact objects, whereas observed
pulsars are, by their definition rotating, and thus we caution a
direct comparison between theory and experiment in this case.\par
We must take further caution when comparing the results of our
numerical calculations with current experimental observations due to
the limitations imposed by the extraordinary challenge of measuring
the radius of an object over galactic distance scales. The current
state-of-the-art measurement techniques used for determining the radii
of these objects (such as X-ray burst timing measurements) are still
undergoing development, and recent analyses of the errors introduced
by the models used to determine the observational data for the most
prominent bounds has shown that much remains poorly
understood~\cite{Bhattacharyya:2009ws}. Until these measurements
become sufficiently discriminating to the point at which we can
exclude particular EOS based on their mass and radius predictions, we
continue to focus our attention on improving our models of the
EOS.\par
We note that the inclusion of the self-energy and medium polarization
Fock terms to nucleonic QHD provides a softening of the EOS which
appears to further lower the maximum mass of compact stellar objects
modelled with such an EOS. While this may appear to be a detriment to
this model, we reserve any conclusions about the validity of such
contributions until we are able to perform full calculations with more
complete additions; in particular the inclusions of hyperons, $\rho$
and $\pi$ mesons, and for modelling QMC. We note that the compression
modulus for Hartree--Fock QHD is in better agreement with the
literature value than Hartree QHD, indicating an improvement to the
original model, though it is overly simplistic to use this single
variable as a test of the model's validity. Further improvements to
the model based on more sophisticated physics should be undertaken
regardless of the experimental constraints.\par
We have shown that the omission of hyperons in the QMC model yields a
transition to a mixed phase of either NJL or MIT bag model quark
matter, as the hadronic EOS is no longer overly soft and the chemical
potential relations for the two phases are such that the transition
can still occur for constituent quark masses. This observation makes
clear that hyperons have a significant role to play in the EOS. We do
however acknowledge that their presence in neutron stars remains
speculative, since no observables are explicitly dependent on the
strangeness content of the system. We admit the possibility that there
may be some unknown mechanism that prevents the production of hyperons
in such a system, and as such we have investigated the possible
effects that this may have on the EOS and stellar solutions.\par
The results presented in Figs.~\ref{fig:QMChybrid} and
\ref{fig:QHDCNFockTOV} indicate that the most sophisticated models
presented in this work (in their current forms) are unable to
reproduce sufficiently massive neutron stars to account for all
observations, notably the largest observed stellar masses. This
appears to be a direct result of the softness of the EOS. This issue
will be explored in a future publication via the inclusion of Fock
terms to QMC, which we have shown to have an effect on the softness of
an EOS, in our case for QHD. We also note that other studies have
shown a strong link between Fock terms and the scalar and vector
potentials~\cite{Krein:1998vc} which indicate that the effects are
non-negligible.\par
Many open questions remain to be investigated in further work,
including further investigations into the effects of Fock terms, and
the density dependence of the bag energy density in the quark phase,
which can be calculated explicitly within the NJL model. The quark
matter models used here are still not the most sophisticated models
available, and further work may involve an investigation of the
effects of color-superconducting quark
matter~\cite{Alford:2007xm,Lawley:2006ps}.\par

%% file: AppendixA_derivations.tex
\chapter{Derivations}\label{sec:derivations}
In this section we will provide some in-depth derivations in order to
provide a more complete understanding of the expressions derived and
their origins. Any conventions used are noted throughout the text
where appropriate, though standard particle physics conventions can
generally be assumed. We present these derivations in context to the
work contained herein, and acknowledge that further extensions may not
be valid for the particular examples shown.\par
\section{Feynman Rules/Diagrams}\label{sec:diagrams}
Although not strictly a derivation {\it per se}, here we present a
summary for one of the most convenient and useful features of Quantum
Field Theory, providing the ability to describe particle interactions
in a diagram by following some simple rules; The Feynman Rules.\par
The diagrams were originally developed for Quantum
Electrodynamics~\cite{Feynman:1949zx}, and were later developed for
QFT in general. These are known as Feynman Diagrams\footnote{Or
  Stuckelberg Diagrams, as \hbox{Murray~Gell-Mann} allegedly preferred
  to call them due to a similar, earlier
  notation~\cite{Stuckelberg:1934}.} and the rules as Feynman
Rules. The diagrams are used throughout this work, and we will outline
the rules governing them here briefly.\par
\begin{center}
\begin{figure}[!t]
\centering \includegraphics[width=0.65\hsize]{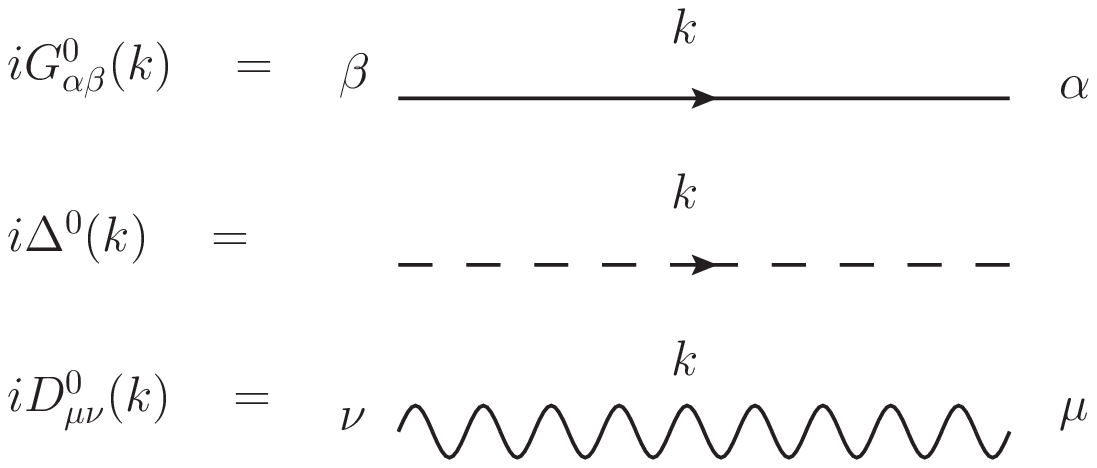}
\caption[Propagators for various types of particles]{Each line
  represents a free propagator for either a fermion (solid line); a
  scalar meson (dashed line); or a vector meson (wavy
  line).}\label{fig:propagators}
\end{figure}
\end{center}

\begin{itemize}
\item To begin with, we first need to determine the particles which
  are to enter and leave the process we are describing. For this
  discussion, we shall limit ourselves to QCD. These are the `external
  legs' of the diagram. The number of external legs will determine the
  number of required momenta. For the case of two external legs, only
  a single momentum is required; due to conservation of momentum,
  what goes in must also come out, and the external legs must
  correspond to `on-shell' particles.
\item For each type of particle, we can represent the propagator (see
  Appendix~\ref{sec:propagators}) by a line. The typical line styles
  are shown in Fig.~\ref{fig:propagators}. Each line introduces a
  propagator into the expression for the diagram. Internal propagators
  may be `off-shell'.
\item All of the lines meet at vertices, and introduce a coupling term
  into the expression for the diagram. For interactions of elementary
  particles, scalar vertices introduce a factor of $ig$ (where $g$ is
  the baryon-meson coupling) and vector particles introduce a factor
  of $-ig\gamma_\mu$. The four-momentum must be conserved at each
  vertex.
\item For each internal momentum corresponding to a loop not fixed by
  momentum conservation, a factor of
  $\displaystyle{\ \frac{1}{(2\pi)^4}\int d^4k\ }$ is introduced to the
  expression for the diagram, and for each closed fermion loop, an
  additional factor of $(-1)$ applies. 
\item Due to the antisymmetry under exchange of fermions, any two
  graphs distinguished only by the exchange of two external identical
  fermion lines must differ by a factor of $(-1)$.
\end{itemize}
Given these rules, diagrams can be easily related to mathematical
expressions, governed by a Lagrangian density, and the appropriate
interactions can be read off. Furthermore, we can consider more
complicated Feynman diagrams to address all higher-order
contributions, such as an addition of gluon exchange between two
external legs.\par

\section{Propagators}\label{sec:propagators}
In order to use the propagators introduced in the previous section, we
need to fully understand their structure and origin. For this purpose,
we provide an overview of the derivation of the scalar propagator.\par
\begin{center}
\begin{figure}[!b]
\centering \includegraphics[width=0.35\hsize]{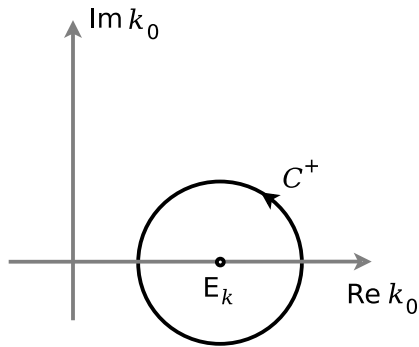}
\caption[Contour integral around a single pole]{Contour integration
  performed counter-clockwise around contour $C^+$ in the complex
  plane of $k_0$ with a pole at real value
  $E_{\vec{k}}$.}\label{fig:ContourInt_1}
\end{figure}
\end{center}

First, we define the four-vector of momentum to be
\be k = k_\mu = (k_0, \vec{k}\, ),
\ee
%
%
%
We can consider a free, positive-energy (frequency) scalar field
$\phi^+$ or negative-energy scalar field $\phi^-$ at a space-time
point $x$ to be expressed in terms of positive-energy creation
($a^\dagger$) and annihilation ($a$) operators (or equivalently,
negative-energy annihilation ($b$) and creation ($b^\dagger$)
operators) as
%
\be \phi(x) = \phi^+(x) + \phi^-(x) = \int \frac{d^3k}{(2\pi)^3}
\frac{1}{\sqrt{2 E_{\vec{k}}}} \left( a(\vec{k}\,) e^{-ik\cdot x} +
b^\dagger(\vec{k}\, ) e^{ik\cdot x}\right)\ , \ee
where $E_{\vec{k}} = \sqrt{\vec{k}^2+M^2}$, and the annihilation
operators are defined by their actions on the vacuum state $|0\ket$;
\be
a(\vec{k}\,)|0\ket = b(\vec{k}\,)|0\ket = 0 \ \ \forall \ \vec{k}\ .
\ee
%
%
%
%
%
%
We can then use the commutation relation for the creation and
annihilation operators
\be \left[ a(\vec{k}), a^\dagger(\vec{k}') \right] =
\delta(\vec{k}-\vec{k}') = \left[ b(\vec{k}),
  b^\dagger(\vec{k}') \right], \ee
to define the positive- and negative-energy propagators
\be i\Delta^\pm(x-y) = \left[ \phi^\pm (x), \phi^{\dagger\; \mp}(y)
  \right], \ee
(where the factor of $i$ has been inserted for convenience), which
satisfy
\be \Delta^-(x-y) = -\Delta^+(y-x). \ee
%
%
%
We may then consider the positive-energy propagator for a Hermitian
field (in which case $\phi^\dagger = \phi$, thus $b=a$ and $b^\dagger
= a^\dagger$.)
\bea \nonumber i\Delta^+(x-y) &=& \left[ \phi^+(x),\phi^{\dagger\;
    -}(y)\right] \\ \nonumber &=& \reci{2 (2\pi)^3} \iint
\left[a(\vec{k}),a^{\dagger}(\vec{k}')\right] \frac{e^{-i k\cdot x}\; e^{i
  k' \cdot y}}{\sqrt{E_{\vec{k}}E_{\vec{k'}}}}\; d^3k\; d^3k'
  \\ \nonumber &=& \reci{2 (2\pi)^3} \iint \frac{e^{-i k\cdot x}\;
    e^{i k' \cdot y}}{\sqrt{E_{\vec{k}}E_{\vec{k'}}}}\;
  \delta(\vec{k}-\vec{k}')\; d^3k\; d^3k' \\
\label{eq:deltaplus}
&=& \reci{2 (2\pi)^3} \int \frac{e^{-i k\cdot(x-y)}}{E_{\vec{k}}}\;
d^3k, \eea
and similarly for the negative-energy propagator such that
\be i\Delta^{\pm}(x-y) = \pm \reci{2 (2\pi)^3} \int \frac{e^{\mp i
    k\cdot(x-y)}}{E_{\vec{k}}}\; d^3k.  \ee
We wish to include the energy components into this definition, but we
also wish to keep the integrations over real numbers. We start by
using contour integration. If we consider a function of the complex
valued $k_0$
\be
\label{eq:k0function}
f(k_0) = \frac{e^{-ik_0(x_0-y_0)}}{k_0+E_{\vec{k}}}, \ee
then we can define this function at $E_{\vec{k}} \in {\mathbb R}$ using
contour integration as shown in Fig.~\ref{fig:ContourInt_1} to be
\be
\label{eq:contint}
f(E_{\vec{k}} ) = \frac{e^{-iE_{\vec{k}}(x_0-y_0)}}{2E_{\vec{k}}} =
\reci{2\pi i}\oint\limits_{C^+} \frac{f(k_0)}{k_0 - E_{\vec{k}}}\;
dk_0. \ee
Eq.~(\ref{eq:contint}) is obtained via Cauchy's Integral Formula which
states that for every point $a$ in the interior of a closed disk $D =
\{z: |z-z_0| \leq r\}$ bounded by a curve $C$,
\be f(a) = \reci{2\pi i} \oint\limits_{C} \frac{f(z)}{z-a}\; dz, \ee
provided that the curve defined by $C$ is taken counter-clockwise.\par
\begin{center}
\begin{figure}[!t]
\centering \includegraphics[width=0.6\hsize]{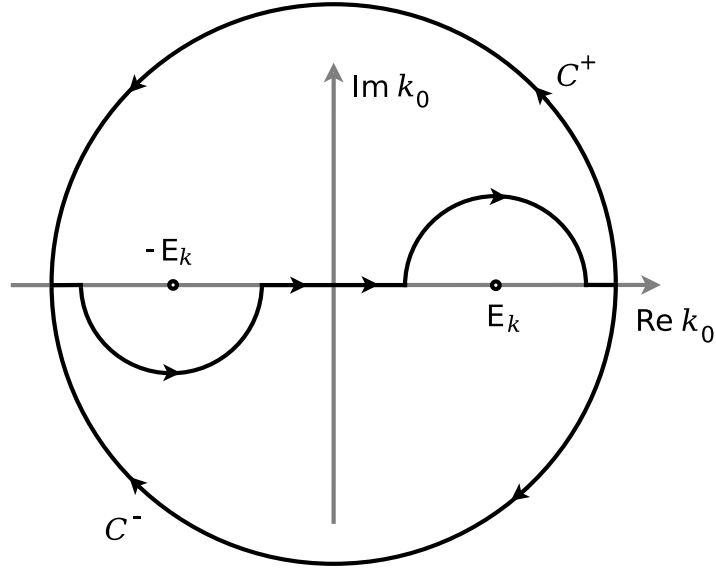}
\caption[Contour integration with poles at $\pm E_{\vec{k}}$]{Contour
  integration performed around contours $C^\mp$ in the complex plane
  of $k_0$ with poles at real values $\pm E_{\vec{k}}$. Note that the
  contour containing $E_{\vec{k}}$ (the lower contour) produces a
  negative value since it is taken clockwise.}\label{fig:ContourInt_2}
\end{figure}
\end{center}

If we consider the on-shell case in which $k_0 = E_{\vec{k}}$ and separate the
temporal components, Eq.~(\ref{eq:deltaplus}) becomes
\bea \nonumber i\Delta^{+}(x-y) &=& \reci{(2\pi)^3} \int e^{i
  \vec{k}\cdot(\vec{x}-\vec{y}\, )}\; \reci{2E_{\vec{k}}}\; e^{-iE_{\vec{k}}(x_0-y_0)}
d^3k \\ &=& \reci{(2\pi)^3} \int e^{i
  \vec{k}\cdot(\vec{x}-\vec{y}\, )}f(E_{\vec{k}})\; d^3k.  \eea
We can then insert the contour integrated form of $f(E_{\vec{k}})$ from
Eq.~(\ref{eq:contint})
\be i\Delta^{+}(x-y) = \reci{(2\pi)^3} \reci{2\pi i} \int e^{i
  \vec{k}\cdot(\vec{x}-\vec{y}\, )} \oint\limits_{C^+} \frac{f(k_0)}{k_0 -
  E_{\vec{k}}}\; dk_0\; d^3k, \ee
and then insert the form of $f(k_0)$ from Eq.~(\ref{eq:k0function}) to
give
\bea \nonumber i\Delta^{+}(x-y) &=& \reci{(2\pi)^3} \reci{2\pi i} \int
e^{i \vec{k}\cdot(\vec{x}-\vec{y}\, )} \oint\limits_{C^+}
\frac{e^{-ik_0(x_0-y_0)}}{(k_0 + E_{\vec{k}})(k_0 - E_{\vec{k}})}\; dk_0\; d^3k
\\ \nonumber &=& \frac{-i}{(2\pi)^4} \oint\limits_{C^+} \frac{e^{-i
    k\cdot(x-y)}}{(k_0)^2 - (E_{\vec{k}})^2}\; d^4k \\ &=& \frac{-i}{(2\pi)^4}
\oint\limits_{C^+} \frac{e^{-i k\cdot(x-y)}}{k^2 - M^2}\; d^4k,
\label{eq:4dprop}
\eea
where the integration over the three-momentum is for $-\infty < k_i \in
{\mathbb R} < \infty$ and the energy is over a contour integral $k_0
\in {\mathbb C}$ such that $k_0 \neq E_{\vec{k}}$. The last line arises from
\be (k_0)^2 - (E_{\vec{k}})^2 = k^2 + (\vec{k}\, )^2 - M^2 - (\vec{k}\, )^2 =
k^2 - M^2. \ee
A similar derivation is possible for the negative-energy propagator
$i\Delta^-(x-y)$, where the only difference will be the contour over
which the integration is performed, in that case $C^-$ to avoid the
point of $k_0 = -E_{\vec{k}}$, but which results in the same
expression as that of Eq.~(\ref{eq:4dprop}). The two integrations
required are illustrated in Fig.~\ref{fig:ContourInt_2}.\par
\begin{center}
\begin{figure}[!t]
\centering \includegraphics[width=0.6\hsize]{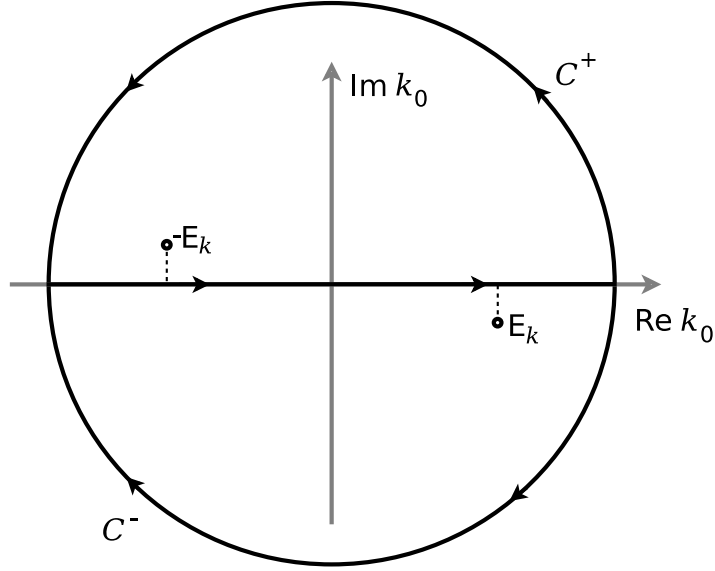}
\caption[Contour integration with poles at $\pm E_{\vec{k}} \mp
  i\epsilon$]{Contour integration performed around contours $C^\mp$ in
  the complex plane of $k_0$ with offset poles at $\pm E_{\vec{k}} \mp
  i\epsilon$.}\label{fig:ContourInt_3}
\end{figure}
\end{center}

To combine the two propagators together, we need to take care of the
contour integrations around both $E_{\vec{k}}$ and $-E_{\vec{k}}$. To
  do this, we can shift both away from the real axis via
\be E_{\vec{k}} \to E_{\vec{k}} - i\eta, \ee
which displaces these points to $- E_{\vec{k}} + i\eta$ and
$E_{\vec{k}} - i\eta$ such that the denominator for the positive
propagator is now
\be (k_0)^2 - (E_{\vec{k}} - i\eta)^2. \ee
Depending on which of $x_0$ and $y_0$ is larger, the exponential in
Eq.~(\ref{eq:4dprop}) will have a different sign. In order to unify
these, we use the time-ordered product to define the Feynman
propagator
\be \Delta_F(x-y) = \bra\Psi_0|T[\phi(y)\phi(x)]|\Psi_0\ket = \left\{ 
\begin{array}{ll} \Delta^+(x-y), \ \ {\rm if}
 & y_0 < x_0 \\ 
\Delta^-(x-y), \ \ {\rm if} & x_0 < y_0 \end{array}
\right. .
\ee
If we define the small parameter $\epsilon = 2\eta E_{\vec{k}}$, neglect terms
of ${\cal O}(\eta^2)$, and enlarge the contour over which we integrate
in the complex plane as shown in Fig.~\ref{fig:ContourInt_3} then the
Feynman scalar propagator can be written as
\be \label{eq:positionspace} \Delta_F(x-y) = \reci{(2\pi)^4}
\int_{-\infty}^{\infty} \frac{e^{-i k\cdot(x-y)}}{k^2 - M^2 +
  i\epsilon}\; d^4k, \ee
which has the advantage that now our integrals are all over real
numbers, and we have a single expression (since we no longer depend on
the contour). The factor of $\epsilon$ is removed after the
integration is performed by taking $\epsilon \to 0$.
\par
We note that Eq.~(\ref{eq:positionspace}) now takes the form of a
Fourier transform, and thus we can write the propagator in either
position-space or momentum-space, as
%
%
%
\be
\label{eq:FourierTransformPropagator}
\Delta_F(x-y) = \volint{k} e^{-ik\cdot(x-y)}\Delta_F(k),
\ee
where we can define the momentum-space propagator using
Eq.~(\ref{eq:positionspace}) as 
\be
\Delta_F(k) = (k^2 - M^2 + i\epsilon)^{-1}. 
\ee
This is the definition of the free scalar field ($\Delta_F(k) \equiv
\Delta^0(k)$) and we can similarly define a vector meson propagator
and a baryon propagator; for a vector field $V$ the free propagator is
\be iD_{\mu\nu}^{0}(y-x) = \bra\Psi_0|T[V_\mu(y)V_\nu(x)]|\Psi_0\ket =
i \volint{k} e^{-ik\cdot(y-x)}D_{\mu\nu}^0(k), \ee
and for a baryon $\psi$,
\be iG_{\alpha\beta}^{0}(y-x) =
\bra\Psi_0|T[\psi_\alpha(y)\bar{\psi}_\beta(x)]|\Psi_0\ket = i
\volint{k} e^{-ik\cdot(y-x)}G_{\alpha\beta}^0(k). \ee
\par
The explicit forms of the momentum-space propagators are
\be \Delta^0(k) = (k_\lambda^2 - m_\s^2 + i\epsilon)^{-1}, \ee
\bea \nonumber D_{\mu\nu}^0(k) &=& \left[-{\rm g}_{\mu\nu} + \frac{k_\mu
    k_\nu}{m_\w^2}\right] (k_\lambda^2 - m_\w^2 + i\epsilon)^{-1}
\\[2mm] &=& -{\rm g}_{\mu\nu} (k_\lambda^2 - m_\w^2 + i\epsilon)^{-1}, \eea
\be G_{\alpha\beta}^0(k) = (\kslash -
M)_{\alpha\beta}\left[(k_\lambda^2 - M^2 + i\epsilon)^{-1} +
  \frac{i\pi}{E^*(\vec{k})}\delta(k_0 - E_{\vec{k}})\theta(k_F - |\vec{k}|)\right],
\ee
using the definitions given later in Eq.~(\ref{eq:HFdefinitions}), and
where the baryon propagator has two parts, the second of which arises
as a result of the immersion in a Fermi sea.\par
%
%
%
%
%
%
%

\section{QHD Equation of State}\label{sec:qhdderiv}
Here we will fully derive the EOS (refer to Section~\ref{sec:EoS}) for
Quantum Hadrodynamics (of which the Quark-Meson Coupling model can be
considered an {\it a posteriori} extension) in more detail than in the
main text. Although we have performed calculations for several
configurations of QHD, the derivation provided here should be a
suitable framework to expand on in order to reproduce all of the
results of this work.\par
%
%
In the derivation of QHD that follows, we consider a collection of
nucleons (protons and neutrons) with strong interactions (which in QCD
are a result of quark-gluon interactions) modelled by $\sigma$,
$\omega$ and $\rho$ mesons (i.e. QHD-II) in a Relativistic Effective
Field Theory.\par
In order to describe the nucleons we combine them into a Dirac spinor
as
\be \psi \equiv \psi_N = \begin{pmatrix}\psi_p\\\psi_n\end{pmatrix}. \ee
We can then write the QHD Lagrangian density with this spinor and
meson terms (as described in Section~\ref{sec:lagrangiandensity}) as
\bea \nonumber {\cal L} &=
&\bar{\psi}\left[\gamma_{\mu}(i\partial^{\mu}-g_{N\omega}\omega^{\mu}-
  g_{\rho}\vec{\rho}^{\; \mu}\cdot\vec{\tau}\; )-(M-g_{N\sigma}\sigma)\right]\psi\\ \nonumber
&&+\frac{1}{2}(\partial_{\mu}\sigma\partial^{\mu}\sigma-m^{2}_{\sigma}\sigma^{2})-
\frac{1}{4}\Omega_{\mu\nu}\Omega^{\mu\nu} -
\frac{1}{4}R^a_{\mu\nu}R_a^{\mu\nu} \\
\label{eq:AppAQHDlag}
&&+\frac{1}{2}m^{2}_{\omega}\omega_{\mu}\omega^{\mu}+
\frac{1}{2}m^{2}_{\rho}\rho_{\mu}\rho^{\mu}+\delta{\cal L} \eea
where the field strength tensors for the vector mesons are
\be \Omega_{\mu \nu} = \partial_{\mu} \omega_{\nu} - \partial_{\nu}
\omega_{\mu}, \qquad R^a_{\mu \nu} = \partial_{\mu} \rho_{\nu}^a -
\partial_{\nu} \rho_{\mu}^a +
g_{\rho}\epsilon^{abc}\rho_\mu^b\rho_\nu^c. \ee
The mass term $M$ should be understood to be a diagonal matrix with
the proton and neutron masses on the diagonal. As discussed in
Section~\ref{sec:su6deriv} the couplings $g$ of the baryons to the
mesons are baryon dependent, with the exception of $g_\rho$ which
appears with the isospin-group dependent $\vec{\tau}$ as defined in
Eq.~(\ref{eq:taus}). \par
The Lagrangian density can also be written in a more familiar form if
we recognize that the vector field terms take the form of the temporal
component of an energy four-vector and hence have units of energy and
act to reduce the energy of the nucleon, whilst the scalar term takes
the form of a scalar mass, and hence acts to reduce the effective mass
of the nucleon. We can observe this better if we define the covariant
derivative as
\be
D_\mu = \del_\mu + ig_{N\w}\omega_\mu  +ig_\rho \rho^a_\mu \tau_a,
\ee
so, neglecting the scalar field for a moment, we have a term that
appears similar to the Dirac equation
\be \left(i\Dslash - M\right) \psi = 0. \ee
\par
The Euler--Lagrange Equations describe the equations of motion for all
particles $\Phi$ involved in the above Lagrangian density,
\be \label{eq:ELEQ} \frac{\partial{\cal L}}{\partial\Phi}-
\partial_{\mu}\frac{\partial{\cal L}}{\partial(\partial_{\mu}\Phi)} =
0. \ee
When we calculate the equations of motion for
Eq.~(\ref{eq:AppAQHDlag}) for the mesons we arrive at a Klein--Gordon
equation for the $\sigma$ field and Maxwell equations for the $\omega$
and $\rho$ fields (by construction, since the free terms in the
Lagrangian density were chosen such that this phenomenology would be
reproduced), thus
\bea \left(\Box+m^{2}_{\s}\right)\sigma &=& g_{N\sigma}\bar{\psi}\psi,
\\ \partial^{\mu}\Omega_{\mu\nu} &=&
g_{N\omega}\bar{\psi}\gamma_{\nu}\psi-
m^{2}_{\omega}\omega_{\nu},\\ \partial^{\mu}R^a_{\mu\nu} &=&
g_{\rho}\bar{\psi}\gamma_{\nu}\tau^a\psi- m^{2}_{\rho}\rho_{\nu}^a.
\eea
Applying Eq.~(\ref{eq:ELEQ}) to the nucleons (and similarly for the
conjugate) we obtain a Dirac equation,
\be \label{eq:derivDirac}
\left[i\!\not\!\partial-g_{N\omega}\!\not\!\omega-g_{\rho}\!\not\!\rho^a\tau_a
  - M + g_{N\s}\s \right]\psi = 0. \ee
\par
We now consider a Mean-Field Approximation (MFA, see
Section~\ref{sec:mfa}) in which the meson fields are described by only
their constant, classical components. We separate the meson fields
into classical and quantum components as
\bea \nonumber \sigma &=& \sigma_{\rm{classical}} +
\sigma_{\rm{quantum}}, \\ \nonumber \omega &=& \omega_{\rm{classical}}
+ \omega_{\rm{quantum}}, \\ \rho &=& \rho_{\rm{classical}} +
\rho_{\rm{quantum}}. \eea
If we now consider vacuum expectation values, the quantum operator
component for the mesons vanishes leaving only a constant classical
component
\bea \nonumber \sigma &\to& \langle \sigma \rangle, \\ \nonumber
\omega^{\mu} &\to& \langle \omega \rangle \equiv \langle \omega_0
\rangle \delta^{\mu 0}, \\ \rho^{\mu} &\to& \langle \rho \rangle \equiv
\langle \rho_0 \rangle \delta^{\mu 0}. \eea
The $\delta^{\mu 0}$ terms for the vector mesons arise due to
rotational invariance as per Section~\ref{subsec:rotational}. Isospin
invariance requires that only the third isospin component of the
isovector meson $\rho_0$ will be non-vanishing, for which we shall
drop the subscript.\par
Applying the MFA For the $\sigma$ field we obtain
%
%
\be \nonumber \left(\Box+m^{2}_{\sigma}\right)\langle \sigma \rangle
\ = \ g_{N\sigma}\langle\bar{\psi}\psi\rangle \ \ \stackrel{\rm
  MFA}{\longrightarrow} \ \ m^{2}_{\sigma}\langle \sigma \rangle \ =
\ g_{N\sigma}\langle\bar{\psi}\psi\rangle \ee
\be \therefore \ \ \langle \sigma \rangle \ =
\ \displaystyle{\frac{g_{N\sigma}}{m^{2}_{\sigma}}\langle\bar{\psi}\psi\rangle}. \ee
For the $\omega$ field we obtain
%
%
\be \nonumber \partial_{\mu}\Omega^{\mu\nu} = \ g_{N\omega}\langle\bar{\psi}\gamma^{\nu}\psi\rangle-
m^{2}_{\omega}\omega^{\nu} \ \ \stackrel{\rm MFA}{\longrightarrow}
\ \ 0 \ = \ g_{N\omega}\langle\bar{\psi}\gamma^{0}\psi\rangle-
m^{2}_{\omega}\langle \omega \rangle \ee
\be \therefore \ \ \langle \omega \rangle \ =
\ \displaystyle{\frac{g_{N\omega}}{m^{2}_{\omega}}\langle\bar{\psi}\gamma^{0}\psi\rangle} \ =
\ \displaystyle{\frac{g_{N\omega}}{m^{2}_{\omega}}\langle\psi^\dag
  \psi\rangle}, \ee
%
%
%
and similarly for the $\rho$ field,
%
%
\be \nonumber \partial_{\mu}R_3^{\mu\nu} = \ g_{\rho}\langle\bar{\psi}\gamma^{\nu}\tau_3 \psi\rangle
-m^{2}_{\rho}\rho^{\nu} \ \ \stackrel{\rm MFA}{\longrightarrow} \ \ 0
\ = \ g_{\rho}\langle\bar{\psi}\gamma^{0}\tau_3 \psi\rangle-
m^{2}_{\rho}\langle \rho \rangle \ee
\be \therefore \ \ \langle \rho \rangle \ =
\ \displaystyle{\frac{g_{\rho}}{m^{2}_{\rho}}\langle\bar{\psi}\gamma^{0}\tau_3 \psi\rangle}
= \ \displaystyle{\frac{g_{\rho}}{m^{2}_{\rho}}\langle\psi^\dag
  \tau_3 \psi\rangle}. \ee
%
%
%
We can define in each of these cases a density associated with the
mean-fields of the mesons. These are the scalar, vector, and
isovector densities,
\bea \label{eq:rhos} \rho_\s \ \equiv &\rho_S &= \
\bra\bar{\psi}\psi\ket, \\ \label{eq:rhow} \rho_\w \ \equiv &\rho_{\rm total}
&= \ \bra\psi^\dagger\psi\ket = \rho_p + \rho_n, \\ \label{eq:rhorho}
\rho_\rho \ \equiv &\rho_{I} &= \ \bra\psi^\dagger\tau_3\psi \ket =
\half\left(\rho_p - \rho_n\right).  \eea
The vector density $\rho_{\rm total}$ and the isovector density
$\rho_{I}$ are the conserved baryon density and isovector density
respectively, the expressions for which can be found using the nucleon
spinor. The closed form of the scalar density $\rho_S \equiv \rho_\s$
will be derived in Appendix~\ref{sec:scalarfield}.\par
Finally, calculating the equations of motion for the nucleons we
obtain
\be \left[i\delslash-g_{N\omega}\langle\omega\rangle\gamma_0-
  g_{\rho}\langle\rho\rangle\gamma_0\tau_3 - M + g_{N\s} \bra\s\ket
  \right]\psi = \left[i\Dslash - M^*\right]\psi = 0, \ee
(and similarly for the conjugate spinor $\bar{\psi}$). We can now
rewrite the QHD Lagrangian density with the surviving mean-field
terms, as
\bea \nonumber\bra{\cal L}\ket &=&
\bar{\psi}\left[i\gamma_{\mu}\partial^{\mu} -g_{N\omega}\langle \omega
  \rangle\gamma_{0} -g_{\rho}\langle \rho \rangle\gamma_{0}\tau_3
  -M^*\right]\psi \\[1mm] \nonumber &&-
\displaystyle{\frac{1}{2}m^{2}_{\sigma}\langle \sigma \rangle^{2}
  +\frac{1}{2}m^{2}_{\omega}\langle \omega \rangle^{2}
  +\frac{1}{2}m^{2}_{\rho}\langle \rho \rangle^{2}} \\[2mm] \nonumber
&=& \bar{\psi}\left[i\gamma_{\mu}\partial^{\mu} -g_{N\omega}\langle
  \omega \rangle\gamma_{0} -g_{N\rho}\langle \rho
  \rangle\gamma_{0}\tau_3 -M^*\right]\psi \\[1mm]
&&-\displaystyle{\frac{g^{2}_{N\sigma}}{2m_{\sigma}^{2}}\rho_{\sigma}^{2}
  +\frac{g^{2}_{N\omega}}{2m_{\omega}^{2}}\rho_{\omega}^{2}
    +\frac{g^{2}_{\rho}}{2m_{\rho}^{2}}\rho_{\rho}^{2}}.\label{eq:MFAlag}
  \eea
Note that the $\sigma$ meson mas term has an opposite sign to the
vector meson terms.\par
For a uniform system, the energy-momentum tensor has the form
\be \langle T_{\mu \nu} \rangle = \left({\cal
  E}+P\right)u_{\mu}u_{\nu}+P {\rm g}_{\mu\nu}, \ee
where the four-velocity for a fluid at rest is $u = (1,\vec{0}\, )$,
and satisfies $u^\mu u_\mu = -1$. We use the standard identity for the
energy-momentum tensor;
\be P = \frac{1}{3} \langle T_{ii} \rangle, \quad {\cal E} = \langle
T_{00} \rangle, \ee
and the Lagrangian form of the energy-momentum tensor
\be T_{\mu \nu} = -{\rm g}_{\mu\nu}{\cal L}+\partial_{\nu}q_i
\frac{\partial {\cal L}}{\partial \partial^{\mu}q_i}, \ee
then insert the MFA Lagrangian of Eq.~(\ref{eq:MFAlag}) to obtain
\be \label{eq:finalemtensor} \bra T_{\mu \nu}\ket =
i\bra\bar{\psi}\gamma_{\mu}\partial_{\nu}\psi\ket -\left(
-\frac{g^{2}_{N\sigma}}{2m_{\sigma}^{2}}\rho_{\sigma}^{2}
+\frac{g^{2}_{N\omega}}{2m_{\omega}^{2}}\rho_{\omega}^{2}
+\frac{g^{2}_{\rho}}{2m_{\rho}^{2}}\rho_{\rho}^{2} \right){\rm
  g}_{\mu\nu}. \ee
Note that there is only one term involving $\partial^{\mu}$ in the
Lagrangian density, and that the equation of motion for the nucleon
(the Dirac equation, Eq.~(\ref{eq:derivDirac})) will prevent any
fermion factors in the term proportional to ${\rm g}_{\mu \nu}$. We
can rearrange this using Eq.~(\ref{eq:derivDirac}) if we expand the
Einstein summation, as
\bea \nonumber \left(i\gamma_\mu \partial^\mu - g_{N\omega} \langle
\omega \rangle \gamma_0 - g_{\rho}\langle \rho \rangle
\gamma_0\tau_3 - M^* \right)\psi &=& 0 \\[2mm]
\nonumber \left(i\gamma_0 \partial^0 + i\gamma_i \partial^i -
g_{N\omega} \langle \omega \rangle \gamma_0 - g_{\rho}\langle \rho
\rangle\gamma_0\tau_3 -M^* \right)\psi &=& 0 \\[2mm]
\nonumber \left(i\gamma^0 \partial^0 - i\gamma^i \partial^i -
g_{N\omega}\langle \omega \rangle \gamma_0 - g_{\rho} \langle \rho
\rangle\gamma_0\tau_3 -M^* \right)\psi &=& 0 \eea
\be \label{eq:rearrangeDirac} \therefore \ \ i\gamma^0 \partial^0 \psi
= \left(i\gamma^i \partial^i + g_{N\omega} \langle \omega \rangle
\gamma_0 + g_{\rho} \langle \rho \rangle\gamma_0\tau_3 +M^*
\right)\psi. \ee
\par
We can now substitute Eq.~(\ref{eq:rearrangeDirac}) into the
expression for the energy-momentum tensor,
Eq.~(\ref{eq:finalemtensor}) to obtain an expression for the energy
density
\bea \nonumber \mathcal{E} &= \ \langle T_{00} \rangle &=
\bar{\psi}\left[i\gamma^{i}\partial^{i} +g_{N\omega}\langle \omega
  \rangle\gamma_{0}+g_{\rho}\langle \rho \rangle\gamma_{0} \tau_3
  +M^*\right]\psi \\
&&-\left( -\frac{g^{2}_{\sigma N}}{2m_{\sigma}^{2}}\rho_{\sigma}^{2}
+\frac{g^{2}_{\omega N}}{2m_{\omega}^{2}}\rho_{\omega}^{2}
+\frac{g^{2}_{\rho N}}{2m_{\rho}^{2}}\rho_{\rho}^{2} \right). \eea
%
%
%
We can furthermore extract a $\gamma_0$ term from
$\bar{\psi}=\psi^{\dagger}\gamma_{0}$ to obtain
\bea \nonumber &{\cal E} &= \psi^{\dagger}\left[i
  \gamma_{0}\gamma^{i}\partial^{i} +g_{N\omega}\langle \omega
  \rangle\gamma_{0}^{2} +g_{\rho}\langle \rho
  \rangle\gamma_{0}^{2}\tau_3 +\gamma_{0}M^{*}\right]\psi \\
\label{eq:A70}
&& \quad -\left( -\frac{g^{2}_{\sigma
    N}}{2m_{\sigma}^{2}}\rho_{\sigma}^{2} +\frac{g^{2}_{\omega
    N}}{2m_{\omega}^{2}}\rho_{\omega}^{2} +\frac{g^{2}_{\rho
   }}{2m_{\rho}^{2}}\rho_{\rho}^{2} \right). \eea
Using Dirac's notation, in which \be \alpha^i = \gamma_0
\gamma^i,\quad \beta = \gamma_0,\quad \gamma_0^2 = I, \quad \partial^i
= (-\vec{\nabla})^i, \ee we can write Eq.~(\ref{eq:A70}) as
\bea \nonumber {\cal E} &=& \psi^{\dagger}\left[-i \vec{\alpha}\cdot\vec{\nabla}
  +g_{N\omega}\langle \omega \rangle+g_{\rho} \langle \rho
  \rangle\tau_3 +\beta M^{*}\right]\psi\\ &&-\left(
-\frac{g^{2}_{N\sigma}}{2m_{\sigma}^{2}}\rho_{\sigma}^{2}
+\frac{g^{2}_{N\omega}}{2m_{\omega}^{2}}\rho_{\omega}^{2}+
\frac{g^{2}_{\rho}}{2m_{\rho}^{2}}\rho_{\rho}^{2}
\right). \eea
We can now separate the interaction terms (in the square brackets) and
we notice that these can be expressed in the same form as the free
meson terms, so we collect these together and find that the signs of
the vector terms are reversed
\bea \nonumber {\cal E} &=& \psi^{\dagger}\left[-i
  \vec{\alpha}\cdot\vec{\nabla} +\beta
  M^{*}\right]\psi+\bar{\psi}g_{N\omega}\gamma_{0}\langle \omega
\rangle\psi +\bar{\psi}g_{\rho}\langle \rho \rangle\gamma_{0}\tau_3
\psi
+\frac{g^{2}_{N\sigma}}{2m_{\sigma}^{2}}\rho_{\sigma}^{2}
-\frac{g^{2}_{N\omega}}{2m_{\omega}^{2}}\rho_{\omega}^{2}
-\frac{g^{2}_{\rho}}{2m_{\rho}^{2}}\rho_{\rho}^{2} \\[2mm] \nonumber
&=& \psi^{\dagger}\left[-i \vec{\alpha}\cdot\vec{\nabla} +\beta
  M^{*}\right]\psi+\frac{g_{N\omega}^{2}}{m_{\omega}^{2}}\rho_{\omega}^2
+\frac{g_{\rho}^{2}}{m_{\rho}^{2}}\rho_{\rho}^{2}
+\frac{g^{2}_{N\sigma}}{2m_{\sigma}^{2}}\rho_{\sigma}^{2}
-\frac{g^{2}_{N\omega}}{2m_{\omega}^{2}}\rho_{\omega}^{2}
-\frac{g^{2}_{\rho}}{2m_{\rho}^{2}}\rho_{\rho}^{2}\\ \nonumber &=&
\psi^{\dagger}\left[-i \vec{\alpha}\cdot\vec{\nabla} +\beta
  M^{*}\right]\psi+\frac{g^{2}_{N\sigma}}{2m_{\sigma}^{2}}\rho_{\sigma}^{2}
+\frac{g^{2}_{N\omega}}{2m_{\omega}^{2}}\rho_{\omega}^{2}
+\frac{g^{2}_{\rho}}{2m_{\rho}^{2}}\rho_{\rho}^{2}\\
\label{eq:unfinishedE} 
&=& \psi^{\dagger}\left[\vec{\alpha}\cdot\vec{k} +\beta
  M^{*}\right]\psi+\frac{g^{2}_{N\sigma}}{2m_{\sigma}^{2}}\rho_{\sigma}^{2}
+\frac{g^{2}_{N\omega}}{2m_{\omega}^{2}}\rho_{\omega}^{2} 
+\frac{g^{2}_{\rho}}{2m_{\rho}^{2}}\rho_{\rho}^{2}, \eea
where we have used the relation $-i\vec{\nabla} = \vec{k}$. To further
simplify the interaction term here we consider solutions to the Dirac Equation
of the form
\be \label{eq:psi} \psi = u(p) e^{-iE^*t}e^{i \vec{p}\cdot\vec{x}}, \ee
where
\be \label{eq:uofp} u(p) = \frac{\!\not\!k +
  M^*}{\sqrt{2E^*}\sqrt{E^*+M^*}}\begin{pmatrix}\chi\\0\end{pmatrix};\quad
  \chi = \begin{pmatrix}1\\0\end{pmatrix} \ \textrm{or} \
 \begin{pmatrix}0\\1\end{pmatrix}, \ee
where we use a normalization factor of
$\displaystyle{\frac{1}{\sqrt{2E^*}}}$. Written out in full,
Eq.~(\ref{eq:uofp}) becomes
\bea \nonumber u(p) &=& \frac{1}{\sqrt{2E^*}\sqrt{E^*+M^*}} \left(
\gamma_\mu k^\mu +M^* \right)
\begin{pmatrix}\chi\\0\end{pmatrix} \\[2mm]
\nonumber &=& \frac{1}{\sqrt{2E^*}\sqrt{E^*+M^*}}
\begin{pmatrix}
(M^*+E^*)I&-\vec{\sigma}\cdot\vec{k}\\ \vec{\sigma}\cdot\vec{k}&(M^*-E^*)I
\end{pmatrix} \begin{pmatrix}\chi\\0\end{pmatrix} \\[2mm]
\nonumber &=& \frac{1}{\sqrt{2E^*}\sqrt{E^*+M^*}}
\begin{pmatrix}(M^*+E^*)\chi\\[1mm]
\vec{\sigma}\cdot\vec{k}\left(\chi\right)\end{pmatrix} \\[2mm]
\nonumber &=& \sqrt{\frac{E^*+M^*}{2E^*}}
\begin{pmatrix}\chi\\[1.5mm]
\frac{\vec{\sigma}\cdot\vec{k}}{{E^*+M^*}}\left(\chi\right)\end{pmatrix}.
\eea
\par
Given the relations
\be (\vec{\sigma}\cdot\vec{k}) = (\vec{\sigma}\cdot\vec{k})^\dagger,
\quad (\vec{\sigma}\cdot\vec{k})^2 = \vec{k}^2, \quad \vec{\alpha} =
\begin{pmatrix}0&\vec{\sigma}\\\vec{\sigma}&0\end{pmatrix},\ee
we can calculate the interaction terms in Eq.~(\ref{eq:unfinishedE})
explicitly;
\bea \nonumber u^\dagger \left(\vec{\alpha}\cdot\vec{k}\right) u &=&
\frac{E^*+M^*}{2E^*}
\left(\chi^\dagger\ \ \frac{\vec{\sigma}\cdot\vec{k}}{E^*+M^*}\,
\chi^\dagger\right)
 \begin{pmatrix}0&\vec{\sigma}\cdot\vec{k}\\\vec{\sigma}\cdot\vec{k}\,
   &0\end{pmatrix}
\begin{pmatrix}\chi\\[1mm]\frac{\vec{\sigma}\cdot\vec{k}}{E^*+M^*}\end{pmatrix} \\[3mm]
\nonumber &=& \frac{E^*+M^*}{2E^*} \left(
\frac{(\vec{\sigma}\cdot\vec{k})^2}{E^*+M^*}\, \chi^\dagger
\ \ (\vec{\sigma}\cdot\vec{k})\chi^\dagger \right)
\begin{pmatrix}\chi\\[1mm]\frac{\vec{\sigma}\cdot\vec{k}}{M^*+E^*}(\chi)\end{pmatrix} \\[3mm]
\nonumber &=& \frac{E^*+M^*}{2E^*} \left(
\frac{(\vec{\sigma}\cdot\vec{k})^2}{E^*+M^*}\, \chi^\dagger\chi \ +
\ \frac{(\vec{\sigma}\cdot\vec{k})^2}{E^*+M^*}\, \chi^\dagger\chi
\right) \\[2mm] &=& \frac{\vec{k}^2}{E^*}, \eea
\bea \nonumber u^\dagger (\beta M^*) u &=& \frac{E^*+M^*}{2E^*}\left(\chi^\dagger
\ \ \frac{\vec{\sigma}\cdot\vec{k}}{E^*+M^*}\, \chi^\dagger\right)
\begin{pmatrix}M^*&0\\0&-M^*\end{pmatrix} 
\begin{pmatrix}\chi\\[1mm]\frac{\vec{\sigma}\cdot\vec{k}}{E^*+M^*}\end{pmatrix} \\[3mm]
\nonumber &=& \frac{E^*+M^*}{2E^*} \left( M^*\chi^\dagger
\ \ -\frac{(\vec{\sigma}\cdot\vec{k})}{E^*+M^*}\, \chi^\dagger \right)
\begin{pmatrix}\chi\\[1mm]\frac{\vec{\sigma}\cdot\vec{k}}{M^*+E^*}(\chi)\end{pmatrix} \\[3mm]
\nonumber &=& \frac{E^*+M^*}{2E^*} \left( M^* -
\frac{(\vec{\sigma}\cdot\vec{k})^2}{(E^*+M^*)^2}\, \chi^\dagger\chi
\right) \\[2mm] &=& \frac{M^{*2}}{E^*}.  \eea
Combining these last two expressions, we can see that
\be \label{eq:udagu} u^\dagger (\vec{\alpha}\cdot\vec{k}+\beta M^*) u
= \frac{(\vec{k})^2+(M^*)^2}{E^*} = E^* = \sqrt{\vec{k}^2+M^{*2}}.  \ee
Furthermore, we see that
\be \label{eq:ubilinears} u^\dagger u = 1, \quad \bar{u}u =
\frac{M^*}{E^*}. \ee
\par
Inserting the results of Eq.~(\ref{eq:udagu}) into Eq.~(\ref{eq:psi}),
and summing over continuous momenta for each of the spin states of
protons and neutrons, we find the expression for the energy density
\be \label{eq:APPE} {\cal E} =
\frac{g_{N\omega}^{2}}{2m_{\omega}^{2}}\rho_{\omega}^2
+\frac{g_{\rho}^{2}}{2m_{\rho}^{2}}\rho_{\rho}^{2}
+\frac{g^{2}_{N\sigma}}{2m_{\sigma}^{2}}\rho_{\sigma}^{2}
+\sum_{p,n}\frac{(2J+1)}{(2 \pi)^{3}}\int_{0}^{k_F} \left(\vec{k\, }^{2}
+M^{*\, 2}\right)^{1/2} d^{3}k, \ee
where $J$ is the spin of the nucleons ($J_{p,n} = \half)$. Similarly,
the equation for the pressure is derived to be
\bea \nonumber P = \ \reci{3}\bra T_{ii}\ket &=&
\frac{1}{3}\psi^{\dagger}\left[-i
  \vec{\alpha}\cdot\vec{\nabla}\right]\psi
+\frac{g_{N\omega}^{2}}{2m_{\omega}^{2}}\rho_{\omega}^2
+\frac{g_{\rho}^{2}}{2m_{\rho}^{2}}\rho_{\rho}^{2}
+\frac{g^{2}_{N\sigma}}{2m_{\sigma}^{2}}\rho_{\sigma}^{2}, \\
\label{eq:APPP}  &=&
\frac{g^{2}_{N\omega}}{2m_{\omega}^{2}}\rho_{\omega}^{2} 
+\frac{g^{2}_{\rho}}{2m_{\rho}^{2}}\rho_{\rho}^{2} -\frac{g^{2}_{N\s}}{2
  m^{2}_{\s}}\rho_\s^{2} +\frac{1}{3}\sum_{p,n}\frac{(2J+1)}{(2
  \pi)^{3}}\int_{0}^{k_F} \frac{\vec{k\, }^2 }{E^{*}} d^{3}k\ .
\eea
These two quantities\emdash the energy density ${\cal E}$ and the
pressure $P$\emdash along with the baryon density $\rho_{\rm total}$
define the EOS for QHD.\par
%

%

%
%
%
\section{Tolman--Oppenheimer--Volkoff Equations}\label{sec:tovderiv}
%
%
The Tolman--Oppenhemier--Volkoff (TOV) Equation (refer to
Section~\ref{sec:stellarmatter}) is derived in General Relativity to
produce a differential relation for the pressure of a perfect fluid as
a function of radius for a sphere of material which is able to sustain
itself against gravitational collapse.\par
To begin with, we define a general metric for a static, spherically
symmetric (non-rotating) star as
\be \label{eq:metric} ds^{2} = -e^{2\Phi}dt^{2} + e^{2\Lambda}dr^{2} +
r^{2}d\Omega^2, \ee
where $\Phi$ and $\Lambda$ are functions of $r$ such that in the
Newtonian limit of $r \rightarrow \infty$, the functions vanish,
i.e. $\Phi\rightarrow 0$ and $\Lambda \rightarrow 0$. Given this, we
can find components of the Einstein Tensor $G_{\mu\nu}$, which is
defined in terms of the Ricci Tensor $R_{\mu\nu}$, and Ricci Scalar
$R$, as
\be G_{\mu\nu} = R_{\mu\nu} -\frac{1}{2}R{\rm g}_{\mu\nu}, \ee
the temporal and radial components of which are
\bea \label{eq:Gtt} G_{tt} &=&
\frac{1}{r^{2}}e^{2\Phi}\frac{d}{dr}\left(r-re^{-2\Lambda}\right), \\[1mm]
\label{eq:Grr} G_{rr} &=& -\frac{1}{r^{2}}e^{2\Lambda}\left(1-e^{-2\Lambda}\right)+\frac{2}{r}\frac{d\Phi}{dr}.
\eea
We can also find the components of the energy-momentum tensor for a
perfect fluid,
\be \label{eq:emtens} T_{\mu\nu} = \left(P+{\cal
  E}\right)u_{\mu}u_{\nu}+P{\rm g}_{\mu\nu} \ee
where $P$ is the pressure of the system, and ${\cal E}$ is the energy
density.\par
Since we are considering a time-like spacetime, we can use the
following relations
\be u_\alpha u^\alpha = -1, \ \ u_i = 0 \ \ \Rightarrow \ \ u_{t}u^{t}
= -1, \ \ \therefore \ {\rm g}^{tt}u_{t}^{2} = -1,\ \ \therefore
\ u_{t}^{2} = e^{2\Phi}, \ee
since we consider a static star, we have no three-velocity components
so $u_{i} = 0$. We can now evaluate the energy-momentum
tensor components to be
\bea 
\label{eq:Ttt}
T_{tt} &=& \left(P+{\cal E}\right)u_{t}u_{t}+P{\rm g}_{tt} = {\cal E}
e^{2\Phi},\quad \\[1mm]
\label{eq:Trr} T_{rr} &=& \left(P+{\cal E}\right)u_{r}u_{r}+P{\rm g}_{rr} = P e^{2\Lambda}.
\eea
\par
The energy-momentum tensor is related to the Einstein tensor via
\be \label{eq:GLT} G_{\mu\nu} = 8 \pi T_{\mu\nu}, \ee
and thus the temporal and radial components are found by inserting
Eqs.~(\ref{eq:Gtt})--(\ref{eq:Grr}) and
Eqs.~(\ref{eq:Ttt})--(\ref{eq:Trr}) into Eq.~(\ref{eq:GLT});
\bea \label{eq:time} G_{tt} &=& 8 \pi \mathcal{E} e^{2\Phi} =
\frac{1}{r^{2}}e^{2\Phi}\frac{d}{dr}\left(r-re^{-2\Lambda}\right), \\
\label{eq:rad} G_{rr} &=& 8 \pi p e^{2\Lambda} =
-\frac{1}{r^{2}}e^{2\Lambda}\left(1-e^{-2\Lambda}\right)+\frac{2}{r}\frac{d\Phi}{dr}.
\eea

The mass within a given radius $r$ is defined by integrating the
energy density, thus
\be M(r) = \int_0^r 4 \pi R^{2} {\cal E} \; dR, \ee
and Eq.~(\ref{eq:time}) already has this form on the right hand side,
so we can divide through by $2 e^{2\Phi} / r^{2}$ and integrate to
obtain a new relation for the mass within radius $r$, which is
\be M(r) = \frac{1}{2}\left(r-re^{-2\Lambda}\right) \int r^{2} dr =
\frac{1}{2}\left(r-re^{-2\Lambda}\right), \ee
which we can rearrange to obtain an expression for the factor that
accompanies the radial component of the metric in
Eq.~(\ref{eq:metric}), thus
\be \label{eq:e2b} e^{2\Lambda} = \left(1 - \frac{2M(r)}{r}
\right)^{-1}.  \ee
%
%
\par
The conservation of local energy-momentum requires that the derivative
of Eq.~(\ref{eq:emtens}) vanishes, and thus
\be T^{\mu\nu}_{\phantom{\mu \nu};\nu} = 0. \ee
Inserting the expression for the energy-momentum tensor of
Eq.~(\ref{eq:emtens}) we have
\be \label{eq:tmunuzero} T^{r \nu}_{\phantom{r \nu};\nu} = [ ({\cal E}
  + P)u^{r}u^{\nu} ]_{;\nu} = ({\cal E} +
P)u^{r}_{\phantom{r};\nu}u^{\nu}+P_{,\nu}{\rm g}^{r \nu} = 0, \ee
where we have taken the $\mu = r$ components since pressure depends
only on the radius, and where $u^{r} = 0$ for a static solution (no
radial velocities). We have also used the fact that ${\rm g}^{r
  \nu}_{\phantom{r \nu};\nu} = 0$. Also note that $P$ and ${\cal E}$
are Lorentz scalar quantities, so the total derivative $P_{;\nu}$
reduces to a partial derivative $P_{,\nu}$.\par
Since the conservation property prescribes that
Eq.~(\ref{eq:tmunuzero}) vanishes, we can multiply both sides by the
metric to lower the $r$ components
\be ({\cal E} + P)u_{r;\nu}u^{\nu}+P_{;r} = 0. \ee
Expanding the covariant derivative in terms of Christoffel symbols
$\Gamma^\alpha_{r\nu}$ gives
\be ({\cal E} + P)u_{r,\nu}u^{\nu}+P_{,r} = ({\cal E} +
P)\Gamma^{\alpha}_{r \nu}u_{\alpha}u^{\nu}. \ee
Once again we use the lack of radial velocity (and acceleration) to
remove the $u_{r,\nu}$ term, thus the remaining terms are
\be P_{,r} = ({\cal E} + P)\Gamma^{\alpha}_{r
  \nu}u_{\alpha}u^{\nu}. \ee
Einstein summation is assumed, and the sums over $\alpha$ and $\nu$
will only have contributions from the temporal components since this
is a static solution, so we then have
\be \label{eq:pcommar} P_{,r} = ({\cal E} + P)\Gamma^{t}_{r
  t}u_{t}u^{t} = -({\cal E} + P)\Gamma^{t}_{r t}. \ee
\par
Slightly aside, the Christoffel symbol can be evaluated from its
definition in terms of the metric
\be \Gamma^{\alpha}_{\mu \nu} = \frac{1}{2}{\rm g}^{\alpha
  \beta}\left({\rm g}_{\beta \mu,\nu}+{\rm g}_{\beta \nu,\mu}-{\rm
  g}_{\mu \nu,\beta}\right), \ee
which reduces to the following if we set $\nu = \alpha$
\bea \nonumber \Gamma^{\alpha}_{\mu \alpha} &=& \frac{1}{2}{\rm
  g}^{\alpha \beta}\left({\rm g}_{\beta \mu,\alpha}+{\rm g}_{\beta
  \alpha,\mu}-{\rm g}_{\mu \alpha,\beta}\right) \\ &=&
\frac{1}{2}{\rm g}^{\alpha \beta}\left({\rm g}_{\beta \mu,\alpha}-{\rm
  g}_{\mu \alpha,\beta}\right) +\frac{1}{2}{\rm g}^{\alpha \beta}{\rm
  g}_{\alpha \beta,\mu}, \eea
where, due to symmetry ${\rm g}_{\beta \alpha,\mu} = {\rm g}_{\alpha
  \beta,\mu}$. Furthermore, the terms in the brackets are
antisymmetric in $\alpha$ and $\beta$, so the bracketed term vanishes
when contracted with the inverse (symmetric) metric $g^{\alpha
  \beta}$. We are then left with
\be \Gamma^{\alpha}_{\mu \alpha} = \frac{1}{2}{\rm g}^{\alpha
  \beta}{\rm g}_{\alpha \beta,\mu} \ee
so if we use the indices we have just derived ($\alpha = t$, $\mu =
r$), we obtain
\be \label{eq:chriss} \Gamma^{t}_{r t} = \frac{1}{2}{\rm g}^{t
  \beta}{\rm g}_{t \beta,r} = \frac{1}{2}{\rm g}^{t t}{\rm g}_{t t,r}
= \frac{1}{2}\frac{-1}{e^{2 \Phi}}(-e^{2 \Phi})_{,r} =
\frac{1}{2}\frac{-1}{e^{2 \Phi}}2 \Phi_{,r}(-e^{2 \Phi}) = \Phi_{,r}
\ee
recalling that the metric is diagonal, thus ${\rm g}^{t\beta}{\rm
  g}_{t\beta} = {\rm g}^{tt}{\rm g}_{tt}$.\par
Inserting the result of Eq.~(\ref{eq:chriss}) into
Eq.~(\ref{eq:pcommar}) we now have
\be P_{,r} = -(\mathcal{E} + P)\Phi_{,r} \ee
which without the comma notation is
\be \label{eq:dpdr} \frac{dP}{dr} = -({\cal E} + P)\frac{d\Phi}{dr},
\ee
where if $P \ll {\cal E}$, Eq.~(\ref{eq:dpdr}) reduces to the
Newtonian result of ${\cal E} \Phi_{,r} = -P_{,r}$ which describes the
balance between gravitational force and the pressure gradient.\par
We can now substitute Eq.~(\ref{eq:e2b}) into Eq.~(\ref{eq:rad}) and
with some rearrangements we can obtain an expression for $\Phi_{,r}$
\be \frac{d\Phi}{dr} = \frac{M(r)+4 \pi r^{3} P }{r\left(r-2M(r)
  \right)}, \ee
which we can then substitute into Eq.~(\ref{eq:dpdr}) to obtain the
Tolman--Oppenheimer--Volkoff Equation
\be \label{eq:OV} \frac{dP}{dr} = -\frac{ \left( P+{\cal E} \right)
  \left(M(r) +4 \pi r^{3} P)\right)}{r(r -2M(r))}. \ee
\par
Finally, if we remove the use of Planck units (for which $\hbar = c =
G = 1$) by dimensional analysis we have
\be \label{eq:TOVnat} \frac{dP}{dr} = \quad -\frac{ G \left( P / c^{2}
  +{\cal E} \right) \left(M(r)+4 r^{3} \pi P / c^{2} \right)}{r(r-2 G
  M(r) / c^{2})}. \ee
%
%

\section{Calculated Quantities of Interest}\label{sec:calcderiv}
Included here are the derivations for some quantities that are used
throughout this work. This will hopefully provide a more detailed
explanation of their origins and their correspondence to the numerical
results.\par
\subsection{Self-Consistent Scalar Field}\label{sec:scalarfield}
The self-consistent mean-scalar-field $\bra\s\ket$ (and hence
effective mass $M^*$ via Eq.~(\ref{eq:effM})) is found such that the
energy density of Eq.~(\ref{eq:APPE}) is minimised with respect to the
scalar density of Eq.~(\ref{eq:rhos}) as
\be \rho_\s = \frac{m_\s^2}{g_{N\s}}\bra\s\ket, \ee
thus, for the case of the nucleons, 
\be \label{eq:intterms} \frac{\del {\cal E}}{\del \rho_\s} = 0 =
\frac{g_{N\s}^2}{m_\s^2}\rho_\s +\frac{\del}{\del \rho_\s}
\sum_{N}\frac{(2J_N+1)}{(2 \pi)^{3}}\int_{0}^{k_{F_N}}\left(\vec{k}^{2}
+M_N^{* 2}\right)^{1/2}d^{3}k\ . \ee
At this point we recall the definition of the effective mass, and
express this in terms of $\rho_\s$
\be M_N^* = M_N - g_{N\s}\bra\s\ket = M_N - \frac{g_{N\s}}{m_\s^2}\rho_\s. \ee
We can now evaluate the integrated terms of Eq.~(\ref{eq:intterms}) as
\bea \nonumber &\displaystyle{\frac{\del}{\del
    \rho_\s}\sum_{N}\frac{(2J_N+1)}{(2
    \pi)^{3}}\int_{0}^{k_{F_N}}\sqrt{\vec{k}^{2} +M_N^{* 2}}}\; d^{3}k
&=-\sum_{N}\frac{(2J_N+1)g_{N\s}}{(2
  \pi)^{3}}\\
\nonumber
&&\times \int_{0}^{k_{F_N}}\ M_N^*\left(\sqrt{\vec{k}^{2} +M_N^{*
    2}}\right)^{-1} \left(\frac{\del \bra\s\ket}{\del \rho_\s}
\right) d^{3}k \\
\nonumber &&=-\sum_{N}\frac{(2J_N+1)g_{N\s}^2}{m_\s^2(2
  \pi)^{3}}\int_{0}^{k_{F_N}}\left(\frac{M_N^*}{E_N^*}\right)d^{3}k, \\
&&
\eea
so the total expression for the scalar density becomes
\be \frac{\del {\cal E}}{\del \rho_\s} = 0 =
\frac{g_{N\s}^2}{m_\s^2}\rho_\s
-\sum_{N}\frac{g_{N\s}^2}{m_\s^2}\ \frac{(2J_N+1)}{(2
  \pi)^{3}}\int_{0}^{k_{F_N}}\left(\frac{M_N^*}{E_N^*}\right)\; d^{3}k,
\ee
which reduces to
\be \rho_\s = \sum_N \frac{(2J_N+1)}{(2
  \pi)^{3}}\int_{0}^{k_{F_N}}\left(\frac{M_N^*}{E_N^*}\right)\; d^{3}k.
\ee
\par
We also note that this expression can be concluded from the bilinears
of Eq.~(\ref{eq:ubilinears}) since the expressions
\be u^\dagger u = 1,\quad \bar{u}u = \frac{M^*}{E^*}, \ee
can be inserted back into Eq.~(\ref{eq:psi}) to calculate the various
densities in Eqs.~(\ref{eq:rhos})--(\ref{eq:rhorho}).
%
%

\subsection{Semi-Empirical Mass Formula}\label{sec:SEMF}
The (Bethe--Weizs\"acker) Semi-Empirical Mass Formula (SEMF) for
nuclei\emdash which is a refined form of the liquid drop model\emdash
describes a binding energy in terms of the number of nucleons $A$, and
atomic number (number of protons) $Z$ as
\be B(A,Z) = a_vA - a_sA^{2/3} - a_c\frac{Z(Z-1)}{A^{1/3}} -
a_{sym}\frac{(A-2Z)^2}{A}+\delta, \ee
where $a_v$ is the volume coefficient, $a_s$ is the surface
coefficient, and $a_{\rm sym}$ is the symmetry coefficient. We also
define $a_c$ as the Coulomb coefficient and $\delta$ as the pairing
term, each defined by
\be a_c = \frac{3e^2}{5r_0}, \quad \qquad \delta = \left\{
\begin{array}{ll}
a_p A^{-1/2}\ , & {\rm for\ even\ } N{\rm -even\ } Z,\\ -a_p
A^{-1/2}\ , & {\rm for\ odd\ }N{\rm -odd\ } Z,\\ 0\ , & {\rm
  for\ odd\ }A,
\end{array} \right.
\ee
where $e$ is the electric charge, $r_0$ is the Coulomb radius constant
which defines the spherical nuclear volume of radius $A^{1/3}r_0$, and
$a_p$ is the pairing coefficient.
These parameters and terms have various empirical derivations; for
example the $Z(Z-1)$ term in the Coulomb term corresponds to Coulomb
contributions of $Z$ protons, subtracting the self-energy contribution
of each of the $Z$ protons, since a single proton should not have a
Coulomb contribution.\par
If we consider the binding energy per baryon $A$
\be \frac{B(A,Z)}{A} = a_v - a_sA^{-1/3} -
\frac{3e^2}{5r_0}\frac{Z(Z-1)}{A^{4/3}} -
a_{sym}\frac{(A-2Z)^2}{A^2}+\frac{\delta}{A}, \ee
then consider the case of nuclear matter, in which we have isospin
symmetry, and thus $Z=N$ (where $N$ is the number of neutrons), the
symmetry term vanishes ($A=Z+N=2Z$), leaving
\be \frac{B(A,Z)}{A} = a_v - a_sA^{-1/3} -
\frac{3e^2}{5r_0}\frac{Z(Z-1)}{A^{4/3}} + \frac{\delta}{A}. \ee
If we now consider the case in which Coulomb contributions are
neglected, as we do in nuclear matter (via setting $e=0$) then the
Coulomb term vanishes, leaving
\be \frac{B(A,Z)}{A} = a_v - a_sA^{-1/3} + \frac{\delta}{A}. \ee
Finally, if we consider the case of infinite nuclear matter neglecting
Coulomb interactions (\mbox{$A\to\infty,$} $e=0$), we find that the
remaining two terms inversely proportional to $A$ vanish and the
binding energy per nucleon becomes simply the volume coefficient
\be \frac{B(A,Z)}{A} = a_v, \ee
which has units MeV. This is the origin of the saturated binding
energy in nuclear matter calculations. The numerical result is the
value obtained when fitting finite nuclei data to the SEMF and
obtaining values for each of the six parameters. At present, several
of these parameter sets exist\emdash hence the wide range of
values\emdash and the set used in this work is that of
Ref.~\cite{Rohlf}.
%
%
\be \frac{B(A,Z)}{A} = a_v = 15.86\ {\rm MeV}.  \ee
The equation for energy per baryon used in this work however takes
into account the rest mass energy of the baryons and defines the
binding as negative, and hence we use
\be \frac{\cal E}{A}-M_N = -15.86\ {\rm MeV}, \ee
or, alternatively, for many baryon species with the possibility of
non-degenerate masses\par
\be \reci{\rho_{\rm total}}({\cal E}-\sum_B\rho_{B}M_B) =
-15.86\ {\rm MeV}.  \ee
%
%

\subsection{Compression Modulus}\label{sec:compressionmod}
The compression modulus is used as a test against experiment, and is
related to the `stiffness' of the EOS. This quantity is a definition,
as the curvature of the energy per baryon at saturation;
\be \label{eq:AppendixKmod} K = \left[ k_F^2
  \frac{d^2}{dk_F^2}\left(\frac{\cal E}{\rho_{\rm
      total}}\right)\right]_{k_{F_{\rm sat}}} = 9 \left[ \rho_{\rm total}^2
  \frac{d^2}{d\rho_{\rm total}^2}\left(\frac{\cal
    E}{\rho_{\rm total}}\right)\right]_{\rho=\rho_0}. \ee
The factor of $9$ arises from using the density rather than the Fermi
momentum, as per Section~\ref{sec:kf}.

\subsection{Symmetry Energy}\label{sec:symenergy}
When we include the $\rho$ meson to the Lagrangian density, we
introduce a term into the energy density that is proportional to
$\bra\rho\ket^2 \propto (\rho_p - \rho_n)^2$, which is thus quadratic
in the deviation from isospin symmetry. We can define this deviation
in terms of a new variable (using just the nucleons here) as
\be t = ( \rho_p - \rho_n ) / \rho_{\rm total}\, ; \ \rho_{\rm
  total} = \rho_p + \rho_n. \ee
We can write the energy density in terms of this new parameter, and
then find the dependence of the energy per baryon on this to be
\bea \nonumber E / A = {\cal E} / \rho_{\rm total} &=&
\reci{2}\left(\frac{g_\rho}{m_\rho}\right)^2 \, t^2 \,\rho_{\rm total} +
\reci{\rho_{\rm total}} \sum_N \frac{(2J+1)}{(2\pi)^3}\int_0^{k_{F_N}}
\sqrt{\vec{k}^2+(M_N^*)^2}\, d^3k \\ \nonumber &=&
\reci{2}\left(\frac{g_\rho}{m_\rho}\right)^2\, t^2 \, \rho{\rm total} +
\reci{\pi^2\rho_{\rm total}} \sum_N \int_0^{k_{F_N}} \vec{k}^2
\sqrt{\vec{k}^2 + (M_N^*)^2} \, dk, \\ && \eea
where the second line is due to the change to spherical coordinates
(for more details, see Appendix~\ref{sec:selfenergyderiv}). If we wish to
know the contribution to the energy due to the isospin symmetry (or
asymmetry), we need to find the term defined as
\be a_{\rm sym} = \reci{2} \left[ \ \frac{\del^2({\cal E}/\rho_{\rm
      total})}{\del t^2} \ \right]_{t = 0}. \ee
We must keep in mind that the integrals do indeed depend on the
isospin symmetry since they depend on the Fermi momenta, which can be
defined as
\be k_{F_n} = k_{F_{\rm sat}}(1+t)^{1/3} \quad ; \ k_{F_p} = k_{F_{\rm
    sat}}(1-t)^{1/3}, \ee
in terms of the saturation Fermi momentum
\be k_{F_{\rm sat}} = \left(\ \frac{3\pi^2 \rho_0}{2}\ \right)^{1/3}, 
\ee
so when we take the first derivative we obtain
\bea \nonumber \reci{2}\left[ \ \frac{\del ({\cal E}/\rho_{\rm total})}{\del t}
  \ \right] &=& \reci{4}\left(\frac{g_\rho}{m_\rho}\right)^2\rho_{\rm
  total}t +
\frac{k_{F_{\rm sat}}^3}{6\pi^2 \rho} \\[2mm] \nonumber &&\times
\left[ \ \sqrt{(t+1)^{2/3} k_{F_{\rm sat}}^2 + M_n^{* 2}}
  -\sqrt{(1-t)^{2/3} k_{F_{\rm sat}}^2 + M_p^{* 2}} \ \right]. \\ &&
\eea
The second derivative produces
\bea \nonumber \displaystyle{\reci{2}\left[ \ \frac{\del^2 ({\cal
        E}/\rho_{\rm total})}{\del t^2} \ \right]} &=&
\reci{4}\left(\frac{g_\rho}{m_\rho}\right)^2\rho_{\rm total} + \frac{k_{\rm
    sat}^5}{18\pi^2 \rho_{\rm total}} \\ \nonumber &&\times \left[
  \ \reci{\sqrt[3]{(t+1)}\sqrt{(1+t)^{2/3} k_{F_{\rm sat}}^2 + M_n^{*
        2}}} \right. \\
&&\left. + \ \reci{\sqrt[3]{(1-t)}\sqrt{(1-t)^{2/3} k_{F_{\rm sat}}^2
      + M_p^{* 2}}} \ \right]. \eea
If we take the limit of $t\to 0$ at saturation ($\rho = \rho_0$) we
obtain
\bea \nonumber a_{\rm sym} &=& \reci{2}\left[ \ \frac{\del^2 ({\cal
      E}/\rho_{\rm total})}{\del t^2} \ \right]_{t=0} \\[3mm]
\nonumber &=&
\left(\frac{g_\rho}{m_\rho}\right)^2\frac{k_{F_{\rm sat}}^3}{12\pi^2}
+\left[ \ \frac{k_{F_{\rm sat}}^2}{3\sqrt{k_{F_{\rm sat}}^2 +
      (M_{n}^{* 2})_{\rm sat}}} + \frac{k_{F_{\rm
        sat}}^2}{12\sqrt{k_{F_{\rm sat}}^2 + (M_{p}^{* 2})_{\rm sat}}}
  \ \right]. \\ && \eea
Alternatively, in the case of charge symmetry in which the proton and
neutron masses are degenerate,
\be a_{\rm sym} = \left(\frac{g_\rho}{m_\rho}\right)^2\frac{k_{F_{\rm
      sat}}^3}{3\pi^2} + \left[ \ \frac{k_{F_{\rm
        sat}}^2}{6\sqrt{k_{F_{\rm sat}}^2 + (M_{N}^{* 2})_{\rm sat}}}
  \ \right], \ee
where now the subscript $N$ refers to the degenerate nucleon mass.\par
It is this that we need to fit to the experimental result of $a_{\rm
  sym} = 32.5~{\rm MeV}$ using the $\rho$ meson coupling $g_\rho$. We
can therefore find the value that satisfies this by rearrangement
\be g_\rho = \left( \frac{3 \pi^2 m_\rho^2}{k_{F_{\rm sat}}^3} \left[
  \ a_{\rm sym} - \frac{k_{F_{\rm sat}}^2}{12\sqrt{k_{F_{\rm sat}}^2 +
      (M_{n}^{* 2})_{\rm sat}}} - \frac{k_{F_{\rm
        sat}}^2}{12\sqrt{k_{F_{\rm sat}}^2 + (M_{p}^{* 2})_{\rm sat}}}
  \ \right] \right)^{1/2} \ee
which only relies on knowing the saturation density (here, $\rho_0 =
0.16~{\rm fm}^{-3}$) and the effective masses at saturation, and thus
varies from model to model.\par
%
%

\subsection{Chemical Potential}\label{sec:chempotderivn}
In this section we will consider Fermi--Dirac statistics\emdash that
of fermions with half-integer spin\emdash in order to carefully define
the chemical potential and associated quantities.\par
We consider the Grand Canonical Partition Function (GCPF), which
describes a grand canonical ensemble in which the system can exchange
both heat and particles with the environment at fixed temperature $T$,
fixed volume $V$, and fixed chemical potential $\mu$.\par
The general expression for a GCPF is
\be {\cal Z}_{\rm GC} = \prod_i {\cal Z}_i = \prod_i \sum_{\{n_i\}}
e^{-\beta n_i (\epsilon_i - \mu)}, \ee
where $\beta=1/kT$, and $\{n_i\}$ is the set of occupation numbers
which satisfy
\be \sum_i n_i = N, \ee
for which the total number of particles is $N$, and for which $i$ runs
over the total number of states. Since we are considering fermions,
the occupation numbers are restricted by the Pauli Exclusion
Principle, and thus are only able to obtain the values 0 and 1, in
which case the GCPF becomes
\be \label{eq:GCPF} {\cal Z}_{\rm GC} = \prod_i \sum_{n_i=0}^1
e^{-\beta n_i (\epsilon_i - \mu)} = \prod_i \left( 1 + e^{-\beta
  (\epsilon_i - \mu)} \right). \ee
\par
The average number of particles in a state is defined in terms of the
GCPF as
\be \langle N \rangle = \frac{1}{\beta}\frac{\partial}{\partial\mu}
\ln {\cal Z}_{\rm GC}, \ee
where the derivative for a natural logarithm dictates that
\be \label{eq:GCPFderiv} \langle N \rangle = \frac{1}{\beta {\cal
    Z}_{\rm GC}}\frac{\partial}{\partial\mu} {\cal Z}_{\rm GC}. \ee
Inserting the expression for the GCPF of Eq.~(\ref{eq:GCPF}) into
Eq.~(\ref{eq:GCPFderiv}) we obtain
\be \langle N \rangle = \frac{1}{\beta {\cal Z}_{\rm
    GC}}\frac{\partial}{\partial\mu} \left[
  \prod_k\left(1+e^{-\beta(\epsilon_k-\mu)}\right) \right]. \ee
Expanding the product notation gives
\be \langle N \rangle = \frac{1}{\beta {\cal Z}_{\rm
    GC}}\frac{\partial}{\partial\mu} \left[
  \left(1+e^{-\beta(\epsilon_1-\mu)}\right)\left(1+e^{-\beta(\epsilon_2-\mu)}\right)\ldots
  \right], \ee
which makes clearer the action of the chain rule for the derivative,
to give
\bea \nonumber \langle N \rangle &=& \frac{1}{\beta {\cal Z}_{\rm GC}}
\left[ \left(\beta e^{-\beta(\epsilon_1-\mu)}\right)\prod_{i\neq
    1}\left(1+e^{-\beta(\epsilon_i-\mu)}\right) \right. \\ && + \left.
  \left(\beta e^{-\beta(\epsilon_2-\mu)}\right)\prod_{i\neq 2}
  \left(1+e^{-\beta(\epsilon_i-\mu)}\right) +\ldots \right]. \eea
Returning to the product notation, the result is
\be \langle N \rangle = \frac{1}{\beta {\cal Z}_{\rm GC}} \sum_{k}
\left(\beta e^{-\beta(\epsilon_k-\mu)}\prod_{i\neq k}
\left(1+e^{-\beta(\epsilon_i-\mu)}\right)\right). \ee
Inserting again the expression for the CGPF, we obtain
\be \langle N \rangle = \frac{1}{\beta} \sum_{k} \frac{\beta
  e^{-\beta(\epsilon_k-\mu)}\prod_{i\neq k}
  \left(1+e^{-\beta(\epsilon_i-\mu)}\right)}
    {\prod_j\left(1+e^{-\beta(\epsilon_j-\mu)}\right)}, \ee
which contracts neatly to
\be \langle N \rangle = \sum_{k} \frac{ e^{-\beta(\epsilon_k-\mu)}} {
  1+e^{-\beta(\epsilon_k-\mu)}} = \sum_k
\frac{1}{1+e^{\beta(\epsilon_k-\mu)}} = \sum_k \bra n_k \ket = \sum_k
n_F(\epsilon_k). \ee
where $\bra n_i \ket$ is the average occupation number also known as
the Fermi--Dirac distribution.\par
Allowing for a continuous distribution rather than discrete sum, we
can write this as
\be \label{eq:FDdist} \langle N \rangle = \int_0^\infty
\rho(\epsilon_k) \ n_F(\epsilon_k) \ d\epsilon_k, \ee
where $\rho(\epsilon_k) = \sum_i \delta(\epsilon_k - \epsilon_i)$ is
the density of states. In the zero temperature limit of $T\to 0$
($\beta \to \infty$), the distribution $n_F(\epsilon_k)$ becomes a
Heaviside step function
\be \label{eq:HSF} n_F(\epsilon_k) \to \theta(\mu-\epsilon_k) =
\left\{
\begin{array}{ll}
0,\ &{\rm if}\ \mu < \epsilon_k \\ 1,\ &{\rm if}\ \mu \ge \epsilon_k
\end{array} \right. ,
 \ee
and thus Eq.~(\ref{eq:FDdist}) becomes
\be \langle N \rangle = \int_0^\infty \rho(\epsilon_k)
\ \theta(\mu-\epsilon_k) \ d\epsilon_k = \int_0^\mu \rho(\epsilon_k)
\ d\epsilon_k, \ee
since the largest value that $\epsilon_k$ can obtain is
$\epsilon_k=\mu$, due to the Heaviside step function,
Eq.~(\ref{eq:HSF}). The lower limit is maintained since we still
require that $\mu \ge 0$. The last level occupied therefore has energy
$\epsilon_{k_{\rm max}} = \mu$, where $\mu$ is defined in
Section~\ref{sec:qftchempot}, and thus the chemical potential is the
energy of a particle at the top of the Fermi sea.\par
%
%
%
%
%
%
%

\subsection{Relation to the First Law of Thermodynamics}\label{sec:firstlaw}
The link between the equation of state for QHD/QMC written as
\be p = \rho_{\rm total}^2 \frac{\del}{\del \rho_{\rm total}}\left( \frac{{\cal
    E}}{\rho_{\rm total}} \right),
\label{eq:appEoS}
\ee
and the first law of thermodynamics,
\be p\, dV = - dE,
\label{eq:appfirstlaw}
\ee
can be seen when we consider that the number of particles $A$ is
constant, and so we can safely use this as a multiplicative factor,
\be p\, dV = - dE \quad \Rightarrow \quad p = - \frac{dE}{dV} = -
\frac{d(E/A)}{d(V/A)},
\label{eq:onA}
\ee
and recall that the particle number density (baryon density) is
denoted by
\be \label{eq:particlenum} \rho_{\rm total} = A/V, \quad \therefore \frac{1}{\rho_{\rm
    total}} = \frac{V}{A}. \ee
We can differentiate Eq.~(\ref{eq:particlenum}) with respect to the
baryon density to obtain
\be \frac{d}{d\rho_{\rm total}}\left( \frac{V}{A} \right) =
\frac{d}{d\rho_{\rm total}}\left( \frac{1}{\rho_{\rm total}} \right) =
- \rho_{\rm total}^{-2} \quad \Rightarrow \quad d\left( \frac{V}{A}
\right) = -\rho_{\rm total}^{-2}\; d \rho_{\rm total}.  \ee
If we substitute this result back into Eq.~(\ref{eq:onA}) we find
\be p = \frac{- d(E/A)}{-\rho_{\rm total}^{-2}\; d\rho_{\rm total}},
\ee
and if we note that the energy density is $\displaystyle{{\cal E} =
  E/V = E\rho_{\rm total}/A}$, we can substitute the differential to find
\be dE = d\left( \frac{{\cal E}A}{\rho_{\rm total}} \right) \quad \Rightarrow
\quad d(E/A) = d\left( \frac{{\cal E}}{\rho_{\rm total}} \right), \ee
and our final result becomes
\be p = \rho_{\rm total}^{2} \frac{d}{d\rho_{\rm total}}\left(
\frac{{\cal E}}{\rho_{\rm total}} \right) = - \frac{dE}{dV}.  \ee
and thus the EOS is related to the first law of thermodynamics.\par
%

\section{Hartree QHD Energy Density}\label{sec:hartreeEderiv}
\begin{center}
\begin{figure}[!b]
\centering
\includegraphics[width=0.35\hsize]{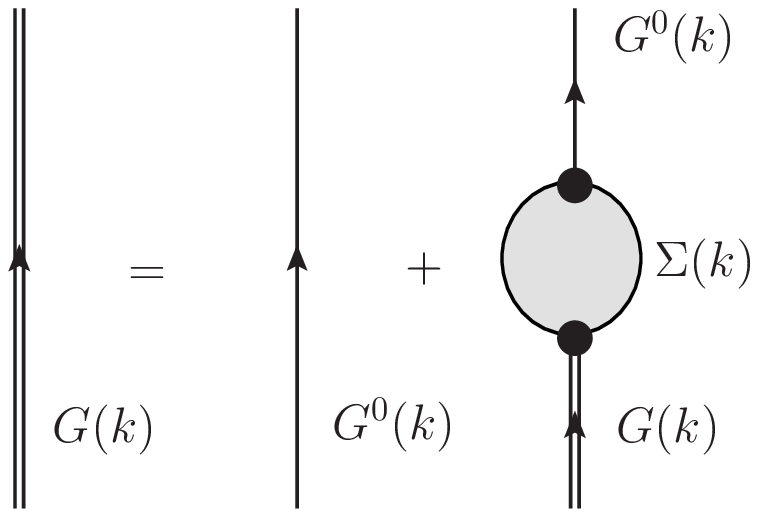}
\caption[Feynman diagram for Dyson's Equation]{Feynman diagram for
  Dyson's Equation as per Eq.~(\ref{eq:dysons}). Here, the double line
  represents $G(k)$; the full, self-consistent dressed baryon
  propagator, and the single line represents $G^0(k)$; the bare baryon
  propagator. $\Sigma(k)$ represents the
  self-energy.\label{fig:secondorderdyson}}
\end{figure}
\end{center}

If we wish to extend the sophistication of the QHD model, we can
include higher-order (in the baryon-meson coupling $g$) terms by using
a perturbative method. The logical next term to include is the
one-loop correction to the self-energy. In order to do this, we must
first formulate Hartree QHD using propagators (see
Appendix~\ref{sec:propagators}) in order to understand how the next
term is introduced. We begin with Dyson's Equation for baryons;
\be \label{eq:dysons} G(k) = G^0(k)+G^0(k)\Sigma G(k), \ee
where $G(k)$ is the exact (dressed) baryon propagator, $G^0(k)$ is the
bare baryon propagator, and $\Sigma$ is the baryon
self-energy. Eq.~(\ref{eq:dysons}) can be represented as a Feynman
diagram, as shown in Fig.~\ref{fig:secondorderdyson}.\par
\begin{center}
\begin{figure}[!t]
\centering \includegraphics[width=0.85\hsize]{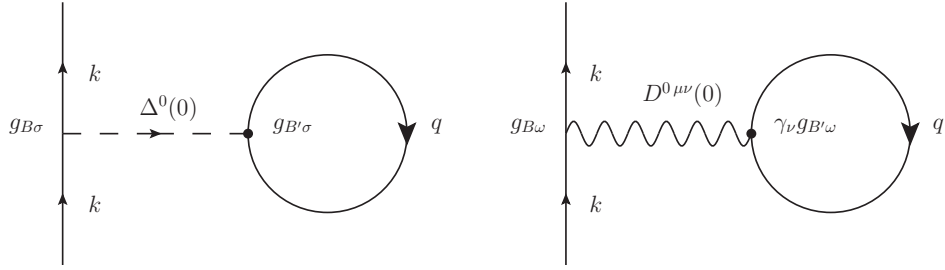}
\caption[Feynman diagram for 2nd order baryon tadpoles]{Feynman
  diagram for second-order self-energies (tadpoles) of
  Eqs.~(\ref{eq:tadpole_s}--\ref{eq:tadpole_v}). \label{fig:secondordertadpole}}
\end{figure}
\end{center}
\begin{center}
\begin{figure}[!b]
\centering
\includegraphics[width=0.8\hsize]{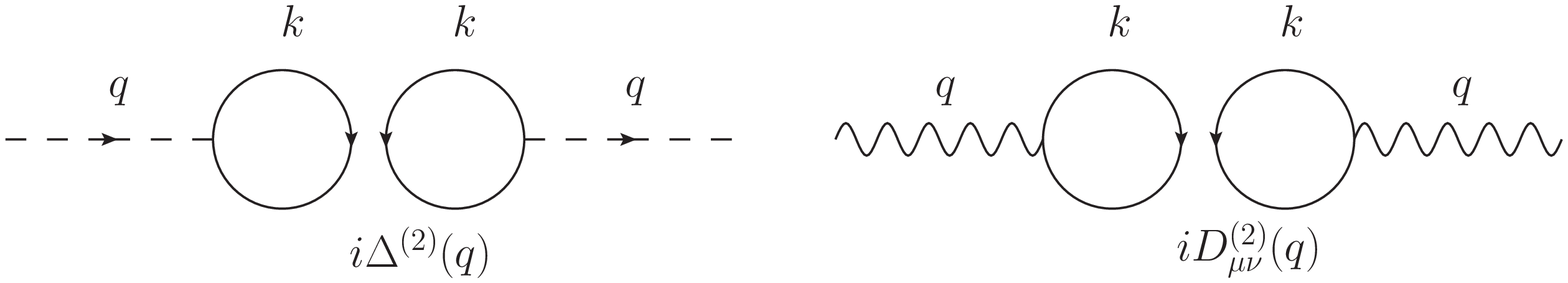}
\caption[Feynman diagram for second-order meson tadpoles]{Feynman
  diagram for second-order meson tadpoles. Note that the momentum is
  conserved, as the loops flow in opposite
  directions.}\label{fig:mesontadpole}
\end{figure}
\end{center}

Up to second-order, this can be written as
\be iG^{(2)}(k) = iG^0(k) + i G^0(k)\Sigma G^0(k), \ee
The second-order (in the coupling $g$) self-energy is written in terms
of a scalar part and a Dirac-vector part (c.f. Eq.~(\ref{eq:LcovSE})
in which we consider the case of Lorentz-covariance)
\bea
\nonumber \Sigma(k) &=& \Sigma^s(k) - \gamma_\mu \Sigma^{\mu}(k)
\\ \label{eq:splitselfenergy} &=&
\Sigma^s(|\vec{k}|,k^0) - \gamma_0\Sigma^0(|\vec{k}|,k^0) +
\vec{\gamma}\cdot\vec{k}\; \Sigma^v(|\vec{k}|,k^0).  \eea
%
%
where the scalar and vector contributions for a single baryon $B$
(with interactions involving $\s$ and $\w$ mesons) are
\bea \label{eq:tadpole_s} \Sigma_{Bs}^{(2)} &=& -ig_{B\s} \sum_{B'}
g_{B'\s} \Delta^0(0) \volint{q} {\rm Tr}\left[ G^0(q) \right]
e^{iq^0\eta}, \\ \label{eq:tadpole_v} \Sigma_{Bv}^{(2)\; \mu} &=&
ig_{B\w} \sum_{B'} g_{B'\w} D^{0\; \mu\nu}(0) \volint{q} {\rm
  Tr}\left[ G^0(q) \gamma_\nu\right] e^{iq^0\eta}, \eea
where the exponential factor ensures that the integrals are finite,
and which can be represented as second-order baryon tadpole diagrams
as per Fig.~\ref{fig:secondordertadpole}.
The second-order tadpole contributions to the meson propagators are
given by
\bea \nonumber i\Delta^{(2)}(q) &=&
i\Delta^0(q)\left\{i(2\pi)^4\delta^{(4)}(q) \left[\sum_{B'} g_{B'\s}
  \volint{k} {\rm Tr} \left[ G^0(k) \right]
  e^{ik^0\eta}\right]^2\right\}\Delta^0(q) \\ &=& (2\pi)^4
\delta^{(4)}(q)\left[ \Sigma_{Bs}^{(2)}\right]^2/g_{B\s}^2
\\ iD_{\mu\nu}^{(2)}(q) &=& (2\pi)^4 \delta^{(4)}(q) \left[ \sum_{B'}
  \Sigma_{Bv}^{(2)\; \mu}\Sigma_{Bv}^{(2)\; \nu}\right]/g_{B\w}^2 \eea
which can be represented as Feynman diagrams as per
Fig.~\ref{fig:mesontadpole}.\par
\begin{center}
\begin{figure}[!b]
\centering
\includegraphics[width=0.85\hsize]{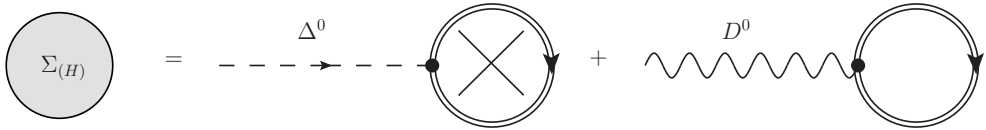}
\caption[Feynman diagram for Hartree self-energy]{Feynman diagram for
  the RHA self-energy, where the double line represents the exact
  (Hartree) propagator.}\label{fig:RHAselfenergy}
\end{figure}
\end{center}

The lowest-order Dyson's equation is not strictly self-consistent,
since the background particles in $\Sigma$ are treated as
non-interacting. However, since the \emph{exact} Green's function can
be expressed as a series containing the proper self-energy,
self-consistency can be achieved, but only if we use the interacting
propagators to determine the self-energy. This condition defines the
Relativistic Hartree Approximation (RHA), thus Dyson's Equation is
written in terms of the Hartree propagators as
\be G^H(k) = G^0(k) + G^0(k)\Sigma_{(H)} G^H(k),
\ee 
where $G^H(k)$ is the Hartree propagator, and $\Sigma_{(H)}$ is the
Hartree self-energy given by
\be \label{eq:HartreeSE} \Sigma_{(H)} = \Sigma_{(H)s} - \gamma_\mu
\Sigma_{(H)v}^{\mu}. \ee
%
%
Eq.~(\ref{eq:HartreeSE}) can be represented as a Feynman diagram as
Fig.~\ref{fig:RHAselfenergy}, where the cross in the scalar tadpole
indicates that it has been renormalized with counterterms, the result
of which is that the free propagator is taken at $k=0$. The expression
for the Hartree self-energy is then
\bea \label{eq:hasCTC} \Sigma_{(H)Bs} &=& ig_{B\s} \sum_{B'}
\frac{g_{B'\s}}{m_\s^2} \volint{q} {\rm Tr}\left[ G^H(q)
  \right]e^{iq^0\eta} + \Sigma_{{\rm CTC}}, \\ \Sigma^{\mu}_{(H)Bv}
&=& ig_{B\w} \sum_{B'} \frac{g_{B'\w}}{m_\w^2} \volint{q} {\rm
  Tr}\left[ \gamma_\mu G^H(q) \right]e^{iq^0\eta}.  \eea
where we note that the bare meson propagator is replaced with the
square of the meson mass. The expression for the scalar self-energy
includes a counterterm $\Sigma_{\rm CTC}$ to render the integral
finite via renormalization. This problem is however overcome easily,
as we will show.\par
\begin{center}
\begin{figure}[!t]
\centering
\includegraphics[width=0.85\hsize]{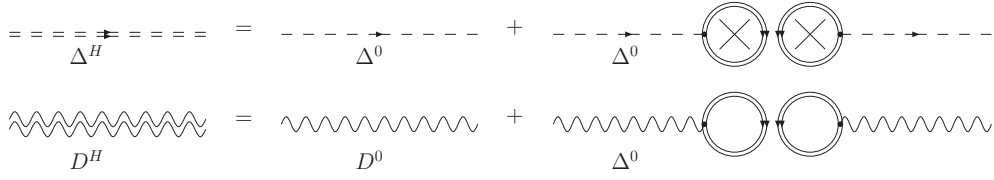}
\caption[Feynman diagram for meson propagators]{Feynman diagram for
  the meson propagators, where the double line represents the exact
  (Hartree) propagator.}\label{fig:RHAmesonpropagators}
\end{figure}
\end{center}

In this RHA case, the meson propagators are given by
\bea \Delta^H(k) &=& \Delta^0(k) -
i(2\pi)^4\delta^{(4)}(k)\left(\Sigma_{(H)Bs}\right)^2 / g_{B\s}^2
\\ D_{\mu\nu}^H(k) &=& D_{\mu\nu}^0(k) -
i(2\pi)^4\delta^{(4)}(k)\left(\Sigma^{\mu}_{(H)Bv}\Sigma^{\nu}_{(H)Bv}\right)
/ g_{B\w}^2.  \eea
and are represented as Feynman diagrams in
Fig.~\ref{fig:RHAmesonpropagators}. The formal solution to Dyson's
Equation gives
\be G^H(k) = \frac{G^0(k)}{1-G^0(k)\Sigma_{(H)}}, \ee
which we can rearrange in order to define the propagator, as
\be \left[ G^H(k) \right]^{-1} = \left[ G^0(k) \right]^{-1} -
\Sigma_{(H)} = \gamma_\mu k^\mu - M - \Sigma_{(H)}, \ee
where $\Sigma_{(H)}$ here represents the average interaction felt by
propagating particles, and is independent of $k$. By considering the
particles to be immersed in a Fermi sea, we can write the Hartree
propagator as a sum of Fermi (corresponding to the antibaryon
contributions) and Dirac (a direct result of the Fermi sea)
components, as
\bea \nonumber G^H(k) &=& G_F^H(k) + G_D^H(k) \\[1mm] \nonumber
&=& (\gamma_\mu k^{*\, \mu} + M) 
\left[ \left( (k_\nu^*)^2 - M^{*\; 2}+
  i\epsilon \right)^{-1} +
  \frac{i\pi}{E^*(k)}\delta(k^0-E(k))\theta(k_F - |\vec{k}|) \right],
\\[2mm]
&&
\eea
where the following definitions have been applied
\be \label{eq:HFdefinitions} \begin{array}{rclcrcl} k^{*\, \mu} &=& k^\mu +
  \Sigma_{(H)v}^\mu, & \quad & E^*(k) &=&  \left( \vec{k^*}^2 + M^{*\; 2}
  \right)^{1/2}, \\[3mm] M^* &=& M + \Sigma_{(H)s}, & \quad & E(k) &=& E^{*}(k)
  - \Sigma_{(H)v}^{0},
 \end{array} \ee
where $E(k)$ is the self-consistent, single-particle energy,
calculated on-shell.\par
Since the $G_F^H$ term corresponds to the sum over all occupied states
in the negative energy sea of quasibaryons, we can drop this term in
the Hartree propagator, thus
\be \label{eq:HartreeProp} G^H(k) = G^H_D(k) = \left( \gamma_\mu
k^{*\, \mu} + M^* \right) \frac{i\pi}{E^*(k)}
\delta(k^0-E(k))\theta(k_F - |\vec{k}|).  \ee
\par
The $G_D^H$ term corresponds to the filled Fermi sea, and produces a
finite MFT result. Thus the counterterms in Eq.~(\ref{eq:hasCTC}) are
of no concern. We may now drop the Hartree subscript, and all
quantities are defined in the mean-field. If we now evaluate the
scalar self-energy contributions we find
\bea \nonumber \Sigma_{Bs} &=& ig_{B\s} \sum_{B'} \frac{g_{B'\s}}{m_\s^2}
\volint{q} {\rm Tr}\left[G_D^H(q)\right] \\ \nonumber &=& - g_{B\s} \sum_{B'}
\frac{g_{B'\s}}{m_\s^2} \frac{(2J_{B'}+1)}{(2\pi)^3} \int_0^{k_{F_{B'}}} 
\frac{M_B^*}{E_B^*(q)} d^3q \\ &\equiv& -g_{B\s} \bra\s\ket.  \eea
and for the vector self-energy contribution we find
\bea \nonumber \Sigma_{Bv}^\mu &=& ig_{B\w}\sum_{B'}
\frac{g_{B'\w}}{m_\w^2} \volint{q} {\rm Tr}\left[\gamma^\mu
  G_D^H(q)\right]\\ \nonumber &=& - g_{B\w} \sum_{B'}
\frac{g_{B'\w}}{m_\w^2} \frac{(2J_{B'}+1)}{(2\pi)^4}
\int_0^{k_{F_{B'}}} d^3q \ \delta^{\mu 0} \\ &\equiv& -g_{B\w}
\bra\w_\mu\ket \delta^{\mu 0}, \eea
where the factor of $J$ accounts for the spin states; 2 for each
species of baryon. This contribution is often accounted for by a
factor of $\gamma=2n_B$ for $n_B$ baryons, since degenerate baryons
result in the same contribution. Since we are summing baryons
individually, and not assuming degenerate masses, we use the factor
$(2J_B+1)$ which also counts spin states, where all the baryons have
$J_B = \half$, thus each baryon contributes a factor of $2$. The Dirac
trace over the baryon propagator exists in order to couple the scalar
propagator to the Dirac propagator. The result of taking the trace for
the scalar self-energy is
\bea \nonumber {\rm Tr}\left[\gamma_\mu k^\mu +M^*\right] &=& {\rm
  Tr}\left[\gamma_0 k^0 - \gamma_i k^i + M^* \right] \\ \nonumber &=& {\rm
  Tr}\left[ \left(\begin{array}{cc} \mathbb{I}&0\\ 0&-\mathbb{I}
\end{array}\right)k^0 - 
\left(\begin{array}{cc} 0&\sigma_i\\ -\sigma_i&0
\end{array}\right)k^i +
\left(\begin{array}{cc} \mathbb{I}&0\\ 0&\mathbb{I}
\end{array}\right)M^*\right] \\
&=& 4M^*\ , \eea
where $\mathbb{I}$ is the $2\times 2$ identity matrix,
$\vec{\sigma}$ are the Pauli matrices (here we are using the Dirac
basis for the gamma matrices) and $M^*$ is the (degenerate) effective
baryon mass.\par
To this point, we have defined the (Hartree) self-energies and the
meson propagators in terms of these self-energies. In order to
calculate the energy density we require a definition; the energy
density is defined by the temporal components of the energy-momentum
tensor.\par
The energy-momentum tensor operator can be expressed in terms of the
Lagrangian density, as per Eq.~(\ref{eq:EML}), and by evaluating this
for the scalar part of the Lagrangian density we find the
energy-momentum operator to be
\be \label{eq:Tprop} \hat{T}_s^{\mu\nu} = -\half \left[ \del_\alpha\s\del^\alpha\s
  -m_\s^2\s^2 \right]{\rm g}^{\mu\nu} + \del^\mu\s\del_\mu\s. \ee

The `physical' energy-momentum tensor for a particle is defined as the
vacuum expectation value of the particle field subtracted from the
ground-state expectation value of the energy-momentum tensor operator,
thus for the scalar mesons, the energy-momentum tensor is
\be \label{eq:physTmunu} T^{\mu\nu}_s =
\bra\Psi|\hat{T}_s^{\mu\nu}|\Psi\ket - {\rm VEV} = \left\{
-i\volint{k}\left[\half(k^2-m_\s^2){\rm g}^{\mu\nu} - k^\mu
  k^\nu\right]\Delta(k) \right\} - \Delta^0, \ee
where the momenta arise from the derivatives above. The energy density
is simply the temporal components of this expression, evaluated with
the appropriate propagator;
\bea \nonumber {\cal E}_s &=& \bra\Psi|\hat{T}_s^{00}|\Psi\ket - {\rm VEV}
\\ \nonumber &=& -i \volint{k} \left[ \reci{2} (k^2 - m_\s^2)g^{00} -
  k^0k^0\right] \\ \nonumber &&\times \left\{\Delta^0(k) -
i(2\pi)^4\delta^{(4)}(k)\left(\Sigma_{H_s}\right)^2/g_{B\s}^2 -
\Delta^0(k) \right\} \\ \nonumber &=& i^2 \volint{k}
(2\pi)^4\delta^{(4)}(k)\left\{\reci{2}(k^2 - m_\s^2)-
(k^{0})^2\right\}\bra\s\ket^2 \\ &=& \reci{2}m_\s^2\bra\s\ket^2 , \eea
where we have used the fact that the ${\rm VEV}$ of a time-ordered
product of operators produces the free propagator
\be {\rm VEV} \propto \bra 0 | T[\phi(x)\phi(y)] | 0 \ket =
i\Delta^0(x-y). \ee
The momentum $k$ vanishes due to the integral over $\delta^{(4)}(k)$
and the contribution to the self-energy from the scalar meson is as
found above. The vector contribution is similarly
\be {\cal E}_v = \bra\Psi|\hat{T}_v^{00}|\Psi\ket - {\rm VEV} =
-\reci{2} m_\w^2 \bra\w\ket^2, \ee
and the baryon contribution is
\bea \nonumber {\cal E}_B = \bra\Psi|\hat{T}_B^{00}|\Psi\ket &=& -i
\sum_B \volint{k} {\rm
  Tr}[ \gamma_0 G_D^H(k) ]k^0 \\
&=& \sum_B \frac{(2J_B+1)}{(2\pi)^3} \int_0^{k_{F_B}}
 E_B^*(k)\, d^3k - \rho_{\rm total}\Sigma_v^0 \eea
where we do not require a ${\rm VEV}$ contribution for the baryon term
because $G_D^H(k) \to 0$ as $k_F \to 0$, and
\be \rho_{\rm total}\Sigma_v^0 = - m_\w^2 \bra\w\ket^2, \ee
which when added to the vector meson contribution changes the sign of
that term. The total energy density for the system of scalar and
vector mesons interacting with baryons for Hartree QHD is therefore
the sum of these components,
\bea \nonumber {\cal E}_{\rm QHD} &=& {\cal E}_B + {\cal E}_v + {\cal E}_s
\\ &=& \sum_B \frac{(2J_B+1)}{(2\pi)^3}\int_0^{k_{F_B}}E_B^*(k)\, d^3k  +
\reci{2}m_\w^2\bra\w\ket^2 + \reci{2}m_\s^2\bra\s\ket^2.  \eea
%
\par
We can now investigate the effects of additional terms in the
self-energy and propagators to extend the sophistication of a
model. By using this perturbative method we are able to identify the
next leading order contributions, and we can include them into our
calculations without changing the method.\par
\section{Hartree--Fock QHD Energy Density}\label{sec:selfenergyderiv}
We can now calculate the effect of introducing the next order term in
the baryon self-energy. To do this, we consider the next likely
Feynman diagram based on the Lagrangian density. Analogous to the
Hartree case, the self-energy contains a tadpole term, and now an
additional exchange contribution
\bea \label{eq:HFSigs} \Sigma_{\s}(k) &=& i g_{B\s} \volint{q}
\left[\sum_{B'} g_{B'\s}  \frac{{\rm Tr}[G(q)]}{m_\s^2}e^{iq^0\eta} +
  \frac{g_{B\s}G(q)}{(k-q)_\mu^2 - m_\s^2 + i\epsilon}\right],
\\ \nonumber \Sigma_{\w}(k) &=& ig_{B\w}\volint{q}
\left[\sum_{B'} g_{B'\w} \frac{{\rm Tr}[\gamma_\mu G(q)]}{m_\w^2} e^{iq^0\eta} +
  \frac{g_{B\w}  \gamma_\mu {\rm g}^{\mu\nu} G(q)  \gamma_\nu}{(k-q)_\lambda^2 -
    m_\w^2+i\epsilon} \right]. \\ && \eea
These self-energy terms depend on the baryon propagator $G$, and in a
fashion similar to that of the Hartree method, we can drop the
$G_F(k)$ components (which correspond to antibaryons), leaving
$G_D(k)$ (the component that arises as a result of the immersion in
the Fermi sea) as the full propagator.\par
\begin{center}
\begin{figure}[!b]
\centering \includegraphics[width=0.8\hsize]{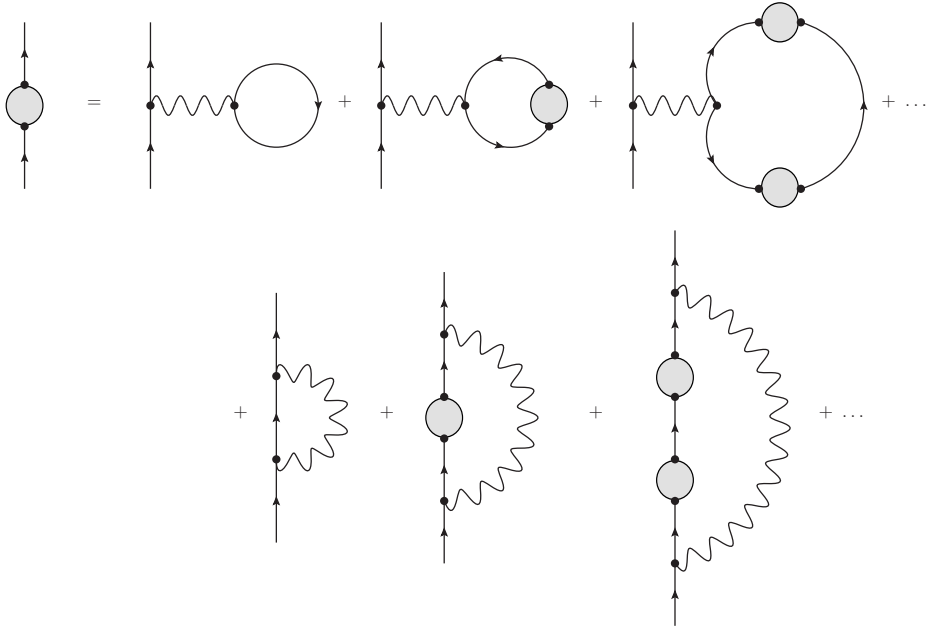}
\caption[Feynman diagram for vector vertices in $\Sigma$]{Summation of
  Feynman diagrams for all possible (vector) interaction terms
  contributing to the Hartree--Fock self-energy. Similar diagrams
  exist for the scalar meson interactions.}\label{fig:alldiagrams}
\end{figure}
\end{center}
Using the expression for $G_D$ of Eq.~(\ref{eq:HartreeProp}), the
self-energies can be separated into terms proportional to the identity
matrix $\mathbb{I}$, and terms proportional to gamma matrices as per
Eq.~(\ref{eq:splitselfenergy}) such that
\bea \nonumber \Sigma_{B} &=& \Sigma_{Bs} - \gamma_\mu \Sigma_B^\mu
\\ &=& \Sigma_{Bs} - \gamma_0 \Sigma_{B0} + \vec{\gamma}\cdot\vec{k}\,
\Sigma_{Bv}. \eea
These contributions can be represented as Feynman diagrams, as shown
in Fig.~\ref{fig:alldiagrams} where we have shown the effect of using
Dyson's equation to illustrate the infinite sum of terms involved.\par
Some of the integrals can then be performed to produce the Hartree
results plus the Fock additions due to the exchange terms in the
self-energy. For the scalar meson $\s$, the component of the
Eq.~(\ref{eq:HFSigs}) proportional to $\mathbb{I}$ is
\bea \nonumber (\Sigma_{\s})_{Bs} &=& ig_{B\s} \volint{q} \left[
  \sum_{B'} g_{B'\s}
  \frac{M_{B'}^*i\pi}{E_{B'}^*m_\s^2}\delta(q^0-E(q))\theta(q_F-|\vec{q}\,
  |)e^{iq^0\eta} \right.  \\ \nonumber &&\left. +g_{B\s}
  \frac{M_B^*i\pi}{E_B^*}D_\s\delta(q^0-E(q))\theta(q_F-|\vec{q}\, |)
  \right] \\[2mm] \label{eq:HFSigSigS} &=& -g_{B\s} \sum_{B'}
\pi\int_0^{q_F}\frac{d^3q}{(2\pi)^4}\frac{M^*}{E^*}\left(
g_{B'\s}\frac{4}{m_\s^2}+g_{B\s}D_\s\right) \eea
we can separate the two terms in Eq.~(\ref{eq:HFSigSigS}) into Hartree
and Fock components, as
\be (\Sigma_{\s})_{Bs} =
\underbrace{-g_{B\s}\sum_{B'}\frac{g_{B'\s}}{m_\s^2}
  \frac{(2J_{B'}+1)}{(2\pi)^3}\int_0^{q_F}\frac{M_{B'}^*}{E_{B'}^*}\,
  d^3q}_{\Sigma_{\rm Hartree} = -g_{B\s} \bra\s\ket} \quad
\underbrace{-\frac{g_{B\s}^2}{2(2\pi)^3}\int_0^{q_F}\frac{M_B^*}{E_B^*}D_\s
  \, d^3q}_{\Sigma_{\rm Fock}}. \ee
We can expand the meson propagator in Fock term carefully, thus
\bea \nonumber \Sigma_{\rm Fock}
&=&-\frac{g_{B\s}^2}{2(2\pi)^3}\int_0^{q_F}\frac{M^*}{E^*} \left[
  (k-q)_\mu^2 - m_\s^2+i\epsilon \right]^{-1} \, d^3q \\ \nonumber &=&
\frac{g_{B\s}^2}{2(2\pi)^3}\int_0^{q_F}\frac{M^*}{E^*} \left[ \vec{k}^2 +
  \vec{q\, }^2 - (E(k) - E(q))^2 - 2 |\vec{k\, }||\vec{q\, } |\cos\theta +
  m_\s^2+i\epsilon \right]^{-1} \, d^3q \\  &=&
\frac{g_{B\s}^2}{2(2\pi)^3}\int_0^{q_F}\frac{M^*}{E^*} \left[ A_\s(k,q) -
  2 |\vec{k\, }||\vec{q\, } |\cos\theta \right]^{-1} \, d^3q, \eea
where we simplify this expression using
\be A_\s(k,q) = \vec{k}^2 + \vec{q\, }^2 - (E(k) - E(q))^2 +
m_\s^2+i\epsilon .  \ee
Currently, this triple integral is over momentum
$\vec{q}=(q_1,q_2,q_3)$ but we can change to spherical coordinates
using the relation
\be \int d^3q = \int_0^\infty dq_1 \int_0^\infty dq_2 \int_0^\infty
dq_3 \to \int_0^\infty q^2 dq \int_0^\pi \sin(\theta) d\theta
\int_0^{2\pi} d\phi. \ee
We can also change the integration parameter $d\theta \to
d(\cos(\theta))$ since $d(\cos(\theta)) = -\sin(\theta)d\theta$
\be \int d^3q \to \int_0^\infty q^2 dq \int_0^\pi \sin(\theta) d\theta
\int_0^{2\pi} d\phi \to \int_0^\infty q^2 dq \int_{-1}^1
d(\cos(\theta)) \int_0^{2\pi} d\phi. \ee
Performing the integral over $\phi$ (since the integrand doesn't
depend on this) we get
\be
\label{eq:spherical}
\int d^3q \to (2\pi) \int_0^\infty q^2 dq \int_{-1}^1 d(\cos(\theta)).
\ee
The integral over $\cos(\theta)$ is possible using the identity
\be \int \left[A - B x \right]^{-1} dx = - \frac{1}{B} \ln | A - B x
|, \ee
and thus we have
\bea \nonumber \Sigma_{\rm Fock} &=& \frac{2\pi
  g_{B\s}^2}{2(2\pi)^3}\int_0^{q_F}\frac{M^*}{E^*} \frac{q^2}{2kq} \ln
\left(\frac{A_\s(k,q) + 2kq}{A_\s(k,q) - 2kq} \right) \, dq
\\ \nonumber &=& \frac{g_{B\s}^2}{16\pi^2k}\int_0^{q_F}\frac{M^*}{E^*}
q \ln \left(\frac{A_\s(k,q) + 2kq}{A_\s(k,q) - 2kq} \right) \, dq
\\ &=& \frac{1}{4\pi^2k}\int_0^{q_F}\frac{M^*}{E^*}
\frac{q}{4}g_{B\s}^2 \Theta(k,q) \, dq, \eea
where we use the definitions
\bea \Theta_i(k,q) &=& \ln \left| \frac{A_i(k,q) + 2kq}{A_i(k,q)-
  2kq}\right|, \\ \Phi_i(k,q) &=& \reci{4kq} A_i(k,q)\Theta_i(k,q) -
1, \\ A_i(k,q) &=& \vec{k}^{\; 2} + \vec{q}^{\; 2} + m_i^2 -
[E(q)-E(k)]^2, \eea
and note that $q = |\vec{q}\, |$, $k = |\vec{k}\, |$. All of these
self-energies are evaluated on shell at the self-consistent
single-particle energies.\par
The total Hartree--Fock self-energy for the $\s$ and $\w$ mesons is
given by
\bea \nonumber \Sigma_{Bs}(k,E(k)) &=& g_{B\s} \sum_{B'}
\frac{-(2J_{B'}+1)}{(2\pi)^3} \frac{g_{B'\s}}{m_\s^2}
\int_0^{k_{F_{B'}}} \frac{M_{B'}^*(q)}{E_{B'}^*(q)} \, d^3q \\ &&+
\reci{4\pi^2k}\int_0^{k_{F_B}}q\; dq \frac{M_B^*(q)}{E_B^*(q)}
\left[\reci{4}g_{B\s}^2\Theta_\s(k,q)-g_{B\w}^2\Theta_\w(k,q)\right],
\eea
\bea
\nonumber \Sigma_{B0}(k,E(k)) &=& g_{B\s} \sum_{B'}
\frac{-(2J_{B'}+1)}{(2\pi)^3} \frac{g_{B'\w}}{m_\w^2}
\int_0^{k_{F_{B'}}} d^3q \\ &&- \reci{4\pi^2k}\int_0^{k_{F_B}}q\; dq
\left[\reci{4}g_{B\s}^2\Theta_\s(k,q)+\reci{2}g_{B\w}^2\Theta_\w(k,q)\right],
\eea
\bea
\Sigma_{Bv}(k,E(k)) &=& - \reci{4\pi^2k}\int_0^{k_F}q\;
dq \frac{q^*}{E^*(q)}
\left[\reci{2}g_{B\s}^2\Phi_\s(k,q)+g_{B\w}^2\Phi_\w(k,q)\right], \eea
%
%
where the terms involving a sum over $B'$ correspond to the Hartree
results. The calculations that include only these Fock contributions
are denoted by the term `Fock1' in our presentation of results. These
are calculated to provide an insight into the effect that these terms
have on the EOS. The energy density can still be calculated as per the
previous section, and although no new terms appear with the addition
of the Fock contribution to the self-energy, the definitions of the
terms appearing in the energy density now include this
contribution.\par
We can however calculate an explicit additional contribution to the
energy density; we recall from the previous section that the energy
density is defined in terms of the physical energy momentum tensor,
which involves propagators, as per Eq.~(\ref{eq:physTmunu}). The
second term that we can include is an additional component to the
interacting meson propagators due to the effect of medium
polarization, in analogy with electric polarization (see
Section~\ref{sec:fockterms}).\par
\begin{figure}[!t]
\centering \includegraphics[width=0.9\textwidth]{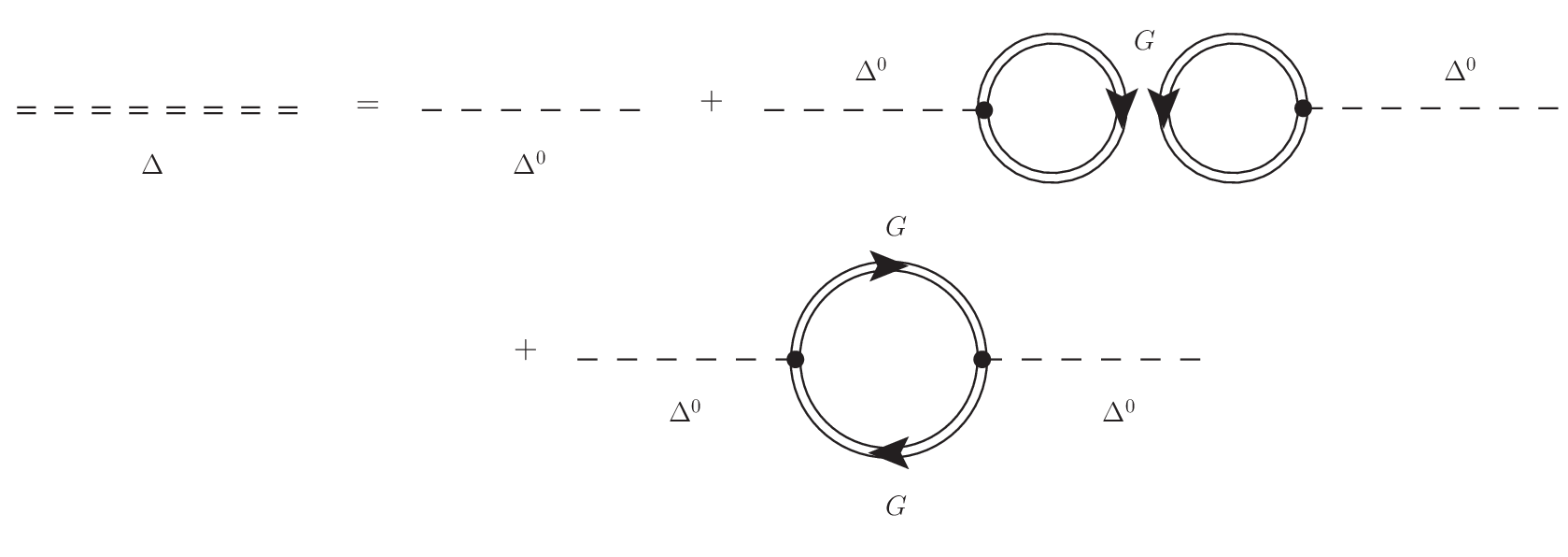}
\caption[3 Terms in the Meson Propagator]{The three leading terms
  contributing to the meson propagator, in this case, the scalar
  meson. The first term corresponds to the bare propagator, the second
  term corresponds to the Hartree contribution, and the third term
  corresponds to the medium polarization term. Similar terms can be
  written for the vector meson.
  \protect\label{fig:3PartMesonPropagator}}
\end{figure}

The Feynman diagram for the medium polarization contribution to the
scalar propagator is given by the third term in
Fig.~\ref{fig:3PartMesonPropagator}, where the first term corresponds
to the bare propagator, and the second term corresponds to the Hartree
contribution.
%
%
The contribution due to medium polarization for the scalar propagator
is given by
\bea \label{eq:Deltaprime} \Delta'(k) &=& \Delta^0 \Pi_\s(k) \Delta^0(k) \\ \Pi_s(k) &=&
-i\sum_B g_{B\s}^2 \volint{q} {\rm Tr}\left[ G_D(q)G_D(k+q)\right], \eea
and for the vector propagator,
\bea D_{\mu\nu}'(k) &=& D_{\mu\lambda}^0 \Pi_\w^{\lambda\sigma}(k)
D_{\sigma\nu}^0(k) \\ \Pi_\w^{\lambda\sigma}(k) &=& -i\sum_B g_{B\w}^2
\volint{q} {\rm Tr}\left[ \gamma^\lambda G_D(k+q)\gamma^\sigma
  G_D(q)\right], \eea
where $\Pi_\s$ and $\Pi_\w^{\lambda\sigma}$ are the medium
polarizations. We can evaluate the physical energy-momentum tensor
using the appropriate propagator which includes this medium
polarization term. For example, the contribution due to baryon $B$ for
the scalar meson is given by
\be \label{eq:physTmunu_2} T^{\mu\nu}_s =
\bra\Psi|\hat{T}_s^{\mu\nu}|\Psi\ket - {\rm VEV} = \left\{
-i\volint{k}\left[\half(k^2-m_\s^2){\rm g}^{\mu\nu} - k^\mu
  k^\nu\right]\Delta(k) \right\} - \Delta^0, \ee
where in this case, $\Delta$ is the three-part propagator in
Fig.~\ref{fig:3PartMesonPropagator}, which we can separate into the
three terms.\par
If we substitute Eq.~(\ref{eq:Deltaprime}) into the temporal components
of the ground-state expectation value in Eq.~(\ref{eq:physTmunu_2}) we
have
\bea \nonumber \bra\Psi|\hat{T}^{00}_s|\Psi\ket &=& -i \volint{k}
\left[\half(k^2-m_s^2){\rm g}^{00} - (k^0)^2\right] \\ \nonumber &&
\times \ \Delta^0(k) \left\{ -ig_{B\s}^2\volint{q}
Tr\left[G_D(q)G_D(k+q)\right] \right\} \Delta^0(k),
\\ && \label{eq:GDkplusQ} \eea
where the Dirac part of the baryon propagator is defined as
\be G_D(k) = \left[\gamma^\mu k_\mu +
  M^*(k)\right]\frac{i\pi}{E^*(k)}\delta(k^0-E(k))\theta(k_F -
|\vec{k}|), \ee
in which we have neglected the vector component of the self-energy,
since it is a $k^{-2}$ term, and thus $k^* = k =
(k^0,\vec{k}\; )$. Rather that working with the $G_D(k+q)$ term, since
all of the above integrals are over $(-\infty,\infty)$, we can shift
the integration variables according to $k+q \to k'$, which can be
rewritten as $k \to k'-q$. Eq.~(\ref{eq:GDkplusQ}) then becomes
\bea \nonumber \bra\Psi|\hat{T}^{00}_s|\Psi\ket &=& -g_{B\s}^2
\volint{k'}\volint{q} \left[\half((k'-q)^2-m_s^2){\rm g}^{00} -
  (k'^0-q^0)^2\right] \\
\label{eq:GDkprime} 
&& \times \ \Delta^0(k'-q) \left\{ {\rm Tr}\left[G_D(q)G_D(k')\right]
\right\} \Delta^0(k'-q), \eea
and so now we can work with the baryon Green's functions in single
variables. The product of these is
\bea \nonumber G_D(q)G_D(k') &=& -\frac{\pi^2}{E^*(q)E^*(k')} \left[
  \gamma^\mu q_\mu \gamma^\nu k'^{\nu} +\gamma^\mu q_\mu M^*(k') +
  \gamma^\nu k'_\nu M^*(q) + M^*(q)M^*(k') \right]. \label{eq:GqGk}
\\ &&\times \ \delta(q^0-E(q))\theta(k_F-|\vec{q}\,
|)\delta(k'^0-E(k'))\theta(k_F-|\vec{k'}|). \eea
Since $\gamma_\mu$ are traceless, and ${\rm
  Tr}[\gamma_aA^a\gamma_bB^b] = 4A\cdot B$, the only terms to survive
in the trace of Eq.~(\ref{eq:GqGk}) are
\bea \nonumber {\rm Tr}\left[ G_D(q)G_D(k') \right] &=& -\frac{4
  \pi^2}{E^*(q)E^*(k')} \left[ q_\mu k'^\mu + M^*(q)M^*(k') \right]
\\ \nonumber && \times \ \delta(q^0-E(q))\theta(k_F-|\vec{q}\,
|)\delta(k'^0-E(k'))\theta(k_F-|\vec{k'}|). \\
\label{eq:GDGD} &&  \eea
Substituting Eq.~(\ref{eq:GDGD}) back into Eq.~(\ref{eq:GDkprime}),
performing the temporal integrals, and applying the $\theta$
functions, we obtain
\bea \nonumber \bra\Psi|\hat{T}^{00}_s|\Psi\ket &=&
\frac{4\pi^2g_{B\s}^2}{(2\pi)^8} \int_0^{k_F} \frac{d^3q}{E^*(q)}
\int_0^{k_F} \frac{d^3k'}{E^*(k')} \left[\half((k'-q)^2-m_s^2) -
  (E(k')-E(q))^2\right] \\ &&\times \ \Delta^0(k'-q) \left[
  q_\mu k'^\mu + M^*(q)M^*(k') \right] \Delta^0(k'-q), \eea
noting that ${\rm g}^{00} = 1$. With the definition of the free scalar
propagator,
\be D_i^0(E(p),\vec{p\, }) = (E(p)^2 - \vec{p\, }^2 - m_i^2)^{-1}, \ee
for which $\Delta^0$ corresponds to $D_\s^0$. We can cancel the term
that was proportional to ${\rm g}^{00}$, leaving
\bea \nonumber \bra\Psi|\hat{T}^{00}_s|\Psi\ket &=&
\frac{g_{B\s}^2}{(2\pi)^6} \int_0^{k_F} \frac{d^3q}{E^*(q)} \int_0^{k_F}
\frac{d^3k'}{E^*(k')} \\ \nonumber && \quad \times \ D_\s^0(k'-q)
\left[\half - D_\s^0(k'-q)(E(k')-E(q))^2\right] \\ &&\times \ \left[
  q_\mu k'^\mu + M^*(q)M^*(k') \right].  \eea
The total energy contribution for all baryons is then the sum of these
terms for all baryons. Further to this, we insert a factor of $(2J_B+1)$
for each baryon to account for spin counting, and an overall factor of
$\half$ due to the symmetry between $q$ and $k'$ which would otherwise
result in overcounting.\par
A similar term exists for the vector meson contribution, in which the
polarization term is
\be \Pi_\w^{\delta\lambda} = -i\sum_B g_{B\w}^2 \volint{q} {\rm Tr}
\left[ \gamma^\delta G_D(k+q)\gamma^\delta G_D(q) \right].  \ee
The combined energy contribution due to medium polarization of these
mesons is therefore
\bea \nonumber &{\cal E}_{\rm med-pol} = & \half \sum_{B}
\frac{2}{(2\pi)^6} \int_0^{k_{F_B}} \frac{d^3q}{E^*(q)}
\int_0^{k_{F_B}} \frac{d^3k'}{E^*(k')} \\ \nonumber && \left\{
g_{B\s}^2 D_\s^0(k'-q) \left[\half - D_\s^0(k'-q)[E(k')-E(q)]^2\right]
\right. \\ \nonumber && \quad \times \ \left[ q_\mu k'^\mu +
  M^*(q)M^*(k') \right] \\ \nonumber &&+ \ g_{B\w}^2 D_\w^0(k'-q)
\left[\half - D_\w^0(k'-q)[E(k')-E(q)]^2\right] \\ && \left. \quad
\times \ \left[ q_\mu k'^\mu - 2M^*(q)M^*(k') \right] \right\} . \eea
In order to perform numerical calculations for this quantity, we must
expand this further. The propagator terms expand as
\bea \nonumber D_\s^0(k'-q) &=& \left( \left[E(k')-E(q)\right]^2 -
(\vec{k}-\vec{q}\, )^2 - m_\s^2 \right)^{-1}\\ &=& \left(
\left[E(k')-E(q)\right]^2 - \vec{k}^2-\vec{q}^{\, 2}+2|\vec{q}\,
||\vec{k'}|\cos(\theta) - m_\s^2 \right)^{-1}, \eea
and owing to the definition of the four-momentum,
\be k^\mu = \left[k^0 + \Sigma^0(k),\vec{k}\right] \stackrel{k^0 \to
  E(k)}{=} \left[E^*(k),\vec{k}\, \right], \ee
the product of the momentum four-vectors is
\bea q_\mu k^{\prime\mu} = E^*(q)E^*(k^\prime) - |\vec{q}\,
||\vec{k^\prime}|cos(\theta), \eea
so once again we need to shift the integration spherical coordinates
according to Eq.~(\ref{eq:spherical}), except that this time the
$d(\cos(\theta))$ integral is non-trivial due to the $\cos(\theta)$
terms.\par
Expanding fully reveals the full expression,
\bea \nonumber {\cal E}_{\rm med-pol} &=& \half \sum_{B}
\frac{(2J_B+1)}{(2\pi)^6} \int_0^{k_{F_B}} \frac{4\pi q^2
  dq}{E^*(q)} \int_0^{k_{F_B}} \frac{2\pi k'^2 dk'}{E_B^*(k')}
\int_{-1}^1 d(\cos(\theta)) \\ \nonumber && \left\{ g_{B\s}^2
D_\s^0(k'-q) \left[\half - D_\s^0(k'-q)[E(k')-E(q)]^2\right]
\right. \\ \nonumber && \quad \times \ \left[ E_B^*(q)E_B^*(k') -
  |\vec{q}\, ||\vec{k'}|\cos(\theta) + M_B^*(q)M_B^*(k') \right]
\\ \nonumber &&+ \ g_{B\w}^2 D_\w^0(k'-q) \left[\half -
  D_\w^0(k'-q)[E(k')-E(q)]^2\right] \\ && \left. \quad
\times \ \left[ E_B^*(q)E_B^*(k') - |\vec{q}\, ||\vec{k'}|\cos(\theta)
  - 2M_B^*(q)M_B^*(k') \right] \right\}.  \eea
Combining the prefactors, we have
\bea \nonumber {\cal E}_{\rm med-pol} &=& \half \sum_{B}
\frac{1}{4\pi^4} \int_0^{k_{F_B}} \frac{q^2 dq}{E_B^*(q)}
\int_0^{k_{F_B}} \frac{k'^2 dk'}{E_B^*(k')} \int_{-1}^1
d(\cos(\theta)) \\ \nonumber && \left\{ g_{B\s}^2 D_\s^0(k'-q)
\left[\half - D_\s^0(k'-q)[E(k')-E(q)]^2\right] \right. \\ \nonumber
&& \quad \times \ \left[ E_B^*(q)E_B^*(k') - |\vec{q}\,
  ||\vec{k'}|\cos(\theta) + M_B^*(q)M_B^*(k') \right] \\ \nonumber &&+
\ g_{B\w}^2 D_\w^0(k'-q) \left[\half -
  D_\w^0(k'-q)[E(k')-E(q)]^2\right] \\ && \left. \quad \times \ \left[
  E_B^*(q)E_B^*(k') - |\vec{q}\, ||\vec{k'}|\cos(\theta) -
  2M_B^*(q)M_B^*(k') \right] \right\} . \eea
Thus the total Hartree--Fock energy density is written as 
\bea \nonumber {\cal E}_{\rm HF} &=& \sum_B \frac{(2J_B+1)}{(2\pi)^3}
\int_0^{k_{F_B}} d^3k \ E(k) + \frac{1}{2}m_\s^2 \bra\s\ket^2 -
\frac{1}{2}m_\w^2 \bra\w\ket^2 \\ \nonumber && + \reci{2}\sum_B
\frac{(2J_{B}+1)}{(2\pi)^6} \int_0^{k_{F_B}} \frac{d^3k}{E_B^*(k)}
\int_0^{k_{F_B}} \frac{d^3q}{E_B^*(q)} \\ \nonumber && \times \left\{
g_{B\s}^2
D_\s^0(k-q)\left[\reci{2}-[E(k)-E(q)]^2D_\s^0(k-q)\right]\right.
\left[k^{*\mu}q_\mu^* + M^*(k)M^*(q)\right] \\ \nonumber && +
2g_{B\w}^2D_\w^0(k-q)\left.\left[\reci{2}-[E(k)-E(q)]^2D_\w^0(k-q)\right]
[k^{*\mu}q_\mu^* - 2M^*(k)M^*(q)]\right\}, \\[2mm] &&\eea
where the double integral is the contribution from the medium
polarization.\par
Calculations performed that include the Fock contribution to the
self-energy as well as this medium polarization energy are denoted by
the term `Fock2' in our presentation of results.\par
\cleardoublepage

%% file: AppendixB_particles.tex
\chapter{Particle Properties}\label{sec:particleprops}
%
%
\begin{center}
 \begin{table}[!h]
 \centering
 \caption[Particle properties]{\protect Particle properties for
   selected leptons, baryons and mesons used in this work, as found in
   Ref.~\cite{Amsler:2008zzb}. Shown here are the quark content (the
   valence quarks that define the particle), the isospin, spin, and
   parity quantum numbers, the mass, Lorentz character, and
   strangeness of each particle.\label{tab:particlesummary}}
  \vspace{4mm}
  \begin{tabular}{||c|c|c|c|c|c||}
  \hline
  \hline
    \\[-4mm]
             &   quarks  &  $I(J^P)$           &   mass (MeV)   &   Lorentz   &   $S$ \\[1mm]
   \hline
   \hline
   \\[-4mm]
  $e^-$          &   fundamental   &  $J=\half$   &   0.51   &   scalar    &   ---  \\[1mm]
  $\mu^- $       &   fundamental   &  $J=\half$   &   105.66   &   scalar    &   ---  \\[1mm]
  $\tau^- $       &   fundamental   &  $J=\half$   &   1776.84   &   scalar    &   ---  \\[1mm]
   \hline
   \hline
   \hline
    \\[-4mm]
  p          &   $uud$   &  $\half(\half^+)$   &   938.27   &   spinor    &   0  \\[1mm]
  n          &   $udd$   &  $\half(\half^+)$   &   939.57   &   spinor    &   0  \\[1mm]
   \hline
    \\[-4mm]
  $\Lambda$  &   $uds$   &  $0(\half^+)$       &   1115.68   &   spinor    &   -1  \\[1mm]
   \hline
    \\[-4mm]
  $\Sigma^+$ &   $uus$   &  $1(\half^+)$       &   1189.37   &   spinor    &   -1  \\[1mm]
  $\Sigma^0$ &   $uds$   &  $1(\half^+)$       &   1192.64   &   spinor    &   -1  \\[1mm]
  $\Sigma^-$ &   $dds$   &  $1(\half^+)$       &   1197.45   &   spinor    &   -1  \\[1mm]
   \hline
    \\[-4mm]
  $\Xi^0$    &   $uss$   &  $\half(\half^+)$   &   1314.83   &   spinor    &   -2  \\[1mm]
  $\Xi^-$    &   $dss$   &  $\half(\half^+)$   &   1321.31   &   spinor    &   -2  \\[1mm]
   \hline
   \hline
   \hline
    \\[-4mm]
  $\s$           &   $\fh$                           &  $0(0^+)$           &   600    &   scalar i-scalar    &   0        \\[1mm]
  $f_0$          &   $\sh$                           &  $0(0^+)$           &   980    &   scalar i-scalar    &   0        \\[1mm]
  $a^0$          &   $(u\bar{u}-d\bar{d})/\sqrt{2}$  &  $1(0^+)$           &   984.7  &   scalar i-vector    &   0       \\[1mm]
   \hline
    \\[-4mm]
  $\w$              &   $\fh$                          &  $0(1^-)$      &   782.65  &   vector i-scalar    &   0    \\[1mm]
  $\phi$            &   $\sh$                          &  $0(1^-)$      &   1019.46 &   vector i-scalar    &   0    \\[1mm]
  $\rho^\pm$        &   $u\bar{d}$/$\bar{u}d$          &  $1(1^-)$      &   775.5   &   vector i-vector    &   0    \\[1mm]
  $\rho^0$          &   $(u\bar{u}-d\bar{d})/\sqrt{2}$ &  $1(1^-)$      &   775.5   &   vector i-vector    &   0    \\[1mm]
   \hline
    \\[-4mm]
  $\eta$       &   $\fh$                           &  $0(0^-)$      &   547.51  &  p-scalar i-scalar   &  0     \\[1mm]
  $\eta'$      &   $\sh$                           &  $0(0^-)$      &   957.78  &  p-scalar i-scalar   &  0     \\[1mm]
  $\pi^\pm$    &   $u\bar{d}$/$\bar{u}d$           &  $1(0^-)$      &   134.98  &  p-scalar i-vector   &  0     \\[1mm]
  $\pi^0$      &   $(u\bar{u}-d\bar{d})/\sqrt{2}$  &  $1(0^-)$      &   139.57  &  p-scalar i-vector   &  0     \\[1mm]
   \hline
   \hline
  \end{tabular}
 \end{table}
\end{center}

\pagebreak

\begin{center}
\begin{figure}[!ht]
\centering
\includegraphics[width=0.7\textwidth]{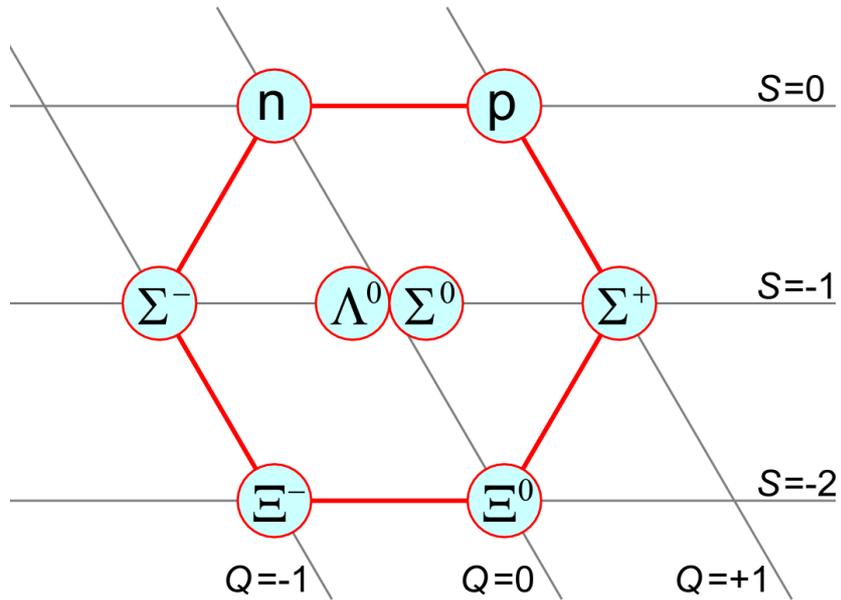}
\caption[Baryon octet]{(Color Online) Diagram of the baryon octet
  showing strangeness and charge. Images for
  Figs.~\ref{fig:BaryonOctet}--\ref{fig:Spin1Meson} have been released
  into the public domain and published on
  wikipedia.org.}\label{fig:BaryonOctet}
\end{figure}
\end{center}
\vfill
%
\begin{figure}[!h]
\begin{tabular}[!ht]{cc}
\begin{minipage}[c]{0.45\textwidth}
\centering
\includegraphics[width=\textwidth]{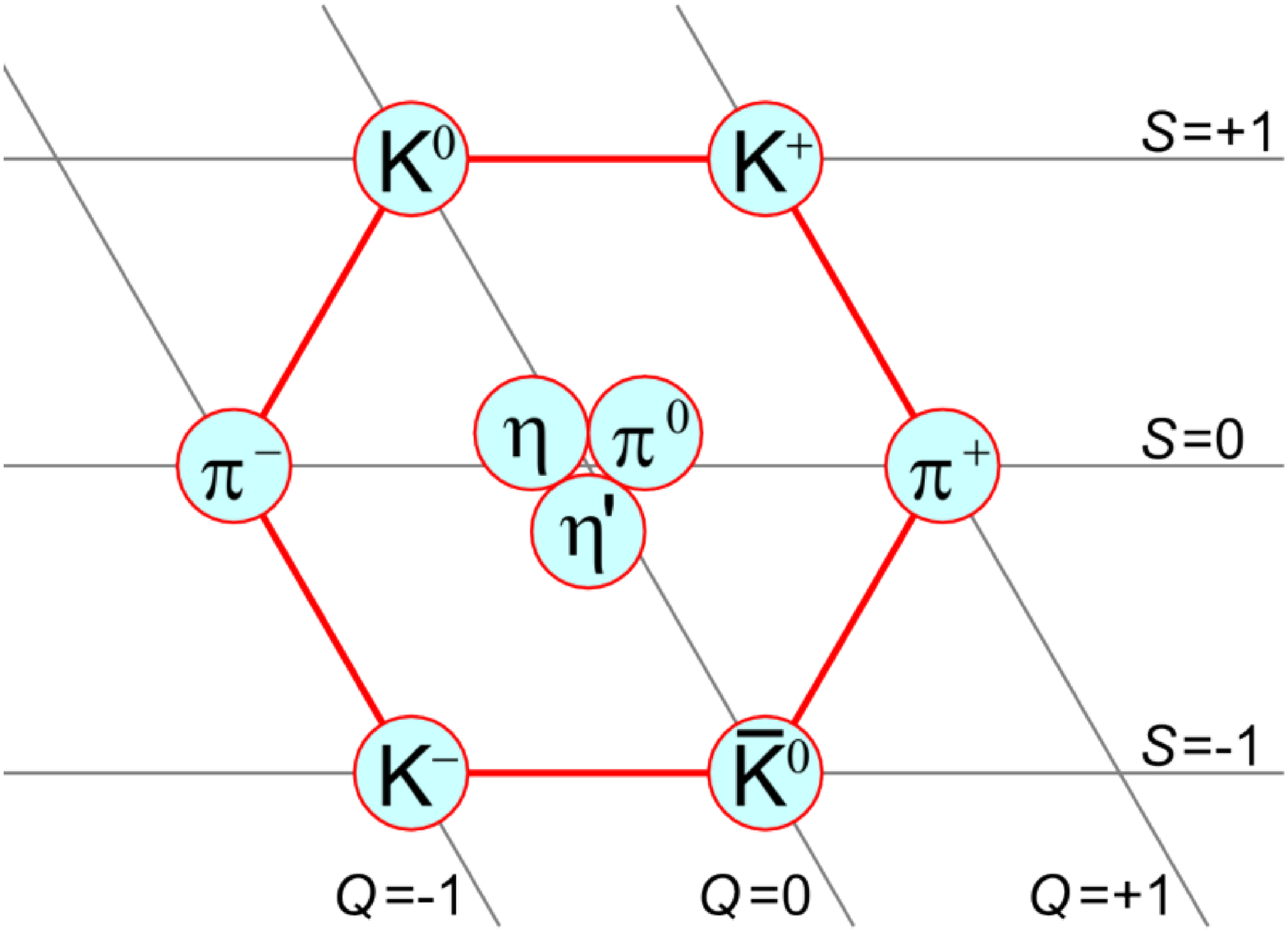}
%
\end{minipage}
&
\begin{minipage}[c]{0.45\textwidth}
%
\centering
\includegraphics[width=\textwidth]{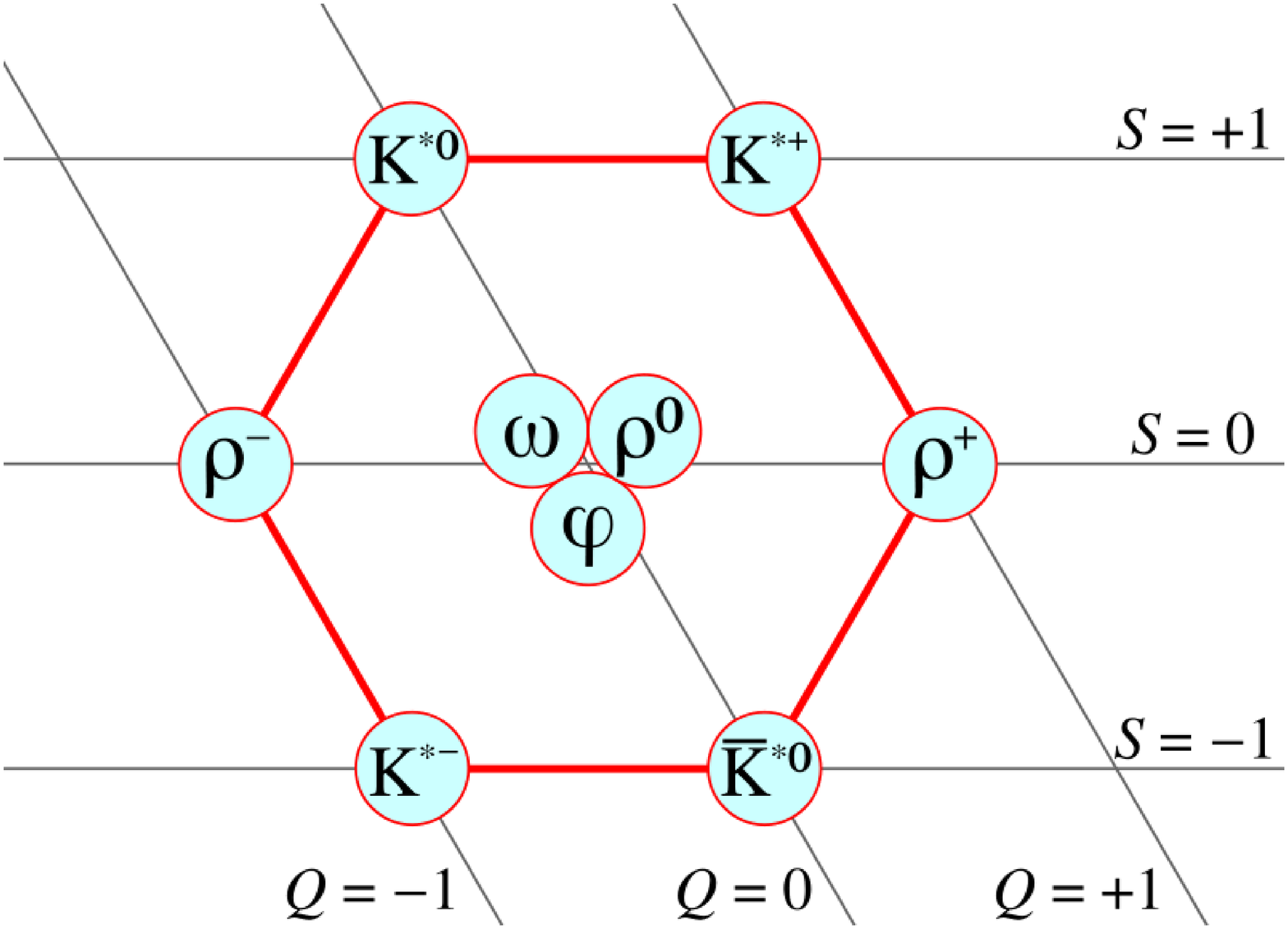}
\end{minipage}
\end{tabular}
\caption[Meson nonets]{(Color Online) Diagrams of the spin-0 and
  spin-1 meson nonets showing strangeness and
  charge}\label{fig:Spin1Meson}
\end{figure}
\vspace{1cm}
%

%% file: AppendixD_papers.tex
\chapter{Relevant Publication By The Author}\label{sec:publications}
The following pages contain a relevant article by the author, in which
we describe the effects of the phase transition from the octet QMC
model to various quark models, and which is published as
\hbox{Phys.~Rev.~C.~79,~045810} (Ref.~\cite{Carroll:2008sv}). The
results presented in that article have not been performed and
published by any other researchers, to the best of the author's
knowledge, and as such constitute a novel contribution to the field.